\documentclass{SCIS2025}
\begin{document}
%\oa
%%%%%%%%%%%%%%%%%%%%%%%%%%%%%%%%%%%%%%%%%%%%%%%%%%%%%%%
%%% Authors do not modify the information below
%%% ×÷Õß²»ÐèÒªÐÞ¸Ä´Ë´¦ÐÅÏ¢
\ArticleType{REVIEW}
%\SpecialTopic{}
%\luntan
\Year{2025}
\Month{}
\Vol{}
\No{}
\DOI{}
\ArtNo{}
\ReceiveDate{}
\ReviseDate{}
\AcceptDate{}
\OnlineDate{}
%%%%%%%%%%%%%%%%%%%%%%%%%%%%%%%%%%%%%%%%%%%%%%%%%%%%%%%
%Architecture, Key Technologies, and Future Directions
%%% title: ±êÌâ
%%%   \title{title}{title for citation}
\title{Distributed satellite information networks: Architecture, enabling technologies, and trends}{Distributed satellite information networks: Architecture, enabling technologies, and trends}

%% Distributed satellite networks: architecture, enabling technologies, and Trends

%%% Corresponding author: Í¨ÐÅ×÷Õß
%%%   \author[number]{Full name}{{email@xxx.com}}
%%% General author: Ò»°ã×÷Õß
%%%   \author[number]{Full name}{}
\author[1,2]{Qinyu ZHANG}{}
\author[1]{Liang XU}{}
\author[1,2]{Jianhao HUANG}{}
\author[1]{Tao YANG}{}
\author[1,2]{Jian JIAO}{jiaojian@hit.edu.cn}
\author[2]{\\Ye WANG}{}
\author[1,2]{Yao SHI}{}
\author[1,2]{Chiya ZHANG}{}
\author[1,2]{Xingjian ZHANG}{}
\author[2]{Ke ZHANG}{}
\author[2]{\\Yupeng GONG}{}
\author[3,4]{Na DENG}{}
\author[3]{Nan ZHAO}{}
\author[5]{Zhen GAO}{}
\author[5]{Shuai WANG}{}
\author[6]{Shujun HAN}{}
\author[2,7]{\\Xiaodong XU}{}
\author[8,9]{Li YOU}{}
\author[8,9]{Dongming WANG}{}
\author[9]{Shan JIANG}{}
\author[8,9]{Dixian ZHAO}{}
\author[10,11]{\\Nan ZHANG}{}
\author[10,11]{Liujun HU}{}
\author[12]{Xiongwen HE}{}
\author[13]{Yonghui LI}{}
\author[8,9]{Xiqi GAO}{}
\author[8,9]{Xiaohu YOU}{}
%% zqy£¬XL£¬HJH£»ZTE£»BUPT£»DaLian£»BeiLi£»DongNan£»PCL; Yonghui Li, Xiongwen He, Jian Jiao
\AuthorCitation{Zhang Q Y, Xu L, Huang J H, et al}

%% ÕÅÇÕÓî£¬ÐìÁÁ£¬»Æ½¨ºÀ£¬ÕÅÐÐ½¡£¬Ê·Ñþ£¬ÕÅ³ÛÑÇ£¬½¹½¡£»
%% ¹¨ÅôÓî£¬ÍõÒ°£¬
%% µËÄÈ£¬ÕÔéª   ´óÁ¬Àí¹¤
%% ¸ßÕò£¬        ±±Àí
%% º«Êé¾ý£¬ÐíÏþ¶«   ±±ÓÊ
%% ÓÈÁ¦£¬¸ßÎ÷Ææ£» ÕÔµÓìÞ£¬ÓÈÐ¤»¢£¬ ¶«ÄÏ
%% ÕÅéª£¬ºúÁô¾ü   ZTE
%% ÀîÓÀ»á£¬Ï¤Äá´óÑ§
%% ºÎÐÜÎÄ£¬

%%% Authors' contribution. Í¬µÈ¹±Ï×
\contributions{These authors contributed equally.}

%%% Address. µØÖ·
%%%   \address[number]{Affiliation, City Postcode, Country}
%\address[1]{Affiliation, City 000000, Country}
%\address[2]{Affiliation, City 000000, Country}
\address[1]{Harbin Institute of Technology (Shenzhen), Shenzhen {\rm 518055}, China}
\address[2]{Pengcheng Laboratory, Shenzhen {\rm 518055}, China}
\address[3]{Dalian University of Technology, Dalian {\rm 116024}, China}
\address[4]{State Key Laboratory of Integrated Services Networks, Xidian University, Xi'an {\rm 710071}, China}
\address[5]{State Key Laboratory of CNS/ATM, Beijing Institute of Technology, Beijing {\rm 100081}, China}
\address[6]{National Engineering Laboratory for Mobile Network Technologies, Beijing University of Posts and Telecommunications, \\Beijing {\rm 100876}, China}
\address[7]{State Key Laboratory of Networking and Switching Technology, Beijing University of Posts and Telecommunications, \\Beijing {\rm 100876}, China}
\address[8]{National Mobile Communications Research Laboratory, Southeast University, Nanjing {\rm 210096}, China}
\address[9]{Purple Mountain Laboratories, Nanjing {\rm 211111}, China}
\address[10]{ZTE Corporation Algorithm Department, Shenzhen {\rm 518057}, China}
\address[11]{State Key Laboratory of Mobile Network and Mobile Multimedia Technology, Shenzhen {\rm 518055}, P.R.China}
\address[12]{Beijing Institute of Spacecraft System Engineering, Beijing {\rm 100076}, P.R.China}
\address[13]{University of Sydney, Sydney NSW {\rm 2006}, Australia}

%%% Abstract. ÕªÒª
\abstract{Driven by the vision of ubiquitous connectivity and wireless intelligence, the evolution of ultra-dense constellation-based  satellite-integrated Internet is underway, now taking preliminary shape.
Nevertheless, the entrenched institutional silos and limited, nonrenewable heterogeneous network resources leave current satellite systems struggling to accommodate the escalating demands of next-generation intelligent applications.
In this context, the distributed satellite information networks (DSIN), exemplified by the cohesive clustered satellites (CCS) system, have emerged as an innovative architecture,
bridging information gaps across diverse satellite systems, such as communication, navigation, and remote sensing, and establishing a unified, open information network paradigm to support resilient space information services.
This survey first provides a profound discussion about innovative network architectures of DSIN, encompassing distributed regenerative satellite network architecture, distributed satellite computing network architecture, and reconfigurable satellite formation flying, to enable flexible and scalable communication, computing and control, fundamentally enhancing network resilience.
The DSIN faces challenges from network heterogeneity, unpredictable channel dynamics, sparse resources, and decentralized collaboration frameworks. To address these issues, a series of enabling technologies is identified, including channel modeling and estimation, cloud-native distributed MIMO cooperation, new waveform design, grant-free massive access, non-orthogonal multicast, distributed phased array antennas, high-speed inter-satellite communication, network routing, and the proper combination of all these diversity techniques. Furthermore, to heighten the overall resource efficiency, the cross-layer optimization techniques are further developed to meet upper-layer deterministic, adaptive and secure information services requirements. In addition, emerging research directions and new opportunities are highlighted on the way to achieving the DSIN vision.}

%A suite of emerging research directions and challenges on the road to making the DSCNs vision a reality
%are discussed
%%% Keywords. ¹Ø¼ü´Ê

\maketitle

%%%%%%%%%%%%%%%%%%%%%%%%%%%%%%%%%%%%%%%%%%%%%%%%%%%%%%%
%%% The main text. ÕýÎÄ²¿·Ö
%%%%%%%%%%%%%%%%%%%%%%%%%%%%%%%%%%%%%%%%%%%%%%%%%%%%%%%
\section{Introduction}
%%%%ÐèÒªÁ½ÕÅÍ¼------·Ö²¼Ê½ÎÀÐÇÌØÕ÷»òÕßÓÅÊÆÍ¼+ÎÄÕÂ½á¹¹Í¼
%%%%½á¹¹×Ü½á
%%%µÚÒ»¶Î£ºÏÈ½éÉÜÎÀÐÇÍ¨ÐÅµÄ·¢Õ¹(´Ó´«Í³µÄµ¥ÐÇÏµÍ³µ½ÏÖÔÚ´ó¹æÄ£ÐÇ×ù½¨Éè)ÒÔ¼°Òý³ö6G¶Ô¿ÕÌìµØº£Ò»Ìå»¯ÍøÂçµÄÉèÏë¡£

%%µÚÒ»¶Î¶ÔÕÕÕªÒª£¬¾¡¿ìÒý³öCCSºÍDSIN£¬ÔÙËµÐÐÒµ·¢Õ¹
Driven by the space-air-ground integrated network (SAGIN), satellite Internet of Things (S-IoT) and satellite-integrated Internet initiatives, the satellite communication (SatCom) system has witnessed a remarkable evolution, transitioning from a standalone large geostationary satellite serving mode to ultra-dense mega-constellation networks such as Starlink, Xingyun and OneWeb, catering to the vision of seamless global coverage and ubiquitous connectivity in the forthcoming sixth generation (6G) \cite{you6G}.
Nevertheless, as the demand for diverse intelligent applications in 6G integration continues to rise, the issue of information barriers inherent in existing independent satellite networks has become increasingly pronounced, rendering these networks incapable of providing globally open, cohesive, and unified information services.
This has raised widespread concern in academia and industry regarding the distributed satellite information networks (DSIN) represented by the cohesive clustered satellites (CCS) system \cite{xu2024semantic,9887918}.

\subsection{The Limitations of Current Satellite Network}
To ensure interoperability and efficient utilization of SatCom, {various key technologies such as advanced multi-satellite multi-beam collaborations, Software-Defined Networking (SDN), on-orbit autonomy, on-orbit computing and intelligent resource management have been proposed to achieve the goal of 6G-enabled DSIN.}
In addition, the standardization efforts for SatCom and Non-Terrestrial Networks (NTN) are gaining momentum, with the active participation of several standardization organizations, including the European Telecommunications Standards Institute (ETSI) and the $3\text{rd}$ Generation Partnership Project (3GPP) \cite{9442378}.
However, current SatCom networks will not meet all the requirements of the future 6G-enabled DSIN. One of the main distinguishing features of the current SatCom systems is the prevalence of using a single satellite to provide emergency communication or Earth Observations (EO) services, which often struggle to offer continuous and robust communication services due to their singular nature and susceptibility to regional disruptions.
Further, the proliferation of diverse satellite systems, including those dedicated to communication, navigation, and remote sensing, has led to a siloed approach where various departments often construct independent ground stations (GSs) to serve their specific domain applications.
This fragmented infrastructure hinders the timely sharing and integrated utilization of massive satellite network business information across different satellite systems.
In response, the future of the satellite networking paradigm is pivoting towards the multi-satellite collaborative shifts, i.e., the CCS system, aiming to create a cohesive platform for integrated acquisition, processing, storage, transmission, and distribution of space-based information. 
Last but not least, the integration of communication, sensing, and computing capabilities for distributed homogeneous or heterogeneous payloads is becoming increasingly critical.
The existing satellite networks are grappling with the challenge of limited orbital and spectral resources, which are unable to scale in line with the growing complexity of space missions and the burgeoning demands of 6G-enabled DSIN. 
The weak functional synergy among diverse payloads, limited perception and computing capabilities, and poor dynamic coordination of heterogeneous resources impede the network from providing task/goal-oriented, information-centric and automation-level intelligent services, which is beyond what current satellite networks can deliver.

The above limitations have spurred the development of DSIN, especially highlighted by the maturation of micro and nanosatellite manufacturing, along with advancements in multi-satellite common orbit control, high-speed inter-satellite communication, and multi-load collaboration technologies.
The typical architecture of DSIN encompasses:
1) the satellite constellations of hundreds to thousands of homogeneous satellites lacking inter-satellite control mechanisms, achieving near-real-time large-scale connectivity and seamless integration;
2) the CCS system of dozens to pairs of homogeneous or heterogeneous satellites in close proximity to form a virtual satellite, built upon orbital control, self-organizing networking and payload synergy technologies, tailored towards various complex space missions to provide flexible, reconfigurable, and resilient space-based information services \cite{xu2024semantic, 9887918}.
These innovative DSIN architectures has significantly enhanced the self-organization and self-healing capabilities of current satellite networks, which have been advanced in notable research projects around the world, i.e., NASA's Earth Observing-1, SpaceX's Starlink, ONION funded by European Union's Horizon 2020, and the Fast, Flexible, Fractionated, Free-Flying (F6) launched by the Defense Advanced Research Projects Agency (DARPA) etc \cite{10400393}.
In this regard, the development of DSIN is no longer a mere concept but an inevitable trend in the evolution of space information technology and 6G networks.

%% ÏÖÓÐµÄ¸ÅÄîÀï£¬ÎÀÐÇ  ºÍ ÎÀÐÇÐÇ×ù¶¼ÊÇ×¨Íø£¬Í¨ÐÅ¡£µ¼º½¡£Ò£¸Ð£¬£»  Î´À´ÊÇÒªÍ¨µ¼Ò£µÈ¹¦ÄÜÒ»Ìå»¯£¬²»Í¬ÎÀÐÇÖ®¼ä»¥Áª»¥Í¨£» ÃæÏòÎ´À´µÄÐÅÏ¢·þÎñÐèÇó£¬ Ìá³öÁË·Ö²¼Ê½ÎÀÐÇÐÅÏ¢ÍøÂçµÄ¸ÅÄî£¬
%% distributed satellites information network: 1) mega constellation, 2)CCS; ÌÖÂÛ¶¨ÒåÇå³þ ·Ö²¼Ê½ÎÀÐÇ£»  %%×ÛÊöµÄ´´ÐÂÐÔ

\subsection{Motivations and Contributions}
The DSIN is anticipated to provide cohesive sharing of heterogeneous resources and ultra-autonomous multi-satellite collaborative networking to handle complex on-orbit operations with unprecedented flexibility and scalability. To achieve this:

1) \textit{Innovative Network Architectures are Imperative}. Among these, current satellite networks, predominantly utilizing transparent satellite mode, reveal inherent limitations in adaptability and operational flexibility.
In contrast, advancing research into \textit{distributed regenerative satellite network architecture}, particularly within DSIN in our survey, is crucial for unlocking enhanced scalability and dynamic traffic management, thus laying the groundwork for next-generation robust and adaptable space information infrastructures.
Nevertheless, this distributed regenerative network architecture poses significant challenges in efficiently scheduling computational resources across widely dispersed satellites and in facilitating effective collaboration among them. In this context, the \textit{distributed satellite computing network architecture} emerges as a promising, pivotal technology, poised to address these challenges and fully unlock the potential of regenerative satellite networks.  %% distributed satellite computing power network  ¸ÄÎª Î´À´µÄÍ¨¸ÐËã¿Ø´æÒ»Ìå»¯ÍøÂç  1118
Moreover, the task-driven, high-volume data generated within DSIN often necessitates dependable satellite formation control and inter-satellite communication amidst complex and variable spatial environments to ensure continuous, deterministic information delivery.
Thus, \textit{reconfigurable satellite formation flying} and collaborative control mechanisms are indispensable for ensuring the success of various formation-flying missions and are discussed in this survey. %%ÀàËÆÓÚµÚ2Ìõ£¬°ÑÄÚÈÝÀïµÄÐ¡½ÚÌâÄ¿Ð±Ìå

2) \textit{New Air Interface and Transmission Technologies are Essential for Achieving High Spectrum Efficiency and Energy Efficiency}. This includes advancements in channel modeling and estimation, cloud-native distributed MIMO cooperation and signal processing, new waveforms, multiple access approaches, multicasting mechanisms, channel coding schemes, phased array antennas, erasure transmission, high-speed inter-satellite communication, network routing, and the proper combination of all these diversity techniques.
Further, another key area of focus is the development of \textit{cross-layer optimization techniques that leverage the dynamic nature of satellite networks, the characteristics of DSIN, and the diverse needs of various intelligent services}. These techniques, including mobility management, resource management, secure communications and testbeds, are crucial for guaranteeing high-deterministic, energy-efficient, ultra-secure and verifiable on-demand services.

\begin{figure}[h]
	\centerline{\includegraphics[width=0.9\textwidth]{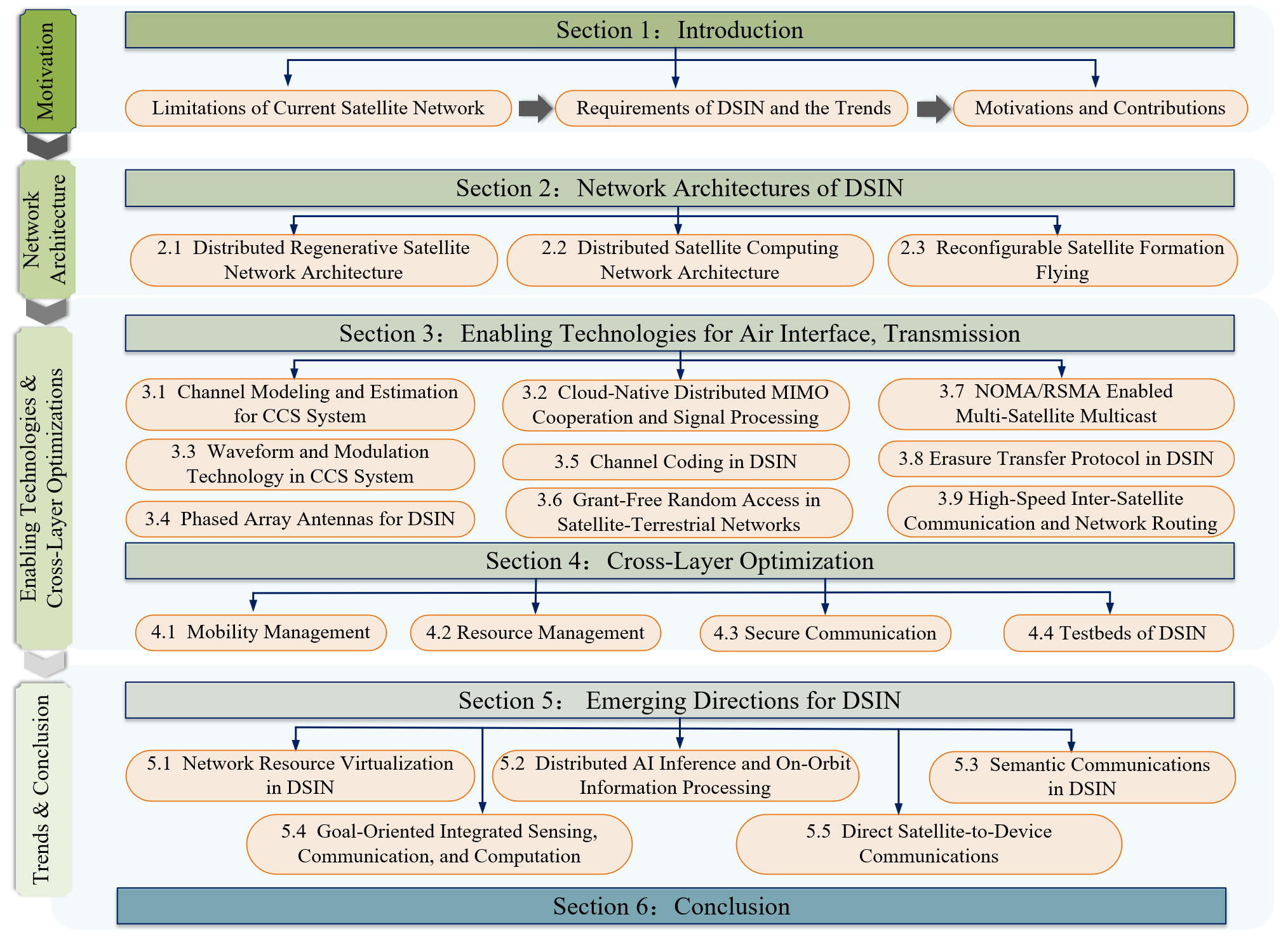}}
	\caption{{Organization of this survey.}}
	\label{fig1}
\end{figure}

3) \textit{A Suite of Emerging Research Directions and Challenges toward Realizing the Vision of DSIN are Discussed in This Survey}.
To provide adaptive resource utilization across dynamically changing network environments, \textit{network resource virtualization (NRV)} technologies combined with resource pooling will be fully exploited.
\textit{Distributed AI inference and on-orbit information processing} technologies will be efficiently combined with multi-satellite collaboration to achieve shared learning, fast inference and optimal decision-making, thus sustaining deterministic and continuous communication in CCS system.
Moreover, to support the emergent paradigmatic transition from bit-level to semantic-level communication, \textit{semantic communications} are anticipated to engender intelligent and concise information-centric services for DSIN.
Further, \textit{goal-oriented integrated sensing, communication, and computation} framework holds the potential to redefine how we utilize limited heterogeneous space-based resources to provide on-demand services for diverse tasks, bridging gaps between sensing, communication, and computation mechanisms to create autonomous and resilient DSIN.
Last but not least, the emerging direction is \textit{direct satellite-to-device communications}, which represents a fundamental milestone in the realization of an integrated satellite-terrestrial framework within DSIN.
By employing innovative on-orbit base station (BS) architecture and sophisticated phased array technology, this approach harmoniously merges SatCom with terrestrial mobile networks and extends a spectrum of communication services such as voice, messaging, and broadband Internet access.
%%ÀàËÆÓÚµÚ2Ìõ£¬°ÑÄÚÈÝÀïµÄÐ¡½ÚÌâÄ¿Ð±Ìå

%An overview of DSINs is shown in Figure \cite{}, where the network architecture, enabling technologies and emerging directions are given.

The organization of this survey is shown in Figure \ref{fig1}.
The limitations of existing satellite networks and the requirements of future DSIN are introduced in Section 1.
In Section 2, the new network architectures of DSIN are presented.
The enabling technologies, including the air interface and transmission technologies are given in Section 3.
The cross-layer optimization techniques are presented in Section 4.
The emerging directions are presented in Section 5.
Conclusions are drawn in Section 6.
The abbreviations in this survey can be found in Appendix A.

\section{Network Architecture}
To better realize the requirements of DSIN, novel network architectures are studied and applied according to network characteristics and specifications in this section, including the distributed regenerative network architecture, distributed satellite computing network architecture and reconfigurable satellite formation flying.

\subsection{Distributed Regenerative Satellite Network Architecture}

\subsubsection{From Transparent to Controllable/Regenerative Payload Architecture}
The satellite with a regenerative payload on board is an emerging trend for space-borne communication. Various system architectures are investigated to enable the flexible network topology (e.g., centralized or distributed) for advanced satellite networks.

\textit{Architecture-1: From transparent to controllable satellite}:
In legacy satellite networks, given the restriction of the satellite platform, the transparent payload for communication is widely deployed, especially for the GEO and MEO systems with following restrictions, e.g.,:
1) Limited processing capability on satellites, which is usually dedicated for satellite control instead of communication payload;
2) The controlling of RF components, e.g., beam and frequency reuse mode, is limited and via the dedicated feeder link along with other information, e.g., satellite movement/gesture adjustment.
Considering the hardware restrictions of satellites, the regenerative satellite payload may have partial or full BS function onboard during its evolution.
Although the satellites with higher capability or lower orbits are popular for the construction of the new satellite networks, re-framing the legacy on-orbit system is still essential considering the additional aspects, e.g., cost and capability of satellite manufacture.

\begin{figure}[t]
	\centering
	\includegraphics[width=0.7\textwidth]{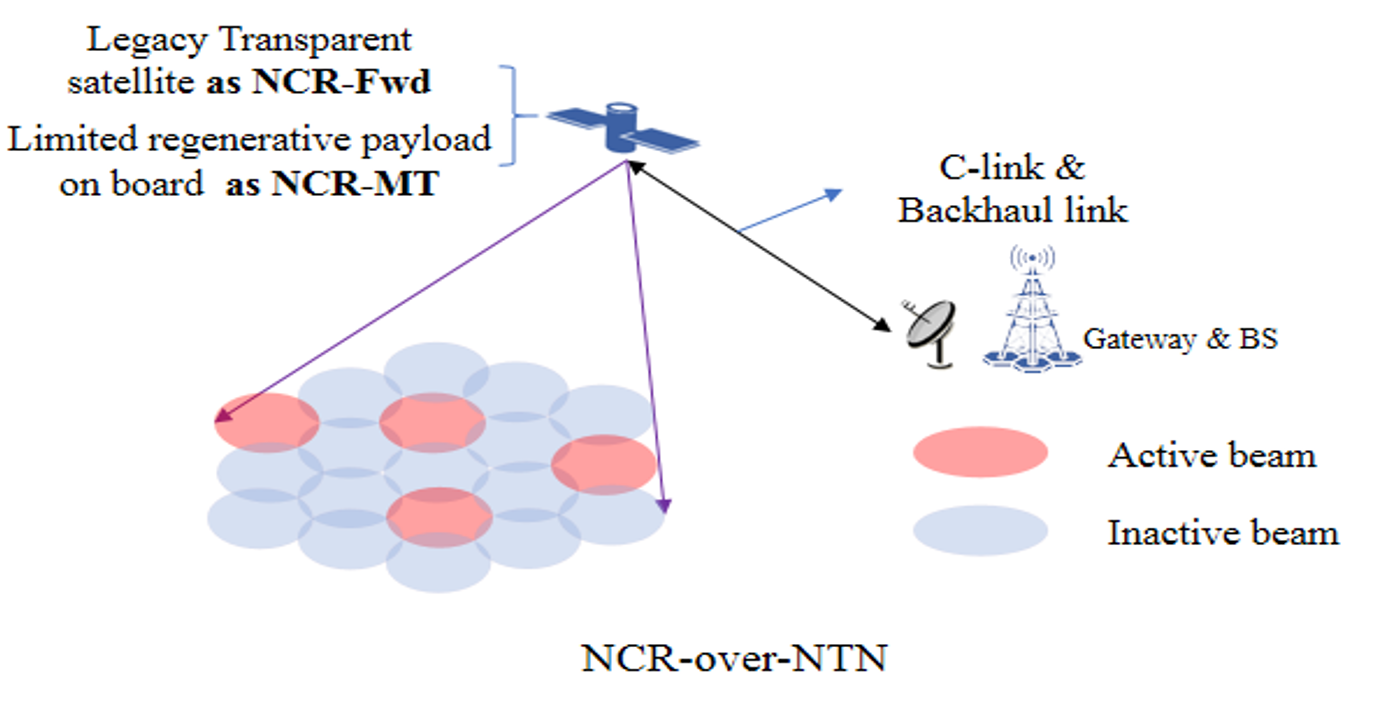}
	\caption{An illustration of controllable payload as evolution of transparent satellite.}\label{controllableSatellite}
\end{figure}

As shown in Figure \ref{controllableSatellite}, the concept of network controllable repeater (NCR) \cite{nan20225g} is introduced to enable the evolution of legacy systems with limited processing capability onboard.
For example, the legacy transparent satellite (i.e., NCR-Fwd), still only implements frequency conversion, a Radio Frequency (RF) amplifier, and beam hopping in both up and down links.
However, additional controlling information will be received by the NCR-MT via the control link from ground BS.
In this way, the coordination direction of the communication system can be well achieved via the centralized control center to optimize the network performance according to the dynamic traffic and scheduling per UEs/Beam footprint.
Moreover, this architecture is more friendly to enable the joint deployment between SatCom and cellular system, e.g., frequency sharing via dynamic turn-on/off the beam or updating the frequency used for each beam.

\textit{Architecture-2: From regenerative payload to full-gNB/CORE onboard}:
Along with the development of mega constellation for LEO satellites, the regenerative satellite has evolved to support more functionalities onboard, which enables more flexible topologies, e.g., CU-DU, full BS onboard with inter-satellite links (ISL) or CORE network onboard.

\begin{figure}[t]
	\centering
	\includegraphics[width=0.7\textwidth]{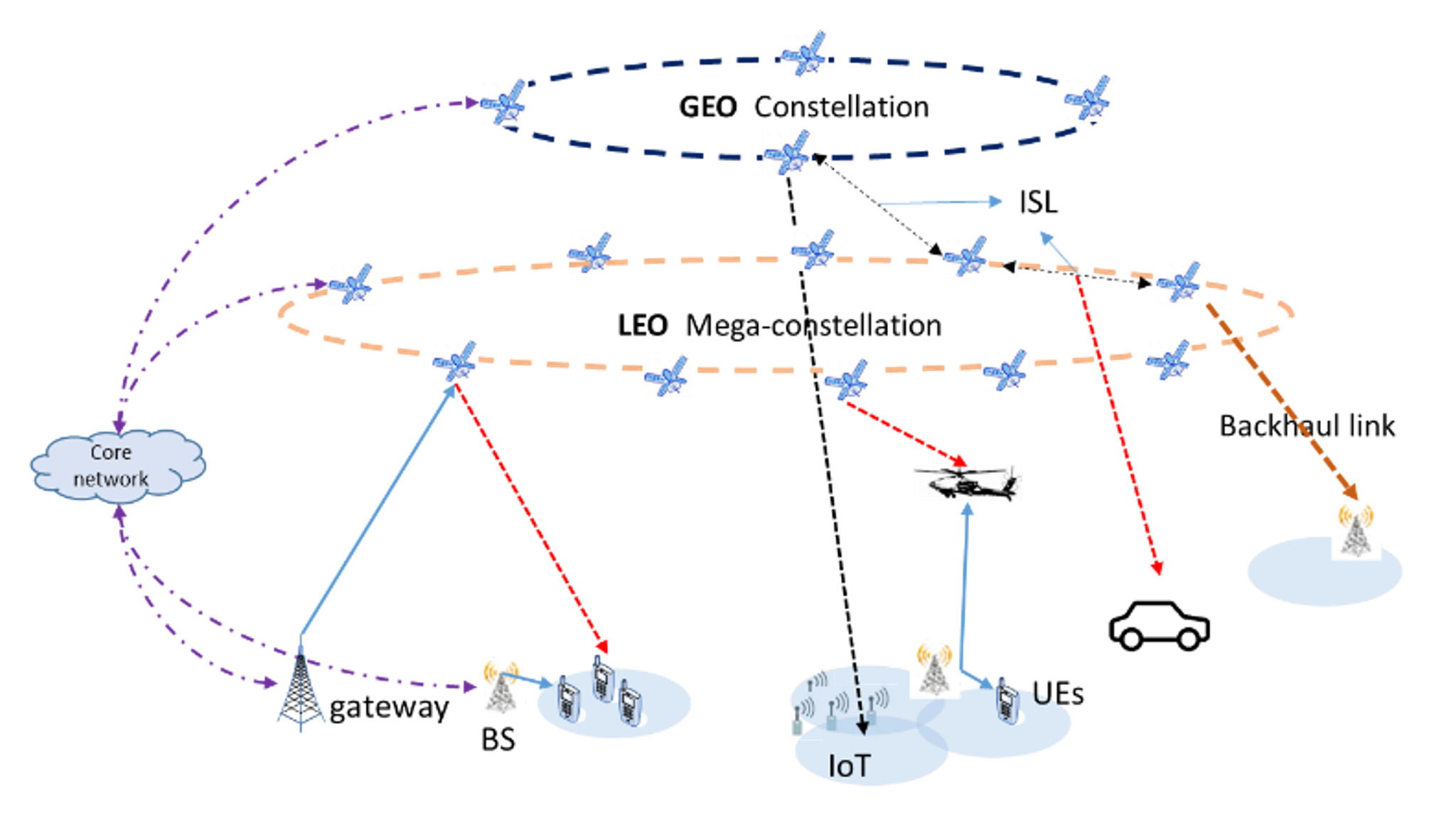}
	\caption{An illustration of flexible network topology based on the regenerative payload.}\label{regenerativeSatellite}
\end{figure}

As the exemplified network shown in Figure \ref{regenerativeSatellite}, with more capable satellites, the system can enable additional applications.
For example, with the ISL, the distributed satellites in DSIN can be deployed to enable complicated coordination onboard.
For instance, enabling certain satellites to function as computation nodes for processing information, such as images, is expected to further reduce latency and alleviate traffic loads during information exchange between satellites and gateways.

Additionally, in \cite{5.5-19}, a group of satellites work together to provide services to the UEs with high reliability since multiple satellites in CCS system can provide the service for the same area, along with backup in case of a satellite failure.
Furthermore, several small and lightweight satellites in CCS system, e.g., CubeSats, equipped with a commercial low gain patch antenna are tuned to create a large equivalent aperture providing a huge gain and a narrow beam.
The CCS system can be arranged in a free-flying configuration (i.e., wireless connected) or tethered configuration (i.e., optically connected).
The Simulation results presented in \cite{10694119} demonstrate that the performance is comparable with the classical and distributed implementations of antenna arrays.

%%1120 Á½¸öÐ¡½Ú£¬È¥µôÖØ¸´µÄÄÚÈÝ
\subsubsection{CU-DU Split Network Architecture}
The ongoing work on 3GPP Rel-19 marks a milestone in the integration of satellite technologies into 5G networks as defined by 3GPP \cite{3gpp.811}, particularly with the inclusion of NTN that feature regenerative, or packet-processing payloads.
The regenerative satellite-based payloads offer greater flexibility and performance by hosting partial or complete functionality of a gNB, an NR BS, and supporting the Xn inter-gNB interface for ISL \cite{10436159}. Nevertheless, due to the limitations in payload, power supply, and heat dissipation on satellite platforms, the processing capability of a single satellite is limited \cite{10648786}.
To deal with this issue, it is crucial to research and develop a distributed, open and intelligent DSIN architecture that can provide more efficient, flexible and smart solutions for a variety of applications, including worldwide communication, electromagnetic spectrum detection, remote sensing applications, and a multitude of other critical missions.
As a promising option, the integration of open radio access networks (O-RAN) with regenerative payloads aligns with the distributed architecture that separates the Central Unit (CU) and Distributed Unit (DU) \cite{10644023}.
As a result, a cutting-edge CU-DU Split~\cite{c1} network architecture can be implemented.
It originates from terrestrial 5G networks, where the CU is tasked with managing non-real-time protocols and services, including Radio Resource Control (RRC) and Packet Data Convergence Protocol (PDCP).
Typically situated in data centers or regional hubs, the CU facilitates efficient resource management and coordination.
Meanwhile, the DU is charged with the real-time processing of wireless access layer protocols, such as Medium Access Control (MAC) and Radio Link Control (RLC).
It is connected to the Radio Unit (RU), often located in close proximity or co-located with the RU to minimize latency.
The CU-DU Split not only fosters the sharing of baseband resources but also paves the way for the slicing and cloudification of wireless access.
This innovative approach ensures effective collaboration between sites in complex DSIN.
In addition, this separation is pivotal for unlocking the full potential of 6G networks in space, where the NTN gateway at the end of the feeder link can function as a router to the core network.
The ISLs are indispensable for regenerative payloads, particularly in Non-Geostationary Satellite Orbits (NGSO) such as LEO and MEO, where the satellites move significantly faster than the Earth's rotation.
This rapid movement necessitates seamless handovers of the feeder link to alternative gateways and the maintenance of service continuity by different space-borne platforms.

For satellite systems, the Satellite Centralized Unit (SAT-CU) serves as the centralized processing unit, enabling efficient sharing of processing capabilities.
By centralizing resource management and complex data processing in the SAT-CU, the system can dynamically allocate resources and quickly adjust task configurations to meet changing demands, while coordinating the collaborative work of multiple Satellite Distributed Units (SAT-DU) in CCS system to enhance the overall efficiency of DSIN.
The SAT-CU, in its role as the control plane and user data anchor, can reduce data interruption delays caused by handovers, further improving the user experience. Meanwhile, the SAT-DU is responsible for preliminary data processing and filtering with high real-time requirements, reducing data transmission delays and improving response speed.
Since SAT-DU only needs to provide RF and limited baseband processing capabilities, with user data being collected and processed centrally by SAT-CU without the need for independent core network connections, the weight, power consumption, and complexity of SAT-DU satellites can be effectively reduced, thereby further reducing the comprehensive cost of satellite network construction and deployment.
This architecture also improves the robustness and reliability of the CCS system, better isolates and handles faults, and reduces the risk of single-point failures.

\begin{figure}[!t]
	\centerline{\includegraphics[width=0.9\textwidth]{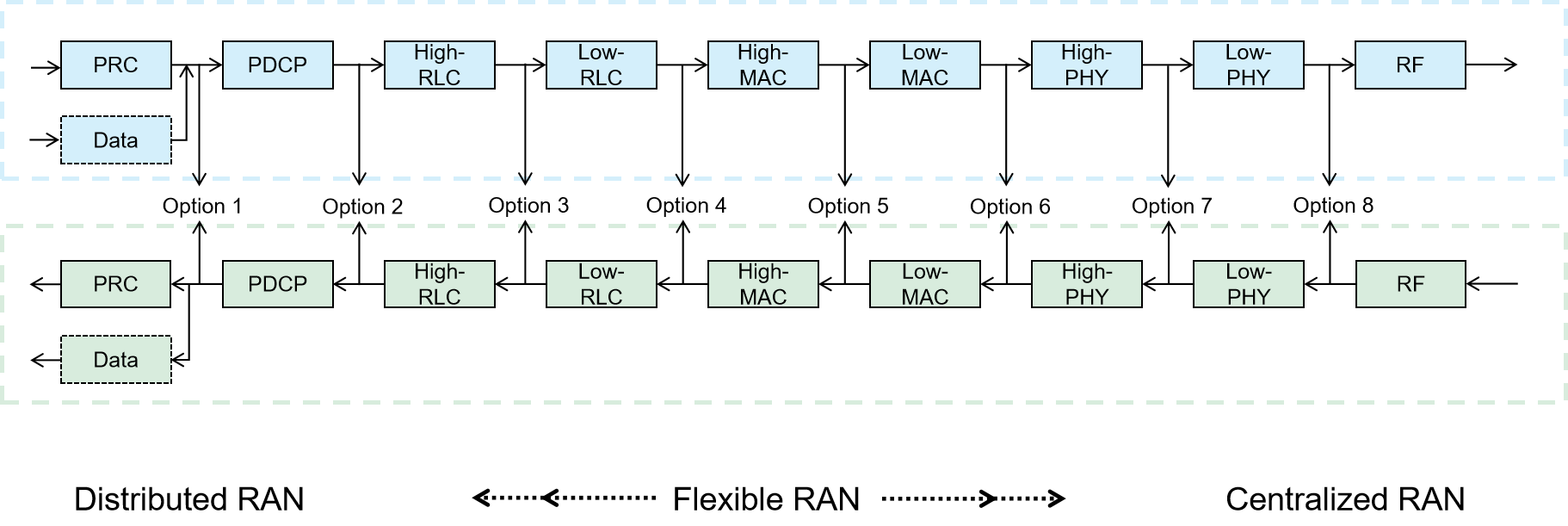}}
	\caption{CU-DU split options.}
	\label{split}
\end{figure}

\hspace{1em}\textit{1) SAT-CU/SAT-DU Split Options.}
When it comes to delineating the roles of the SAT-CU and SAT-DU, it is essential to explore different CU/DU functional partitioning methods tailored to diverse use cases.
Factors to consider include network topology and coverage, latency and bandwidth requirements, computing and storage resources, network load and traffic patterns, security and privacy, cost and complexity, and the extent of collaboration (such as joint scheduling, joint reception, and joint transmission).
Specific partitioning can refer to 3GPP TS 38.401~\cite{3gpp.38.401} and TR 38.801~\cite{3gpp.38.801}.
3GPP provides 8 functional split options as shown in Figure \ref{split}, varying based on the distribution of responsibilities between the CU and DU.
Here is an overview of the common CU-DU split options:

\begin{itemize}
    \item \textit{High-layer Split}: In high-layer split configurations, the CU manages upper-layer functions such as RRC and PDCP, while the DU oversees lower-layer tasks like RLC, MAC, and Physical (PHY) layers.
        This setup effectively centralizes key control functions within CU, reducing the processing requirements at DU.
        High-layer splits are advantageous for scenarios where the DU has limited processing power, or when centralized control and streamlined maintenance are priorities.
        However, the reliance on a centralized control structure can lead to increased latency, making high-layer splits less suitable for applications that require real-time responsiveness.
        Additionally, this configuration may impose significant bandwidth demands on the fronthaul link, potentially impacting efficiency in bandwidth-constrained environments.
    \item \textit{Mid-layer Split}: Mid-layer splits, which divide RLC functions between the CU and DU, offer a balanced approach to functional partitioning.
    By handling the PDCP in CU while assigning RLC, MAC, and PHY to DU, mid-layer splits manage latency more effectively than high-layer splits, while maintaining efficient control and resource allocation.
    Certain mid-layer variations, such as Options 3a and 3x, further refine the division of RLC tasks, with retransmission functionalities assigned differently to meet specific performance objectives.
    This distribution provides flexibility in controlling retransmission and segmentation operations, improving error handling close to the user end.
    While mid-layer splits can optimize latency and processing, they introduce complexity in managing distributed RLC functionalities, requiring careful coordination to ensure seamless operation.
    The variable processing load across CU and DU in dynamic DSIN conditions also adds to the challenges of mid-layer configurations.

    \item \textit{Low-layer Split}: Low-layer splits assign the most high-layer functionalities, such as MAC scheduling and Forward Error Correction (FEC), to the DU, effectively decentralizing processing tasks.
        By placing these real-time functions closer to the user, low-layer splits significantly reduce end-to-end latency, making them ideal for latency-sensitive applications.
        For instance, mission-critical applications such as satellite-based disaster response and real-time navigation services benefit from this approach, as it allows for immediate data processing at DU, minimizing the end-to-end latency.
        Additionally, low-layer splits facilitate scalability, enabling decentralized processing across broad coverage areas.
        However, these configurations may require DUs with substantial processing power and advanced capabilities, increasing deployment costs and power consumption. The power constraints are a concern in satellite networks, especially in remote locations where low-layer splits could be less feasible.

\end{itemize}

Selecting the optimal SAT-CU/SAT-DU split configuration requires a nuanced understanding of the tradeoffs between latency, bandwidth, and processing resources \cite{9946423}.
High-layer splits are often well-suited for S-IoT applications where low-cost DU units can efficiently manage large-scale device connectivity without the need for real-time response.
In contrast, mobile satellite services, which require a balance between latency and data integrity, may benefit from mid-layer splits.
By distributing control and error-correction functions across the CU and DU, mid-layer configurations can offer enhanced reliability and efficiency, making them a practical choice for applications where moderate latency and bandwidth demands coexist.
For critical applications requiring near-instantaneous response, low-layer splits present the most viable option.
By handling MAC and FEC functionalities at the DU, low-layer splits provide the freshness information needed for mission-critical applications, albeit with higher infrastructure demands.

\hspace{1em}\textit{2) Interfaces.}
In future DSIN architecture, seamless communication between network elements is crucial to achieving efficient data flow and resource management.
High flexibility in positioning network elements introduces challenges in interface design and standardization.
The F1 and E1 interfaces, which connect the CU and DU, play a central role in managing these connections.
The F1 interface further divides into F1-C (Control Plane) and F1-U (User Plane) components, allowing the CU and DU to manage control signalling and data transmission independently.
This division ensures efficient separation of control and data functions, enabling adaptive scaling of control and user data processing according to network conditions.

The E1 interface, connecting CU-CP and CU-UP, allows for further modularization within the CU.
By enabling control and user plane functions to optimize independently, the E1 interface ensures flexibility in managing network signalling and data flow, enhancing resource utilization.
Designing these interfaces to handle transmission delays between satellite and ground elements is critical.
Given that satellite networks often experience varying levels of latency and intermittent network coverage, interface protocols must accommodate these dynamics while ensuring high interoperability and compatibility between components.
The 3GPP standards provide a baseline for interface design in CU/DU split systems, promoting consistency in functionality and data transmission even as network elements vary.

In conclusion, by employing functional, service-based, and physical splits, network operators can optimize latency, bandwidth, and processing resources based on specific service requirements.
The design of interfaces such as F1 and E1 further supports seamless integration, ensuring compatibility across distributed elements and enhancing the overall performance of DSIN.
As SatCom continues to evolve, the SAT-CU/SAT-DU architecture will play a crucial role in enabling high-performance, adaptable satellite networks capable of meeting the growing demands of global connectivity.
Future research could explore adaptive split configurations that respond dynamically to changes in network conditions, pushing the boundaries of efficiency and responsiveness in DSIN.

\subsection{Distributed Satellite Computing Network Architecture} 
With the development of satellite technology, the available computing, communication, sensing and storage resources of satellites are increasingly abundant.
To reduce the data transmitted back to the ground for processing and improve information delivery efficiency, particularly in tasks such as pre-processing and target recognition of remote sensing images, the requirements of intensive computation tasks executed at satellites have become more urgent.
The consensus among industry and academics is that by leveraging the regenerative capabilities of satellites, deploying computational power nodes on satellites represents a promising solution \cite{Wang2023Satellite}.
{In this context, distributed satellite computing has become a common approach for onboard large-scale and massive computation tasks, offering benefits such as high reliability, fault tolerance, scalability, and fast computation. However, distributed satellite computing faces significant challenges. On the one hand, the computing nodes in the distributed satellite network must exchange numerous intermediate results with each other to compute the final result, which greatly increases communication overhead, especially in high-dynamics, high-mobility LEO satellite networks. On the other hand, distributed satellite computing is executed by a large number of computing nodes, which may have varying computing and networking resources. As a result, there can be straggling nodes—computing nodes that run slower than others, unintentionally increasing the overall time required to complete the computing tasks.
To address these challenges, the concept of coded distributed computing, which combines coding techniques and distributed computing, has been proposed by \cite{8051074} and has recently garnered significant attention. By employing coding-theoretic techniques, the distributed computing framework encodes the subtasks for the computing nodes, enabling the master node to recover the final result from partially finished nodes, thus mitigating the effects of stragglers \cite{9463425}. Furthermore, coded distributed computing is highly compatible with NFV, mobile edge computing \cite{9933792}, and collaborative deep reinforcement learning (DRL) \cite{10638529}, which facilitates unified resource and service management within SAGIN.
The network topology in DSIN is characterised by strong dynamic properties, with asymmetric node computing capabilities, making it difficult to apply AI methods designed for terrestrial networks to DSIN. To address this, \cite{10638529} proposes a collaborative DRL paradigm for satellite-terrestrial networks to achieve intelligent multi-orbit spectrum and computing resource management. This paradigm employs a DRL algorithm combining graph convolutional networks and federated learning (FL) to analyze and infer network topology features, guiding the communication and computing resource management of distributed satellites. This enables efficient adaptive computing offloading while avoiding model retraining due to node mobility or failures.
Additionally, the integration of data centers into satellite networks has also attracted attention \cite{9887918} for achieving low latency, flexible networking, and efficient resource utilisation. Moreover, by deeply integrating cloud computing, the cloud-native satellite computing network can shape DSIN into a distributed, elastic, and horizontally scalable satellite system composed of interrelated microservices, isolating state in a minimal number of stateful components \cite{10433234}.
However, how to efficiently schedule the heterogeneous resources at distributed satellites and make those satellites located in a wide area collaborative is still an urgent problem that needs to be solved.}

\begin{figure}[t]
\centering
\includegraphics[width=0.9\textwidth]{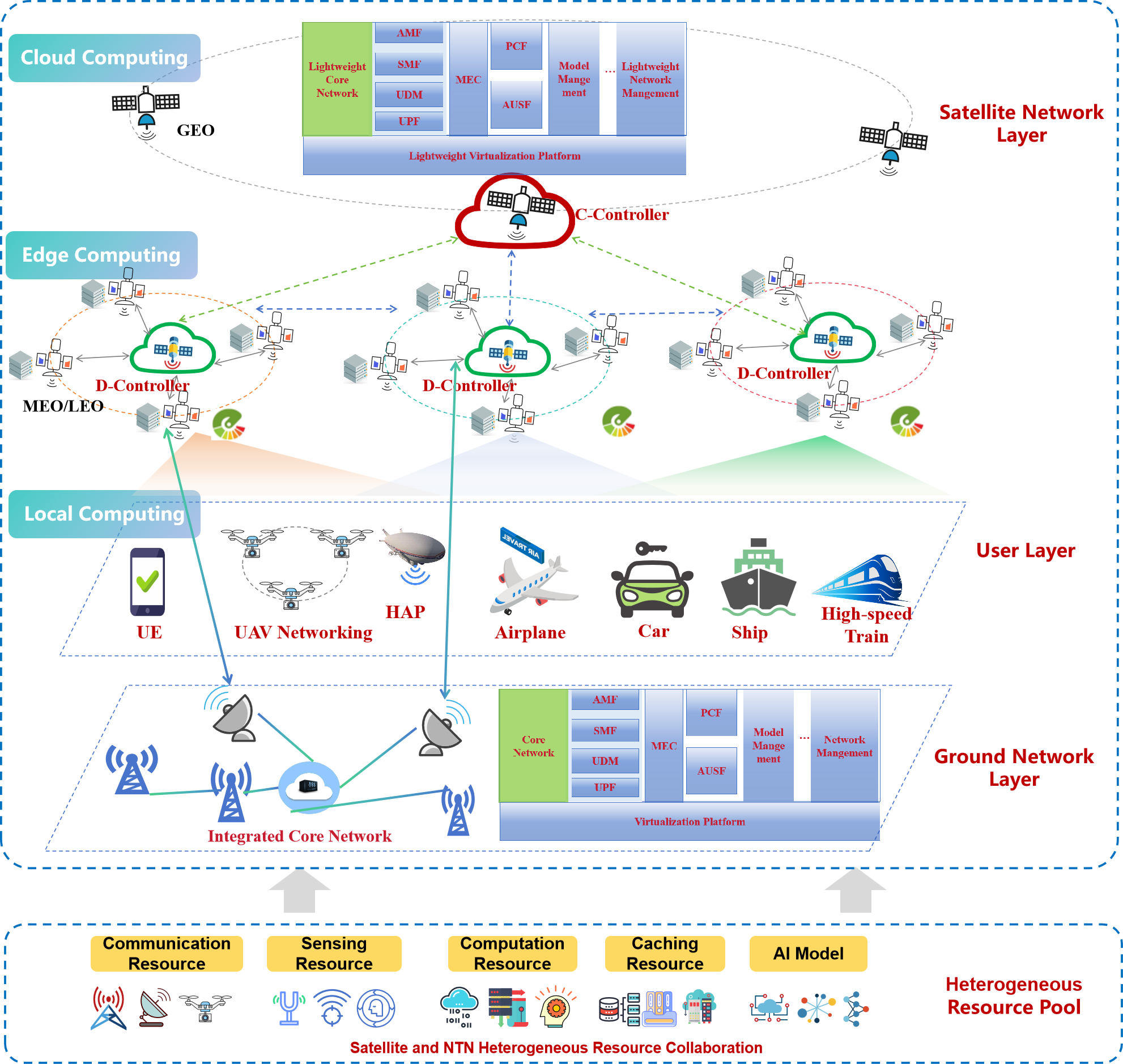}
\caption{{Distributed satellite computing network architecture for DSIN.}}
\label{Computing}
\end{figure}

To solve the above issues, we propose a distributed satellite computing network architecture as shown in Figure \ref{fig1}.
Exploiting the characteristics of GEO, MEO, LEO, and terminals, we design a Cloud-Edge-Local collaborative computing architecture.
In the satellite network layer, the GEO provides cloud computing services due to its abundant computing power and stability, MEO and LEO provide edge computing services with their advantages in quantity and distance to the Earth's surface.
The terminals in the user layer provide local computing services using their various computing power capabilities.
To enhance network performance through LEOs' cooperative transmissions, satellite clusters are proposed in \cite{Jung2023Satellite}.
Inspired by \cite{Jung2023Satellite}, we consider there are some MEOs/LEOs form a CCS system, in which the leader satellite plays the role of the distributed controller (D-controller) in the MEO/LEO satellite layer, while GEO works as the center controller (C-controller), and the follower satellites cooperate with transmission and computation.
To make the onboard service forwarding or migrate between different satellites, deploying the CORE network function on the satellite network layer is a promising solution. Therefore, we consider an integrated core network function to be entirely deployed at the ground networks layer, and the lightweight core network function is deployed at GEO in the satellite network layer as shown in Figure \ref{Computing}.

\subsubsection{Key Technologies of Satellite Computing Deployment}

\begin{itemize}
\item \textit{Lightweight Virtualization Technology of Satellite}:
Virtualization technology abstracts core network functions from distributed satellite hardware, allowing these generated core network functions to operate independently of physical infrastructure.
Lightweight techniques such as network element function trimming, function fusion and reconstruction, and interface protocols are lightweight and employed to achieve satellite-based lightweight core networks \cite{Pan2024Architecture}.
The deployment set of satellite network element functions can include access and mobility management function (AMF), session management function (SMF), unified data management (UDM), user plane function (UPF), policy control function (PCF), and authentication server function (AUSF).
Moreover, the functions of MEC, model management, and lightweight network management are considered to better process intelligent tasks.
Due to the rapid changes in satellite-ground topology, satellite-based computing nodes face significant time delays when making decisions.
Integrating the computational resources of satellite-based distributed nodes through virtualization technology, computation tasks in the network are split into several microservices, and then executed on the computing units of CCS system, efficiently reducing the computational burden on the individual satellites.

\item \textit{Edge Computing Function of Satellite Division Strategy}
Structured functional division of satellite onboard edge computing is essential for optimizing space-based data processing capabilities, enabling modular and collaborative task execution in satellite edge computing.
To improve the capability of satellite in-orbit service and to organize the intelligent satellites' edge computing nodes, the functional division of onboard edge computing is strategically organized into a layered architecture, and segmented into 4 distinct layers  \cite{Wang2023Satellite, 9442378}: Resource layer, virtual abstraction layer, system service layer, and application layer.
In the 4-layer architecture, services and applications with varying requirements demand close collaboration among system components\cite{Wang2023Satellite}.
Vertical collaboration involves coordination across hierarchical levels, harnessing the unique capabilities of each to optimize resources and enhance system performance.
Horizontal collaboration facilitates the distribution of computing tasks and service composition among computing modules at the same level, which is essential in multi-service scenarios or computing fusion.
Integrated collaboration combines both vertical and horizontal approaches, integrating computing resources, data, and services across layers and entities.
This comprehensive model is crucial for achieving maximum efficiency and functionality in DSIN, and enabling the development of more robust and intelligent satellite computing power technologies \cite{Wang2024End}.

\item \textit{Onboard Intelligent Fusion of Multi-source Satellite}
With the evolution of satellite technology, the received EO data are more precise and have a wider range of information with improved spatial and temporal resolution.  Multi-source satellite data fusion can enable effective complementarity to EO data, eliminate conflicts and uncertainties between data, and obtain more accurate and reliable information than a single data source.
Pre-processing, feature extraction, and fusion algorithm are three key steps critical for achieving high-quality fusion results.
Based on the information abstraction level, the fusion algorithm can be classified into three levels: Data, features, and decisions \cite{Gao2023Onboard}.
Meanwhile, AI has the potential to revolutionize the onboard information fusion processing method.
Currently,  AI-based methods have been proposed to correlate data from different sensors, DL has become increasingly prevalent in feature extraction and fusion algorithms, and FL has been also used to improve the performance of satellite data in-orbit fusion \cite{Li2024FedFusion}.
These AI-based algorithms adapt their strategies based on feedback, offering a dynamic approach to data fusion that is sensitive to the nuances of multi-source satellite data.

\end{itemize}

\subsubsection{Distributed Computing Power Collaboration}
\begin{itemize}
\item \textit{Computing Task Migration Mechanism}
The collaboration of satellite and ground computing power can better assist in completing large-scale data processing tasks.
On one hand, in satellite remote sensing data processing, utilizing onboard computing power for preliminary data processing can reduce the amount of data that needs to be transmitted to ground stations.
On the other hand, by employing task migration between the satellite network layer and the user layer, it becomes feasible to offload tasks from remote terrestrial devices to satellites.
However, the limited payload capacity of satellites poses constraints on onboard computational resources, making it challenging to handle computationally intensive tasks effectively.
The computing task migration mechanism mainly includes cloud-edge, edge-local, and edge-edge collaborations.
In addition, imbalances in workloads among satellites result in higher queuing delays for tasks on heavily loaded satellites, while under-utilization of computing resources occurs on lightly loaded satellites.
The evolution of large-scale satellite constellations has emphasized the importance of collaborative task processing among multiple satellites to formulate a CCS system to address the inadequacies in individual satellite computational resources \cite{Xi2023Multi}.
Satellites engaged in satellite-ground computational migration not only participate in collaborative computations across CCS system but also serve as pivotal scheduling decision units for task scheduling and resource allocation.
Migrating tasks with high queuing delays to less loaded satellites benefits load balancing.
Tasks partially offloaded from devices can be seamlessly transferred within CCS systems, thereby enabling a more flexible task migration approach.

\item \textit{Multi-dimensional and Heterogeneous Resource Collaborative Management}
Currently, the independent networking of satellites and ground networks results in low flexibility in information interaction and significant differences in network characteristics, failing to meet various business requirements \cite{Cui2023Latency}.
Additionally, different operations of satellite and ground networks often lead to isolated systems, causing inefficiency and resource waste.
These issues underscore the urgent need for a collaborative management framework for multi-dimensional heterogeneous resources between satellites and ground networks to ensure optimal resource utilization and seamless interoperability, providing users with perception, communication, computing, and caching services \cite{Gong2024Intelligent}.
Several key technologies, such as MDMA \cite{Zhang2023MDMA},  can support the development of collaborative management of multi-dimensional heterogeneous resources between satellites and ground networks.
However, the collaborative management of heterogeneous resources must meet the requirements of abstraction, flexibility, and scalability.
Abstraction simplifies diverse resources into manageable entities, achieving unified resource sharing and reducing overall costs despite underlying heterogeneity. Flexibility allows the system to adapt dynamically to varying workloads and resource requirements.
Scalability ensures the system can expand and integrate new resource types or technologies as they develop.
In addition, to reasonably allocate the entire network resources among DSIN and NTN networks, it is essential to take into account the complexity of computing tasks, and the computing and transmission capabilities of DSIN and NTN, guaranteeing the match between heterogeneous resources and diverse tasks regardless of their source or characteristics.
\end{itemize}

\subsection{Reconfigurable Satellite Formation Flying}
The functionality of CCS system relies on the cooperation among multiple satellites. The number of available satellites within a specific space determines the system's service performance in that area. To address sudden and intense functional demands, satellites need to remain clustered within a small area. When facing widespread and uniform functional demands, a symmetric and regular satellite configuration is often required. Depending on the spatial scale and control method, existing satellite system configurations are generally categorized into three types: Satellite formations, satellite swarms, and satellite constellations.

As a key advancement in DSIN, satellite formation flying (SFF) plays a pivotal role in modern space exploration and applications to maintain a targeted orbit configuration with desired relative separation and orientation between multiple spacecraft.
The concept of SFF was first introduced in the 1970s to leverage multiple satellites to conduct interferometric infrared synthetic aperture imaging tasks traditionally handled by single, large satellites \cite{8293791}.
A specific satellite formation is %a specific geometry configuration 1202
 composed of multiple satellites distributed on the same or adjacent orbits, functioning as an extended ``virtual spacecraft'' and avoiding the technical and financial challenges of building one satellite of equivalent size.
For a specific SFF, the member satellites are relatively close to each other, with fixed relationships or dynamical relationships that can be represented by linearized equations, and the space-relative positions are commonly expressed using Cartesian coordinate parameters.
Early satellite formations are primarily employed in Synthetic Aperture Radar (SAR) applications, such as the US Techsat-21 program, France's Cartwheel program, and Germany's Pendulum project \cite{yin2018review}.
Due to their configuration flexibility, satellite formations are widely utilized in Earth environment monitoring missions, such as the ESA Swarm, QB-50, and GRACE projects.
In space science experiments, many projects such as TPF and LiSA have been launched, and satellite formations for gravitational wave detection projects like Tianqin and Taiji are under construction in China.

The SFF evolved from simple dual-satellite configurations to more complex architectures involving dozens or even hundreds of small satellites.
A satellite swarm (or cluster) is a broad concept.
In general, any satellite system in which multiple satellites collaborate to perform tasks can be referred to as a satellite cluster.
To distinguish from other concepts, the ``satellite swarm" discussed in this subsection typically consists of multiple satellites with similar orbital parameters and close spatial proximity \cite{8293791}.
The concept of ``satellite clusters'' emphasizes extending the functionality  of system on a spatial scale, without requiring the satellites to maintain stable relative positions.
The functionality of a satellite swarm can be realized by ensuring a certain number of satellites within a specified spatial range.
In remote sensing applications, projects like OLFAR, SWIFT, and KickSat propose the deployment of large numbers of satellites as observation arrays to improve remote sensing performance.
Remote sensing satellite swarms such as FISC and Flock enable frequent global observation updates.
%In deep-space exploration, satellite swarm projects such as ANTS and Breakthrough Starshot have provided new approaches for asteroid belt and exoplanet explorations.

A ``satellite constellation'' is a satellite system composed of multiple satellites distributed in several identically shaped orbits according to a specific pattern. Generally, satellites in a constellation are evenly distributed along their orbits, with nearly equal spacing between orbital planes \cite{yin2018review}.
The concept of satellite constellations emphasizes a balanced spatial distribution of satellite functionality.
The relative positions of satellites within a constellation are typically characterized by orbit parameters, and the use of homologous orbits ensures a stable configuration, thereby reducing the need for frequent maintenance.
Satellite constellations are commonly employed for global or latitudinally distributed communication and navigation services.
Since the 1970s, various countries have progressively developed mid- and high-orbit navigation constellations such as GPS, BeiDou, GLONASS, and Galileo, low-orbit communication constellations such as Iridium and Global Star, as well as inclined elliptical orbit constellations such as Molniya and O3B.
Large-scale LEO Internet constellations, represented by Starlink, OneWeb, Kuiper, and Telesat, have also been extensively deployed.
The aforementioned satellite systems all use fully functional satellites as their basic units.
Additionally, there is a distribution concept where satellite subsystems are separated and perform formation flying in space. %, which is beyond the scope of this paper.

\begin{figure}[t]
\centering
\includegraphics[width=0.9\textwidth]{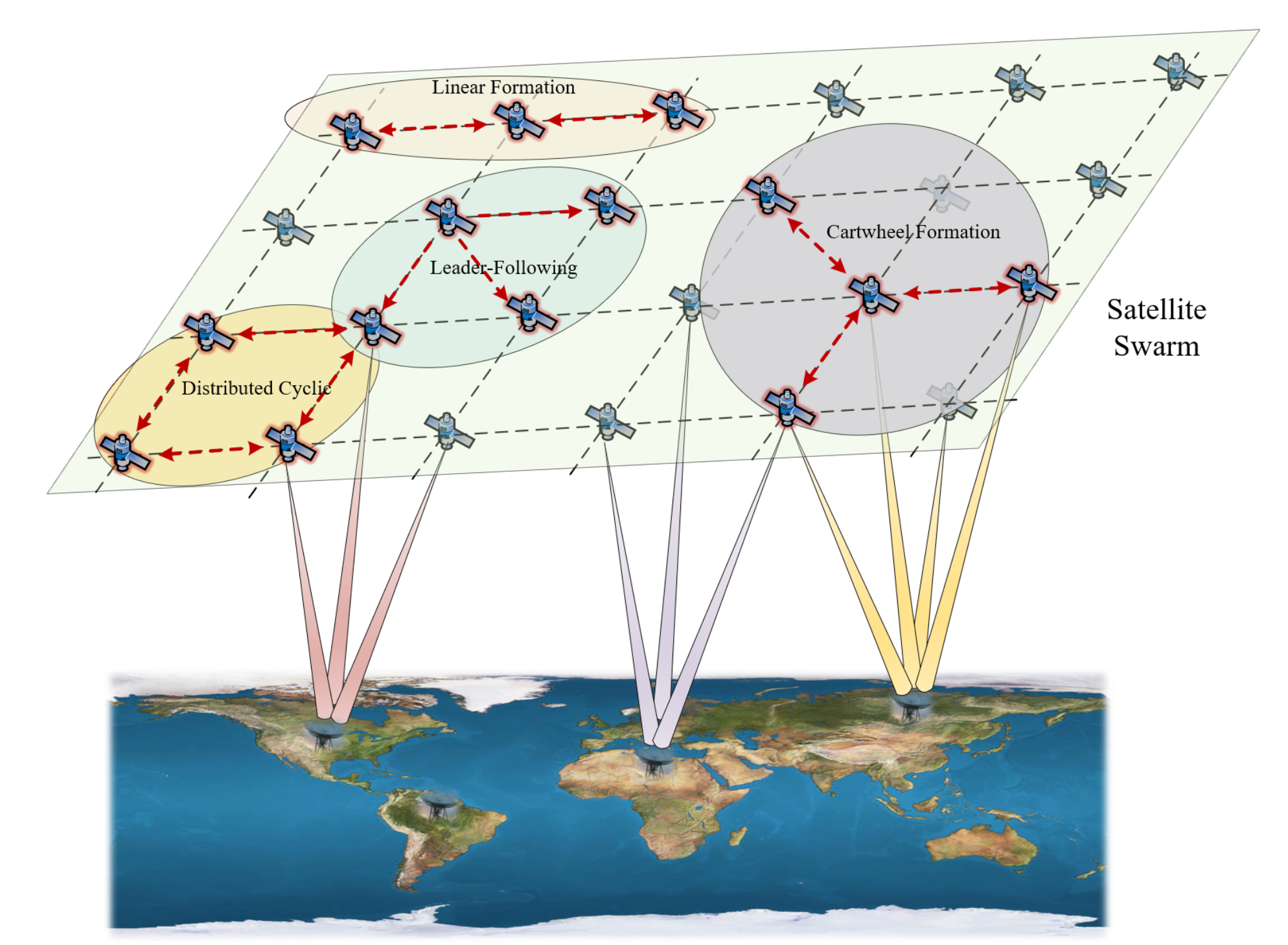}
\caption{An illustration of various satellite formations in satellite swarms}
\label{satelliteformation}
\end{figure}

To meet the demands of different scenarios, CCS systems require scalable and reconfigurable capabilities for configuration adjustments.
In near-circular orbit formation flying scenarios, the relative motion of satellites can be described by the linearized Hill equation (also known as the C-W equation) \cite{clohessy1960terminal}.
Building on this, \cite{tschauner1964optimale} proposed the T-H equation through variable substitution, which is a linear time-varying equation describing elliptical orbits via the eccentric Anomaly.
The aforementioned linearized equations are valid only for formation scenarios where the inter-satellite distance is less than 20 km.
{However, in non-near-circular orbits and conservative perturbation environments, obtaining the analytical solutions of these equations is challenging, rendering them unsuitable for relative configuration design or the construction of edge information hubs \cite{10679984}, but more suitable for motion control when combined with closed-loop control algorithms such as PID, sliding mode, optimization, or robust control methods.}
In addition, driven by the growing need for large-scale, the DSIN is envisioned to provide continuous, real-time services from deep space exploration to planetary orbit application fields \cite{wang2016satellite}, such as the high-orbit high-resolution optical EO, space target attachment, electromagnetic force formation and Confederacy space system design, etc.
As such, the development of satellite formation configurations has evolved from simple relative position maintenance to complex multi-satellite cooperation, and further to intelligent and adaptive control.
However, these expanded formations introduced new challenges in terms of communication constraints, incomplete information, and the need for advanced coordination strategies to maintain operational efficiency.

Various approaches to satellite formation control have been explored in the literature, each offering unique advantages and addressing specific challenges associated with maintaining and controlling satellite configurations in space, as shown in Figure \ref{satelliteformation}.
One of the foundational approaches to formation control is the leader-follower method.
In this configuration, one or more satellites serve as ``leaders," setting the trajectory for the rest of the satellites, or ``followers," to track \cite{2.3.2}.
The followers adjust their position and orientation relative to the leader, thus achieving a coordinated formation.
The Leader-following methods can be implemented in various structures, including single-leader, multi-leader, and virtual-leader formations, which offer different levels of flexibility and robustness.
However, a key limitation of the leader-following approach is its dependence on the leader. If the leader experiences a malfunction or significant perturbations, the entire formation may become destabilized, posing risks to the mission \cite{zou2012neural}.

In constraint to centralized formation control, the behavior-based formation control method represents a more decentralized approach \cite{balch1998behavior}.
Instead of relying on a designated leader, this method assigns specific behaviors to each satellite, such as collision avoidance, formation keeping, and reconfiguration. Through these individual behaviors and local control rules, the overall formation is maintained in a coordinated manner.
The core challenge in behavior-based control lies in designing effective behavior-coordination mechanisms that ensure global formation objectives are achieved.
For example, satellites may have to balance conflicting requirements, such as maintaining formation while avoiding potential collisions \cite{2.3.4}.
Behavior-based methods are particularly advantageous for large formations with multiple interacting units, as they allow for greater adaptability and autonomy.
However, the decentralized nature of this approach can make it difficult to achieve the high levels of precision required in certain missions, as the formation's overall stability depends on the interaction of individual satellite behaviors.

Moreover, the virtual structure formation method takes a different approach by treating the formation as a single ``rigid body'' \cite{2.3.5}.
Each satellite maintains a fixed position within a virtual structure, such as a wheel, line, or grid.
This method is effective in achieving high precision in maintaining formation shape and alignment.
For example, wheel formations involve satellites distributed around a circular or ring-like path, providing 360-degree coverage ideal for tasks like surveillance and communication relays.
Other configurations include linear formations, which arrange satellites in a straight line, suitable for continuous EO and synchronized orbit tracking, and rectangular or grid formations, where satellites form a structured grid pattern for broad-area coverage.
Triangular formations and cubic or spherical formations are also common in virtual structure control, offering three-dimensional coverage for complex space missions that require multi-directional observations or high spatial resolution.
Additionally, honeycomb formations follow a hexagonal grid structure, maximizing spatial efficiency and are often used in communication networks and environmental monitoring due to their dense, uniform coverage.
Arrow or V-formations are another variant, typically used in applications that require directional alignments, such as clustered navigation or coordinated maneuvering, as they reduce aerodynamic drag (for atmospheric applications) and optimize the communication link among satellites.
Beyond these conventional configurations, recent research has also introduced optimization and predictive control techniques into satellite formation control.
For instance, optimal control methods aim to minimize fuel consumption or time delay, which is essential for deep-space missions with limited resources.
Model predictive control (MPC) allows satellites to adjust their trajectories based on predictions of future system states, providing adaptive and efficient control for complex formations with high maneuverability demands.

As satellite networks become more distributed, the importance of reconfigurable SFF (RSFF) cannot be overstated.
Traditional formations are often designed as static structures, optimized for a specific set of mission parameters.
However, in distributed satellite networks, each satellite may experience varying orbital constraints, mission objectives, and environmental conditions, requiring the formation to adapt dynamically to maintain operational effectiveness \cite{2.3.6}.
The RSFF allows satellites to perform signal synchronization, change relative positions, adjust their orientations, or even reallocate tasks among themselves, offering unparalleled flexibility for mission planning and execution \cite{2.3.7,9761868}.

Several research efforts have focused on developing the necessary technologies to achieve RSFF.
One of the most prominent approaches is the use of graph theory to model inter-satellite communication and control \cite{graph}.
By representing satellites as nodes in a graph and their communication links as edges, researchers have developed algorithms to optimize formation control under various constraints, including limited communication bandwidth, time-varying network topologies, and external disturbances.
Another key area of research involves the development of formation control strategies based on artificial potential functions.
These methods model satellites as particles moving within a potential field generated by their relative positions and target locations.
By adjusting the forces acting on each satellite, the formation can be dynamically reconfigured to achieve the desired geometries while avoiding collisions or other hazards. This approach has proven effective in both simulation and experimental settings, particularly for missions involving close-proximity operations.
Moreover, the control methods for swarm configurations share similarities with satellite formation control methods.
However, for CCS system already in the desired configuration, frequent control task execution leads to resource wastage.
Therefore, an event-triggered mechanism can be designed to apply control only when configuration errors exceed the predefined threshold.
Event-triggered control can effectively reduce both communication frequency among members and energy consumption for orbit maintenance.
Currently, event-triggered control has been increasingly utilized for maintaining swarm system configurations, specifically in research areas such as first-/second-order integral multi-agent systems \cite{li2014event}, directed/undirected topologies \cite{lu2017distributed}, centralized/distributed triggering mechanisms \cite{garcia2014cooperative}, fixed/switched topologies \cite{li2014event}, linear/nonlinear systems \cite{zhu2014event}, with/without time delays \cite{fan2014semi}, and input saturation \cite{seuret2016lq}.

Further, developing decentralized control architectures that distribute decision-making across the DSIN is essential to improve resilience and flexibility.
The DSIN should incorporate adaptive algorithms capable of responding to environmental changes and communication delays, enabling autonomous reconfiguration without relying on a central command unit.
Moreover, hybrid centralized and decentralized control architectures use centralized control for high-precision tasks and decentralized strategies for adaptive and resilient reconfiguration, enabling the balance between precision and robustness.
For inter-satellite distances exceeding 20 km, orbit maneuver requirements can be addressed based on the Lambert problem.
The Lambert problem is a typical two-point boundary value problem that determines the maneuver trajectory given the spacecraft's initial and final positions and the transfer time \cite{sconzo1962use}.
Considering the univariate nature of the Lambert problem, recent studies have derived transfer time equations using parameters such as flight path angle \cite{nelson1992alternative}, eccentricity \cite{avanzini2008simple}, and terminal velocity \cite{zhang2020terminal}, and proposed iterative solution methods. Furthermore, maneuver theory has been improved by incorporating factors such as environmental perturbations \cite{yang2015homotopic}, thrust direction constraints \cite{zhang2020reachable}, and multi-revolution orbit solutions \cite{thorne2015convergence}.
The last but not least, researching algorithms that optimize energy consumption during formation reconfiguration is also crucial.
These algorithms could focus on minimizing fuel use while considering the long-term sustainability of the mission.
Techniques such as cooperative energy management and predictive energy budgeting can be explored to extend the operational lifespan of satellite constellations.

By adopting a phased configuration control method at different spatial scales, the CCS systems can be restructured according to service demands, choosing system configurations and control maintenance strategies that best fit the task requirements.
This enables the design of highly flexible, scalable, and open network architectures for DSIN, along with the corresponding functional partitioning strategies.
The future of RSFF lies in the continued integration of advanced control algorithms, high-precision sensors, and scalable communication networks.
One promising avenue is the application of machine learning techniques to predict and adapt to environmental disturbances in real time.
By incorporating predictive models into the control loop, satellite formations can proactively adjust their configurations to maintain optimal performance.
Another important direction is adopting a phased configuration control method at different spatial scales, which allows the CCS system to be restructured according to service demands and to choose system configurations and control maintenance strategies that best fit the task requirements.

\section{Enabling technologies}
In order to achieve spectrum- and energy-efficient communication, as well as high timeliness and reliability, a variety of new air interface and transmission technologies will be used in the DSIN, especially in the CCS system.
In this section, we will first introduce the air interface technologies at the physical and link layers, including channel modeling and estimation, new waveforms, distributed antennas, channel coding methods, multiple access approaches, and multicasting mechanisms, followed by network and transport layer technologies, involving erasure transfer protocols, distributed routing, and congestion control.

\subsection{Channel Modeling and Estimation for CCS System}
As the dawn of 6G communication technologies beckons, the crucial role of DSIN in achieving uninterrupted global connectivity has become more pronounced than ever \cite{6G}.
Within this evolving landscape, the creation of precise channel models and the enhancement of channel estimation methods have become pivotal to the advancement of these systems.
Channel models are indispensable for capturing the intricacies of satellite-ground communication links, offering a fundamental insight into their performance across diverse operational scenarios.
Concurrently, channel estimation stands as a cornerstone technology, facilitating the real-time acquisition of channel state information, which is indispensable for fine-tuning signal transmission and guaranteeing the efficiency and dependability of DSIN \cite{Sate}.
Despite extensive research into terrestrial communication channel modeling and estimation, the unique challenges posed by satellite-to-ground channels in CCS system demand a closer look.
In this subsection, we delve into the channel modeling and estimation within the realm of CCS system, and explore the latest advancements and present a comprehensive review of the current research landscape.

\subsubsection{Channel Model}
In CCS system, the channel model is essential to understand and predict the behavior of signals propagating between satellite-to-satellite and satellite-to-user links.
We delve into the two primary categories of the channel model, the satellite-to-satellite and user-satellite channel model, focusing on the user-satellite channel model \cite{test1}.

\begin{figure}[t]
	\centering
	\includegraphics[width=0.8\textwidth]{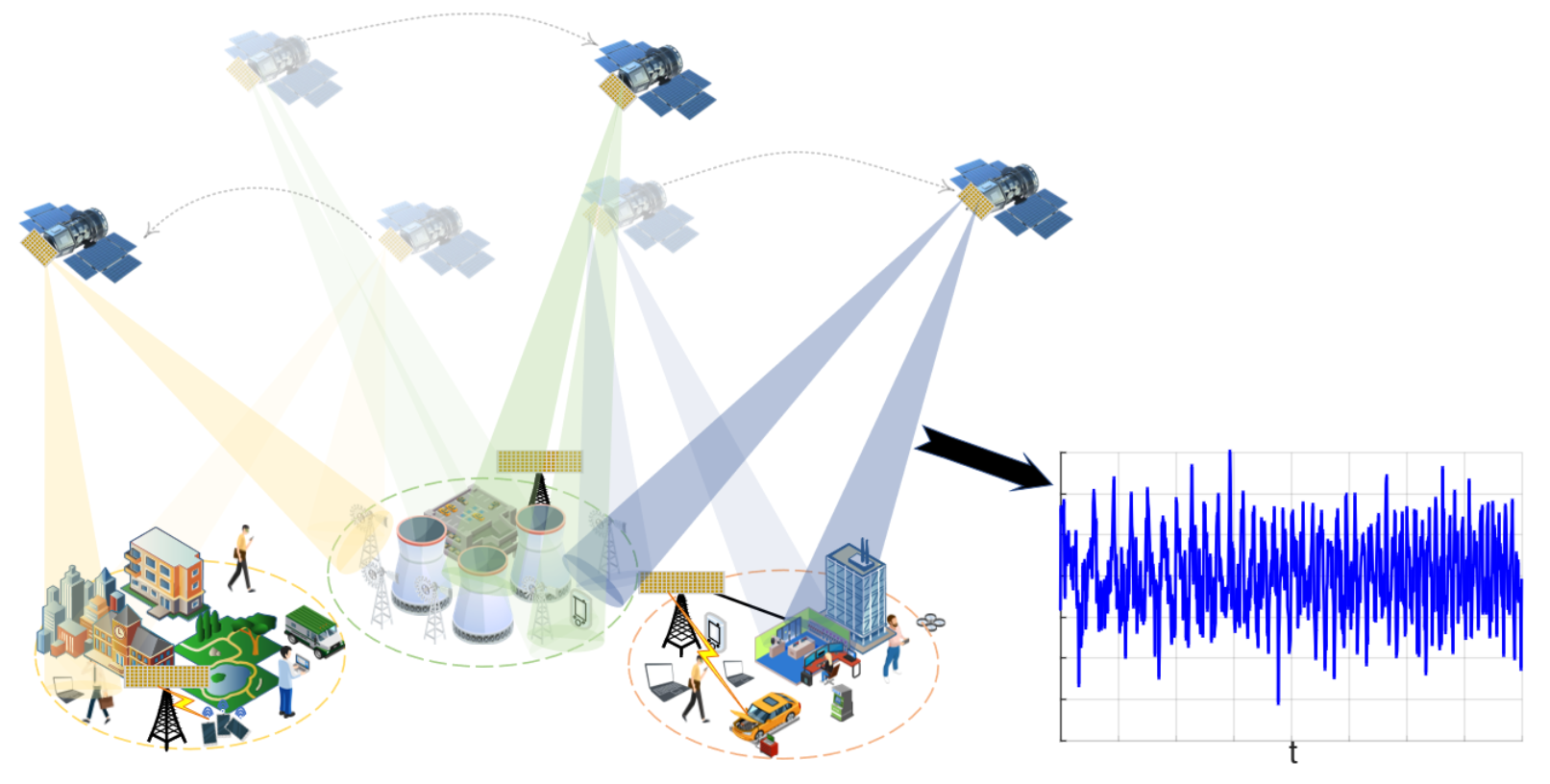}
	\caption{An illustration of channel model in CCS system.}\label{channel}
\end{figure}

The satellite-to-satellite channel model, which concerns the communication links between satellites, is relatively straightforward compared to user-satellite modeling.
This simplicity arises because space-to-space communication does not involve the complexities of the Earth's atmosphere or terrestrial obstacles.
As shown in Figure \ref{channel}, in satellite-to-satellite channel models, the primary considerations include the relative positions and movements of the satellites, the distances between them, and the characteristics of the transmission medium, which is typically the vacuum of space.
The main challenges in the satellite-to-satellite channel model involve accounting for the dynamic nature of CCS system, including the effects of gravitational forces, satellite orbit perturbations, and the need for precise alignment of communication beams.
However, the relative ease of this model allows researchers to focus on optimizing the performance of ISL without the added complexities of atmospheric and terrestrial interference.
The user-satellite channel model is where the real challenge lies.
This type of model ensures reliable communication between satellites and ground stations \cite{ISL}.
The user-satellite channel model is more challenging than the satellite-satellite channel model, primarily due to many factors, including the influence of the Earth's atmosphere, interference from terrestrial obstacles, and the dynamic nature of the communication environment.
User-satellite channel model methods mainly include geometric stochastic models and machine learning (ML)-based models.

The geometric stochastic model is designed with rational assumptions regarding the distribution of obstacles and scatterers near the receiving user.
It abstracts the effective scatterers within the environment into one or more models characterized by distinct geometric forms, systematically distributing them across the environment's abstracted geometric structure.
In \cite{SG}, the studies undertook the channel model of a multi-satellite communication system, postulating that the satellites are uniformly dispersed across the surface of a sphere at a specified altitude above the Earth. Using the geometric relationship between the satellite and the Earth, the research deduced the correlations among the elevation angle, azimuthal angle, and the distance between the satellite and the Earth. Although the geometric stochastic model has high universality and low computational complexity, it has low modeling accuracy in specific scenarios.
To have a more accurate channel model of CCS system, much research is needed to consider more factors.

In CCS wireless communication environments, the channel conditions are more complex, the operating frequency bands are higher, and the mobility of the terminals is further enhanced.
In such scenarios, ML's self-learning and predictive capabilities are particularly suitable, as they can extract critical features from highly complex channel data.
Moreover, by thoroughly training the channel model, we can enable the model to better adapt to the rapid changes in channel states in high-dynamic scenarios, thereby significantly improving the generalization ability and performance of the model.
The authors in \cite{AICM} integrate ML with traditional channel modeling techniques, presenting a novel approach for channel modeling in LEO satellite systems.
It underscores the critical role of radio channel forecasting in improving link quality in the face of growing atmospheric impairments.
In recent years, machine learning-based satellite-Earth channel modeling has seen rapid advancement, with many models being proposed, showcasing the technology's robust learning capacity and proficiency in handling vast datasets.
The long short-term memory (LSTM)-based modeling has garnered significant attention, with superior prediction accuracy demonstrated in several studies.
However, the predictive accuracy of current ML-based models is contingent upon the richness of the training dataset and demands substantial computational resources, which limits the applicability of many models to specific scenarios or frequency bands, as noted in \cite{WCM}.

\subsubsection{Channel Estimation}
Channel estimation is an essential aspect of CCS systems, and is a critical issue that directly affects the performance and reliability of the communication link. This entails estimating the impact of the channel through which a transmitted signal traverses in a wireless environment.
Conventionally, the channel effect is encapsulated in a channel state information (CSI) block in modern communication systems.
While conventional methods like Minimum Mean Square Error (MMSE) are employed for CSI estimation, they often entail high computational costs and may not always align with the demands of real networks.
Furthermore, obtaining timely CSI information gets more challenging due to extended propagation delays and fast-changing propagation environments in CCS systems \cite{zCE}.

The long propagation delays and high mobility inherent in SatCom channels contribute to increased channel estimation errors, while the large number of antennas adds significant overhead.
In CCS systems, utilizing statistical CSI (sCSI) has proven to be a more practical approach \cite{Li}.
This method effectively addresses the challenges of obtaining instantaneous CSI (iCSI) and significantly reduces the computational burden on satellite payloads by allowing for fewer updates to transmission strategies. 
Additionally, incorporating an effective pilot design can further enhance channel estimation, providing a robust solution to these challenges \cite{PCE}.

Various techniques have been developed to measure SatCom channels.
In \cite{CSCE}, Wang et al. introduce a method known as adaptive random-selected multi-beamforming estimation, which focuses on estimating geometric-based millimeter-wave Multiple-Input Multiple-Output (MIMO) CSI.
They create a geometry-based channel model for multi-satellite applications within high-throughput satellite systems.
A notable feature of this method is its application of compressive sensing, which estimates the combination of the transmit beamformer on the satellite side and the receiving combiner on the user side, organized randomly across several time slots.
Taking advantage of the sparsity in the angle domain, this approach significantly reduces the number of measurements needed for precise channel estimation.

ML-based methods are increasingly being adopted as a promising alternative for channel estimation.
This channel estimation can be framed as a supervised learning problem by using various channel features as inputs, including distance, time delay, received power, azimuth angles of arrival (AoA) and departure (AoD), elevation angle, root mean square (RMS) delay spread, and frequency, with CSI serving as the output labels.
In \cite{AICE}, the reciprocity property of downlink and uplink channels in time division duplexing (TDD) systems is explored, allowing the downlink channel to be estimated from uplink CSI using an LSTM-based deep learning model.
Lu et al. in  \cite{MCE} propose a channel estimation method utilizing a fundamental deep learning architecture in multi-satellite, multi-ground environments.
{In \cite{8979256}, a ML-based CSI prediction method is applied in a massive MIMO communication scenario, using convolutional neural networks to extract temporal channel correlation features. In \cite{9439942}, an LSTM-based predictor is used to address the problem of channel ageing in LEO satellite communication systems. Meanwhile, several DL-based joint CSI prediction methods are incorporated in downlink precoding schemes\cite{9000850,10550141}. A DL-based quantized phase hybrid precoder is introduced to enhance spectral efficiency in \cite{9000850}. Furthermore, taking channel prediction errors into account, \cite{10550141} proposes a deep learning (DL)-based joint channel prediction and multibeam precoding scheme, which achieves significant gains in robust CSI acquisition and uplink transmission performance, even under high Doppler shifts and long propagation delays.

Although AI-enabled channel estimation methods provide a favourable trade-off between generalisation and performance-complexity, they lack interpretability in decision-making. To address this issue, \cite{10368353} proposes a novel explainable AI (XAI)-based channel estimation scheme to offer detailed and reasonable interpretability of deep learning (DL) models.
In summary, AI-enabled channel estimation methods have demonstrated unique advantages in enhancing the reliability and robustness of satellite-terrestrial communications. However, the significant mismatch between the learning capabilities of existing AI models and the constraints of onboard satellite systems remains a challenging issue. Therefore, guiding AI-enabled channel estimation methods toward lightweight and universally reconfigurable solutions will be a key focus for future CCS system.}

% \subsection{Distributed MIMO Cooperation for CCS System}
\subsection{Cloud-Native Distributed MIMO Cooperation and Coordinated Signal Processing}
% \subsubsection{Cloud-Native Distributed MIMO Cooperation}
The deployment of large-scale satellite constellations, comprising thousands or even tens of thousands of satellites, has garnered considerable attention from industry and academia. In such systems, multi-satellite MIMO can reduce the reliance on a more significant number of antennas while improving data throughput \cite{XZY1}. One approach utilizes multiple non-collaborating satellites to independently apply MIMO techniques. Alternatively, cooperation between satellites, through data exchange or CSI, enables joint tasks such as transmission and resource allocation \cite{multisatellitemimo, XZY2}. This cooperation approach maximizes the benefits of the MIMO technique, significantly enhancing overall system performance. In addition, in CCS system, collaborative reception and transmission among multiple satellite nodes can achieve uplink reception diversity gain and downlink coherent beamforming gain, respectively. In theory, joint transceiver processing can achieve optimal performance.
Scalable distributed cooperative baseband signal processing has been extensively studied in cell-free massive MIMO systems \cite{r3x-1}. The authors in \cite{LEODMIMO} apply the concept of cell-free massive MIMO to satellite communications, and both theoretical and simulation results demonstrate that this approach can significantly enhance system capacity. However, the cell-free massive MIMO in \cite{LEODMIMO} requires the deployment of a centralized processing unit in space, which is highly challenging to implement.

\begin{figure}[t]
\centering
\includegraphics[width=0.5\textwidth]{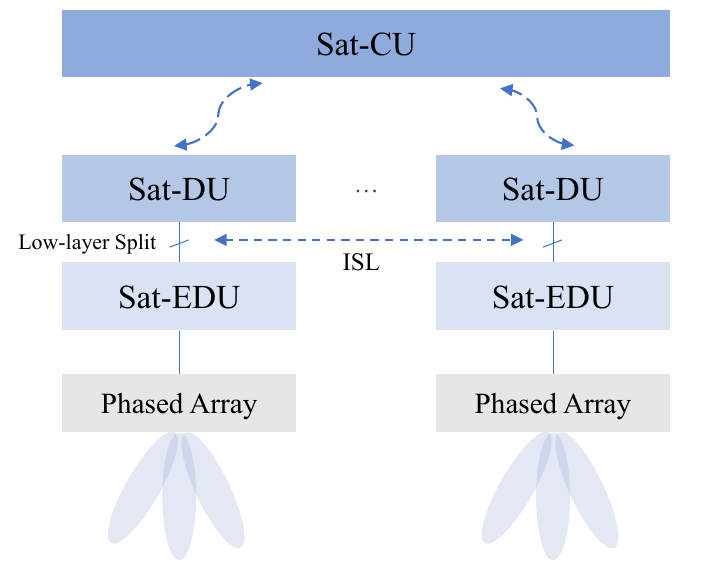}
\caption{Cloud-native baseband coordination.}
\label{fig-cloud-native}
\end{figure}

Based on Option 7 in Figure \ref{split}, the authors in \cite{r3x-3} propose a new low-layer physical splitting scheme to enable scalable distributed baseband signal processing, further evolving into a cell-free wireless access network. Introducing the new low-layer physical layer splitting scheme into the CCS architecture enables cloud-native collaboration of baseband computing power, thereby reducing the complexity of implementing cooperative transmission.
As shown in Figure \ref{fig-cloud-native}, the edge distributed units (Sat-EDU) implements distributed baseband transceiver processing, the Sat-DU performs high-layer physical layer processing and upper-layer processing for anchor users, while the Sat-CU handles anchor user selection, user-to-node and beam association, and other control functions.

% However, implementing distributed baseband processing onboard also faces significant technical challenges, primarily including the realization of distributed transceivers and phase synchronization among nodes.
However, the implementation of distributed MIMO cooperation and on-board coordinated baseband signal processing in CCS system faces significant technical challenges, {primarily involving the realization of distributed transceivers, as well as synchronization and calibration among nodes.}
Specifically, when the clock or local oscillator signals are generated locally at each satellite, attaining precise synchronization and calibration in terms of absolute phase, frequency, and time becomes extremely challenging \cite{9761868}. Also, distributed cooperative baseband signal processing is frequently hampered by the limitations of on-board processing capabilities and inter-satellite
interaction capacity. Consequently, there is an urgent need to develop robust, efficient, and resource-conserving synchronization and calibration techniques to bolster the functionality of multi-satellite MIMO systems.
In addition to the above challenges, distributed MIMO proves to be of vital importance for the effective management of the vast amounts of data generated by such systems. Through the utilization of decentralized processing nodes, each satellite is enabled to independently process signals and subsequently share the results with neighboring satellites. This approach not only diminishes the overreliance on centralized data processing, but also mitigates communication overhead. Moreover, cloud-native architectures take the distributed MIMO cooperation and coordinated baseband signal processing to a new level by furnishing scalable and flexible capabilities that can dynamically adapt to the ever-evolving needs of the network \cite{cloud}.

\subsubsection{Uplink Distributed Receiver}
When applying the cell-free wireless access network of reference \cite{r3x-3} to LEO networks, each node is equipped with a large-scale phased array, and both Sat-EDU and Sat-DU are deployed on LEO nodes. The system adopts a user-centric approach for node association, and the nodes serving users include a primary service anchor node and multiple secondary service nodes. For uplink reception, the receiver of each node independently performs multi-user detection \cite{Li}. The detected signals are sent to the primary service node Sat-DU of the user through the high-speed ISL for combined detection, then perform higher-layer physical layer processing such as demodulation and decoding. As shown in Figure \ref{fig-cloud-native}, uplink distributed cooperation relies on ISL. Compared to the link between Sat-DU and Sat-CU, the link between Sat-EDU and Sat-DU requires the exchange of detected modulated symbol data, placing higher demands on ISL capacity.

\subsubsection{Synchronization, Calibration and Downlink Distributed Beamforming}
Signal propagation delays, determined by satellite altitude and other parameters, typically range from tens to hundreds of milliseconds. This significant transmission delay imposes stricter timing management requirements for uplink and downlink signals. Moreover, signals from multiple satellites often cover the same area, enabling a single terminal to be served by multiple satellites simultaneously. However, the propagation delay differences and Doppler frequency offsets between various satellites and user terminals are significantly larger than those in terrestrial communication systems. This makes it challenging to achieve synchronization between satellites and terminals akin to synchronization and calibration in terrestrial systems. Managing synchronization and calibration between satellites and terminals to avoid interference between downlink transmissions from satellites and uplink transmissions from terminals is a critical challenge in multi-satellite cooperative transmission.

The synchronization and calibration methods on the terrestrial are mainly divided into hardware calibration and over-the-air (OTA) calibration. Hardware calibration requires additional reference antennas, which have been extensively studied in TDD massive MIMO \cite{synchronization}. OTA calibration requires no additional hardware and is realized through the transmission of reference signals between the remote radio unit (RRU) or between RRU and UE, including self-calibration and UE auxiliary calibration.  For example,  the authors in \cite{synchronization1}  propose a trunk-based calibration method, but the calibration time is long. The authors in \cite{synchronization2} respectively adopt different synchronization methods in WiFi distributed MIMO experiments. The authors in \cite{synchronization4} implement the system based on the OpenAirInterface (OAI) platform and propose a fast calibration method. 

For downlink transmission, the primary serving node distributes information to the secondary nodes, and each node independently performs beamforming \cite{Li}. As mentioned above, multi-node coherent cooperative transmission relies on phase synchronization between nodes. Time-frequency synchronization between multiple satellite communication nodes is the core issue for downlink coherent transmission. Phase synchronization between analog beams of multiple distributed phased arrays are affected by the following factors: (1) time-frequency synchronization between multiple nodes; (2) consistency calibration between multiple analog beams of multiple nodes.
Unlike terrestrial systems, satellite nodes typically direct their beams toward the ground, and phased arrays often lack an OTA link, making it challenging to achieve inter-satellite self-synchronization. A feasible method involves ground terminal-assisted phase synchronization and tracking. For TDD systems, terminal-assisted calibration and phase tracking can support coherent cooperative transmission among multiple phased arrays. For frequency-division duplex (FDD) systems, phase synchronization among multiple phased arrays can be achieved through terminal measurement, feedback, and phase tracking.

Terminal-assisted phase synchronization involves periodic measurement and feedback. However, the significant delays and Doppler frequency offsets between the terminal and multiple nodes result in substantial time overhead for measurement and feedback. Therefore, achieving high-precision phase synchronization imposes higher requirements on the local oscillator precision of the phased arrays and the accuracy of Doppler frequency offset estimation.
For downlink coherent cooperative transmission, phase synchronization errors among the analog beams of multiple phased array nodes are unavoidable. Thus, designing robust distributed cooperative digital precoding is essential \cite{DMIMO,r3x-5}.

From the above analysis, it can be observed that satellite systems differ significantly from terrestrial systems, rendering the synchronization and calibration technologies of terrestrial systems inapplicable to CCS systems. Satellite synchronization continues to face numerous challenges, such as achieving precise synchronization and calibration in terms of absolute phase, frequency, and time, especially when the clock or local oscillator signals are locally generated in each satellite. Moreover, the constrained resources of satellite systems further complicate these issues. It is necessary to develop more robust, efficient and resource-saving synchronization and calibration techniques to support MIMO cooperation and coordinated baseband signal processing in CCS system.

\subsubsection{Cloud-Native Distributed MIMO}
In DSIN, LEO satellites move at high speed along a predetermined orbit, which continuously changes their position relative to the UTs. As a result, UTs must frequently switch links between different LEO satellites to maintain a stable network connection. This complex switching process spans two critical layers: the link and network layers. At the link layer, the primary goal is to seamlessly transfer the communication link from one satellite to another within the UTs line of sight, akin to constructing an invisible yet indispensable bridge. The network layer, however, poses a more significant challenge. During a satellite handover, when the UT connects to a new ``home" satellite network, higher-level protocols such as transmission control protocol (TCP) and user datagram protocol (UDP) must migrate swiftly and seamlessly to the UTs' new internet protocol (IP) address. This ensures that ongoing data transmissions remain uninterrupted. Unfortunately, the visible window of any single LEO satellite to a UT lasts only a few minutes, necessitating frequent handovers. This results in high signaling overhead and many issues, including reduced throughput, processing delays, data forwarding congestion, and lagging location updates. These challenges severely degrade network performance, compromise spectrum utilization, and negatively impact the user Quality of Service (QoS).

Distributed MIMO technology, which has demonstrated remarkable success in terrestrial communications, offers promising solutions for LEO satellite networks. In terrestrial-based scenarios, distributed MIMO leverages multiple access points in a collaborative, cellular-free manner to achieve high spectral efficiency, superior power efficiency, and excellent network flexibility \cite{DMIMO}. Now, this cutting-edge technology is poised to revolutionize LEO satellite networks. By capitalizing on ultra-dense satellite constellations, ultra-fast ISLs, and line-of-sight (LoS) connectivity with terrestrial UTs, distributed MIMO is expected to unlock unprecedented communication performance. 
In parallel, integrating cloud-native with LEO satellite networks has injected new energy into the development of distributed MIMO systems \cite{LEODMIMO}. 

% Cloud-native platforms based virtualization infrastructure, with their robust data storage, computation, and resource integration capabilities, serve as a ``cloud brain" for satellite networks \cite{cloud}. When massive amounts of satellite data converge, cloud-native is able to manage a series of micro-services running on the cloud with high feasibility to achieve rapid processing and insightful analytics \cite{10234306}, allowing distributed MIMO systems to monitor real-time network traffic patterns and dynamically adapt to fluctuating user demands.

{Cloud-native refers to a set of technologies that decompose applications into microservices and package them into lightweight containers for deployment and orchestration across various servers \cite{10433234}. The cloud-native satellite cluster can be seen as a distributed, elastic, and horizontally scalable system, serve as a ``cloud brain" for satellite networks \cite{cloud}. The cloud-native satellite cluster consists of interrelated onboard microservices that isolate state in a minimal number of stateful components. As a cloud-native CCS system, it adopts practices such as microservices, containerisation, and orchestration to enable agility, scalability, and rapid development and deployment of satellite-integrated Internet applications.
To set up a cloud-native distributed satellite cluster, the monolithic system must first be decomposed into self-deployable, function-specific microservices \cite{10433234}, which can communicate with one another through lightweight messaging protocols over ISL. 
Subsequently, each microservice is packaged into a container using virtualisation technologies such as Docker, KubeEdge, and Kubernetes. Each container is then orchestrated into an integrated system for functionality, with automatic configuration.
Through the pooling and unified orchestration of computing power at satellite edge nodes, the cloud-native CCS system can dynamically allocate resources to each node for distributed signal processing and collaborative computation \cite{10234306}.
In this regard, cloud-native distributed MIMO efficiently leverages fragmented satellite node resources for cluster node status monitoring and information awareness. Based on the information gathered, it facilitates distributed collaborative transmission \cite{9183752}.
A cloud-based cell-free distributed massive MIMO system is investigated by \cite{9401305}, addressing challenges related to synchronization, calibration, and real-time baseband processing in 5G NR. General-purpose multi-core CPUs are employed for baseband signal processing. 

In addition, when LEO satellite clusters perform collaborative transmission towards the ground, frequent connection handovers occur, which slow down the cluster service restart process and reduce satellite node resource recovery efficiency. On one hand, the cloud-native CCS system ensures consistency of satellite cluster information across nodes through a distributed communication protocol, while periodically updating the health status of cluster nodes to ensure the rapid dissemination and proactive exclusion of faulty node information. On the other hand, by relying on the service migration and rapid resource release and recovery capabilities of the cloud-native CCS system, the cloud-native distributed MIMO can quickly rebuild container services for new satellite nodes and initialize satellite cluster information, thereby enabling flexible and robust signal processing and satellite-terrestrial transmission.}

\subsection{Waveform and Modulation Technology in CCS System}
The waveform is the shape of signal as it propagates through the physical medium, typically represented by its distribution over time or space.
It can also be abstracted to other domains such as frequency, code, or index.
Since the waveform directly affects the effective Signal-to-Noise Ratio (SNR) and Spectral Efficiency (SE) of CCS systems, flexible and efficient waveform design, along with its corresponding modulation techniques, is a key focus in DSIN.

\subsubsection{Current Waveform and Modulation Technology}
\hspace{1em}\textit{1) Orthogonal Frequency-Division Multiplexing.}
Orthogonal Frequency-Division Multiplexing (OFDM) is one of the most widely used modulation technology, which has been adopted by many wireless standards such as IEEE 802.16, 4G Long-Term Evolution (LTE), 5G New Radio (NR) and Wi-Fi \cite{3.2-1}.
In particular, OFDM transforms a wideband fading channel into a set of parallel narrowband flat-fading sub-channels.
By appropriately selecting the number of subcarriers, it ensures that the bandwidth of each sub-channel is smaller than the coherent bandwidth of the transmission channel, thus preventing frequency-selective fading and avoiding Inter-Symbol Interference (ISI).
However, conventional OFDM with rectangular shaping suffers from high sidelobes in frequency-domain and stringent criteria of orthogonality between subcarriers \cite{3.2-2}, which limits the flexibility of its waveform configuration and makes it difficult to adapt to the diverse services emerging in 5G.

In response, although OFDM is still used in 5G, several variants are proposed during the standardization process to achieve overlapping orthogonal design.
These schemes are mainly categorized into subband filtering and subcarrier filtering in multicarrier systems.
On one hand, subband filtering includes Universal Filtered Multi-Carrier (UFMC) \cite{3.2-3} and Filtered OFDM \cite{3.2-4}.
The former yields a good frequency-domain localization but may suffer from severe ISI due to the absence of Cyclic Prefix (CP).
In contrast, the latter offers subband filtering in a flexible manner to meet user requirements and uses the CP to mitigate ISI.
On the other hand, subcarrier filtering involves Filter Bank Multi-Carrier (FBMC) \cite{3.2-5} and Generalized Frequency Division Multiplexing (GFDM) \cite{3.2-6}.
The former provides a certain degree of flexibility through the selection of arbitrary pulse shaping filters and their parameters, whereas the absence of CP leads to similar issues as UFMC.
The latter, based on independent block modulation, offers a flexible frame structure but has higher requirements for time synchronization

In addition to filtering, windowing also plays a crucial role in waveform shaping.
For example, a windowed OFDM scheme proposed in \cite{3.2-7} can smooth the transitions between adjacent symbols by adding additional prefixes and suffixes, thereby reducing the out-of-band radiation caused by the rectangular pulse shaping.
Overall, these techniques effectively prevent potential Inter-Carrier Interference (ICI), ISI, and Adjacent Channel Interference (ACI), at the cost of increased transceiver complexity, delay, and reduced SE.
\begin{figure}[t]
	\centering
	\includegraphics[width=0.9\textwidth]{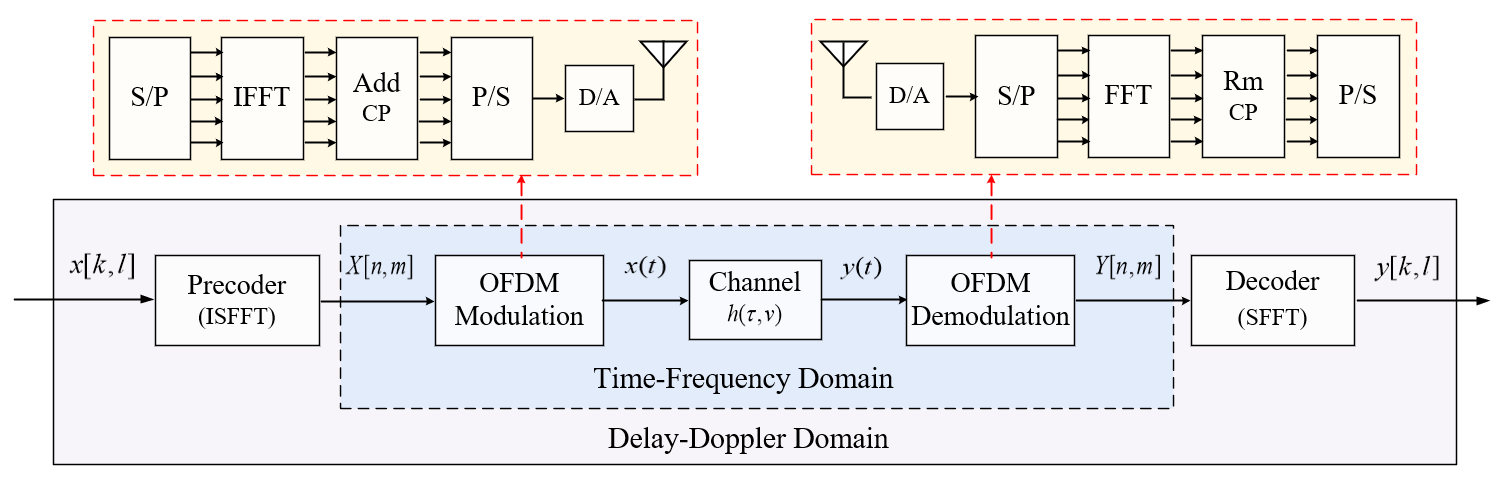}
	\caption{{An illustration of OTFS architecture compatible with OFDM system.}}\label{OTFS}
\end{figure}

\textit{2) Orthogonal Time Frequency Space.}
Current OFDM and its variants can only support high data rates and service quality in low- and moderate-speed mobility environments, while their performance is limited in LEO satellites of DSIN.
The reason is that the OFDM-based modulation process occurs in the Time-Frequency (TF) domain, making it highly susceptible to the Doppler effect caused by the high mobility of LEO satellites, which can easily disrupt the orthogonality of the OFDM waveform and lead to severe ISI and ICI \cite{3.2-8}.
To address this issue, a direct solution is to increase the CP ratio in OFDM to preserve orthogonality, although it results in a significant decrease in SE.
Another approach is to predict the Doppler shift based on the ephemeris and perform pre-compensation, but this method requires tracking each OFDM subcarrier individually, which brings significant complexity and hardware costs, further burdening the limited payload of LEO satellites \cite{3.2-9}.

Recently, a novel waveform modulation technique for high-speed mobility scenarios, called Orthogonal Time Frequency Space (OTFS), has been proposed \cite{3.2-10,3.2-11}.
OTFS is a two-dimensional modulation technique that represents transmitted signals in the Delay-Doppler (DD) domain.
By utilizing the Symplectic Finite Fourier transform (SFFT) and its inverse transform ISSFT, it converts the highly time-varying TF channel model into a sparse, slow-varying channel model in the DD domain, thus averaging out the rapid channel dynamics caused by satellite movement.
This method takes full advantage of the complete diversity offered by time and frequency selective channels, resulting in much higher SE compared to OFDM, with smaller subcarrier spacing and reduced CP overhead.
It also achieves a lower Peak-to-Average Power Ratio (PAPR), as the ISFFT spreading operation enhances the maximum energy per bit.
As a result, OTFS is considered a promising waveform and modulation technology for 6G and has been included in discussions at the 3GPP meetings \cite{3.2-12,3.2-13}.

The OTFS modulation also presents several significant advantages.
First, OTFS can be implemented by embedding an ISFFT precoding and corresponding SFFT decoding module within the existing OFDM framework, allowing for seamless integration with current OFDM systems \cite{3.2-14} and ensuring architectural compatibility with technologies like LTE as shown in Figure \ref{OTFS}.
Second, the reduced time variability of channels in the DD domain enhances the practicality and robustness of OTFS, while also lowering the overhead and complexity associated with physical-layer adaptation.
Moreover, the inherent sparsity of satellite-to-terrestrial channels is also beneficial for the design of low-complexity OTFS receivers \cite{3.2-15}.
Last but not least, the compact representation of DD channels in OTFS enables efficient packing of reference signals, which provides robust support for large-scale MIMO applications and facilitates the integration with multiple access techniques, such as non-orthogonal multiple access (NOMA) \cite{3.2-16}.
In summary, OTFS is a promising technology for enabling coordinated waveform design within CCS system in DSIN.

\subsubsection{Challenges and Development of OTFS}
There are still some open issues and challenges in OTFS modulation that require further investigations, in addition to the advantages mentioned above.

\textit{1) Interference Management.}
If the time and frequency resolution of the signal is not inversely related to the channel's Doppler shift and delay in OTFS, it can result in a loss of channel sparsity and lead to ICI in the DD domain, also known as Inter-Doppler Interference (IDI).
In response, researchers have proposed several solutions to enhance the sparsity of channels and mitigate the impact of IDI, such as performing cross-domain iterative detection and estimation jointly \cite{3.2-17}, executing windowing based on water-filling power allocation in the TF domain \cite{3.2-18}, and applying block-based joint detection and IDI elimination \cite{3.2-19}.
However, these schemes all come with the cost of increased receiver complexity.
Unlike conventional OTFS receiver designs, recent works have indicated that the Successive Interference Cancellation (SIC)-based MMSE detector can achieve superior interference cancellation with lower complexity, even in the presence of imperfect CSI \cite{3.2-20,3.2-21}.

Similarly, the RF impairments at the OTFS transmitter can also cause interference, with the presence of In-phase and Quadrature Imbalance (IQI) leading to Mirror Doppler Interference (MDI) in the DD domain \cite{3.2-22, 3.2-23}.
Different from the classical OFDM schemes, IQI does not cause saturation in the Bit Error Rate (BER) performance of OTFS; 
Instead, it only diminishes the channel diversity gain.
Simulations in \cite{3.2-24} evaluate the performance of OTFS under various RF impairments and demonstrate that the increase in pilot power and the number of Doppler bins, as well as the application of windowing, can enhance its interference resilience.
Further, due to the difficulty of implementing complex nonlinear receivers on computing- and energy-constrained LEO satellites, machine learning techniques, leveraging the predictability of Doppler shift (i.e., satellite movement), can be used with well-trained agents to reduce the complexity of interference detection and cancellation.

\textit{2) Channel Estimation.}
Recent researchers have proposed several channel estimation algorithms for OTFS systems to fully leverage the sparsity of the DD domain, including methods based on impulse surrounded guard symbols \cite{3.2-25}, embedded pilots \cite{3.2-26}, modified Compressive Sensing (CS) matrix \cite{3.2-27}, and Sparse Bayesian Learning (SBL) frameworks \cite{3.2-28}.
In contrast to the above algorithms that only make use of pilot symbols, the schemes that incorporate data symbols can further enhance estimation performance.
For example, the OTFS channel estimation method with superimposed pilots proposed in \cite{3.2-29} utilizes the sum-product algorithm (SPA) for data detection, and then performs data-assisted channel estimation that achieves higher SE.
Expanding on the use of data symbols as ``virtual pilots", the authors in \cite{3.2-30} further exploit the sparsity of the DD domain within an SBL framework by applying Variational Bayesian Inference (VBI) to estimate the DD channel vector, thereby achieving reduced complexity while maintaining channel estimation accuracy.

It is noteworthy that, although several studies have proved that collaborative downlink transmission in DSIN can substantially enhance the SE \cite{3.2-31, 3.2-32}, the parallel transmission of multiple data streams may cause a loss of channel sparsity in the DD domain, which requires more pilot resources to maintain channel estimation accuracy and significantly increases overhead.
Therefore, the key lies in designing a novel channel estimation method that can strike a tradeoff between cost and accuracy for DSIN.
The scheme proposed in \cite{3.2-33} introduces a simultaneous pilot-based aggregate channel estimation algorithm to improve channel estimation in CCS systems, and designs a three-stage peak-searching correlation-based method to handle fractional Doppler estimation.
%However, there are only a few works studying OTFS-based multi-satellite communication, and more in-depth research is needed.

\textit{3) MIMO-based OTFS.}
MIMO technology, with its extensive spatial degrees of freedom, low cost, and high integration, has accelerated research on OTFS beamforming and transmission efficiency, which enables further enhancements in the SE and diversity performance of current OTFS systems \cite{3.2-34}.
One of the main challenges for the MIMO system is the high processing complexity at the receiver.
For instance, the Message Passing Algorithm (MPA) proposed for MIMO-OTFS systems in \cite{3.2-35} provides excellent performance but also incurs high computational complexity due to their nonlinear nature, especially in channels with high Doppler shift and high-order modulation.
Moreover, spatial correlation can degrade the BER performance of MIMO-OTFS systems, particularly when receiving algorithms are designed without accounting for antenna correlation and Inter-Antenna Interference (IAI) \cite{3.2-36}.

To address these issues, recent studies have proposed several linear detection algorithms for MIMO-OTFS systems, based on Maximum Ratio Combining (MRC) \cite{3.2-37}, MMSE \cite{3.2-38}, Least Squares Minimum Residual (LSMR) \cite{3.2-39} and others, to fully exploit the two-dimensional sparsity in the DD domain and reduce computational complexity.
Further, these schemes all utilize Whitening Transformation (WT), a technique commonly used in MIMO systems to modify its channel matrix, such as Cholesky decomposition, to mitigate the spatial correlation between transceivers.
Considering that the methods mentioned above do not eliminate the spatial correlation at the receiver compared to the transmitter, the authors in \cite{3.2-40} designed a whitening filter to further improve the BER performance of MIMO-OTFS systems with receiver correlation.

Furthermore, researchers have recently explored the integration of MIMO-OTFS with Grant-Free Random Access (GFRA) technology for more complex multi-user detection scenarios.
The authors in \cite{3.2-41} introduced a two-dimensional pattern coupling hierarchical prior in SBL, combined with a generalized approximate MP algorithm, to improve the utilization of 2D burst block sparsity in the GFRA channel matrix, which arises from the three-dimensional structural sparsity of MIMO-OTFS in the DD domain.
In \cite{3.2-42}, the investigation is extended to multi-satellite MIMO-OTFS systems, where a joint scheme for device identification, channel estimation, and symbol detection in collaborative multi-satellite GFRA is proposed, and a distributed approach is adopted to offload computational tasks to edge satellites, thereby reducing the burden on individual satellites. {In addition, it is important to note that OTFS modulation ensures inherent stability of the wireless channel in the DD domain \cite{3.2-43, 3.2-44}}, which greatly facilitates the design of robust physical layer security schemes \cite{3.2-45}.

%security of satellite communications faces serious challenges due to the long distance and openness of satellite-terrestrial links and the limited computational resource of satellite platforms.

%\textit{4) Secure Communication.} %% ºÏµ½°²È«ÄÇÒ»Ð¡½Ú
%The security of satellite communications faces serious challenges due to the long distance and openness of satellite-terrestrial links and the limited computational resource of satellite platforms.
%In this case, OTFS modulation is considered a promising technology for enhancing secure communication because of the inherent stability of wireless channel in the DD domain, which facilitates the design of physical layer security schemes.
%For example, secret-key generation based on CSI in the DD domain requires low complexity, while eavesdroppers face difficulty in obtaining the time-varying CSI in the TF domain \cite{3.2-43}.
%Moreover, recent studies on OTFS-based uplink transmission \cite{3.2-44} and OTFS-based downlink beamforming \cite{3.2-45} have also demonstrated that OTFS  modulation can enhance security performance in LEO satellite communication scenarios with high mobility.
%Despite this, the investigation on OTFS-based security mechanisms is still limited, with current researches mainly focusing on simple end-to-end transmission scenarios.
%Overall, the OTFS-based secure communication in our DSIN requires further in-depth exploration.

\subsection{Phased Array Antennas for DSIN}
\begin{figure}[t]
	\centering
	\includegraphics[width=0.9\textwidth]{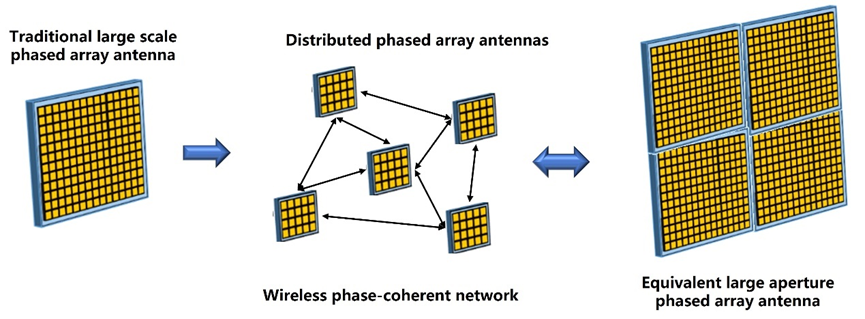}
	\caption{Traditional large scale phased array antennas vs. distributed phased array.}\label{antennaFig1}
\end{figure}

The transition from large, single-satellite communication platforms to networks of relatively small, coordinated satellite platforms marks an important advancement in SatCom \cite{mudumbai2009distributed}.
The distributed phased array antennas on these platforms can form a large equivalent aperture and enhanced flexibility in platform spacing, formation configurations, and network topology, enabling adaptable system configurations, which offer several advantages, including improved data throughput, and great resilience to interference \cite{nanzer2017open}. 
By utilizing the collective capabilities of multiple small phased array antennas, the CCS system can dynamically adapt to mission changes and provide flexible coverage over vast areas, while minimizing the impact of individual platform failures \cite{9761868,nanzer2021distributed,rashid2023high}.
Compared to traditional phased array antennas, CCS systems allow subarrays to be deployed across different coordinated satellite nodes, as shown in Figure \ref{antennaFig1}.
This approach mitigates the challenges of mounting large-scale phased array antennas on a single satellite, simplifying design and dramatically reducing the cost of satellite production and launch.
To sum up, the distributed phased array antennas can enhance overall performance through phase-coherent collaboration in CCS system.
However, deploying distributed phased arrays introduces new design constraints and technical challenges, such as the limited payload capacity of small satellite platforms, precise synchronization between subarrays, multi-beam beamforming and fast scanning.
Addressing these challenges is crucial for unlocking the full potential of distributed phased array systems in large-scale DSIN.

\subsubsection{High-Performance Silicon-based Integrated Phased Array Antennas}
\begin{figure}[t]
	\centering
	\includegraphics[width=0.5\textwidth]{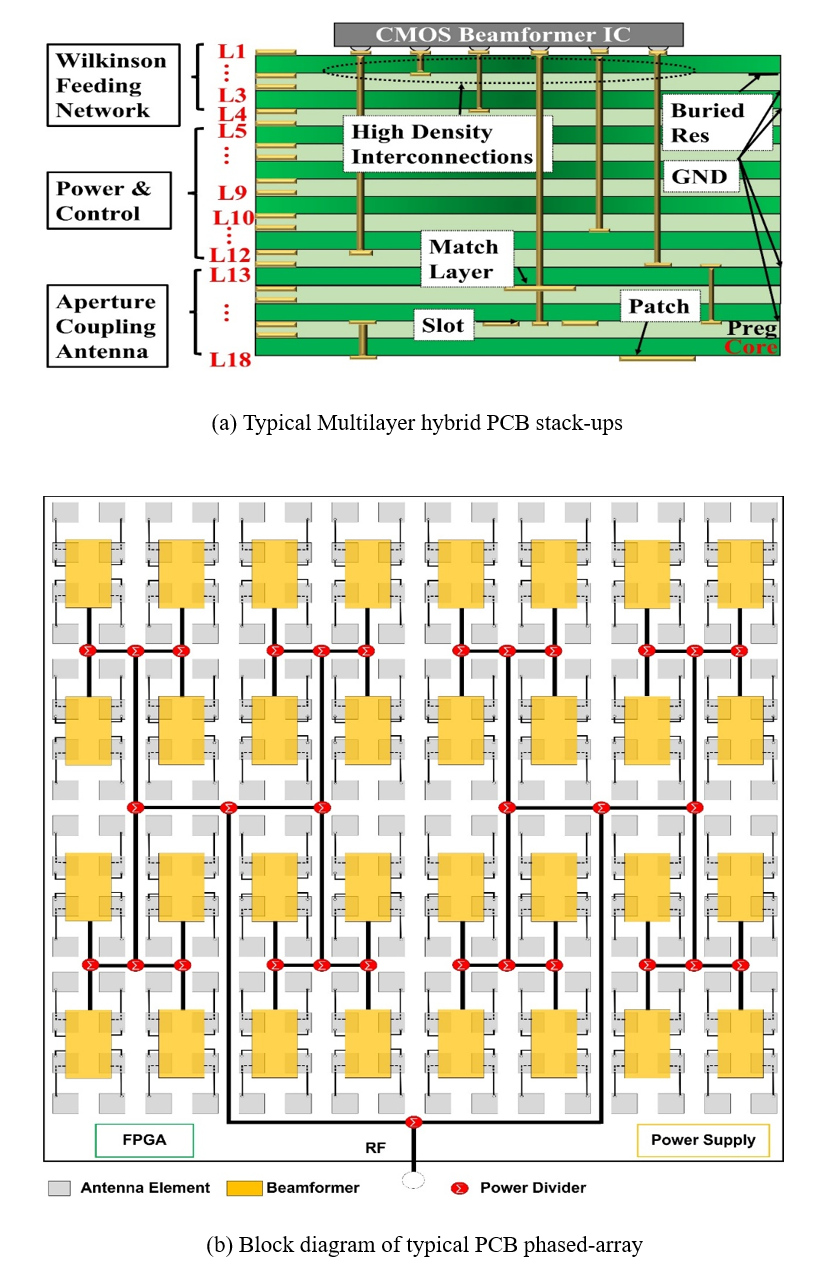}
	\caption{Typical Multilayer hybrid PCB stack-ups and block diagram of typical PCB phased-array.}\label{antennaFig2}
\end{figure}

The development of distributed phased array antennas involves two key considerations.
First, the limited payload capacity of distributed satellite platforms necessitates highly integrated phased array antennas that deliver high output power, maintain low noise levels, and operate with low power consumption.
Second, the demand for cost-effective designs to support the deployment of amounts of distributed phased array antennas across numerous satellites.
The emergence of large-scale phased array antennas started in the late 1970s with the development of phased-array radars for missile early warning systems.
Since then, phased array antenna technology research has been predominantly concentrated on military applications. Although the advantages of phased arrays in communications and radar are well-established, their use in industrial and commercial products has been limited due to the high cost.
Over the years, notable efforts have been dedicated to reducing the cost, weight, and size of phased array systems.
Recent advances in silicon-based CMOS process and PCB technologies, coupled with the increasing demand for millimeter-wave satellite communication applications, have spurred growing interest in developing low-cost, high-efficiency integrated phased array antennas \cite{zhao2021millimeter,zhao2024w,liu202224}.
Si-based CMOS process offers key advantages, including high yield and reliability, which, compared to compound semiconductors, significantly reduce the cost of active channels in phased arrays, making large-scale distributed systems feasible \cite{zhao2021design}.
Furthermore, heterogeneous hybrid PCB technology allows the integration of large-scale microstrip antennas, RF circuits, control circuits, power supply circuits, and phased array chips onto a single PCB at millimeter-wave frequencies, as shown in Figure \ref{antennaFig2}.
This replaces traditional brick-and-tile phased array architectures, leading to enhanced antenna integration while reducing size, weight, and profile, thus enabling satellite-based phased array applications.
The PCB-based approach also streamlines production by eliminating complex micro-assembly steps, increasing throughput, and reducing manufacturing costs \cite{luo2020scalable,fu2023low}. However, Si-based CMOS technology still lags behind compound semiconductors in terms of key parameters, such as breakdown voltage, power density, electron mobility, and thermal conductivity.
As a result, Si-based CMOS devices show inferior performance in RF applications, particularly in the millimeter-wave spectrum, in terms of output power and noise figure. To overcome these limitations and improve the performance of individual channels in the phased array system, hybrid CMOS-GaAs packaging is a promising solution \cite{zhao2022k}.

\subsubsection{Distributed Phased Array Synchronization}
The objective of synchronization in distributed phased arrays is to ensure that signals from multiple nodes reach the target location with precise phase alignment and accurate timing, thereby achieving constructive interference to maximize signal power \cite{mudumbai2007feasibility}.
For received signals, precise alignment of phase and timing across the distributed array is necessary to enable coherent processing.
The techniques used to estimate and adjust these phase and timing discrepancies vary depending on whether the system is in receive-only mode or engaged in transmission.
On one hand, for receive-only configurations, data-driven methods can be used to estimate and correct phase alignment during post-processing, though this may introduce some latency.
On the other hand, coherent distributed transmission is more complex, as it requires real-time alignment of the electrical states of each antenna element in the array. This process includes ensuring frequency synchronization across all nodes, calibrating internal phase delays, and correcting phase and timing differences caused by varying node positions \cite{coleri2002channel}.

\subsubsection{Multi-Beam Beamforming Mechanism}
\begin{figure}[t]
	\centering
	\includegraphics[width=1.0\textwidth]{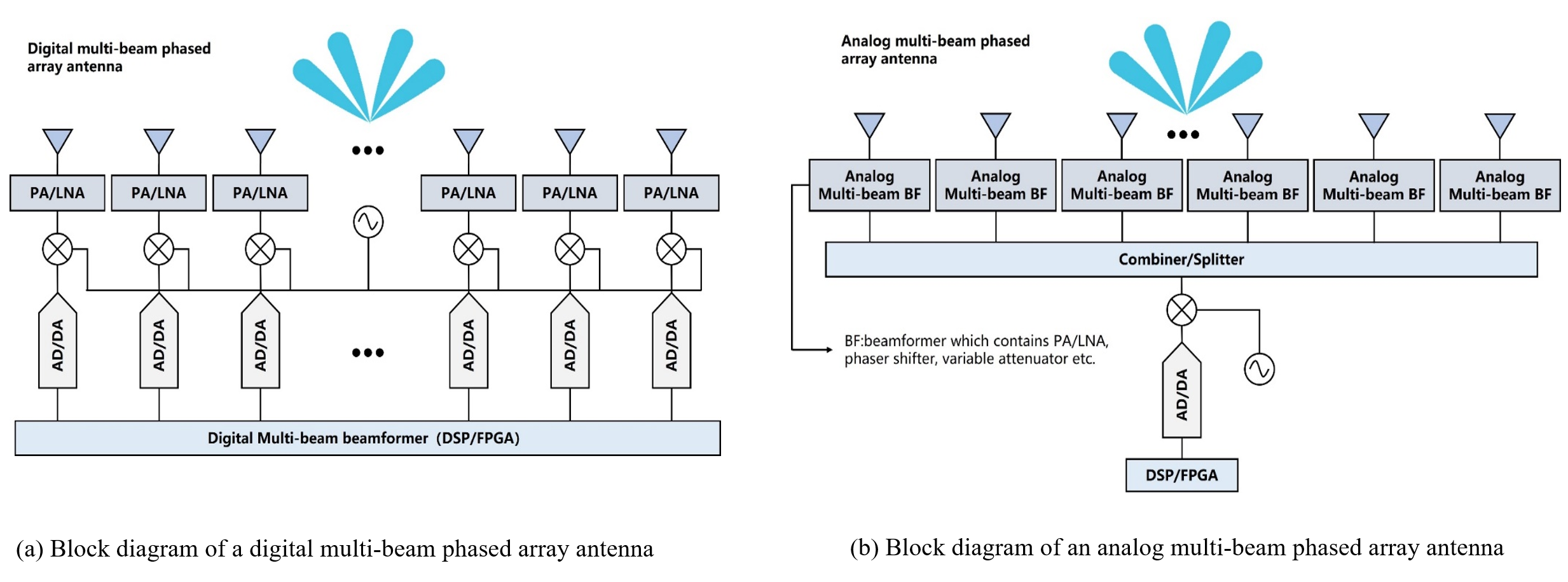}
	\caption{Block diagrams of digital multi-beam phased array antenna and analog multi-beam phased array antenna.}\label{antennaFig3}
\end{figure}

In CCS systems, the multi-beam capability of phased array antennas is crucial, as it enhances system spectral efficiency through beam diversity \cite{yang2024k,yeh2020multibeam}.
As shown in Figure \ref{antennaFig3}, there are two implementation methods for multi-beam phased array systems: digital phased arrays and analog phased arrays \cite{hu2020orthogonal,he2021review}. In a digital phased array, the signal received by each antenna element is amplified, filtered, down-converted, digitized at an intermediate frequency (IF), and digitally down-converted to create a zero-IF digital signal. Each antenna element's zero-IF digital signal is then sent to an array signal processing subsystem, where the desired multi-beams are formed.
However, the digital phased array approach requires down-conversion and digitization of signals for each antenna element.
Given that hundreds of channels may be needed to achieve the necessary antenna gain, this system requires a large amount of equipment, resulting in high power consumption and significant costs \cite{hu2018digital, yu2019multibeam, jang201816}.
In an analog phased array, the signal received by each antenna element is amplified and split into multi-path, and each path is independently weighted in amplitude and phase in the analog domain, with delay compensation and synthesis to ultimately form the multi-beam.
All beams are then down-converted and sent to the baseband processing subsystem \cite{peng2020ka}.
The greatest advantage of the analog phased array lies in its greatly lower cost compared to digital phased arrays, as well as its high level of technical maturity.
When the number of beams is four or fewer, multiple beams can be efficiently realized using standard analog phased array components, offering excellent cost-effectiveness \cite{talisa2016benefits}.
To overcome the hardware limitations of fully digital beamforming, hybrid beamforming can also be utilized.
In hybrid beamforming systems, a large number of antenna elements are connected to a limited number of RF chains through a network of phase shifters.
In certain scenarios, hybrid beamforming can achieve spectral efficiency comparable to that of fully digital systems \cite{sohrabi2016hybrid, wan2021performance}.

\subsection{Channel Coding in DSIN}

Channel coding is essential for ensuring the transmission reliability of our DSIN \cite{3.4-1}.
With the advancements in 5G and the potential applications of 6G, future communication scenarios will become increasingly diverse, presenting both challenges and opportunities for the design of channel encoding and decoding.
For instance, the large- and moderate-length coding schemes with high throughput and reliability are more suitable to the Enhanced Mobile Broadband (eMBB) in 5G, which tolerates higher delays but demands greater channel capacity.
On the other hand, Ultra-Reliable and Low Latency Communications (URLLC) require extremely low latency and higher reliability, making short- and moderate-length coding schemes that excel in low complexity, low delay, and ultra-high reliability more appropriate \cite{3.4-2}.
Table \ref{tab-code} illustrates a comparison between different channel coding and decoding techniques for our DSIN.

\begin{table}[h]\scriptsize
	\centering
	\caption{Channel coding used in air interface of DSIN}
	\label{tab-code}
	\begin{tabular}{m{2cm}|m{2cm}|m{5cm}|m{5cm}}
		%\rowcolor{morelightGreen}
		\hline
		\multicolumn{4}{l}{\textbf{State-of-the-Art Channel Coding Techniques}} \\
		\hline
		\noalign{\vskip 2pt}
		\hline
		\textbf{Code} & \textbf{Decoder} & \textbf{Advantages} & \textbf{Disadvantages} \\
		%\rowcolor{lightGreen}
		\hline
		\vspace{9pt} TBCC codes \vspace{7pt} & Viterbi decoding & Fast encoding and favorable rate performance with very low BLER at short block lengths  & Low BLER requires high-order memory, but the decoding complexity increases exponentially with the memory order \\ [9pt]
		%\rowcolor{morelightGreen}
		\hline
		\vspace{14pt} Polar codes \vspace{11pt} & SC decoding, SCL decoding, CA-SCL decoding & Fast encoding and decoding with simple algorithms, can achieve very low BLER at short block lengths with SCL decoding, no sign of error floor  & The decoding with sequential structure introduce delay, a large list is required to achieve low BLER, the list decoding is not practical for low-power terminals \\ [14pt]
		%\rowcolor{lightGreen}
		\hline
		\vspace{14pt} Turbo codes \vspace{11pt} & Iterative soft decoding, MAP decoding & Fast encoding and decoding with parallel structure, can achieve near-Shannon limit performance at large block lengths & The iterative decoding introduce delay, exists an error floor at the high SNR region, poor performance at moderate and short block lengths \\ [14pt]
		%\rowcolor{morelightGreen}
		\hline
		\vspace{14pt} LDPC codes \vspace{11pt} & BP decoding, MS decoding, IMS decoding & Fast and efficient decoding with parallel structure, excellent BLER performance at large block lengths  & Exists error floor at short block lengths and low code rate, the BP decoder is sub-optimal  at short block lengths, the modulation granularity is not flexible  \\ [14pt]
		%\rowcolor{lightGreen}
		\hline
		\noalign{\vskip 2pt}
		\hline
		\multicolumn{4}{l}{\textbf{Universal decoder design}} \\
		\hline
		\noalign{\vskip 2pt}
		\hline
		\textbf{Encoder} & \textbf{Decoder} & \textbf{Advantages} & \textbf{Disadvantages} \\
		\hline
		\multirow{6}{2cm}{Any linear encoding scheme}
		& \vspace{11pt} GRAND algorithm \vspace{9pt} & can achieve very low BLER at short block lengths and high code rate with acceptable complexity  & The decoding complexity is high at low code rate and moderate block lengths, the BLER performance is sensitive to noise patterns  \\ [13pt]
		%\rowcolor{morelightGreen}
		\cline{2-4}
		& \vspace{18pt} OSD algorithm \vspace{16pt} & can achieve near ML performance at short block lengths without an error floor, exhibits strong practical feasibility & The decoding complexity is increases exponentially with the order especially at moderate and large block lengths, the low-complexity improved methods introduce many adjustable parameters  \\ [18pt]
		%\rowcolor{lightGreen}
		%\rowcolor{lightGreen}
		\hline
	\end{tabular}
\end{table}

\subsubsection{State-of-the-Art Channel Coding}
The challenges and potential coding technologies for various code lengths under the transmission requirements of our DSIN for future 6G networks are summarized as follows:

\textit{1) Short-Length Coding Schemes.}
In the short codes with lengths below 1024 bits, the competitive coding schemes include: the classic algebraic code such as Bose-Chaudhuri-Hocquenghem (BCH) and Reed-Solomon (RS) codes \cite{3.4-3}, the convolutional code \cite{3.4-4} that can achieve efficient encoding and decoding through trellis structures, and the Polar code \cite{3.4-5}, which have been standardized for the control channel in 5G NR.
Currently, various types of classic algebraic codes have been fully optimized, while convolutional codes, represented by Tail-biting Convolutional Code (TBCC), have been replaced by the higher-performance Polar codes in the 5G standard.
Thus, we focus on the development of Polar code, which can achieve near-capacity transmission by applying Successive Cancellation (SC) decoding to polarized sub-channels created through merging and splitting operations on $i.i.d.$ binary memoryless channels, where some sub-channels approach noiselessness and others become fully noisy under complete polarization.

Motivated by the construction of the Polar code, it is evident that its code length is restricted to powers of two, making it challenging to meet the rate-adaptive requirements in various applications.
To address this limitation, a widely adopted solution is to use the original Polar code as a base code, and then adjust its length by removing certain codewords.
Techniques like puncturing \cite{3.4-6} and shortening \cite{3.4-7} are typically employed, with puncturing being more suited for low-rate scenarios and shortening better matching high-rate requirements.
As a result, the 3GPP adopts a rate-adaptive scheme for Polar code that integrates puncturing, shortening, and repetition in the current 5G standard \cite{3.4-8}.

\textit{2) Large- and Moderate-Length Coding Schemes.}
In the large and moderate codes with lengths over 1024 bits, notable coding schemes include Turbo code \cite{3.4-9} and Low-Density Parity-Check (LDPC) code \cite{3.4-10}.
Thanks to the near-Shannon limit decoding performance at large block lengths and high-throughput parallel decoding capabilities, the LDPC code has already been standardized for 5G data channels.
The Turbo code, however, has been replaced by LDPC codes in 5G due to the inherent error floor and relatively high decoding delay in iterative processes.
Moreover, the concatenated BCH and LDPC coding scheme is used in the Digital Video Broadcasting-Satellite-Second Generation (DVB-S2) standard for satellite broadcasting, achieving near-Shannon capacity and excellent spectral efficiency in complex satellite-terrestrial channels.
According to the Consultative Committee for Space Data Systems (CCSDS), the outer code in DVB-S2 employs Quasi-Cyclic LDPC (QC-LDPC) code, which offers reduced computational complexity and effective parallel processing performance \cite{3.4-11, 3.4-12}.

To support the demand for higher data throughput in future networks, conventional serial coding faces high complexity at large block lengths, while parallel coding imposes significant hardware and power constraints.
Consequently, coupled LDPC code \cite{3.4-13} has become a key candidate scheme for handling ultra-long data stream transmission, constructed by introducing coupling constraints across multiple independent LDPC code blocks.
Representative schemes involve Spatially-Coupled LDPC (SC-LDPC) code \cite{3.4-14} and staircase LDPC code \cite{3.4-15}, each offering unique advantages for high-throughput streaming.
SC-LDPC code can be constructed by coupling multiple independent LDPC code blocks into a chain using the matrix expansion-based method \cite{3.4-16} or protograph-based method \cite{3.4-17}.
Staircase LDPC code is a type of product code with a generalized coupling structure.
It introduces additional encoding constraints between adjacent code blocks to ensure that each row and column represents a codeword, which leads to a lower error floor.
However, conventional staircase LDPC code with a fully coupled structure faces the issue of fixed re-encoded length and code rate once the component codes are determined.
Thus, several studies proposed a partially coupled LDPC coding scheme to address this problem, allowing for rate-adaptive code designs with minimal loss in coding gain \cite{3.4-18, 3.4-19}.

\textit{3) Distributed Coding within Multi-Satellite Cooperation.}
The above coding schemes may encounter conflict resolution issues in multi-satellite cooperative transmission within the CCS system.
In such cases, code-domain multiple access is a promising technology in the CCS system, which mainly includes $T$-fold multiple access \cite{3.4-20} and Lattice-Code Multiple Access (LCMA) \cite{3.4-21} schemes.

The $T$-fold multiple access scheme using concatenated codes can reliably recover a certain number of conflicting information with low complexity.
Specifically, when the number of conflicting sources does not exceed $T$, the outer code can correctly decode the messages of each source from the superimposed codewords without errors.
However, the decoding complexity of both the $T$-fold codebook and the outer code increases significantly with code length, especially when the outer code is Polar code rather than LDPC code, rendering it appropriate only for short block length transmissions \cite{3.4-22}.

In contrast, the lattice code maps messages onto lattice points with power constraints and achieves rate limits approaching $\log(1+\text{SNR})$ through the design of nested lattice structures \cite{3.4-23}.
Therefore, the LCMA scheme no longer distinguishes signals from interference.
Instead, it approaches the limits of multiple access capacity and multiplexing rate by exploring the optimal mapping between multi-user aggregated signal structures and lattice structures, combined with the advantages of fast parallel processing and low-complexity single-user decoding, which also enables it to handle a wide range of block lengths.
Nevertheless, the spreading sequences of lattice code cannot adapt to time-varying channels in real-time, and the modulation schemes must match the encoding alphabet of lattice code \cite{3.4-24}, which both limit its performance in diverse variable-length data services under complex channel constraints within our DSIN.

\subsubsection{Decoder Design}
The decoder can be classified into two types: the dedicated decoder designed for a specific coding scheme, and the universal decoder that can adapt to any linear coding scheme.

\textit{1) Dedicated Decoder.}
Polar codes can be efficiently decoded using the SC algorithm, but its performance may degrade due to incomplete channel polarization at short and moderate block lengths.
In response, the SC List (SCL) algorithm \cite{3.4-25} is proposed to enhance decoding performance, which can achieve a Block Error Rate (BLER) lower than $10^4$ with a list size greater than 32 for 5G Polar code.
Further enhancement in the SCL decoding performance can be obtained by integrating Cyclic Redundancy Check (CRC) bits, but at the cost of increased decoding complexity \cite{3.4-26}.
Moreover, to address the extra delay introduced by the serial SC decoding, several works suggest parallel processing of certain special nodes in SC decoding to reduce decoding delay, such as Belief Propagation (BP)-based parallel decoding for Polar code \cite{3.4-27}.

The BP decoding algorithm commonly used for LDPC code involves nonlinear functions in the computation of check nodes, resulting in high implementation complexity.
Therefore, to meet the high-speed transmission requirements of 5G LDPC codes, several simplified alternative algorithms have been proposed, such as the widely used Min-Sum (MS) algorithm \cite{3.4-28} and its various improvement schemes (collectively referred to as IMS) \cite{3.4-29}, which also come with a considerable performance loss, particularly in scenarios involving short and moderate block lengths and low code rate.
Further reduction in the complexity of the LDPC decoder while ensuring decoding performance is still necessary to make it a realistic solution for the DSIN with Ubiquitous diversified services.

\textit{2) Universal Decoder.}
The Maximum Likelihood (ML) algorithm offers excellent BLER performance, but its huge decoding complexity makes it impractical for satellite platforms with limited payload.
Similarly, most of the near-ML performance universal decoding algorithms, such as Guessing Random Additive Noise Decoding (GRAND) \cite{3.4-30} and Ordered Statistics Decoding (OSD) \cite{3.4-31}, also face these challenges.
More precisely, while GRAND supports high-rate linear block codes effectively, it incurs very high decoding complexity for codes with moderate to low rates.
Leveraging the assumption that errors are most likely to occur at the least reliable positions of received symbols, OSD identifies the Most Reliable Basis (MRB) and flips a small number of bits within the MRB to regenerate a list of candidate codewords, from which it selects the best one as the output.

Researchers have recently proposed several methods to reduce their complexity.
In the case of OSD algorithm, its decoding complexity mainly stems from the re-encoding and Gaussian Elimination (GE) steps, which make its improvements broadly focus on two approaches: one involves avoiding the execution of GE in the OSD process \cite{3.4-32}, while the other introduces stopping or skipping criteria to reduce the number of Test Error Patterns (TEP) evaluated during the search for the optimal candidate codewords \cite{3.4-33}.
There are still however challenges that require resolution.
First, the above improved OSD algorithms help reduce decoding complexity in several ways, but they also require extensive parameter tuning, which complicates the process of achieving optimal BLER performance.
Second, most OSD-based algorithms operate in iterations and lack effective methods for parallel decoding.
For these issues, several studies suggest that the complexity can be further reduced through techniques such as segmentation-discarding method \cite{3.4-34}, effective tree-based search algorithms \cite{3.4-35}, and cascaded BP decoding \cite{3.4-36}.

\textit{3) Joint Decoding of Parallel Multi-Stream Reception.}
Existing code-domain multi-satellite cooperative transmission technologies, including the MPA decoder for Sparse Code Multiple Access (SCMA) scheme \cite{3.4-37}, the BP-based $q$-ary decoder in LCMA scheme \cite{3.4-38}, and the Joint Decoder (JD) used in $T$-fold multiple access scheme \cite{3.4-39}, all suffer from high decoding complexity.
For instance, the decoding complexity of JD is approximately $O(mT\log^2T\log\log T)$, where $m$ represents the message length.
Moreover, since the signal quality is heavily influenced by the complicated and volatile satellite-terrestrial channels, further research on distributed decoding schemes that adapt to such communication conditions and achieve high-reliability decoding of superimposed coded signals in DSIN, especially in CCS system is necessary.

%\subsection{Satellite-ground many-to-many authorization-free random access}
\subsection{Grant-Free Random Access in Satellite-Terrestrial Networks}
\begin{figure}[!h]
		\centering
		\includegraphics[width=0.8\textwidth]{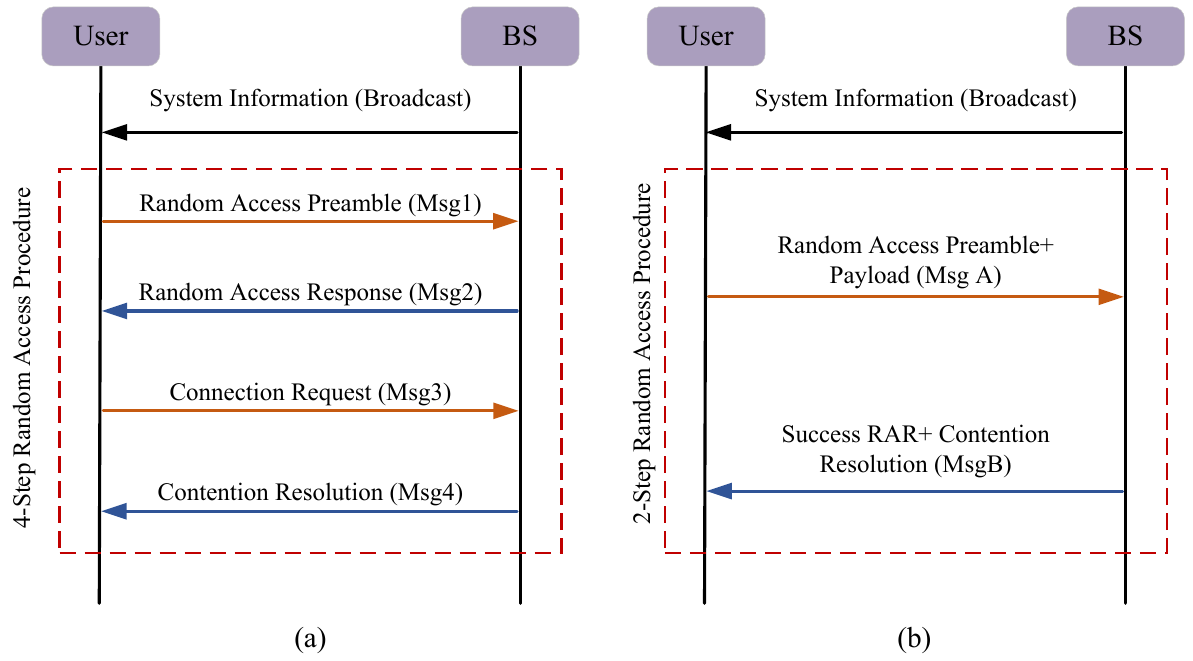}
		\caption{(a) The four-step random access procedure in 4G LTE/5G NR; (b) The two-step random access procedure introduced in 5G NR.}
		\label{GFRA}
	\end{figure}

Random access and multiple access technologies are essential to establish communication links in satellite-terrestrial networks \cite{KeIoTJ}.
The traditional four-step random access procedure is depicted in Figure \ref{GFRA}(a).
Initially, a user terminal randomly chooses a preamble from a set of sequences and transmits it as Msg1 over the physical random access channel to the BS.
Upon detecting the transmitted preamble, the BS replies with a random access response (RAR), i.e., Msg2, which includes the random access preamble identifier (RAPID), a cell-radio network temporary identifier (C-RNTI), and the timing advance (TA) information.
Once the user receives Msg2 within a specified time window, it sends Msg3 to the BS, containing a connection request and terminal identifier for contention resolution. If multiple users employ the same preamble, the BS may not be able to decode Msg3 correctly.
If successful, the BS transmits a contention resolution message, Msg4, which embeds the user's identifier.
The terminal confirms successful access to the BS if it can correctly parse its identifier; otherwise, a new access schedule is attempted.

To decrease access delay, 3GPP Release 16 introduced a two-step random access process, illustrated in Figure \ref{GFRA}(b).
This approach merges the preamble sequence from Msg1 with the payload data from Msg3 into MsgA and merges the RAR from Msg2 with the contention resolution information from Msg4 into MsgB, reducing the interactions between the user and the BS.
Nevertheless, both the four-step and two-step access methods depend on signaling interactions between the BS and terminals, classifying them as granted-based access. Besides, the existing 3GPP framework primarily uses orthogonal preamble sequences, which inevitably result in sequence collisions as the number of access terminals increases, reducing the probability of successful access.
	
In contrast to terrestrial cellular communication, terrestrial-satellite communication (TSC) typically experiences high propagation delays.
Grant-free random access permits users to transmit pilot and data payload directly without BS coordination, thus satisfying fast access requirements in high-delay scenarios. Moreover, non-orthogonal pilot sequence allocation can reduce the pilot collision probability, thereby enhancing multi-user capacity within limited time-frequency resources.
Therefore, grant-free NOMA presents an effective strategy for TSC.
	
In the context of grant-free access, the wide geographic coverage of satellite-based BSs leads to significant differential delays among users within the same service area, further complicated by high Doppler frequency shifts associated with relative movement.
Consequently, time-frequency synchronization and pre-compensation are essential before transmission to ensure.
Unlike grant-based access where the BS calculates and sends TAs to users based on preamble sequences, grant-free access permits multiple users to independently perform timing advances and frequency offset pre-compensation by utilizing broadcast signal from the BS, which enhances efficiency.
When terrestrial terminals have precise ephemeris information and positioning capabilities, they can compute the relative distance and velocity between the user and the satellite, facilitating time-frequency offset pre-compensation for synchronization.
If user terminals lack positioning capabilities or cannot access ephemeris data, the method leveraging downlink synchronization signal block for time-frequency offset estimation, as proposed in \cite{21Location}, can convert differential time-frequency estimates into the locations of either user terminals or satellites to achieve synchronization.

In NOMA scenarios, compressive sensing-based techniques for active user detection and channel estimation can effectively leverage user activity sparsity to reduce the required pilot length.
To ensure optimal sparse recovery performance, designing non-orthogonal pilot sets with low correlation is crucial for minimizing user interference.
A method for constructing pilot sets based on Golay sequences is proposed in \cite{21Golay}, which successfully minimizes pilot correlation and the Peak-to-Average Power Ratio (PAPR) of spreading codes in multicarrier transmission systems.
This approach provides lower PAPR while maintaining robust sparse recovery performance compared to random binary, Gaussian, pseudo-random, and Zadoff-Chu (ZC) sequences. Authors in \cite{24mask} present a deterministic scheme for non-orthogonal pilot construction using a discrete Fourier transform (DFT) matrix with a mask.
This method can generate $\mathcal{O}(L^3)$ non-orthogonal sequences within a length-$L$ sequence and offers theoretical guarantees on inter-sequence correlation bounded by $\mathcal{O}(\frac{1}{\sqrt{L}})$.
This enlarges the pilot set size and outperforms Gaussian random pilots by reducing collision probability.
In \cite{23diff}, a composite preamble construction method is proposed.
By combining orthogonal ZC sequences with multiple ZC root sequences with different phase rotations, this approach extends the pilot set, thereby differentiating users selecting the same orthogonal pilot.
{An iteratively constructed pilot sequence is proposed in \cite{10108023} using conjugate gradient descent and space projection, subject to the constraint on the PAPR. The pilot sequence proposed in \cite{10108023} can achieve up to 47\% and 28\% lower coherence compared with the binary Golay sequence and the traditional ZC sequence, respectively. Moreover, a Low-Correlation-Zone Periodic Sequence (LPS) is designed for the superimposed pilot structure in \cite{10246293}. The average auto-correlation and cross-correlation of the LPS is $\mathcal{O}(1/L)$, which is $1/\sqrt{L}$ times lower than that of the multi-root ZC sequence, and results in a reduced decoding failure probability in satellite-based massive access systems.
It is worth noting that the expansion of the non-orthogonal pilot sequence set not only alleviates access conflicts but can also be applied to pilot index modulation \cite{9686735} and superimposed transmission \cite{10246293}, thus enhancing spectral efficiency.}

Regarding active user detection and channel estimation for LEO satellite uplink access, authors in \cite{20IoTJ} address satellite-terrestrial link channel modeling, incorporating phase shifts and channel impairments.
A Bernoulli-Rayleigh message-passing algorithm is proposed, which leverages the sparse activity of ground users to reduce pilot overhead.
Acknowledging the wideband transmission needs of terminal devices and the high-delay, fast-varying characteristics of satellite-terrestrial links, \cite{zCE} introduces an OFDM symbol repetition mechanism.
This approach addresses residual time and frequency offsets in LEO satellites, proposing an enhanced variance state propagation algorithm for joint active user detection and channel estimation.
To accommodate rapidly changing satellite-terrestrial channels, \cite{23ZXY_TWC} applies OTFS modulation for massive multi-user access, delineating a two-stage process for joint active user detection and channel estimation followed by data detection.
{However, previous studies have primarily focused on Global Navigation Satellite System (GNSS)-based solutions, which are unsuitable for power-limited Time-Sensitive Communications (TSC). In response to this, \cite{10839280} proposes a joint design for device identification, channel estimation, and symbol detection in LEO satellite-enabled GFRA systems, without relying on GNSS assistance. This approach uses OTFS modulation with a message-passing-based algorithm to handle large differential delays and Doppler shifts at the satellite receiver.}
Additionally, algorithms based on block coordinate descent, tensor Bayesian learning, and approximate message passing are also extensively employed for compressed sparse recovery problems and can be effectively utilized in uplink signal estimation and detection for TSC.

To meet the extensive data transmission demands of terrestrial users, implementing NOMA for overload transmission is an effective strategy to enhance multi-user capacity.
According to \cite{24LDS}, a code-domain NOMA scheme based on low-density signatures (LDS) is designed.
This scheme exploits the structured sparsity of transmitted signals and user activity to develop a Gaussian-approximated message-passing-aided sparse Bayesian learning algorithm, achieving superior bit error rate and false alarm rate performance compared to existing approximate message-passing algorithms.
In \cite{24AFDM}, a SCMA scheme using affine frequency division multiplexing is proposed.
This approach modulates sparse codewords onto chirp subcarriers and discusses in detail the design of sparse codes, chirp rate selection, and receiver algorithms.
The method demonstrates significantly better performance than OFDM-based SCMA in high mobility scenarios, underscoring its potential application in multi-user access scenarios for LEO satellites.

    \begin{figure}[!h]
    	\centering
    	\includegraphics[width=0.7\textwidth]{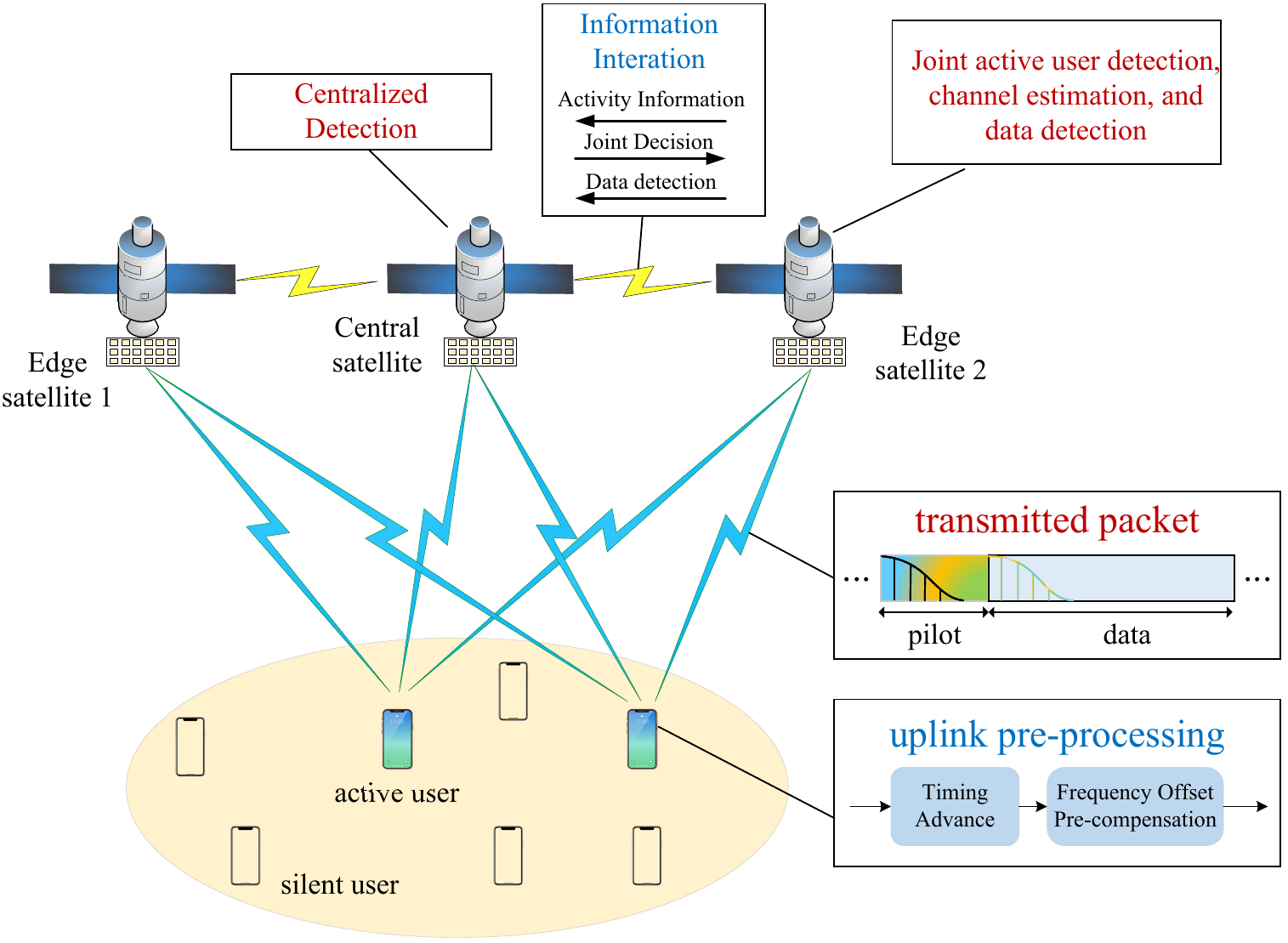}
    	\caption{Diagram of multi-satellite cooperative grant-free random access.}
    	\label{MultiSatGFRA}
    \end{figure}

{Another research branch of GRRA focuses on improving access timeliness. It is worth noting that large-scale user device access conflicts, along with access failures and the occurrence of multiple retransmissions, can lead to the accumulation of long queue buffers at IoT devices, significantly degrading the timeliness of information \cite{10492466}, which can be quantified using the Age of Information (AoI) metric \cite{3.7-27}. In \cite{9766119}, the instantaneous AoI evolution of each device is traced through Markov analysis to derive the expression for the average AoI of the access system. A grant-free age-optimal random access protocol is then proposed to dynamically adjust the number of access slots, optimizing the system's average AoI while enhancing the maximum throughput.
Furthermore, considering that satellite networks need to accommodate service devices with diverse timeliness requirements, \cite{10412105} proposes the unequal timeliness protection massive access scheme for mission critical communications in S-IoT. This scheme aims to minimize the average AoI for user groups with different timeliness demands, ensuring reliable access under their respective reliability constraints.}

In multi-user TSC, a constellation comprising multiple satellites extends coverage and design flexibility beyond a single satellite system.
This configuration offers spatial diversity and multiplexing gains owing to multiple observations in the spatial domain.
As depicted in Figure \ref{MultiSatGFRA}, in complex scenarios where multiple users simultaneously access multiple satellites, precise multi-channel estimation strategies, and multi-user detection mechanisms with multi-satellite collaborative data demodulation strategies are vital for achieving reliable massive access.
Authors in \cite{19MSC_RA} propose a novel multi-satellite collaborative user detection scheme for traditional slotted ALOHA-like random access methods.
By modeling the communications between multiple users and multiple satellites as an equivalent distributed multiple-input multiple-output (MIMO) system, the proposed multi-satellite cooperative MIMO detection algorithm significantly enhances the success detection probability of collided data packets when compared to the single-satellite scheme.
The authors in \cite{KeJSAC} introduce an advanced multi-satellite cooperative scheme for active user detection and data detection.
By capitalizing on the observation diversity gain provided by multiple satellites, this scheme substantially improves the accuracy of active user detection and effectively mitigates the performance degradation in data detection caused by spatial channel correlation among active users.
In addition, \cite{3.2-42} proposes a MIMO-OTFS modulation-based multi-satellite collaborative random access scheme.
Initially, it derives users' initial channel estimates at each satellite through a message-passing algorithm and subsequently aggregates feedback information from multiple edge satellites to achieve enhanced activity detection, and data detection outcomes.
Moreover, the proposed approximate expectation propagation method can be executed either centrally by a single satellite or distributed across multiple satellites. Notably, the distributed algorithm requires only two rounds of information exchange to achieve performance levels comparable to those of centralized processing.

\subsection{NOMA/RSMA Enabled Multi-Satellite Multicast}
%%%%%%%%%%%%ÕâÀïÐèÒª¼ÓÉÏÊ±Ð§NOMA/RSMAÏà¹ØÑÐ¾¿
\subsubsection{NOMA Enabled Multi-Satellite Multicast}
\begin{figure}[t]
	\centering
	\includegraphics[width=1.0\textwidth]{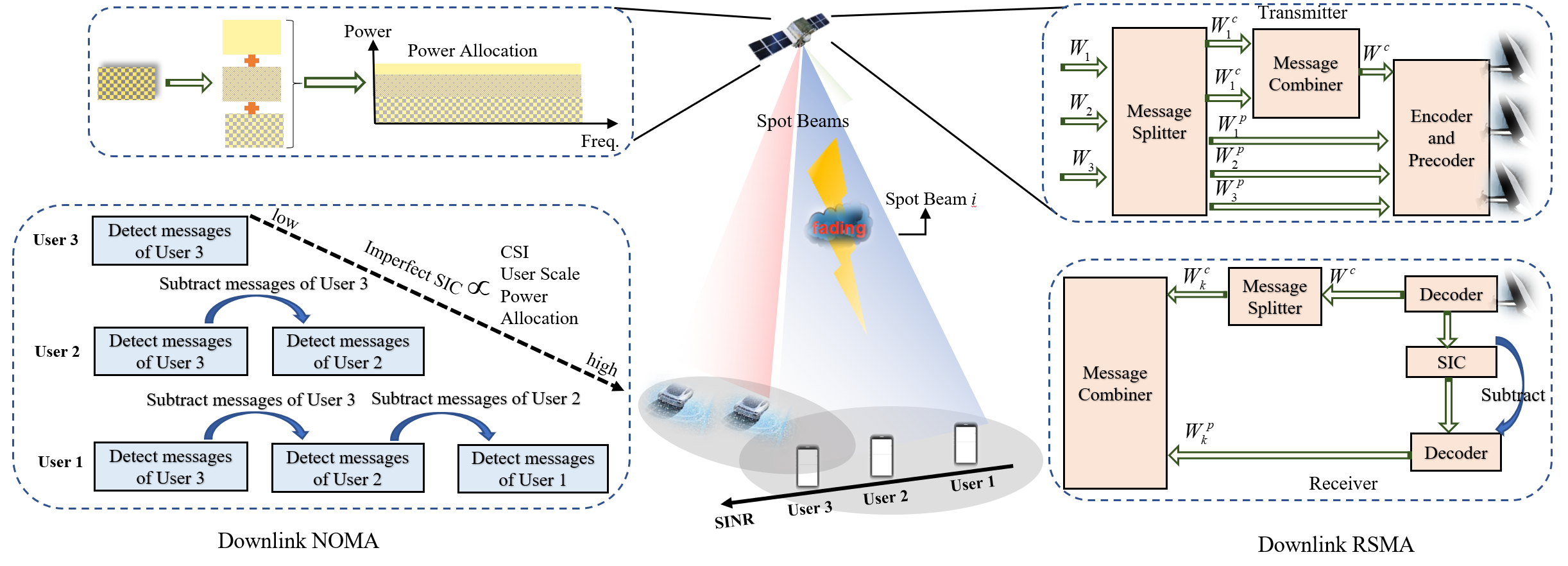}
	\caption{An illustration of downlink NOMA and RSMA enabled satellite communication system.}\label{NOMA_RSMA}
\end{figure}

The rapid evolution of mobile networks towards 5G and beyond has driven the need for highly efficient multiple-access techniques capable of handling the ever-increasing demand for wireless connectivity. 
Among these techniques, the NOMA stands out as a promising solution due to its ability to support multiple users on the same frequency and time resources by exploiting power domain multiplexing \cite{7973146}.
Unlike traditional orthogonal access schemes such as TDMA and FDMA, NOMA enables simultaneous access for multiple users, making it particularly well-suited for scenarios where resources are scarce, such as SatCom.
As NOMA evolves, the integration of NOMA into SatCom systems, particularly in downlink multicast communication of CCS system, presents a significant opportunity to enhance service quality, improve spectral efficiency, and support a larger number of users \cite{8957684}. As shown in Figure \ref{NOMA_RSMA}, by enabling superposition transmission, the NOMA operates by allowing multiple users to share the same time-frequency resources, with user separation achieved through the power domain, thus enhancing the overall system capacity and the overall spectral efficiency of satellite networks. This is particularly beneficial for CCS system, which often serve a wide geographical area with diverse user demands.

Research on NOMA in satellite multicasting has seen significant progress. 
Early works focus on the theoretical performance of NOMA in satellite channels, demonstrating its superiority over orthogonal multiple access (OMA) in terms of user capacity \cite{8357810}, spectral efficiency \cite{7906532} and user fairness \cite{8957684}. 
Subsequent studies delve into the practical aspects of implementing NOMA in satellite systems, including the design of precoding techniques to mitigate inter-beam interference \cite{9115278}, the development of user scheduling algorithms to optimize resource allocation \cite{7560605}, the improvement of information freshness to support time-critical services \cite{10336741}.
The application of NOMA in satellite multicasting has also been explored in the context of high-throughput satellite (HTS) systems, where the combination of NOMA with advanced beamforming techniques has been shown to significantly improve the throughput performance \cite{kodheli2020satellite}. Moreover, the potential of NOMA in cognitive radio-inspired satellite networks has been investigated, where NOMA principles are applied to enhance spectrum sharing between primary and secondary users \cite{perez2019non}.
Recent research has also addressed the challenges of limited feedback and imperfect CSI in NOMA-enabled SatCom systems, proposing robust designs to ensure reliable communication in the presence of channel uncertainties \cite{chu2021robust}. Additionally, the application of NOMA in multi-beam satellite systems has been explored, aiming to improve the sum capacity throughput by using the joint beamforming and power allocation design \cite{lin2019joint}. Other studies have examined the integration of NOMA with other advanced satellite technologies, such as cognitive radio \cite{Liu2021Spectrum} and DRL \cite{jiao2023age}. Cognitive radio enables dynamic spectrum sharing between satellites and terrestrial networks, while DRL can be used to optimize resource allocation in NOMA-enabled CCS system.
Over time, the focus shifted towards more complex satellite architectures, such as LEO constellations and CCS system. In particular, multi-satellite systems, where multiple satellites collaborate to provide downlink services to a common set of users, have become a key area for NOMA deployment. The transition to multi-satellite systems necessitates new approaches to address inter-satellite interference, and optimize decentralized resource allocation \cite{Zhao2024Dynamic} under the constraint of asynchronous cooperative communication \cite{Zhao2023Multi}.

While NOMA has shown great potential in enhancing downlink multicast communication for DSIN, several challenges remain. One of the key challenges is managing the complexity of resource allocation in DSIN. As the number of satellites in a CCS system increases, the coordination of power allocation, beam hopping, and user scheduling becomes more complex. It is necessary to develop more efficient algorithms for managing these resources in large-scale NOMA-enabled DSIN. 
{{A DL-based MIMO-NOMA framework is proposed in \cite{9093213} to maximize the sum data rate and energy efficiency. As an evolution of NOMA, hybrid NOMA offers superior compatibility, enabling effective coexistence between OMA and NOMA. The optimality of downlink hybrid NOMA in the two-user case has been invalidated in \cite{10824682}. Furthermore, a scalable hybrid NOMA scheme is proposed in \cite{10283937} to jointly control power allocation and bandwidth overlap in the presence of both imperfect CSI and SIC, thereby improving the sum rate in a spectrally efficient manner.
Additionally, leveraging the spectral efficiency advantages of NOMA, Reconfigurable Intelligent Surfaces (RIS)-aided NOMA systems have garnered significant attention for maximizing the communication success ratio, while meeting the diverse QoS requirements of users and the dynamic energy constraints at the RIS \cite{10185599}. In particular, the RIS-aided NOMA scheme can significantly enhance the sum rate under finite-block length coding by jointly optimising user power allocation and RIS beamforming \cite{10103152}.}}
Moreover, most of the existing PHY signal processing methods are limited to the study of two-user NOMA systems with limited satellites. Extending those solutions to large-scale networks with multiple satellites in a cluster or a swarm is of great importance.
Another challenge is the management of inter-satellite interference. In CCS systems, satellites often operate in close proximity, leading to increased interference. Techniques such as multi-agent deep reinforcement learning (MADRL) enabled cooperative NOMA and beamforming can be seen as a good avenue for future research to mitigate the interference issue. In addition, as a common consumption in existing research works, perfect SIC processing may lead to overestimating the performance of NOMA-enabled CCS system, which can be addressed in future research.

\subsubsection{RSMA Enabled Multi-Satellite Multicast}
The integration of rate-splitting multiple access (RSMA) with satellite communication systems has emerged as a promising approach to enhance spectral efficiency and energy efficiency, particularly in multi-satellite downlink broadcasting CCS system. The RSMA is a novel NOMA scheme that has been gaining traction in the SatCom domain. Unlike traditional NOMA, which focuses on power domain multiplexing, the RSMA splits the messages into common and private parts, allowing for more flexible interference management as shown in Figure \ref{NOMA_RSMA}. This approach has been shown to outperform conventional schemes in terms of spectral efficiency and energy efficiency, especially in multi-beam, multi-group multicast scenarios.

The primary distinction between the RSMA and the aforementioned NOMA lies in the message splitting strategy. The RSMA and NOMA both aim to maximize spectrum efficiency but differ fundamentally in their approaches. While the NOMA relies on successive interference cancellation (SIC) to decode signals from multiple users, the RSMA treats part of the interference as noise and decodes another part via SIC, which results in greater flexibility and robustness.
This ability to combine noise cancellation and SIC allows RSMA to outperform NOMA, especially in scenarios involving imperfect CSI or complex satellite networks where multiple downlink beams may overlap. The flexibility of RSMA in handling various levels of interference makes it superior to NOMA for CCS system, particularly in high-interference regimes where NOMA's performance degrades due to the complexity of interference management across multiple beams. An exhaustive survey of the RSMA combined terrestrial communication literature from both the information-theoretic and communication-theoretic perspectives is presented by \cite{mao2022rate}.

Inspired by the advantages of RSMA-aided multigroup multicasting in terrestrial networks, the deployment of RSMA in the realm of DSIN is intriguing and shows great potential.
Specifically, in multi-satellite downlink broadcasting, the RSMA enables managing inter-beam and inter-satellite interference effectively, which is a significant advantage over the NOMA.
Several key technologies make RSMA a feasible option for CCS system: \textit{1) Massive MIMO}: The RSMA leverages multi-antenna systems to deliver high spectral efficiency and robust communication links \cite{mishra2021rate}. This is particularly useful in DSIN, where beamforming and spatial multiplexing are critical. The practical applicability of the RSMA in downlink multi-antenna communications is demonstrated by utilizing the physical layer design and link-level simulations \cite{9217326}.
\textit{2) Intelligent Reflecting Surfaces (IRS)}: The RSMA can be combined with IRS to dynamically control signal propagation, improving coverage and reducing interference in DSIN. As a result, the capacity of SatCom can be improved significantly by integrating the RSMA and IRS, as explored by \cite{de2022rate}.
\textit{3) Precoding and Beamforming}: The RSMA requires sophisticated precoding techniques to optimize the transmission of common and private streams across satellite networks. This ensures minimal interference between beams and users, enhancing the overall system performance. The statistical beamforming and common stream separation based on RSMA are incorporated to maximize the minimum user rate under the total power budget of the transmitter \cite{9420034}, which demonstrates RSMA's robustness against the inaccuracy of CSIT and its superiority over SDMA in terms of explicit max-min fair rate gain.
In addition, some works have also begun to integrate RSMA with AI-driven optimization algorithms to manage complex satellite constellations dynamically, opening up new possibilities for automated satellite network management and enhanced performance under variable conditions \cite{zhuo2023ris}. Joint optimization of resource allocation and power control for RSMA-based LEO satellite-terrestrial networks is investigated in \cite{Huang2024} by employing DRL to maximize energy efficiency. Further, the concept of distributed RSMA has opened new avenues for research in DSIN.
To enhance spectral efficiency and manage interference in complex satellite networks, \cite{xu2024distributed} introduced a distributed RSMA approach for multi-layer satellite systems.
More recently, the RSMA-enabled multi-satellite multi-beam communication systems have been investigated. \cite{8385504} addresses the enhancement of sum-rate in multi-beam satellite systems by employing rate-splitting precoding, enabling simultaneous transmission of public and private frames, and proposes a low-complexity design that demonstrates significant performance gains.
The RSMA is also proven to be promising for addressing practical challenges such as feeder link interference and imperfect CSIT in multigateway multibeam satellite systems \cite{9684855}.

Despite its promising capabilities, several challenges remain for the deployment of RSMA in DSIN: \textit{1) Complexity of SIC in Large Networks}: While the RSMA simplifies SIC compared to the aforementioned NOMA technology, the complexity can still increase in large-scale satellite constellations, requiring further research into low-complexity decoding algorithms. \textit{2) Interference Management}: Managing interference across a CCS system with overlapping beams remains a significant challenge. Future work should focus on optimizing RSMA for highly dynamic satellite networks where user mobility and environmental factors can create varying interference patterns. \textit{3) Integration with Emerging Technologies}: The integration of RSMA with emerging technologies such as AI, removable antennas, and ultra-massive MIMO presents both opportunities and challenges. Research is needed to ensure these technologies can work together seamlessly in large DSIN.

%%%%%%%%%%%%%%%%%%%%%%%
%[0] RSMA-Enabled Aerial RIS-aided MU-MIMO System for Improved Spectral-Efficient URLLC
%[1] Rate-Splitting Multiple Access for Multi-Antenna Broadcast Channels with Statistical CSIT
%[2] Distributed Rate-Splitting Multiple Access for Multilayer Satellite Communications
%[3] Energy Efficiency of Rate-Splitting Multiple Access, and Performance Benefits over SDMA and NOMA
%[4] Rate-Splitting Multiple Access for Downlink Multi-Antenna Communications: Physical Layer Design and Link-level Simulations
%[5] Performance Analysis for RSMA-Empowered STAR-RIS-Aided Downlink Communications
%[6] Two-Step Adaptive Grouping Access Based on RSMA for Multibeam Satellite System

\subsection{Erasure Transfer Protocol in DSIN}
%intermittent spatial and temporal connectivity,
In contrast to relatively stable terrestrial networks, the dynamic formation configurations, non-trivial round-trip propagation delays, and significant long-distance transmission losses in the DSIN substantially impact and constrain the performance of conventional terrestrial communication protocols, such as the widely used Transmission Control Protocol/Internet Protocol (TCP/IP), which may misinterpret the packet loss under poor CSI as an indication of network congestions \cite{3.7-1, 3.7-2}.
To tackle the issues that TCP/IP encounters in space communications, various solutions have been proposed at both the transport layer and other layers.
The most widely adopted schemes are the CCSDS protocol suite represented by the Space Communication Protocol Standard (SCPS) \cite{3.7-3}, and Delay/Interrupt Tolerant Networking (DTN) protocol suite, represented by the Bundle protocol and Licklider Transmission Protocol (LTP) \cite{3.7-4}.
Currently, the SCPS protocol has been downgraded to historical status, while the LTP protocol, benefiting from its no-handshake operation and delayed acknowledgement-based retransmissions, is widely applied in the satellite-terrestrial and ISLs characterized by long-distance and high-interruption probability, and has been adopted within the CCSDS protocol suite \cite{3.7-5}.

\begin{figure}[t]
	\centering
	\includegraphics[width=0.9\textwidth]{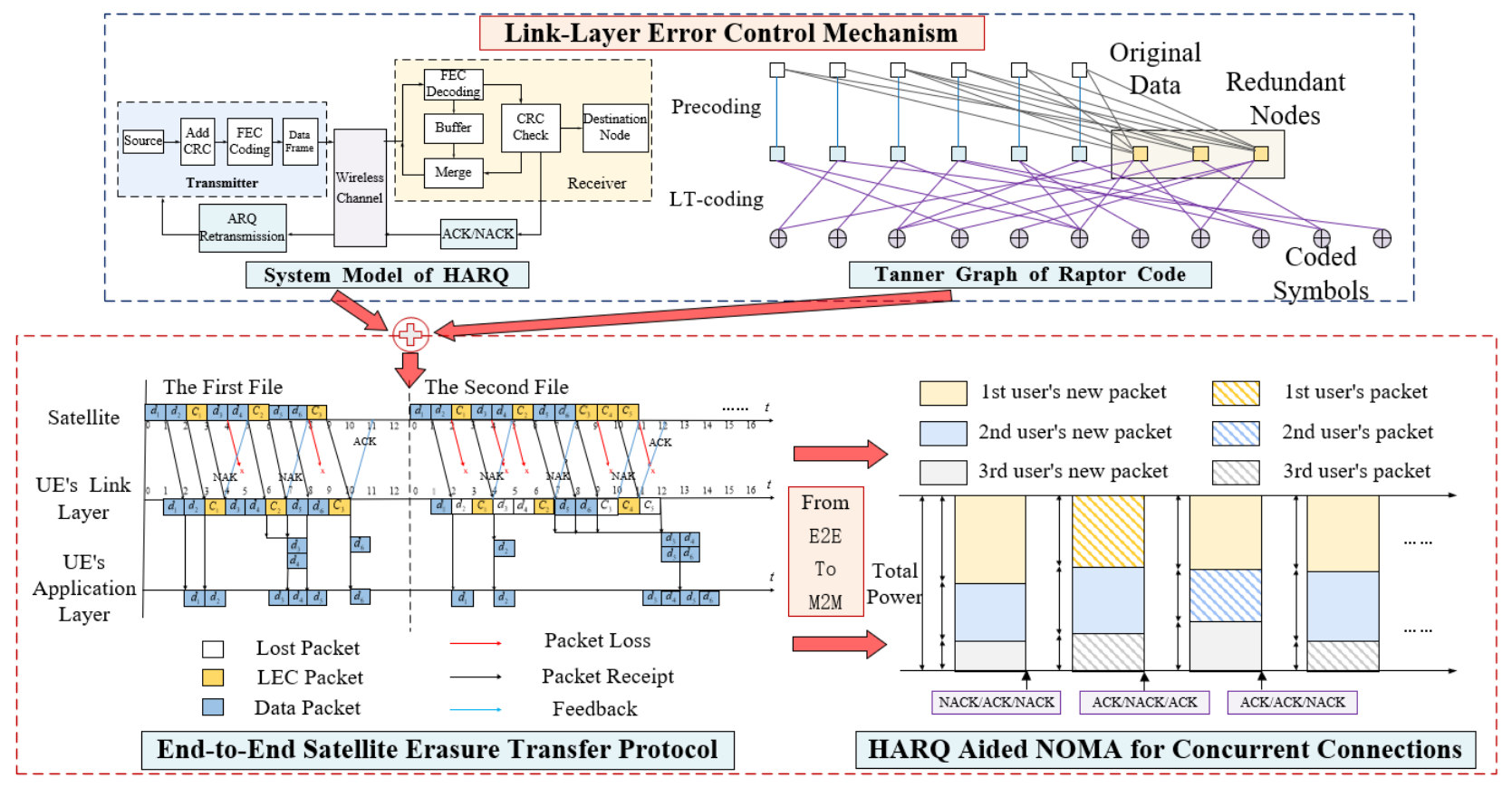}
	\caption{A overview of the introduced satellite erasure transfer protocols.}\label{Erasure}
\end{figure}

Achieving freshness and reliable information delivery required by mission-critical applications with the LTP protocol has long been a core area of research in SatCom, where reliability serves as a critical prerequisite for ensuring timely delivery.
However, the Physical Layer (PHY) Forward Error Correction (FEC)
approaching the Shannon limit still cannot fully ensure reliable packet-level transmission over satellite-to-terrestrial channels.
Indeed, to maintain a low BER and even BLER level, these FEC schemes typically require a lower coding rate, which results in limited throughput and compromised timeliness \cite{3.7-6, 3.7-7}.
Consequently, link-layer error control methods, particularly retransmission and redundancy mechanisms, have been extensively explored as alternative solutions to achieve cost-effective error-free reception. The key techniques of these two mechanisms and their application in satellite transfer protocols are illustrated in figure \ref{Erasure}.

\subsubsection{Current Link-Layer Error Control Techniques}
\hspace{1em} \textit{1) Low-Delay Retransmission Mechanism.}
The classic retransmission mechanism of LTP protocol, Automatic Repeat Request (ARQ), enables the retransmission of uncorrectable frames through feedback from the receiver to the transmitter, but multiple retransmissions can significantly amplify the high-delay effects of long-distance transmissions \cite{3.7-8}.
A more advanced error control method is Hybrid ARQ (HARQ), which combines ARQ with FEC to improve reliability by enabling both frame correction and retransmission of uncorrectable frames.
In contrast to Type I HARQ, which discards the uncorrectable received frames and performs memoryless decoding for each single encoded frame at the receiver, Type II HARQ employs frame combining to improve the utilization of transmitted frames and enhance detection performance.
It can be further classified into two types: Chase Combining (CC) \cite{3.7-9}, which retransmits identical coded frames, and Incremental Redundancy (IR) \cite{3.7-10}, which retransmits additional parity bits.

Given that throughput, outage probability, and the number of average retransmissions for conventional HARQ have been extensively studied in terrestrial networks \cite{3.7-11}, this subsection focus on the timeliness issues of HARQ in SatCom.
More precisely, the feedback-based retransmission mechanism in standard HARQ renders it a high-delay scheme in the context of long-distance transmission.
To minimize long delay, various enhancements to the HARQ protocol have been proposed, such as Fast HARQ \cite{3.7-12} and Dynamic HARQ \cite{3.7-13}.
The Fast HARQ reduces delay by omitting certain feedback signals and allowing the receiver to decode only when the estimated overall channel gain is sufficiently high, while the Dynamic HARQ adaptively adjusts the number of retransmissions for each packet based on the decoding condition of the previous one.
Further taking into account the delayed feedback, a novel HARQ protocol called Early HARQ has been proposed to mitigate this issue by introducing predictive mechanisms \cite{3.7-14}.
Specifically, the receiver predicts the decoder outcome before actual decoding by comparing its BER estimate with an empirically calculated threshold and then generates early Acknowledgment (ACK)/non-ACK (NACK) or uncertain feedback to reduce feedback delays.
Moreover, the authors in \cite{3.7-15} introduced machine learning to improve the early HARQ, and demonstrated that the enhanced decoding prediction can significantly reduce estimate errors and redundant retransmissions compared to the empirical threshold-based scheme.

\textit{2) Link-Layer Erasure-based Redundancy.}
The Long Erasure Codes (LEC) specification \cite{3.7-16} published by CCSDS is an effective erasure scheme, which utilizes a packet-level LEC to alleviate packet loss at the PHY, i.e., an opportunity to recover packets that failed to decode at the PHY.
Typical LEC schemes, such as RS code, LDPC code, and fountain code can reconstruct the original information from any $k$ received packets by adding $m$ redundant erasure packets to $k$ original data packets using their respective coding methods.
Since link-layer RS codes are mainly used in storage systems and LDPC codes are more commonly used for PHY error correction, we focus on fountain code with the advantages of rate adaptability, no feedback retransmission required, and low encoding/decoding complexity \cite{3.7-17}.
From this perspective, fountain code is also a promising scheme for satellite communications featuring time-varying channels, long feedback delays, and limited payload of space segments.

Luby Transform (LT) code was the first to implement the concept of fountain code \cite{3.7-18}, which generates an arbitrary number of encoded packets by randomly selecting and XORing data packets from the original data, and allows the receiver to decode the complete information once a sufficient number of encoded packets are received.
To address the issue of unnecessary redundant packets generated by the random selection of data packets in LT code, which increases transmission overhead and reduces decoding efficiency, researchers have proposed Raptor code by adding concatenated precoding on top of LT code \cite{3.7-19}.
Compared to LT code, Raptor code not only reduces the number of encoded packets required for decoding but also decreases the decoding complexity at the receiver, and has been standard for DVB Handheld and 3GPP Multimedia Broadcast/Multicast Service (MBMS) \cite{3.7-20}.
To improve the performance of Raptor code, the authors in \cite{3.7-21} introduced the feedback mechanism of HARQ to optimize the degree distribution function, and proved that the feedback Raptor codes can achieve linear complexity with only a slight increase in decoding overhead.
%To improve the performance of Raptor code, the authors in \cite{3.7-21} and \cite{3.7-22} proposed a encoding mechanism that dynamically adjusts the connections between check nodes and variable nodes, as well as a parallel soft decoding algorithm based on multiple soft iterative cycles, both designed to enhance decoding efficiency.
Moreover, the rate adaptability of Raptor code allows flexible allocation of redundant packets across different nodes or transmission paths, which means that even if some nodes fail, the original data can still be recovered \cite{3.7-22}.
Therefore, Raptor code is highly adaptable for the DSIN, where the distributed Raptor code can greatly improve decoding probability and system fault tolerance, as well as reduce feedback delay.

\subsubsection{Satellite Erasure Transmission Scheme}
%\textit{3) Satellite Erasure Transmission Scheme.}
Network Coding (NC) applies a similar concept to LEC at the network layer, which enhances the transmission efficiency and reliability by forwarding data packets that are linear combinations of the original ones.
Random Linear NC (RLNC), which selects coding coefficients randomly over a finite field, can theoretically achieve the maximum network flow \cite{3.7-23, 3.7-24}.
To lower the encoding and decoding complexity of NC, the authors in \cite{3.7-25} proposed an excellent RLNC scheme that integrates fountain code to achieve both low computational complexity and high transmission efficiency, in which the outer code is typically a matrix-based fountain code, and the inner code uses RLNC to encode data within the same block.
Leveraging the advantages of the above LEC scheme, where the receiver can recover the $k$ lost original packets with high probability and low decoding complexity when it successfully receives $k$ NC packets, the authors in \cite{3.7-26} proved that the NC aided HARQ (NC-HARQ) enables energy-efficient transmission in satellite-to-terrestrial downlink communication.

Further, recent studies have explored the timeliness of NC-HARQ erasure protocols, which further incorporate AoI as a measure of information freshness \cite{3.7-27}.
The authors in \cite{3.7-28} investigated the NC-HARQ with limited or no feedback retransmission for satellite communications, showing that limited retransmission optimises delay, while no retransmission yields improved AoI.
An excellent erasure scheme named adaptive NC-inserted HARQ was proposed in \cite{3.7-29} to further improve timeliness, where NC packets are dynamically inserted into the data stream based on delayed CSI to accelerate packet recovery at the receiver.
The above scheme was extended to a dual-hop satellite relay scenario in \cite{3.7-30}, aiming to achieve optimal AoI and delay through a combination of adaptive CSI-based relay modes with adjustable LEC packet insertion intervals.
Further enhancement on SatCom can be obtained through cross-layer optimization of error control methods.
For example, the authors in \cite{3.7-31} proposed a dual-layer coding scheme that jointly optimizes link-layer LEC and physical-layer FEC.
By balancing redundancy between the LEC and FEC, i.e., finding their optimal coding rate, this erasure transfer protocol enables age-critical and reliable transmission in satellite networks.
Similarly, designing a cross-layer optimization framework between the link and transport layers is another promising research direction.
The benefit of this approach is that the TCP variants and LTP protocols can regard the satellite-terrestrial channel as a transparent transmission
path by leveraging link-layer error control to mitigate packet loss \cite{3.7-32}.
In this case, many congestion control algorithms can be applied to the satellite network effectively, further enhancing the actual transmission performance.

It is worth noting that the above schemes are designed for point-to-point transmission and are not applicable in DSIN with multiple points at least on one end.
In response, a satellite transmission scheme that combines HARQ, cooperative communication, and NC, is proposed in \cite{3.7-33}, which enables concurrent multi-stream data transmission through cooperation between multiple user antennas, thus reducing retransmissions and increasing throughput.
In more complex scenarios involving multiple users or satellites, recent studies suggest integrating HARQ with NOMA to achieve age-critical, high-throughput and reliable transmission in multi-node concurrent connections.
The outage performance \cite{3.7-34}, diversity order \cite{3.7-35}, average throughput \cite{3.7-36}, and BLER performance \cite{3.7-37}, have been thoroughly analyzed in the finite block length for HARQ aided NOMA (HARQ-NOMA) system.
The obtained closed-form expressions, approximations, or bounds of the above metrics show that the HARQ-NOMA system can significantly improve the performance of conventional NOMA schemes by adjusting the power level of users during retransmissions to reduce the number of attempts.

Recent works have further analyzed and optimized the AoI of the HARQ-NOMA system, where the results showed that HARQ can mitigate information ageing due to retransmissions by enhancing the reliability of NOMA \cite{3.7-38}.
Moreover, to further improve transmission efficiency, the authors apply DRL for the HARQ-NOMA system in \cite{3.7-39} to derive an age-optimal scheme integrating intelligent power allocations and retransmission decisions.
In summary, most studies on HARQ-NOMA systems lack the application of link-layer LEC, and the erasure transfer protocols for DSIN still need further investigation.

\subsection{High-Speed Inter-Satellite Communication and Network Routing}
The proliferation of LEO satellite constellations has ushered in a new era of DSIN, promising extensive coverage, unprecedented connectivity and intelligent informative services. At the core of this paradigm shift is the inter-satellite high-speed communication technology, efficient routing of data and congestion control through the constellation, which demands innovative solutions to address the dynamic and complex nature of these networks.

\subsubsection{High-Speed Inter-Satellite Communication}
The DSIN requires high-speed inter-satellite communication to achieve global ubiquitous coverage, but the increasing demand for inter-satellite communication bandwidth, coupled with the instability of free-space channels, presents new challenges in DSIN. As the communication capacity required for satellites and various spacecraft grows exponentially, the current communications based on RF are becoming insufficient to meet the sharply rising demand for capacity \cite{samy2022space}. 
Inter-satellite Free Space Optical (FSO) communication has emerged as a promising alternative or complementary solution to traditional RF systems, owing to its vast unlicensed bandwidth and the ability to transmit data at exceptionally high rates over long distances \cite{5.5-2}. Moreover, in contrast to conventional RF-based wireless systems, the narrow and directional nature of the laser beams used in FSO communications offers enhanced security, reduced power consumption, and immunity to electromagnetic interference. Constellation systems such as Starlink, OneWeb, Kuiper, and Telesat have already adopted inter-satellite FSO communication as one of their core transmission links. 
Nevertheless, despite the great potential of FSO communication for DSIN, its performance suffers from various limitations and challenges. On the one hand, the FSO communication is susceptible to the detrimental effects of atmospheric turbulence, including beam-wandering-induced pointing errors, beam scintillation, and attenuation caused by weather conditions such as haze, snow, fog, and clouds \cite{samy2022space}. Specifically, pointing errors induced by beam wandering and misalignments between receivers and transmitters can severely impact the reliability of the link. To address these challenges, the hybrid FSO/RF scheme within satellite networks presents a promising solution, leveraging the complementary nature of both technologies to enhance link quality without causing interference between the FSO and RF links \cite{le2022link}.
Relay-assisted transmission via high-altitude platforms (HAPs) offers another promising solution to enhance the usability of FSO links. For example, unmanned aerial vehicles (UAVs) can be deployed between satellites and GSs to establish a structured space-air-ground FSO network within DSIN, providing high reliability and high-speed transmissions \cite{}. This approach has already been explored in \cite{arum2020review} and has been validated by NASA's Laser Communications Relay Demonstration (LCRD) mission, which successfully transmitted data from GEO at a rate of 1.2 Gb/s. However, the performance of FSO communication significantly deteriorates as the propagation distance increases, particularly in environments with strong turbulence.

On the other hand, the successful implementation of long-distance FSO communications in DSIN largely hinges on the performance of the pointing, acquisition, and tracking (PAT) system \cite{kaymak2018survey}. The PAT system is critical for ensuring accurate alignment and maintaining stable communication links over vast distances, especially in the presence of atmospheric disturbances and dynamic satellite movements. 
FSO communications can mitigate pointing errors, acquire incoming light signals, and maintain stable links during missions by continuously monitoring system-wide performance metrics, such as received signal power and Strehl ratio, and dynamically adjusting correction elements like gimbals, mirrors, or adaptive optics. Pointing refers to the process of aligning the transmitter within the receiver’s field-of-view (FOV), while signal acquisition involves aligning the receiver with the arrival direction of the beam \cite{moon2024pointing}. Tracking ensures the ongoing maintenance of both pointing and signal acquisition throughout the optical communication link between satellites.
To achieve high data rates through concentrated light intensity and the extended reach of narrow beams, various PAT mechanisms have been proposed, including gimbal-based, mirror-based, gimbal-mirror hybrid, adaptive optics, liquid crystal, RF-FSO hybrid, and other PAT approaches, as detailed in \cite{kaymak2018survey}. With the decreasing size of satellites, UAVs, and airborne micro-stations, PAT mechanisms are anticipated to become smaller and less complex, necessitating innovative ATP designs. For example, mirror-based PAT mechanisms struggle to maintain a stable optical link between a GS and an LEO satellite travelling at 7.5 km/s for more than 1 millisecond.
In this context, developing an agile PAT system is both essential and challenging to ensure the stability and reliability of FSO links in DSIN, where network outages are either impermissible or must be anticipated proactively. To explore the feasibility of FSO communications in near-Earth 6G NTNs, a baseline PAT system for vertical FSO links equipped with conventional detectors and actuators has been introduced in \cite{moon2024pointing}. This system demonstrates improved robustness and agility through the incorporation of angle-of-arrival (AoA) estimation and retroreflectors, showcasing significant performance gains.
Moreover, current laser terminal control typically relies on a one-to-one mechanical beam-linking approach, which results in prolonged access times. Additionally, since the beam direction is fixed to a single trajectory, this method is limited to point-to-point communication and cannot establish links with multiple terminals simultaneously, falling short of the requirements for efficient inter-satellite networking.
To address this limitation, researchers have proposed an optical spherical head array \cite{2019USAConference}, which enables electronically programmable control of optical head directions to achieve rapid beam angle adjustments. This approach ensures swift angle switching with high resolution, presenting significant potential for advancing laser-based inter-satellite networking in the future.

Further, in DSIN, small satellites can significantly enhance information fusion and high-speed transmission capabilities through distributed cooperation in communication and computation. However, achieving precise synchronization in absolute phase, frequency, and time remains a formidable challenge when the clock or local oscillator signals are generated locally at each distributed node, which has received plenty of attention in the literature \cite{sundararaman2005clock, kodheli2020satellite}. 
As previously discussed for distributed MIMO collaboration, the GSs must maintain synchronized clocks with sub-nanosecond accuracy to support bandwidths of several hundred MHz, thereby ensuring symbol-level synchronization. Moreover, the relative motion of satellites in DSIN, along with varying and long distances, further complicates signal synchronization in diverse space science tasks such as remote sensing, inter-satellite ranging, and relative positioning. 
A non-collaborative CFO estimation technique is proposed in \cite{9348867} to address synchronization challenges in distributed mMIMO mobile communication systems. 
Additionally, methods such as closed-loop, open-loop, master-slave, and consensus-based approaches have been suggested for implementing frequency and phase synchronization, as detail reviewed in \cite{9761868}. Furthermore, in recent years, ML techniques have garnered widespread academic interest in addressing various synchronization issues in end-to-end communication systems, including frame synchronization, compensation for sampling frequency/time offsets, and phase noise characterization. In this regard, the authors in \cite{wu2019deep} propose a CNN-based synchronization model with a softmax activation function to detect the actual position of a frame header, while the authors in \cite{wang2013learning} employ multi-instance learning to solve frame synchronization problems across different frequency ranges without requiring additional modifications. However, due to the limited on-orbit computational power, ML models may struggle with poor convergence rates, especially under varying transmission conditions. Therefore, there is a pressing need to investigate optimal model synchronization technologies among small satellites in a DSIN to improve the accuracy of ML training models and expedite training times.

\subsubsection{Distributed Network Routing}
The pursuit of ubiquitous global internet connectivity has driven the development of satellite constellations, which aim to provide seamless coverage across the globe. With the increasing number of satellites in orbit, the routing of data within these constellations has become a critical area of research.
To achieve efficient routing in satellite networks, researchers in the space industry are delving into network routing design, predominantly decentralized and centralized, to offer choices for the multi-satellite cooperative backhaul. The centralized routing approaches are mostly computed on the ground and distributed to the satellites. Early routing algorithms, such as Dijkstra's shortest path algorithm, were adaptations of centralized routing algorithms and were not designed for the dynamic nature of satellite networks \cite{kodheli2020satellite}. The stale information on the traffic status can result in sub-optimal routing algorithms and valueless routing updates in CCS systems.
The ongoing advancement of space-based networking technologies, including space-based routing and ISL, is critical for meeting the substantial data transmission requirements associated with S-IoT services. It also spurred the transition of satellite constellation routing from centralized to distributed approaches.
The distributed routing enables to calculation of the routes in orbit as the satellites are typically equipped with onboard computation and storage resources, thus guaranteeing the timely utilization of the latest traffic queuing status.
Distributed, heterogeneous satellite constellations that integrate communication, navigation, and remote sensing capabilities are now recognized as a pivotal development trajectory for satellite networks \cite{qi2021using}. These networks, characterized by high-velocity LEO satellites and frequent ISL changes, are inherently dynamic, particularly within multi-layer LEO topologies. This dynamic nature presents significant challenges for fully distributed routing at the network layer, necessitating innovative inter-layer link establishment and maintenance strategies \cite{dong2023novel,hu2022software,jiang2023software}.

\begin{figure}[t]
	\centering
	\includegraphics[width=0.9\textwidth]{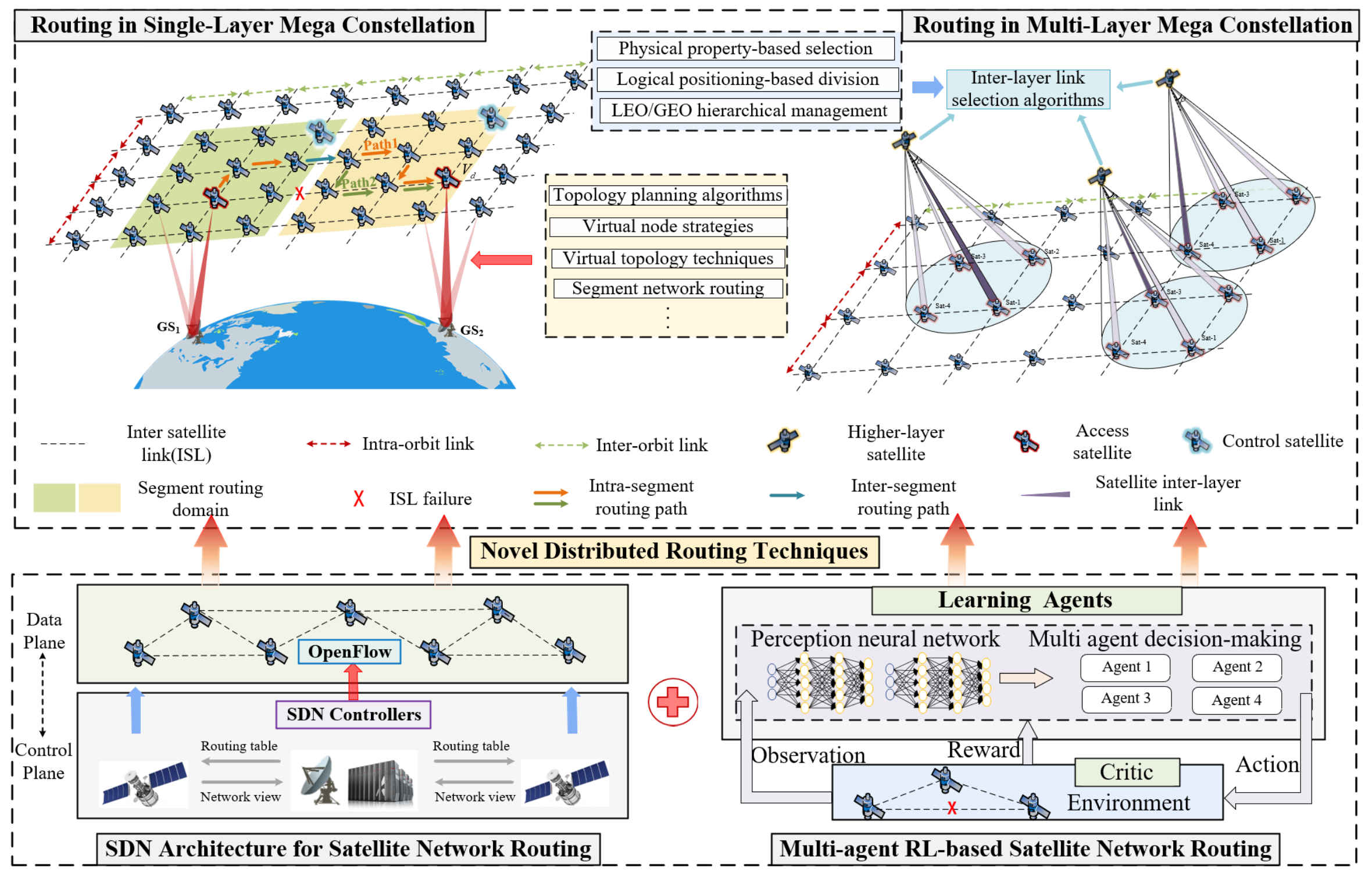}
	\caption{A overview of distributed network routing and congestion control for DSIN.}\label{routing}
\end{figure}

To address these complexities, researchers have proposed a variety of adaptive routing strategies that emphasize topological flexibility and low-complexity modeling. For instance, the inter-layer link algorithms which designed for inclined orbital constellations choose to prioritize link selection based on physical properties such as distance and angular velocity \cite{qi2021using}. Virtual node strategies, which divide satellite constellations into geographically bounded virtual regions, further mitigate dynamic network topology issues by assigning domain controllers within each region \cite{dong2023novel}. By introducing domain controllers, virtual node methods enable to reduction of the impact of topological fluctuations and facilitate manageable sub-network structures. In addition to virtual nodes, virtual topology techniques provide a structured approach by dividing network timeliness into discrete snapshots, thereby maintaining a manageable database of topological states for each time slot \cite{chen2019topology}. This approach reduces the demand on satellite storage resources while enabling efficient, snapshot-based routing \cite{wang2023reliability}. A notable example of such virtual topology applications includes the hybrid use of virtual topology and virtual node methodologies, which, despite their efficiency, can still present computational challenges when inter-layer node counts are substantial \cite{jiang2023software}. An alternative approach involves multi-hop route stability estimation, where link interruption probabilities are calculated using stochastic geometry to ensure robust connections through multi-hop routes \cite{wang2023reliability}.
Routing in multi-layer satellite networks also benefits from algorithmic approaches that integrate logical positioning within LEO constellations, allowing each satellite to communicate exclusively with neighbouring layers. This cross-layer design, leveraging geographic divisions, minimizes signaling overhead and maintains load-balancing \cite{hu2022qos,roth2022distributed}. Back-pressure routing strategies, such as Distance-based Back-Pressure Routing (DBPR), assign link weights according to distance, thereby favouring non-congested, shorter paths to the destination \cite{deng2022distance}. 
These strategies have proven effective in fulfilling key performance criteria, such as high throughput and low latency that are often difficult to achieve concurrently with traditional algorithms.

As DSIN grows in scale and complexity, future routing strategies must be tailored to meet the needs of satellite-integrated Internet applications. Strategies grounded in SDN principles offer a promising framework for these networks by leveraging global network awareness to optimize routing paths and reduce the computational cost associated with link selection in real-time \cite{kumar2021fybrrlink}. For example, the fybrrLink QoS-aware routing algorithm, designed for GEO/LEO satellite networks, uses SDN's global perspective to expedite inter-layer link selection by combining modified Bresenha and Dijkstra algorithms, thus significantly reducing optimal route computation times \cite{kumar2021fybrrlink}.
Further, AI techniques are being explored to predict network conditions and optimize routing dynamically.
Recent research has focused on leveraging ML, particularly reinforcement learning (RL), to address the challenges of routing in LEO satellite constellations. Q-learning, a type of RL, has been proposed for distributed routing in LEO satellite constellations \cite{soret2024q} This approach models the routing problem as a multi-agent Partially Observable Markov Decision Process (POMDP), where each satellite interacts only with its neighbours. The proposed Q-learning solution demonstrates comparable delays to centralized algorithms under steady-state conditions, increased supported traffic load without congestion, and minimal signaling overhead among satellites.
Another significant contribution is the work on robust beam-to-satellite routing strategies for mega-constellations \cite{10623789}. Aiming to minimize end-to-end latency and maximize the supported traffic load, strategies are proposed to address the challenges of routing in the presence of highly imbalanced traffic and dynamic network topology.

Moving forward, researchers anticipate the need for multi-constraint, integrated sensing-computation-routing mechanisms tailored for partitioned satellite networks. These mechanisms are expected to facilitate adaptive, rapid re-routing by calculating optimal paths based on ISL attributes for each hop, enabling network state awareness and further enhancing resilience and adaptability in multi-layer, heterogeneous DSIN.

\subsubsection{Congestion Control}
The dynamic nature of DSIN introduces unique challenges in traffic management and congestion control. On the one hand, in DSIN, a large amount of signaling backhaul will result in severe local network congestion and aggravate network control load, which further worsens the control delays \cite{ji2021mega}.
On the other hand, the unprecedented demands of real-time, high-volume data transmission in DSIN necessitate advanced multi-hop relaying and satellite-based processing to optimize data flow. Characterized by high-bandwidth and long-latency ISL, these networks resemble ``long fat networks'' (LFNs) and require carefully tailored flow control protocols to effectively manage both congestion and link idleness. Current end-to-end flow control algorithms typically operate through two primary phases: link state assessment and rate adjustment \cite{wei2006fast,ahmad2019enhancing}.

During link state assessment, protocols monitor status metrics such as packet loss and delay, enabling dynamic evaluation of network conditions. This real-time monitoring allows for adaptive decision-making, optimizing network performance and ensuring efficient data transmission by adjusting to fluctuating network conditions. Packet loss detection algorithms, including TCP Reno and TCP Cubic, perform well in stable link environments with low packet loss but face challenges in LEO satellite networks, where frequent handovers introduce elevated packet loss \cite{deutschmann2023cubic}. Conversely, delay-sensitive algorithms like TCP Vegas and Fast TCP respond to increases in round-trip time (RTT) by preemptively reducing transmission rates, mitigating packet loss through early congestion avoidance \cite{claypool2021comparison,liu2022effects}. The above approaches aim to maximise bandwidth utilisation without causing congestion, keeping the link in a near-threshold state upon stabilisation. However, as arrival rates approach service rates, high bandwidth utilisation can compromise data timeliness, with the effects being particularly pronounced in latency-sensitive applications \cite{liu2022effects}.

Recent advances, inspired by the AoI-sensitive ACP+ algorithm \cite{9484567,9562140}, indicate that leveraging Age of Information (AoI) metrics can provide timely congestion management across large-scale satellite constellations \cite{chiu1989analysis}. By adjusting transmission rates based on end-to-end AoI variation, this approach enhances the timeliness and reliability of flow control in LFNs. Rate adjustment methods often employ Additive Increase Multiplicative Decrease (AIMD) mechanisms, which converge sources to comparable rates over time, dynamically balancing throughput across sources. While Multiplicative Increase Multiplicative Decrease (MIMD) methods allow faster adjustment, they are generally less fair \cite{chiu1989analysis}. In fact, each adjustment mechanism impacts network stability and convergence. For instance, AIMD protocols like TCP Reno struggle to quickly reach target congestion windows in LFNs, limiting convergence speed \cite{10228914}. TCP Hybla attempts to address this by comparing actual RTT with a predefined standard to fine-tune window increments, though slow convergence remains a constraint under LFN conditions \cite{claypool2021comparison}. By contrast, TCP Cubic introduces a nonlinear window growth function to expedite rate stabilization, although its aggressive scaling risks incurring congestion \cite{deutschmann2023cubic}.
Future research should prioritise the development of flow control protocols that strike a balance between data freshness and efficiency while adapting to the frequent link handovers and sudden traffic surges characteristic of mega-constellations. Such advancements are crucial for ensuring robust, real-time performance in next-generation satellite networks, enabling them to meet the demands of emerging, latency-sensitive applications.

\section{Collaborative Cross-layer Optimization}

\subsection{Mobility Management}
Mobility management can be categorized into location management and handover management, where the former focuses on tracking and updating the real-time location information of Mobile Terminals (MTs), while the latter ensures uninterrupted service continuity as an MT switches between two access nodes.
We then elaborate on the overall architecture of mobility management and its specific schemes for the above two components in the following.

\subsubsection{Mobility Management Architecture}
Consider that the dual mobility arising from the high-speed movement of CCS system, and the uneven distribution of mobile terminals, along with the highly overlapping satellite coverage areas that cause frequent handovers, can significantly increase handover delay, signaling overhead, and decision-making burden in DSIN \cite{4.1-1}.
These issues are intensified in the real world by the limited number and fixed locations of GSs equipped with Mobility Management Functions (MMFs), as well as the unavoidable non-trivial backhaul delays across satellite-terrestrial nodes.
In such cases, a large amount of mobility management signaling must be relayed through inter-satellite links with many hops to a few GS, leading to severe network congestion, increased management delays, and excessive signaling overhead \cite{4.1-2}.
Moreover, in the NTN architecture proposed by 3GPP with gNB functional split that allows for a tradeoff between costs and payloads, the two main network functions involving MMFs, i.e., the gNB-CU-CP and Access and Mobility Management Function (AMF), are still deployed on GSs, which makes their feeder links struggle to meet the delay and capacity requirements necessary for the mobility management of a large number of satellites \cite{4.1-3}.
Therefore, a mobility management architecture with dynamic MMF configuration is crucial for ensuring service continuity and timely management in DSIN.
\begin{figure}[t]
	\centering
	\includegraphics[width=0.9\textwidth]{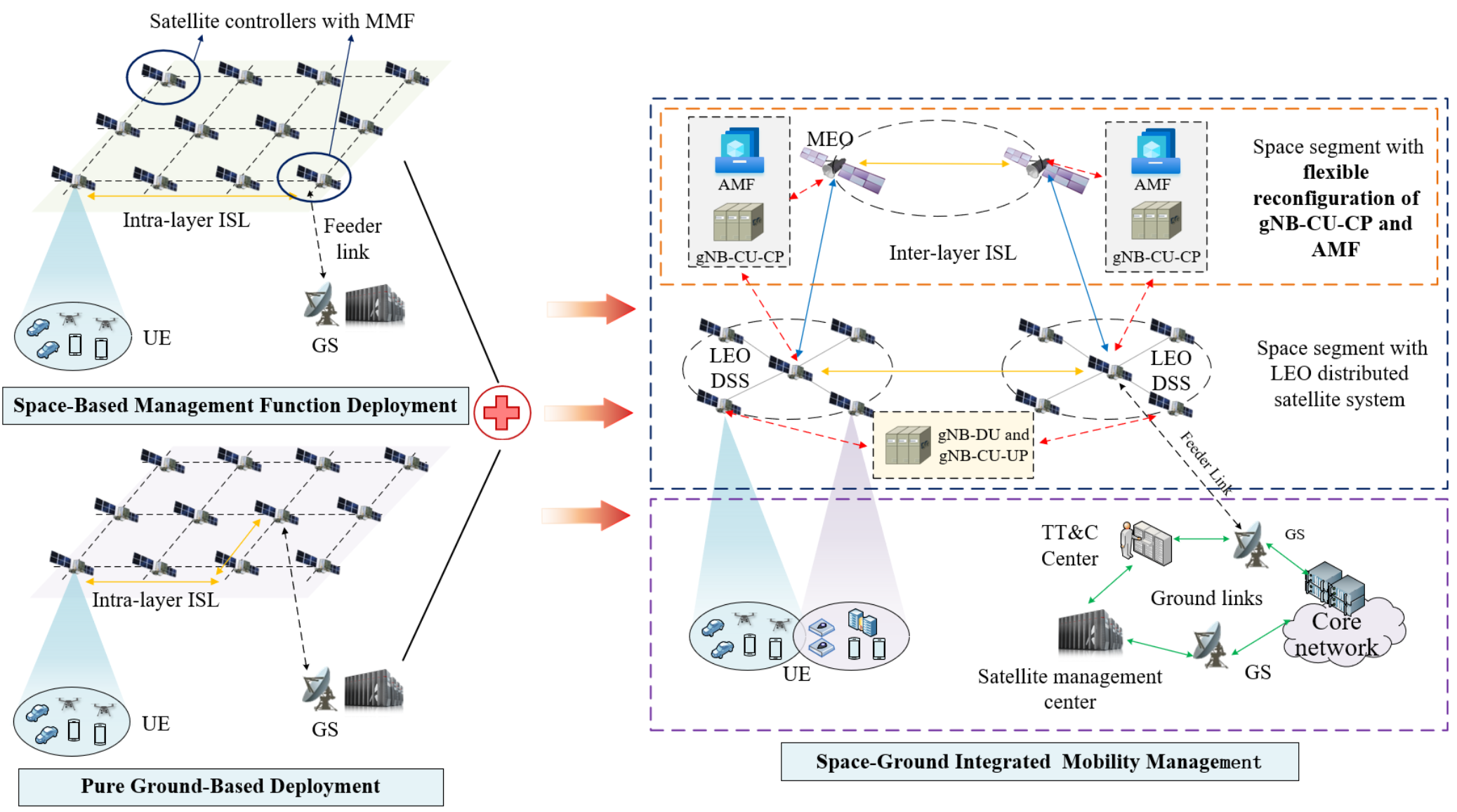}
	\caption{A network view of SGIMM under the 3GPP NTN architecture with gNB functional split.}\label{mobile}
\end{figure}

Conventional mobility management architectures of LEO satellite systems are primarily divided into two parts: Pure Ground-based Deployment (PGD) \cite{4.1-4} and Space-based Management Function Deployment (SMFD) \cite{4.1-5}.
In the PGD architecture, which relies on GSs for mobility management, the number of GSs is significantly insufficient compared to the satellite networks because of the geographical and policy factors, which makes it challenging to improve or even ensure the delay and overhead associated with mobility management.
The SMFD architecture performs mobility management through satellite controllers equipped with MMFs.
However, the satellites are required to continuously interact to gather information from across the network, which not only consumes data transmission resources but also creates a heavy control load.
In addition, the number of controllers needs to be dynamically adjusted to the network scale.

To address these shortcomings, a novel architecture named Space-Ground Integrated Mobility Management (SGIMM) was proposed to integrate the advantages of the above two approaches \cite{4.1-6, 4.1-7}.
It introduces non-LEO satellites as management nodes in the space segment, characterized by wider coverage, higher capacity, and fewer requirements for deployment quantity, and enables them to collaborate with GSs to perform mobility management, achieving a separation of transmission and management both physically and logically.
On this basis, we can further apply the functions and interfaces defined by the 3GPP NTN architecture as shown in Figure \ref{mobile}, where the Medium Earth Orbit (MEO) satellites and GSs are deployed with AMF and gNB-CU-CP for user location management and handover decision-making, respectively, while each LEO satellite is equipped with gNB-DU and gNB-CU-UP for accessing and signaling transmission \cite{4.1-8}.
This approach is equally applicable in DSIN, in which the management nodes of MTs can be further switched among the leader satellites in LEO parts of DSIN, the MEO satellites and GSs based on the tradeoff between delay and overhead.

Moreover, the SGIMM architecture will also cause lots of issues, such as the triple mobility due to MEO satellites, along with the additional delay and overhead caused by storage and content migration during the handovers among management nodes.
In response, several studies suggest that distributed RL algorithms can be employed to address the relative mobility among MTs, GSs, and satellites, enabling ground MTs to intelligently select the optimal ground/satellite management node with minimal handover and migration delay \cite{4.1-9}.
Further, the authors in \cite{4.1-10} proposed a Flexible and Distributed Mobility Management Architecture (FDMMA) based on the SGIMM, including an overview of its network configuration, communication protocols, functional implementation, and handover procedures.
The lightweight handover decision and distributed MMF configuration designed for the FDMMA offer improved handover delay and reduced signaling overhead.

Besides, further research is needed on the implementation of location management and handover strategies in our DSIN.
On one hand, due to the time-varying topology and increased density of satellite systems, location management requires more frequent updates and paging for real-time tracking and recording, which imposes high demands on overhead and delay.
On the other hand, the exponential growth of access requests, ultra-dense cellular coverage, and diverse types of services considerably complicate the handover triggering conditions and increase the selection of access nodes, straining the computational capabilities of MMFs and impeding the seamless handover \cite{4.1-11}.

\subsubsection{Handover and Location Management}
The feasible schemes for these two types of management techniques regarding the above challenges are discussed in this subsection.

\textit{1) Distributed Handover Schemes.}
Handover schemes in mobility management fall into two main categories: beam handover and satellite handover, where the former is triggered when MTs move across boundaries between neighboring beams of a satellite, and the latter occurs as MTs move across the coverage areas of adjacent satellites.
The early research on handover schemes primarily focuses on beam handover since the number of satellites in conventional satellite systems is limited.
Moreover, the channel allocation strategies \cite{4.1-12} that ensure new channels are available during handovers and the handover guarantees \cite{4.1-13} that prevent call drops or blockages throughout the handover process are the prime issues in managing beam handover requests.
With the growing density of satellite networks, the frequency of handovers between satellites has increased considerably, prompting a shift in recent research toward the design of satellite handover schemes.

Graph theory is a typical approach for implementing satellite handover, where each satellite's coverage period is treated as a node, and the handovers between satellites are represented as directed edges \cite{4.1-14, 4.1-15}.
In this framework, the handover process can be described as selecting an optimal path from all available handover paths.
Further taking into account the MT request distribution, available channel state, satellite coverage time and other QoS requirements, the constructed graph can be utilized to find the optimal handover handover in diverse communication scenarios \cite{4.1-16}.
However, these handover schemes rely on centralized decision-making based on global state information, leading to excessive complexity and overhead in DSIN with numerous satellites and MTs.

Distributed RL, such as multi-agent Q-learning, is viewed as a potential solution to the above issues.
By deploying well-trained agents at various nodes of DSIN, each satellite can derive localized strategies that simultaneously satisfy payload constraints and minimize handover costs with low computational complexity \cite{4.1-17, 4.1-18}.
Nevertheless, due to the limited state-action pairs stored in the Q-table, most RL-based handover schemes are only suitable for simple scenarios with a limited number of satellites.
In more complex DSIN, it is necessary to employ more advanced DRL algorithms to design effective handover schemes.
A MT-driven Deep Q-Network (DQN) based handover scheme was developed in \cite{4.1-19}, in which a centralized agent deployed at the management satellite node for training the DQN, and each MT individually makes its handover decisions based on the parameter disseminated from the trained node.
Adopting a similar centralized-training and distributed execution approach, the authors in \cite{4.1-20} proposed a multi-agent fingerprints-enhanced double deep Q-network to address the handover issues under burst traffic scenarios with delay constraint.
Since the above handover schemes are all designed under the static propagation conditions, the authors in \cite{4.1-21} develop a multi-agent successive hysteretic DQN algorithm to address the handover problem that involves throughput requirement of MTs and load-balancing demand of satellites in time-varying satellite-Terrestrial channel.

\textit{2) Dynamic Location Management.}
Location Management (LM) primarily involves the Location Area (LA) (i.e., tracking area) design, location update and paging, where the size of LA is crucial, as it affects the frequency of signaling interactions during updates and paging.
Specifically, a larger LA may result in delayed updates of MT location information, while a smaller one may lead to excessive updates and paging overhead \cite{4.1-22, 4.1-23}.
as a result, a dynamic adaptive LA scheme is proposed in \cite{4.1-7} to achieve an optimal tradeoff between LA division and updates/paging overhead.
The scheme enlarges the LA size for high-speed MTs to reduce overall LM overhead and shrinks the LA size in high call-traffic areas for faster user access while keeping the number of paged satellites manageable in large-scale satellite networks.

In an IP-based network, the Mobile IP version 6 (MIPv6) \cite{4.1-24} is a classical management scheme introduced by the Internet Engineering Task Force (IETF), where an MT only binds to a new IP address upon handover to maintain its TCP connection with an access point or GS and minimize the impact of MT location changes.
As communication networks become more complex, the IETF introduces enhancements such as Proxy MIPv6 (PMIPv6), Fast Handover for MIPv6 (FMIPv6), and Hierarchical MIPv6 (HMIPv6) to strengthen the location management performance of IP-based network \cite{4.1-25}.
Moreover, since these centralized LM schemes face high update overhead and limited scalability due to frequent handovers with increasing MTs and satellites, the IETF is working on developing LM solutions for future satellite networks.

Seamless IP diversity-based Generalized Mobility Architecture (SIGMA) is a promising LM solution for DSIN, where an MT can continue using its previous IP address while obtaining a new one, with the timing for switching to the new IP address and deleting the old one predicted based on the deterministic movement path of satellites \cite{4.1-26}.
However, SIGMA does not account for the significant signaling overhead induced by frequent satellite handovers under high mobility conditions. To mitigate the link management (LM) cost, two dynamic LM approaches have been proposed.
The first approach is a dual-location area link management (LM) scheme \cite{4.1-27}, where a Satellite Location Area (SLA) is introduced for the positioning of LEO satellites. By combining the SLA with the User Location Area (ULA), this scheme adapts to the varying mobility patterns of LEO satellites and mobile terminals (MTs), effectively reducing the frequency of global updates.
The second approach is a Virtual Attachment Point (VAP)-based link management (LM) scheme \cite{4.1-28}, in which mobile terminals (MTs) first establish a connection to a fixed logical point, the VAP. From there, MTs access the satellite network via this designated point.
In such a case, the AMF only track the MT's location relative to the VAP, along with the VAP's location relative to the satellite, which can mask the mobility of satellites from MTs.

Further, to enhance the scalability of the existing LM schemes, several researchers propose distributed solutions that offer advantages like near-optimal path selection, reduced workload distribution, and improved handover performance \cite{4.1-29}, which can be viewed as a potential approach for DSIN.
Virtual MIPv6 (VMIPv6) is a typical distributed LM scheme \cite{4.1-30}, where a Virtual Agent Cluster (VAC) comprising a group of LEO satellites within a specific LA collaboratively manages the mobility of MTs within its Virtual Agent Domain (VAD).
When MTs perform handovers within the same VAD, they only change their local addresses and retain their global addresses, which helps reduce LM overhead and delay.

\subsection{Resource Management}
The proliferation of SatCom, particularly with the advent of satellite-integrated Internet and the anticipated 6G services, underscores the criticality of resource management and allocation mechanisms in DSCN. For instance, satellite constellations operating in LEO provide reduced latency and improved bandwidth for applications such as real-time telemedicine and autonomous vehicle coordination. However, the dynamic dual mobility nature of satellites and users, especially in LEO and medium Earth orbit (MEO) based satellite networks, requires advanced resource allocation strategies to address both scalability and flexibility in resource distribution across hundreds or thousands of satellites \cite{kodheli2020satellite}.
Moreover, the aggregated resource management for satellite edge computing highlights the need for efficient resource allocation to support the computational and storage demands at the network edge \cite{9500539}.
This need underscores the critical role of resource management mechanisms in meeting the complex requirements of DSCN.
In general, the DSCN faces the following core challenges in resource management:

\textit{1) Scarcity and Uneven Distribution of Resources.}
On the one hand, traditional resource management solutions do not fully exploit the regenerative capabilities of next-generation satellite systems, which offer flexibility in bandwidth, power control, and onboard data processing. To address this, research has ventured into areas such as dynamic bandwidth allocation, edge caching optimization, and onboard signal predistortion for digital transparent satellites \cite{10436159}. On the other hand, exacerbated by the rapid growth in the number of satellites and the diverse nature of satellite missions, the scarcity of available frequency spectrum becomes a critical bottleneck \cite{jia2020intelligent}. Techniques such as cognitive radio (CR) have been proposed to enable dynamic spectrum sharing among satellite networks, minimizing interference and maximizing spectrum efficiency. However, implementing CR in CCS system poses unique challenges due to the need for real-time interference management across multiple satellites. It is important to note that with the increasing density of satellite constellations or clusters, managing interference, both intra-system and inter-system, becomes a significant challenge. Hence, intelligent resource allocation algorithms that can predict and mitigate interference are essential for improving spectrum efficiency.

\textit{2) Hysteresis Effect.} The DSCN exhibits a hysteresis effect, where resource scheduling lags behind the update of resource status information, resulting in low resource utilization efficiency. To tackle this, researchers have introduced resource scheduling algorithms based on DRL, which can adapt to complex and dynamic environments, addressing the mismatch problem of traditional scheduling methods. The DRL algorithm is widely used to address the joint allocation of sub-channels and power in multi-beam SatCom systems \cite{deng2022dynamic}, offloading between terrestrial and satellite networks \cite{cao2022edge}, demonstrating the innovative significance of DRL in heterogeneous terrestrial-satellite communication networks.
Additionally, in the field of distributed resource management in DSIN, MADRL has been increasingly utilized to address the complex challenges of coordinating and optimizing resources across multiple agents (satellites). By employing MADRL, researchers have been able to develop innovative solutions that address the challenges of dynamic heterogeneous resource allocation, task offloading, and service allocation in satellite networks. The integration of MADRL with other techniques and frameworks has shown promise in enhancing the performance and efficiency of satellite communication systems, and it is expected to play a crucial role in DSIN.
For instance, a hierarchical cross-domain satellite resource management framework is analysed by employing MADRL to guide the collaborative work of multiple satellites within a domain \cite{he2023hierarchical}, enhancing the scheduling capabilities of multi-domain satellite systems.

\textit{3) Edge Computing and Network Slicing.} Edge computing has been introduced to reduce latency by bringing processing closer to the data source. Distributed caching can alleviate bandwidth demands by storing data at or near satellite locations. The utilisation of MADRL in optimizing resource allocation in complex satellite-terrestrial networks allows for enhancing the efficiency of edge computing in complex satellite-terrestrial networks \cite{jia2022joint}. Moreover, network slicing strategies for real-time applications in large-scale satellite networks is crucial for managing network resources efficiently in satellite networks \cite{guo2024network}. Further, the use of digital twins to enhance resource slicing in LEO satellite networks is explored, emphasizing the importance of robust and adaptive resource management strategies for the dynamic nature of SatCom \cite{he2024digital}.
However, these technologies are still in their infancy within satellite networks, and their deployment across globally dispersed satellite constellations remains a challenging endeavor. Issues such as data synchronization, cache coherence, and global scalability hinder the full-scale adoption of these methods in real-world applications.
  
Looking ahead, the future DSIN will require innovative approaches to enhance resource management efficiency. One promising direction is the development of integrated resource management for terrestrial-satellite systems, which can optimize the use of resources across both domains. Additionally, research into intelligent resource management for satellite and terrestrial spectrum shared networking is essential \cite{10371362}, focusing on spectrum sensing, prediction, and allocation to improve spectrum efficiency. Moreover, in the distributed satellite network architecture, blockchain-based frameworks for decentralized resource management could enable autonomous resource allocation among CCS system, enhancing flexibility and reducing the need for centralized control. By providing a secure, distributed ledger, blockchain can facilitate transparent and verifiable transactions, improving the reliability and scalability of resource sharing. Further, future DSIN would benefit from the development of adaptive algorithms capable of real-time resource allocation adjustments to handle dynamic demand fluctuations. These algorithms should be designed to operate in high-mobility environments, allowing networks to dynamically adjust to the changing conditions of satellite constellations without compromising efficiency. The last but not the least direction is multi-orbit satellite collaboration, where collaborative resource management strategies among satellites in different orbits, such as LEO, MEO, and GEO can be obtained. This significantly enhances resource utilization by enabling satellites to share resources based on their specific capabilities and operational constraints, resulting in improved network resilience and performance.

\subsection{Secure Communications}
\begin{figure}[t]
\centering
\includegraphics[width=0.55\textwidth]{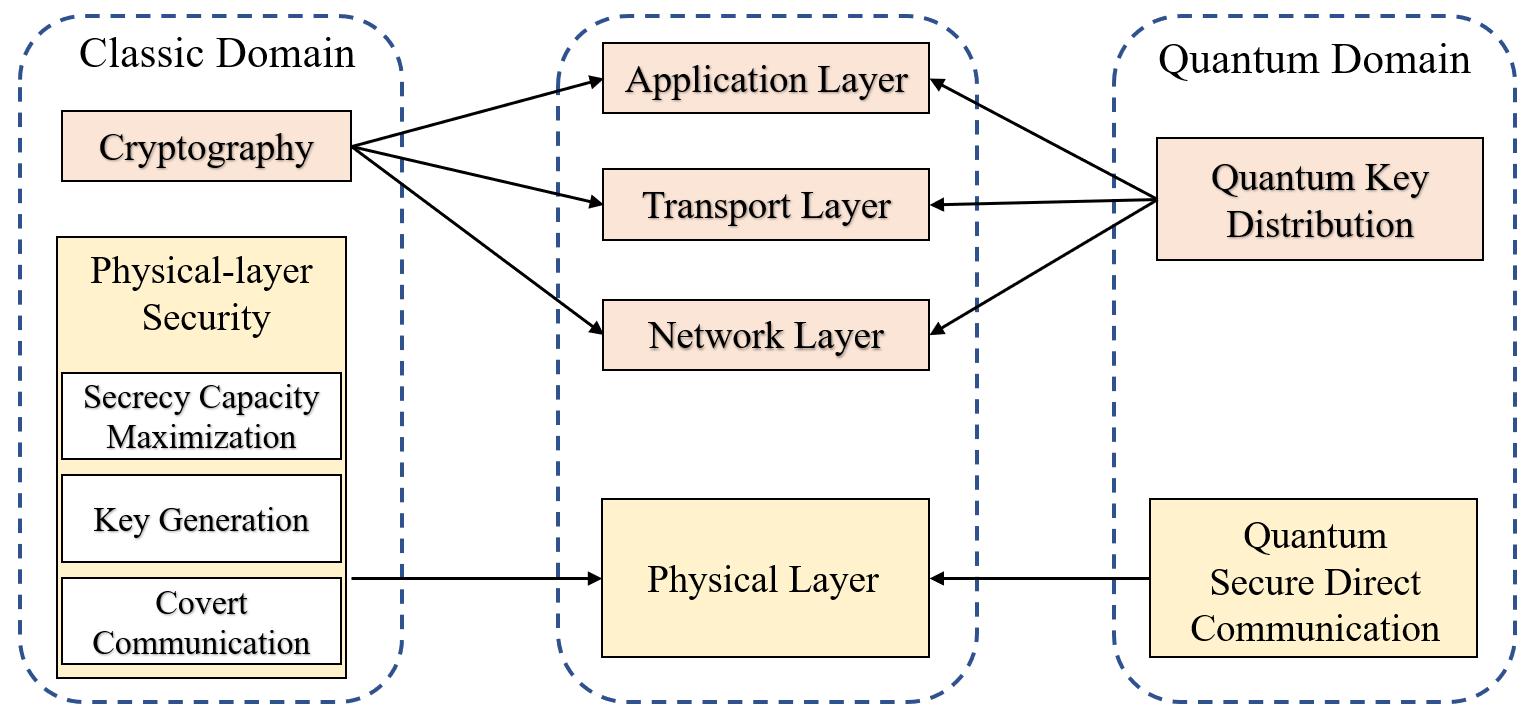}
\caption{Different information security solutions at different open system interconnection layers}
\label{SecCom}
\end{figure}

With the rapid development of communication industry, the exponential increase in confidential and sensitive data over wireless links yields increasingly prominent security concerns. Compared with terrestrial communication networks, the electromagnetic broadcast characteristics of satellite networks enable longer transmission distances and broader wireless coverage, making these links more susceptible to eavesdropping, jamming and unauthorized access. Moreover, the three-dimensional wide-area coverage of satellite networks allows malicious nodes to conduct illegal activities from any position in space, air or ground, and it is challenging to detect and cope with malicious nodes in the passive listening mode. Additionally, Resolution No. 35 of the 2019 World Radiocommunication Conference requires satellite operators to submit actual deployment parameters of their satellite systems, which, however, facilitates malicious nodes to predict satellite orbits and positions and prepare for potential attacks or eavesdropping. Consequently, DSIN faces severe challenges in ensuring secure information transmission. To ensure the security of information transmission, three mainstream solutions are classic cryptography, physical-layer security, and quantum-domain security, which constitute multi-level security mechanisms and provide comprehensive security at the physical layer, network layer, transport layer, and even application layer, as shown in Figure \ref{SecCom}.

Cryptography uses public/private keys to encrypt/decrypt confidential information, rendering it undecipherable without the key \cite{diffie1976new}. The security essence of key-based cryptosystems lies in that it is difficult for devices with limited computational power to carry out mathematical operations to decipher the transmitted data. The cryptography-based security solutions operate at the network or upper layers and can be easily transplanted to satellite communication systems \cite{ingemarsson1981encryption}, where the limited on-board storage and processing capabilities need to be prudently considered. With the growth in device computational power and advancements in quantum computing technology, the capability to decipher information will be significantly improved. A possible way for the security enhancement is to increase the key length, which can temporarily alleviate the pressure of cracking brought by the enhancement of classical computation and cryptanalysis capabilities. However, the longer the key, the lower the efficiency of the encryption algorithm becomes, resulting in slower encryption, decryption and key distribution speeds. This motivates us to apply more lightweight and efficient algorithms to traditional cryptographic methods and also reveals that cryptography alone is insufficient to guarantee secure information transmission in DSIN.

To cope with this issue, it is promising to move the mechanisms for secure information transmission down to the physical layer directly through physical signal transformation. Based on the concept of information theory security, physical-layer secure communication technologies exploit the differences between legitimate and eavesdropping channels to achieve complete secrecy at the physical signal transmission level, effectively preventing the third parties from intercepting information through eavesdropping channels \cite{liu2017physical}. Existing methods for secure transmission at the physical layer are mainly divided into three categories. The first one is to increase the rate disparity between legitimate and eavesdropping channels by using techniques such as information coding, beamforming, artificial noise and relay cooperative jamming, thereby enhancing the secrecy capacity \cite{han2022challenges}. 
The second one is to generate inherently random key pairs for physical-layer key encryption of the transmitted data by utilizing the reciprocity, randomness and spatial uniqueness of wireless channels \cite{zeng2015physical}. 
In this regard, the uniqueness of channel information ensures that the transmitted data cannot be deciphered even if malicious nodes enhance their computational capabilities. Although these methods protect the information contents from being deciphered, malicious nodes can still detect communication behavior, locate the sources of radiation and launch physical attacks, thereby posing a security threat to legitimate nodes. To achieve a higher level of secure physical-layer transmission, the third one is the emerging covert communication technology. It typically hides the behavior of information transmission by mimicking environmental noise or other natural signals. Therefore, the transmitted signal appears to be part of the electromagnetic environment and the information is transmitted without drawing the attention of unauthorised entities \cite{chen2023covert}. The key to information covertness lies in the various uncertainties in the end-to-end physical transmission, such as the signalling uncertainty, noise uncertainty, fading uncertainty and interference uncertainty. In \cite{Yu2024covert}, a covert satellite communication scheme over an overt channel by randomizing the variance feature of the transmitted Gaussian signalling to realise a positive covert rate, i.e., increasing the uncertainty at the transmitted signal. The interference uncertainty inherently in large-scale LEO satellite networks is utilised to enhance the covertness in \cite{feng2024covert}, where a two-stage Stackelberg game is used to model the conflict dynamics between the adversarial BS and the satellite network. A Stackelberg equilibrium is achieved to reveal the trade-off between transmission reliability and covertness in leveraging co-channel interference. Since physical-layer secure communications depend on how to transform the physical signal, the information security can be further enhanced via effectively combining physical layer transmission with advanced channel coding, massive multiple input multiple output (mMIMO), Tera Hertz technologies, etc.

Although cryptography-based and physical-layer solutions have the capability to ensure communication security, there is always a certain possibility of information leakage. Instead, due to the fundamental principles of quantum mechanics, quantum communications provide a promising approach to realize unconditionally secure communications even when malicious nodes have unlimited computational power \cite{pan2024evolution}. Specifically, the no-cloning theorem states that an unknown quantum state cannot be perfectly copied and thus prevents the eavesdroppers from copying the transmitted quantum states. Quantum superposition means that information is encoded in quantum states that can be in a combination of `0' and `1' simultaneously, and the detection of these states collapses the superposition, making it impossible to extract the full information without knowing the correct measurement basis. Furthermore, quantum entanglement shows that any detection on one entangled particle instantaneously affects the state of the other, thereby making the entanglement suitable for secure communications. To this end, numerous quantum cryptographic or quantum communication protocols have been proposed, which can be classified into four main branches: quantum key distribution (QKD), quantum teleportation, quantum secret sharing and quantum secure direct communication (QSDC). Different from conventional cryptography, QKD utilizes the laws of quantum physics to distribute unconditional secret keys between a pair of legitimate parties rather than the classic wireless exchange, and the messages are encrypted by the agreed secret keys and then transmitted over a classic wireless channel. The QKD systems have been extended to ranges of hundreds of kilometres through optical fibers \cite{yin2016measurement} and thousands of kilometres in satellite systems of Micius satellite for quantum science experiments \cite{liao2017satellite}. QSDC encodes the secret message directly into quantum states and then allows the direct transmission of secret messages through a quantum channel without establishing a shared secret key first. It should be noted that QKD systems have become commercially available, while QSDC is still an active area of research and development, and its practical implementation is still in its early stages.

To sum up, the three mainstream security solutions constitute the built-in fundamental security capabilities in future DSIN. The requirements for embedded security capabilities in DSIN can be addressed by integrating these fundamental security capabilities in 6G and making the corresponding adaptive improvements. Hence, the functions of the collection, control, and isolation are provided. On this basis, DSIN can also leverage distributed FL technology to realize decentralized secure and trustworthy mechanisms, constructing a secure and reliable intelligent distributed satellite network. Consequently, the constructed network satisfies the differentiated security requirements of various service scenarios, enhances the autonomous security capabilities of network communications, and establishes a measurable and evolvable intrinsic security protection system.

\subsection{Testbeds of DSIN}
%Distributed satellite systems comprise constellations of multiple small satellites working in tandem to achieve objectives that traditionally would have required large, expensive, and singular satellites. These systems are highly beneficial for applications such as global internet connectivity, disaster management, environmental monitoring, and defense communication. Their scalability, cost-efficiency, and flexibility make them indispensable in modern satellite operations.

One of the key challenges faced by DSIN is ensuring reliable and efficient communication between satellites and ground stations, as well as inter-satellite communication. 
As satellite systems become more complex, the testbeds of DSIN are a critical infrastructure that facilitates the development and validation of advanced space communication technologies. As the space industry continues to evolve, several platforms have been developed to incorporate orbital simulations, inter-satellite communication, ground-station connectivity, and dynamic topology changes for SatCom systems.
These platforms are essential for assessing the performance of communication protocols under varying environmental conditions. 
Historically, a variety of satellite network simulators in the market such as iTrinegy's Network Emulators and the DataSoft Satellite Network Simulator, or in OpenSource projects like the Satellite Network Simulator 3 (SNS3), OPSAND and Real-Time Satellite Network Emulator, have been widely used for network simulations \cite{kodheli2020satellite}. However, the advent of DSIN requires testbeds that can handle the complexity of multi-satellite networking and their associated communication protocols. The development of these testbeds has evolved from simple, single-satellite simulations to sophisticated systems that can simulate entire satellite constellations.

The international application status of these platforms is marked by several key developments and trends. {For instance, the integration of SDN in satellite networks has been a significant step forward.} 
The SDN allows for the separation of the data plane and control plane, enabling centralized management of network resources and dynamic reconfiguration in response to varying traffic demands and network conditions \cite{quan2017enhancing}. This approach not only reduces the computational burden on individual satellites but also optimizes the use of network resources, catering to the high-dynamic and heterogeneous nature of satellite networks. {As a cost-effective SatCom solution for 5G}, the virtualisation of SatCom network functions is validated in the SAT5g project \cite{SAT-5G}, ensuring compatibility with the 5G SDN and NFV architecture. To address the challenge of analyzing and evaluating integrated space, aerial, and ground networks, a SAGIN simulation platform which integrates multiple network protocols, node mobility, and control algorithms is developed in \cite{cheng2020comprehensive}, optimizing the network functions such as access control and resource orchestration in a combined centralized and decentralized manner. While the flexibility and scalability brought by SDN/NFV technologies are beneficial for heterogeneous network architecture, the issues of deploying the SDN controllers in DSIN and coordinating the actions among the controllers require careful investigation. As a viable option, cloud/edge computing and network slicing can be incorporated into resource isolation, cloud services, computation offloading, and edge caching, {fundamentally meeting the varying requirements of ever-increasing services and applications.}
In addition, the development of testbeds like the DVB-RCS2/S2 has been instrumental in the design and validation of next-generation satellite systems \cite{6934571}. These testbeds provide a distributed environment that mimics the operational characteristics of real satellite networks, allowing researchers and engineers to experiment with various protocols, routing algorithms, and network configurations in a controlled setting.

A critical component of testbeds for DSIN is the constellation network simulation. The constellation network simulation includes orbital simulation, topology modeling, and link parameter analysis \cite{9814560}. Orbital simulation is essential for analyzing satellite link performance and designing communication topologies. By predicting satellite positions and simulating their operational behavior, these platforms enable accurate modeling of communication links and the dynamic nature of DSIN.
Using high-precision orbital models such as the Keplerian and SGP4 models is important for the DSCNs to ensure accurate simulations. The Keplerian model is used for scenarios with low precision requirements \cite{spiridonov2022small}, while the SGP4 model is a more advanced numerical integration model that accounts for factors such as Earth's nonspherical shape and the gravitational influence of the Sun and Moon \cite{10560107}. The use of these models allows researchers to simulate various scenarios, including LEO constellations and MEO satellites.
Furthermore, network topology construction based on simulation tools like NS3 allows the creation of different types of constellations, such as polar, Walker, and hybrid constellations, which are then visualized in 3D. This provides a robust framework for evaluating how satellite movement affects communication links and network performance. Specifically, the platform can simulate changes in the routing topology over time and evaluate metrics such as packet delay, packet loss, and bandwidth utilization.

%\item {Multi-satellite collaboration simulation and performance evaluation}:
%In the design of experimental platform for the DSCNs, various simulation tools can be leveraged to simulate orbit design, satellite or device mobility, and data communications. For instance, VISSIM is a powerful tool for traffic simulation and analysis by utilizing the embeded world map and GUI controller to simulate and display real world scenarios. STK provides highly comprehensive parameters and data of existing LEO, Medium Earth Orbit (MEO), GEO, and High Earth Orbit (HEO) satellite systems
%Therefore, how to design a unified experimental platform for DSCNs is an important yet challenging. issue.
Distributed computing resources for managing complex simulations are necessary for DSIN. Modern testbeds is envisioned to employ a distributed architecture where the simulation and computation modules are separated. This architecture allows for the efficient allocation of resources across distributed nodes, enabling to simulate large-scale satellite constellations.
The platform utilizes a central control center that manages task allocation and monitors resource usage, ensuring efficient execution of simulation tasks. By leveraging distributed computing resources, the platform can concurrently calculate satellite parameters such as orbit, attitude, link budgets, and topology changes, resulting in higher computational efficiency.
Moreover, simulation modules must support multiple communication protocols, including satellite-to-ground, inter-satellite, and intra-satellite communication protocols. These modules access real-time computation results from the distributed nodes and display them to the users, enabling the real-time monitoring of network performance. The inclusion of collaborative satellite scenarios further enhances the platform's utility by supporting tasks such as cooperative satellite computation, data transmission coordination, and interference management.
The simulation module is designed to cover four key multi-satellite collaboration scenarios: computational collaboration, transmission collaboration, interference coordination, and data injection coordination. For instance, when individual satellites face computational limitations or high latency, a multi-satellite computational collaboration strategy is used. This strategy leverages a grid model to track changes in the satellite network topology, selecting a central node as the computational coordinator. This node processes collaboration requests and returns a list of suitable satellites based on criteria such as maximum computational power and shortest transmission distance, enabling efficient inter-satellite collaboration.

Additionally, the testbed supports the configuration of distributed computational collaboration simulations, providing performance evaluations for these scenarios. For handling large volumes of remote sensing data, the system employs multi-link data transmission strategies, where data is split and transmitted in parallel through multiple channels, significantly reducing latency. The testbed also includes distributed multi-path data transmission simulations, supporting large data downlink tasks and enabling efficient evaluations of remote sensing images delivery.
Dynamic beamforming capabilities are used to allocate and manage frequency spectrum resources between satellites, optimizing the use of limited spectrum and avoiding interference. The interference management can achieve large-scale network construction and flexible spectrum allocation.
A testbed for MU-MIMO precoding in multi-beam satellite systems is designed by \cite{9148757} to demonstrate the feasibility of spatial multiplexing with full frequency reuse for video transmission across two co-located GEO satellites.
To handle large-scale computational model data injection and synchronization, a combination of SDN-enabled centralized and distributed routing mechanisms can be used to ensure rapid synchronization of ground network data. The platform supports the configuration of injection tasks, enabling collaborative simulation and performance evaluations of data injection across multiple satellites. In the beyond 5G NTN-terrestrial networks, the SDN-based testbed has been proposed to monitor the substrate network with traffic statistics and apply routing decisions, which allows for managing the procedure proactively to minimize traffic losses \cite{10154319}.

%\item {Integrated ground and ob-orbit testing}:
The need for integrated ground and on-orbit testing of various communication protocols should also be highlighted in the testbed of DSIN. Ground testing systems are designed to emulate the satellite environment by software and utilize four key components: space simulation systems, ground simulation systems, testing systems, and auxiliary equipment. This allows for functional verification of network devices and interoperability testing, ensuring that the technology meets specified standards.
On-orbit testing, on the other hand, involves deploying the satellite in a real-world environment, performing computing or inference tasks \cite{qiao2024orbit}, embedding a virtual network \cite{minardi2022virtual}, and evaluating its performance under actual conditions. This process verifies whether the communication protocols, technical standards, and system performance align with expectations. Furthermore, data obtained from on-orbit tests are used to calibrate the ground-based testing systems, enabling iterative refinement and optimization of DSIN.

\section{Emerging Directions}
In this section, we focus on the evolving trends in DSIN and the associated technical challenges, examining key aspects such as network resource virtualisation, distributed AI-enabled on-orbit information processing, semantic communications, direct satellite-to-device communications, and goal-oriented integration of sensing, communication, and computation. These elements are crucial for advancing DSIN to support a wide range of applications.
\subsection{Network Resource Virtualization in DSIN}
\begin{figure}[t]
	\centering
	\includegraphics[width=0.7\textwidth]{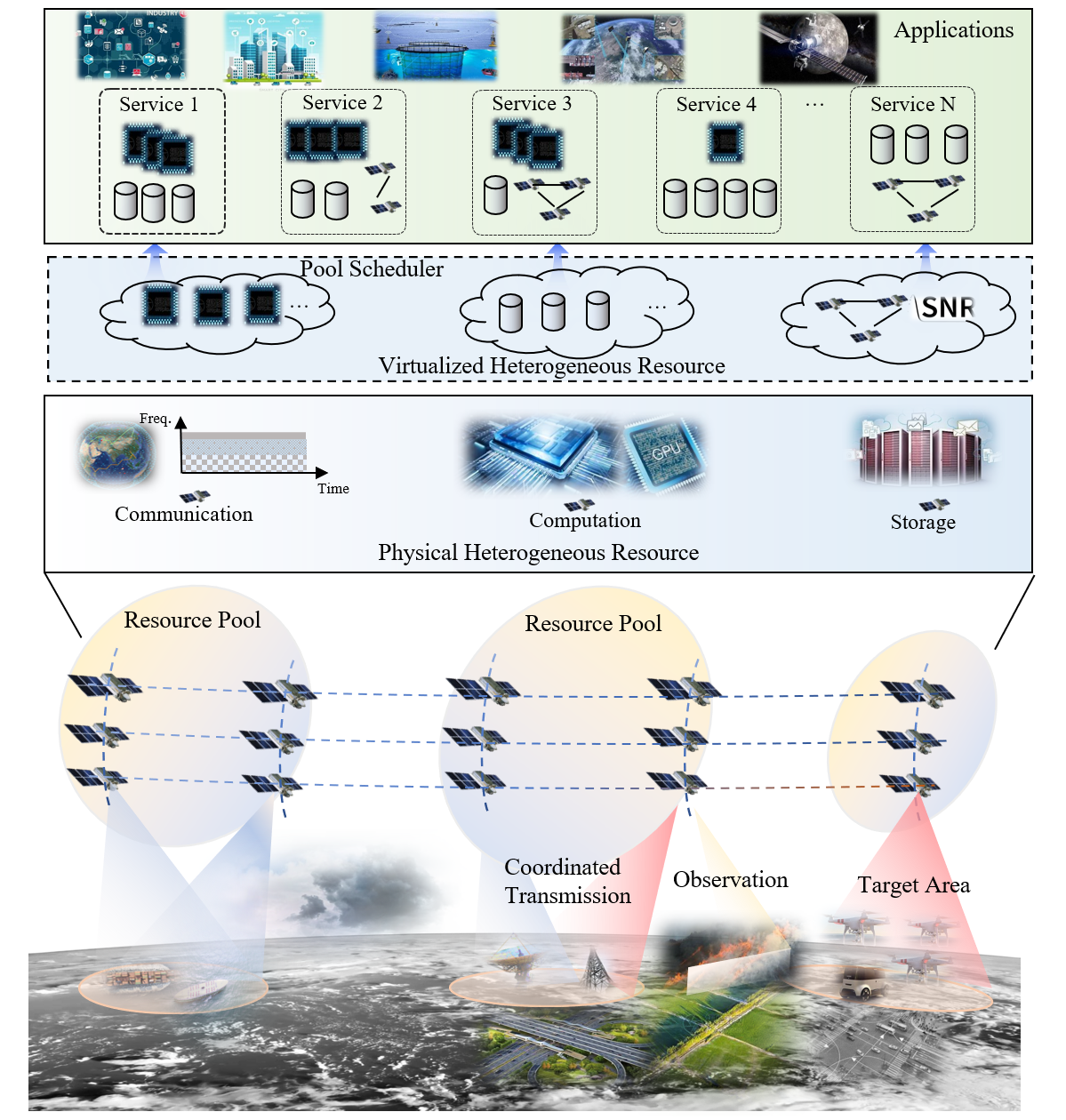}
	\caption{An illustration of resource pooling as a foundation for enhanced NRV.}\label{NRV}
\end{figure}

Resource constraints such as limited computational power, bandwidth, and storage present significant hurdles in the DSIN.
Unlike traditional terrestrial networks, satellite networks operate in constrained environments where heterogeneous resources must be managed with high precision \cite{9887918}.
%This limited-resource environment often results in coarse-grained resource allocation across various resource, i.e., computation, communication, and storage etc, thus leading to inefficiencies and underutilization.
Moreover, the fragmented and unbalanced resources across multiple satellites significantly hinder resource utilisation in DSIN, particularly when accommodating diverse intelligent applications or tasks.
For instance, urgent tasks, such as disaster monitoring, demand higher prioritization and faster resource deployment compared to routine data collection missions. However, the current satellite network lacks the adaptability to prioritize and scale resources based on these differences, leading to either under-allocation or over-allocation of resources and ultimately reducing energy efficiency.
As a promising technology to deal with the issue of resource management in DSIN, the network resource virtualization (NRV) abstracts space-air-ground physical resources to create virtualized resource entities and corresponding heterogeneous resource set for diverse services \cite{10236381}, enabling flexible and efficient resource orchestration, as shown in Figure \ref{NRV}.

%Through virtualization, DSIN can achieve benefits like fault isolation, efficient resource utilization, load balancing, and enhanced security \cite{10236381}. Additionally, virtualization facilitates the intelligent migration and management of virtualized heterogeneous space-air-ground resources, addressing fault tolerance and ensuring reliable, scalable on-demand resource services for emerging intelligent satellite-integrated Internet applications.
%As a promising approach, the network resource virtualization (NRV) abstracts and pools physical resources, enabling flexible and efficient allocation within complex, distributed systems.
%This concept emerged the need to optimize resource utilization across dynamically changing network environments.
%%The resource management features of future DSINs will include the following aspects: 1) frameless coverage definition; 2) users always
%% focused at the coverage center; 3) flexible serving set construction; 4) eliminating edge effects and the BS, UAV or satellite boundary; 5) eliminating the handover of users; 6) centralized signal processing.
In NRV-enabled DSIN, heterogeneous resources can be dynamically scheduled and prioritized to address the areas of greatest need, thereby optimizing key performance indicators across multiple satellites.
{Specifically, based on the distributed satellite computing network architecture presented in Section 2.2, the DSIN can decentralize the computation, storage, and networking
resources of satellites across different spatial environments, and connects them via ISL
to form a collaborative working framework. This architecture allows for the flexible allocation of satellite, ground, and edge network resources, enabling efficient information collection, processing, computation, storage, and distribution. It is particularly well-suited to meet the massive, intelligent information service demands of the future, which will be driven by the vision of the Internet of Everything.} 

Despite the potential of NRV, several critical challenges hinder its full integration into DSIN, the foremost being \textit{the lag effect in resource allocation caused by long-distance communication}. The physical distances between satellites introduce latency issues, potentially undermining the effectiveness of NRV, particularly in time-sensitive applications. Research efforts have explored multi-layer architectures and edge computing solutions to reduce latency \cite{10275860}.
By enabling data processing closer to the source, edge computing helps mitigate latency. In satellite networks, LEO satellites can function as edge devices, processing data locally before transmission. However, while edge computing alleviates some latency issues, it only provides a partial solution and demands substantial onboard processing capabilities, posing a challenge for smaller satellites with limited resources. Moreover, achieving
resource virtualization across heterogeneous satellite platforms presents challenges in forming a unified, interoperable resource scheduler.

As a foundation technology for enhanced NRV, the \textit{resource pooling} concept is proposed by \cite{xuRP} to achieve uniform resource management.
The resource pooling technology aggregates heterogeneous satellite network resources, i.e., processing power, bandwidth, and storage, into a centralized, shared pool accessible by any satellite.
The multi-dimensional optimization problem is typically NP-hard, but recent advances in meta-heuristic methods, such as genetic algorithms (GA), ant colony optimization (ACO), and simulated annealing (SA), have made it possible to solve many of these problems \cite{xu2013resource}.
Benefiting from this advance, we can utilize the NRV technology to meet the unique requirements of different tasks, allowing for task-oriented resource prioritization. For example, a real-time disaster monitoring mission could be allocated a higher share of computational and bandwidth resources than routine environmental observation tasks, thus ensuring faster data processing and transmission for time-sensitive applications.
To achieve resource pooling, physical pooling must be implemented to eliminate barriers between homogeneous and heterogeneous satellites, enabling a shared communication channel that facilitates resource sharing across users, instances, and time scales. In addition, logical pooling can be utilized to aggregate fragmented resources into a cohesive virtual pool in a software-defined manner. 
Specifically, this approach facilitates fine-grained and flexible resource management, enabling resources to be dynamically partitioned and allocated based on task requirements and swiftly released and reclaimed upon task completion. 
Such rapid allocation and deallocation mechanisms significantly improve resource flexibility and availability. 
It is worth noting that the edge computing combined resource pool architecture has shown promise in optimising resource sharing for satellite-ground cooperative tasks \cite{zhu2020two}.
However, a significant challenge in establishing a resource pool of clustered satellites lies in the dynamic topology changes caused by orbital movements, which result in instability. The core issue revolves around managing the capacity of the resource pool and dynamically selecting satellites in CCS system to ensure efficient resource sharing \cite{qin2022application}.

Another critical issue lies in the limitations of multidimensional heterogeneous resources in DSIN, which are confined to different network layers, lacking interoperability and unified global coordination.
To address this, {the virtualized hybrid satellite-terrestrial systems (VITAL) project introduces the network functions virtualization (NFV) and SDN technologies into satellite networks}, enabling flexible and resilient resource management in satellite-ground networks \cite{shenxuemin}.
The SDN facilitates dynamic resource management in heterogeneous satellite networks. By dynamically deploying SDN controllers in LEO satellite constellations, it becomes possible to effectively address network traffic fluctuations caused by variations in user geographic locations and time zones \cite{8647843}.
Moreover, the SDN/NFV technologies in SAGIN have been proven to effectively transform the network from a connection-based model to a service-based model, offering on-demand resource allocation and service customisation \cite{8460139}.
However, the inherent resource dynamics and service uncertainties in DSIN render traditional scheduling methods unsuitable for decision-making tasks requiring high efficiency and rapid response, such as service function chain deployment and mapping.
Reinforcement learning (RL), as a self-learning and adaptive decision-making approach, has been explored in SDN/NFV networks to some extent \cite{ku2017study}.
Nevertheless, how to apply RL to integrate lightweight SDN/NFV for enabling adaptive and scalable NRV in DSIN remains relatively scarce, highlighting a gap in the current research.

%
%\begin{itemize}
%  \item \textit{Integrating SDN and NFV to Enable Adaptive and Scalable NRV}: The synergy between SDN and NFV can enable a more adaptive, scalable NRV system within DSINs. SDN could dynamically route data across satellite nodes based on real-time resource availability, while NFV would provide virtualized network functions that adapt to changing satellite configurations and workloads. Future, with a focus on reducing power and bandwidth requirements, the lightweight SDN and NFV design specifically optimized for satellite environments can also be considered as promising directions.
%  \item \textit{Advanced Network Slicing with AI for Customized Service Provision}: Network slicing can be further enhanced with AI technologies to create dynamic, predictive models that optimize resource allocation for specific applications in DSCNs. For instance, AI-driven algorithms could predict the resource needs for applications like Earth observation or military surveillance, adjusting network slices accordingly. This would enable a more efficient allocation of NRV, enhancing the system's overall service capacity.
%  \item \textit{Robust Security Models Using Blockchain and Distributed Ledger Technology (DLT)}: Security challenges in NRV can be addressed by integrating blockchain or DLT to ensure transparency and tamper-resistance in resource allocation. By logging all resource allocation and usage events on an immutable ledger, satellite networks can enhance accountability and prevent unauthorized access.
%\end{itemize}

\subsection{Distributed AI Inference and On-Orbit Information Processing}
Given the constrained transmission capacity of satellite-to-ground links, traditional architectures reliant on GS struggle to address the growing demands for massive data return \cite{pachler2021updated,chen2021analysis}.
Moreover, the limited resources of a single satellite render it impractical to process all received data independently.
Consequently, leveraging on-orbit computing and storage resources across DSIN becomes essential for the real-time processing of vast sensor data collections.
{As outlined in Section 2.2, the distributed satellite computing network architecture integrates the available computing, communication, sensing, and storage resources of distributed satellites into a virtual resource pool for unified management. Onboard intelligent fusion of multi-source satellites is achieved by managing multi-dimensional and heterogeneous resources, enabling collaborative distributed computing power.}
Distributed AI technologies, such as federated learning \cite{mcmahan2017communication}, transfer learning and split learning \cite{wu2023split}, can be effectively employed within CCS system in DSIN to facilitate on-orbit information processing.
Moreover, With the rapid development of AI, large language models (LLMs) have become a star technology in the field of natural language processing. However, the inference process of LLMs often faces significant challenges in terms of computational resources and time costs, especially for on-orbit intelligent applications in satellite networks. To address this issue, distributed AI inference technology has emerged. Distributed AI inference involves breaking down the large language model inference task, which would typically be executed on a single machine, into multiple sub-tasks that can be processed in parallel across multiple nodes with limited computational power. This approach enables the efficient utilisation of computational resources from distributed and edge nodes, significantly improving on-orbit information processing and inference speed, while also reducing inference costs. During the distributed inference offloading process, managing a large number of computational resources, including CPUs, GPUs, and others, is crucial. To accelerate the model inference speed, NRV and distributed satellite computing network architecture, as mentioned in Section $2.2$ and $5.1$, can be employed to achieve compact and efficient computational power collaboration and task priority scheduling. Furthermore, in response to the dynamic network characteristics of DSIN, such as frequent topology changes and narrow communication time windows between the satellites and GSs, distributed incremental inference emerges as a promising technique. By learning the implicit relationships between various tasks or model information, it abstracts a distributed inference model with basic inference capabilities. This approach avoids full-scale learning and model updates, thereby accelerating model inference speed while reducing communication overhead.

The research on distributed AI in DSIN involves two architectures: Synchronous architecture and asynchronous architecture.
Synchronous architecture requires all satellites to update simultaneously to ensure state consistency, making them suitable for tasks that require real-time coordination.
However, they may suffer from reduced efficiency due to factors such as network latency.
In contrast, the asynchronous architecture allows satellites to update independently, providing flexibility and effectively accommodating delays and node failures, but the inconsistency in update pacing may increase coordination complexity.

\begin{figure}[t]
	\centering
	\includegraphics[width=0.7\textwidth]{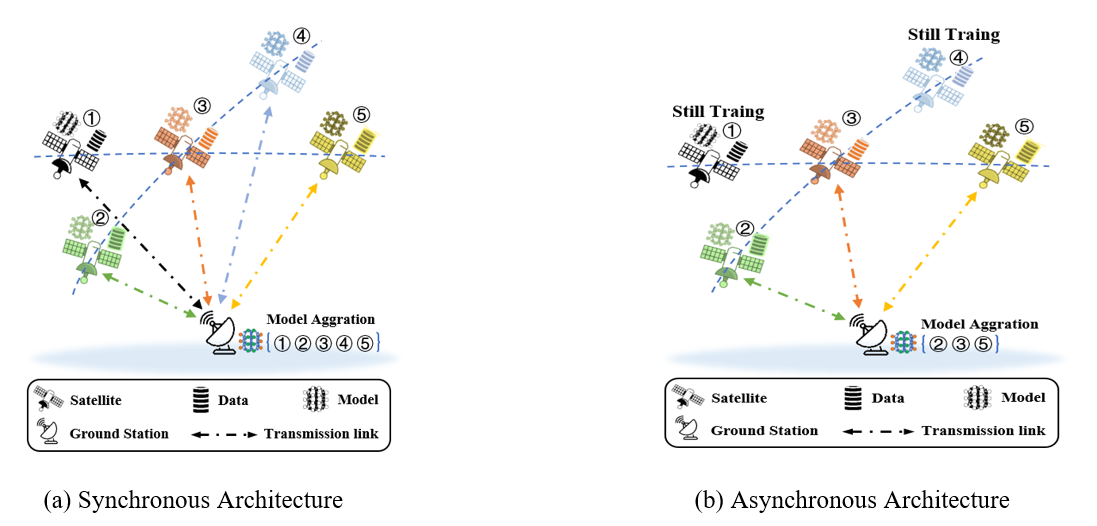}
	\caption{An illustration of synchronous and asynchronous architectures of distributed AI for on-orbit information processing.}\label{Architecture}
\end{figure}

In synchronous architecture, each satellite is equipped with edge computing capabilities and acts as a typical client group of traditional on-orbit distributed architectures.
Each satellite can train on its locally relevant data and, after several rounds of local training, obtain a high-performance local model.
This model is then sent to the GS for global model aggregation at the same time, as illustrated in Figure \ref{Architecture} (a).
Similar to the naive idea of FedAvg \cite{mcmahan2017communication}, after multiple global rounds of model training and aggregation, the CCS system ultimately generated a unified model suitable for global inference.
FedProx \cite{li2020federated} addresses each satellite's significant differences in data distribution by introducing a loss regularization term that utilises the current local model and the previous global aggregated model, ensuring that any onboard model does not deviate too far from the global model parameters.
FedNova \cite{wang2020tackling} treats the adjustment of learning rates and weight decay parameters as a joint optimization problem.
In each global round, after the server receives all local gradients, it calculates the global gradient and sends it back to each satellite, allowing them to adaptively adjust their learning rates and weight decay parameters to optimize local model performance.
To reduce the risks of single-point failure and communication congestion, the CCS system can perform local model updates through communication between co-orbital satellites and cross-orbital exchanges, enabling satellites to collaboratively train models without a central server \cite{10092560}.

The advantages of deploying distributed AI in synchronous architecture lie in the simplicity and intuitiveness of the algorithms.
This makes them easy to implement and manage while ensuring that all satellites are at the same state during the training process. However, a significant drawback is the low global efficiency, as faster nodes must wait for slower ones, leading to resource wastage in CCS system. Additionally, delays or failures from any single node can impact overall performance.
As the number of satellites in the CCS system increases, the waiting time for model aggregation will also grow.
Therefore, synchronous architecture has poor adaptability to constellations with varying onboard resources.

In asynchronous architecture as shown in Figure \ref{Architecture} (b), the clustered satellites in CCS system cannot update or aggregate at the same time.
Therefore, there is a greater need to focus on the impact of single satellite model lag on global aggregation.
Previous research has extensively explored optimizing local training and global model aggregation.
By proposing an improved asynchronous federated learning method to enhance robustness in heterogeneous environments, this approach focuses on utilizing the predictable availability of satellites and introduces a new communication protocol and algorithm framework to improve the efficiency of the training process \cite{razmi2022ground}. Considering the deployment of distributed AI orchestrated by an external constellation parameter server, a novel communication scheme has been proposed that leverages inter-satellite links within the orbit, the predictability of satellite movements, and partial aggregation methods to significantly reduce training time and communication costs \cite{razmi2022boardhuiyi}.
FedSpace \cite{so2022fedspace} dynamically schedules model aggregation, leveraging the deterministic and time-varying connectivity of satellite orbits and Earth's rotation.
An asynchronous satellite system AsyncFLEO \cite{10021101} that utilizes high-altitude platforms as parameter servers, addressing both idle waiting in synchronous FL and the model staleness issues caused by lagging satellites.
An asynchronous distributed algorithm called FedGSM \cite{10304545}, introduces a compensation mechanism that leverages the deterministic and time-varying characteristics of satellite orbits to mitigate the adverse effects of gradient staleness.
The proposed asynchronous distributed algorithm called FedBuff \cite{Nguyen2021FederatedLW}, utilizes buffered asynchronous aggregation to address scalability and privacy issues in cross-device FL.

Compared to synchronous architecture, asynchronous architecture has the advantage of improving computational efficiency in CCS system, as each satellite can independently update its model, reducing global wait time and allowing all satellites to adapt to varying computational capabilities and network delays.
When some satellites fail, other satellites can continue to update, which may accelerate the convergence of the global model in certain scenarios.
However, the drawbacks of asynchronous architecture include instability in the global state.
The independence of the satellites can lead to inconsistencies in the global updates, resulting in fluctuations in the performance of the global model.
Additionally, frequent asynchronous updates may increase network burden and communication overhead.
The algorithms involved in asynchronous updates are also more complex, requiring greater technical investment to design effective asynchronous mechanisms.

For distributed on-orbit information processing, both synchronous and asynchronous architecture face significant technical challenges and research opportunities.
%The communication connections in dynamic and heterogeneous satellite networks are extremely complex.
Due to the relative motion between satellites, the constellation topology changes rapidly, making it difficult to maintain stable and efficient inter-satellite links. Optimizing resource allocation among satellites is crucial. Each satellite has limited and heterogeneous computational, energy, and storage resources, necessitating the deployment of effective algorithms for real-time task and resource allocation \cite{10042025}.
%In a decentralized scenario as shown in Figure \ref{Architecture}(b), it is essential to label each satellite in the constellation, perform layered pooling of critical node knowledge, and manage the scheduling of numerous nodes appropriately.

\subsection{Semantic Communications in DSIN}
The 6G-enabled DSIN are expected to provide ubiquitous intelligent services with stringent QoS requirements in dynamic network environments, which prompts a shift from conventional architectures that focus on high transmission rates between transceivers to new architectures centered on intelligent connectivity of everything.
The resulting extensive scales of data which require a massive amount of bits to represent, impose a huge burden and bottleneck on communication systems that handle only bit-level reconstruction, especially on DSIN with limited transmission power, storage space and computing resources \cite{5.3-1}.
In response, a novel paradigm known as semantic communication is inspired as a potential technology to break the bottleneck by utilizing the semantics and content of data to optimize the utilization of communication resources \cite{5.3-2,5.3-3}.
{For example, when CCS system collaboratively executes a remote sensing task, each satellite extracts semantic information about the target in orbit and transmits it to the ground station or device. This effectively reduces the volume of data transmitted over the link and improves transmission timeliness. In addition, in poor channel conditions, semantic communication proves more robust to fading than traditional bit-level reliable transmission methods, offering greater resilience for DSIN \cite{10485510,10486973}. In this sense, semantic communication can be viewed as a promising development direction for the future of DSIN.}

\begin{figure}[t]
	\centering
	\includegraphics[width=0.9\textwidth]{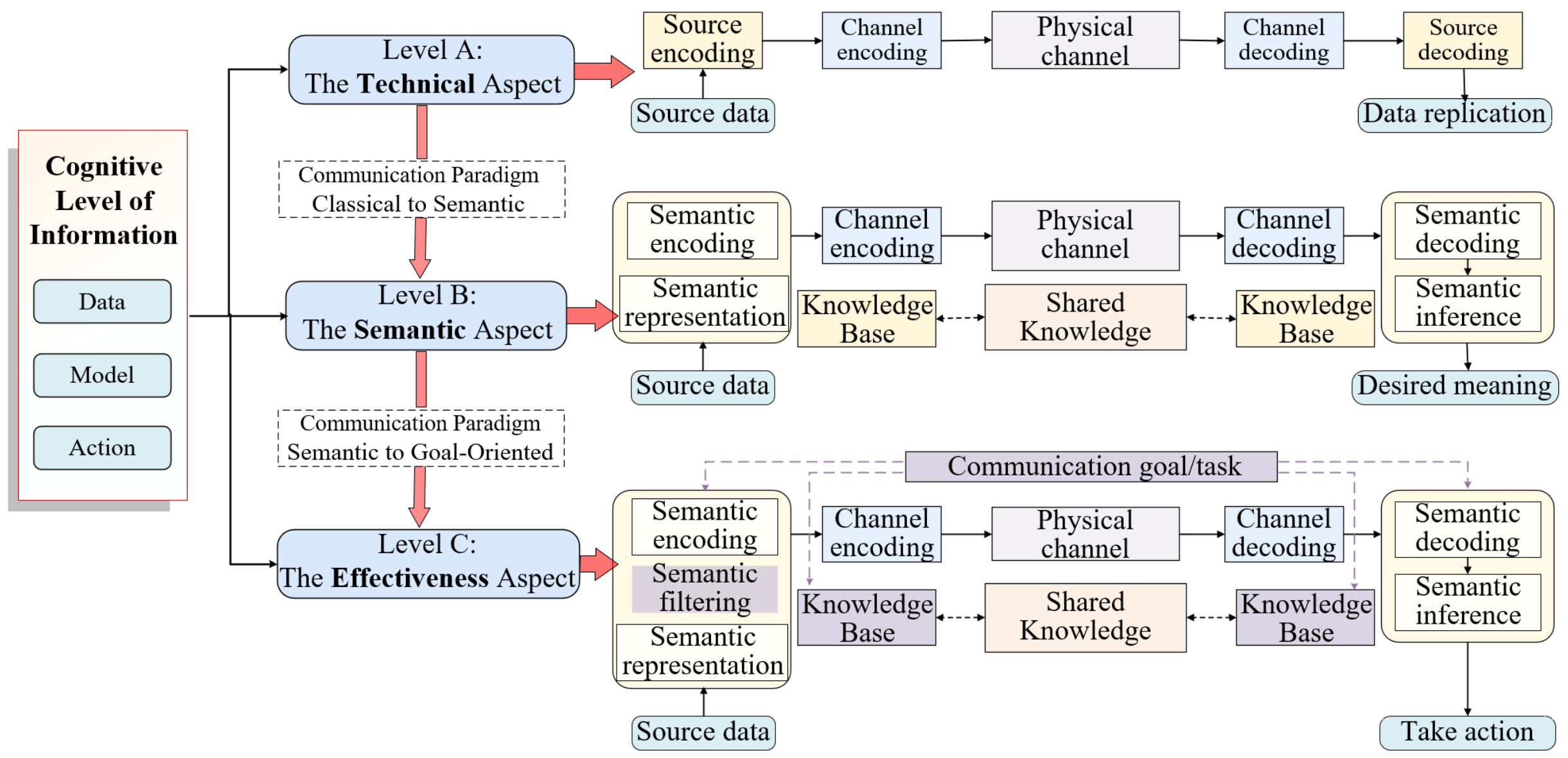}
	\caption{An illustration of general communication system models for three communication levels.}\label{semantic} %%1119Í¼ÒªÌåÏÖÓïÒåÍ¨ÐÅÓëDSINµÄ¹ØÏµ
\end{figure}

Specifically, motivated by the three communication levels identified by  Shannon and Weaver \cite{5.3-4}:
\begin{itemize}
	\item Level A addresses the technical problem, which answers ``How accurately can the symbols of communication be transmitted?"
	\item Level B tackles the semantic problem, and asks ``How precisely can the transmitted symbols convey the desired meaning?"
	\item Level C solves the effectiveness problem, where ``How effectively do the received symbols affect conduct in the desired way?" is the core.
\end{itemize}

Semantic communications in DSIN enable all participants to achieve Level B and Level C communication with minimal overhead by leveraging advanced AI technology \cite{5.3-5}.
This approach ensures that only the most valuable and contextually relevant information is transmitted to applications, efficiently distilling insights from massive datasets and delivering them to their destinations at the right time, as illustrated in Figure \ref{semantic}.
As a result, it serves as a key technology for ensuring high QoS service in DSIN via extracting, transmitting, and evaluating critical semantic information.
Moreover, the advancement of semantic communication and the construction of DSIN are mutually reinforcing.
On one hand, the distributed computing and AI networks facilitated by space-terrestrial collaboration support the large-scale deployment of semantic communication systems \cite{5.3-6}.
On the other hand, semantic communication significantly enhances network performance, particularly in hardware-constrained satellite networks, which effectively broadens the applications and versions of DSIN.
Currently, the works on semantic/effectiveness-empowered communications primarily follow two approaches: The application-centric engineering implementations and the universal representation of information importance.

\textit{1) Application-Centric Engineering Implementations.}
This approach stems from attempts to leverage DL at the physical layer for representing underlying data meanings in the absence of a rigorous theoretical framework for semantic communications \cite{5.3-7}.
In this case, the works in this vein focus on the architecture design and engineering implementation of semantic compression and extraction.
An early prototype of end-to-end semantic communication is Deep Joint Source-Channel Coding (Deep JSCC), in which textual sentences are encoded into fixed-length bit streams within a simple channel environment \cite{5.3-8}.
Following this, several researchers have developed semantic communication systems tailored for image \cite{5.3-9} and speech \cite{5.3-10} transmission based on this framework, with the goal of achieving efficient reconstruction of image and speech data at the receiver, respectively.
In addition, for task-specific applications, semantic communication systems must be capable of extracting task-relevant information at the transmitter and performing correct decision-making relying on the limited received information at the receiver.
For instance, a speech recognition-oriented semantic communication system has been developed in \cite{5.3-11}, where the receiver can reduce the volume of data from the transmitter by converting the received speech signals into text, and then applying speech synthesis to reconstruct the original signal.
Similarly, to achieve accurate image recognition with minimal overhead, the author in \cite{5.3-12} proposed a classification-oriented semantic communication system, which can accurately identify various objects based on the transmitted image features, rather than the full image, under limited bandwidth and power constraints.

Moreover, given the practical necessity of gathering multi-modal data from diverse users/devices and fusing them at the receiver, a Transformer-based framework with high representational capacity can be employed to unify the transmitter structure across various tasks \cite{5.3-13,5.3-14}.
This approach enables the extension from single-user, single-modal systems to multi-user, multi-modal scenarios, such as integrating text and image information for complex tasks like Visual Question Answering (VQA).
Furthermore, the complexity of multi-modal multi-tasks and the variability of transmission environments raise the need for stronger generalization in semantic communication systems, which drive the advancements in Knowledge Base (KB)-based approaches \cite{5.3-15}.
In typical satellite networks, such as onboard remote sensing processing, KB-sharing mechanisms are widely integrated into semantic encoding modules to enhance the utilization of ground-based remote sensing knowledge and expert insights which can sustain long-term semantic alignment between transceivers.
Therefore, the KB-assisted semantic extraction further strengthens the ability of satellites for semantic representation across multi-sources data and multi-modal tasks in complex space-terrestrial channels \cite{5.3-16}.

%\subsubsection{Universal Representation of Information Importance}
\textit{2) Universal Representation of Information Importance.}
This approach focuses on the etymological meaning of semantics, i.e., the importance or priority of information \cite{5.3-17}.
From this perspective, the priority is not determined by classical information entropy but rather relies on the semantics conveyed by the information, where the semantic metrics are no longer confined to specific applications outlined in the first approach, such as the semantic distance in text or Structural Similarity Index (SSIM) in images \cite{5.3-18};
Instead, they serve as a more comprehensive multidimensional metric that reflects the overall performance of the network.
Within such a paradigm, the semantic communication systems are capable of recognizing high-priority information and adjusting resource allocation accordingly.
For example, a packet is given more priority when its destination has not been updated for a while in the AoI-based framework \cite{5.3-19}.
However, the assessment of information importance in AoI relies on an assumption that fresher messages always contain more valuable information, which makes it overlook the differing temporal requirements in various applications and fail to capture how the content of information impacts the task execution.
In response, the Value of Information (VoI) is born to address the above problems, typically denoted by a non-linear penalty function of AoI, i.e., $f({\rm AoI}(t))$ \cite{5.3-20}.
Thus, on one hand, VoI can capture the varying sensitivity to freshness across different applications by mapping age penalty to linear, exponential, or logarithmic functions \cite{5.3-21}.
Nevertheless, the selection of penalty functions completely relies on empirical setups, and thus some efforts aim to derive the function stringently, such as the mutual information-based construction \cite{5.3-22}.
On the other hand, it can reflect the differences in information content related to the transceiver state $(X_t,\hat{X}_t)$ through a content-aware error penalty function $g(X_t,\hat{X}_t)$, such as Mean Squared Error (MSE) or threshold error \cite{5.3-5}.
It is worth noting that the error function can also be regarded as a kind of nonlinear AoI.

Moreover, since the above metrics only focus on one attribute of information at a given time, several semantic metrics that combine multiple attributes have been proposed.
One of the most notable examples is Age of Incorrect Information (AoII) \cite{5.3-23}, which quantifies the utility reduction from asynchronous duration between transceivers by simply integrating $f({\rm AoI}(t))$ with $g(X_t,\hat{X}_t)$ \cite{5.3-7}.
On this basis, by further incorporating a cost function $C$ dependent on real-world constraints like bandwidth limits and energy consumption, the authors in \cite{5.3-24} propose a three-dimensional Utility of Information (UoI) evaluation framework that can simultaneously encompass timeliness, accuracy, and energy efficiency of information.
Further, UoI expands the error function $g(X_t,\hat{X}_t)$ to capture not only packet-level physical process mismatches but also the application-layer semantic states required for extraction and recovery, such as KBs \cite{5.3-25}.
In this regard, the improvement both broadens the application scenarios of UoI and initiates a preliminary integration with the first approach.

Overall, the current semantic-empowered communication, particularly in the context of DSIN, still faces several challenges for future investigations: \textit{1) Research in Semantic Information Theory:} The shortcomings in foundational theoretical research have led to issues such as the lack of interpretability and explainability in semantic extraction, as well as an unclear relationship between semantic rate and Shannon bit rate, which make the real implementation of semantic communication systems on DSIN challenging.
\textit{2) Flexible Access and Handover:} The inherent flexibility of satellites, especially those in CCS systems, requires constant adaptation of communication modes in response to real-time environmental changes, which imposes more demands on the generalization capabilities of semantic communication systems, including the time and resources needed for model retraining and the continual expansion of KBs in dynamic environments.
\textit{3) Inconsistent KB and Sharing Costs:} The information acquisition from the broad coverage of DSIN will create inconsistencies in KBs across different nodes, leading to mismatched semantic networks.
However, frequent updates to maintain semantic network alignment amid constantly expanding KBs are extremely time- and resource-consuming.

\subsection{Goal-oriented  Integrated Sensing, Communication, and Computation}
\begin{figure}[t]
	\centering
	\includegraphics[width=0.9\textwidth]{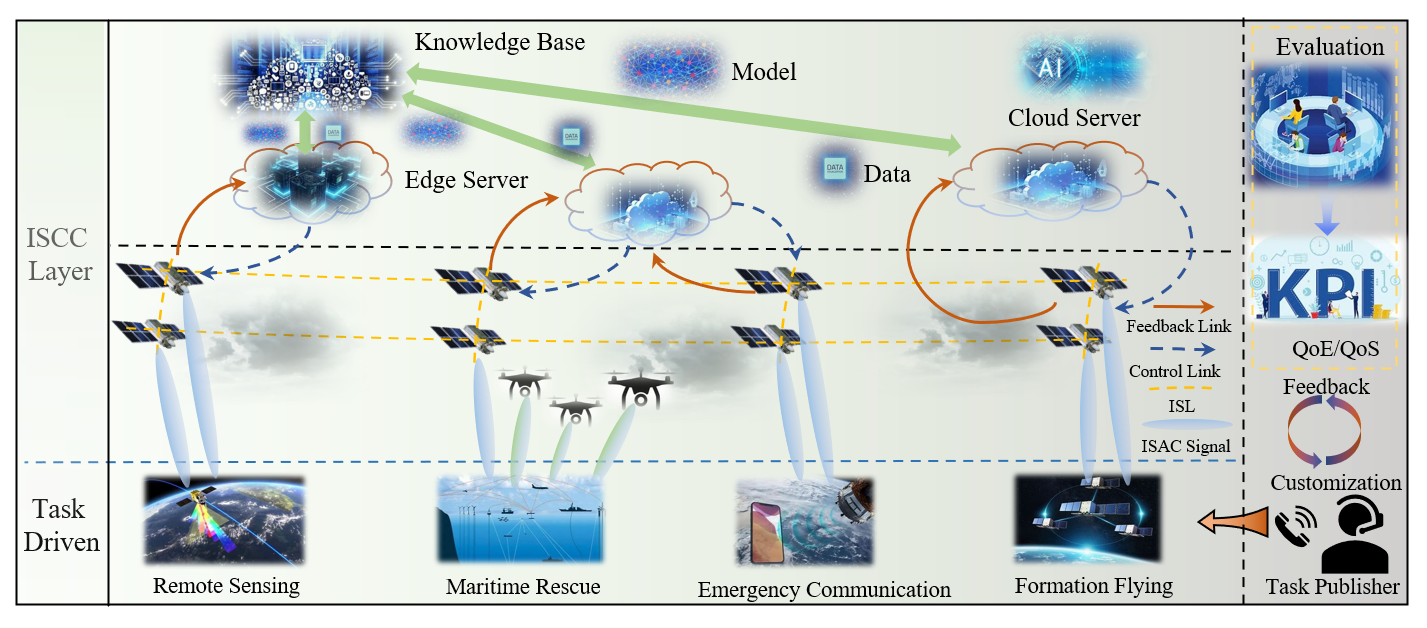}
	\caption{{An illustration of goal-oriented ISCC for DSIN.}}\label{ISCC}
\end{figure}

%In recent years, the rapid advancement in satellite communication technologies has led to a paradigm shift toward highly integrated systems that simultaneously perform sensing, communication, and computation tasks. This shift has been driven by the growing demands for more efficient resource utilization, real-time decision-making, and enhanced system intelligence. The integration of these three functionalities, ISCC, provides new opportunities to optimize the use of bandwidth, energy, and computational resources for DSINs.
As DSIN continues to advance, a diverse range of communication-assisted applications is emerging, including remote sensing, disaster management, and environmental monitoring, each requiring integrated sensing and computational capabilities alongside distinct communication performance metrics \cite{10480327}.
For instance,
%environmental monitoring systems demand ultra-low latencies, typically within 1 to 5 milliseconds, combined with exceptional transmission reliability of 99.999\% or higher for critical scenarios.
%Additionally,
emerging applications such as on-orbit object detection and tracking necessitate not only reliable communication capabilities but also sophisticated computing and sensing functionalities, underpinned by the satellite network infrastructure.
As a transformative concept within DSIN, the integrated sensing, communication, and computation (ISCC) has begun to attract increasing attention from both the academic and industrial sectors \cite{you2024ubiquitous,zhang2024joint}, and has recently been adopted among the key usage scenarios for IMT-2030/6G by the radio communication division of the International Telecommunication Union (ITU-R) \cite{kaushik2024toward}.
While recent studies have proposed several approaches to address these challenges of ISCC in satellite networks, i.e., ISCC in air-ground integrated networks \cite{fei2023air} and S-IoT \cite{10480327}, goal-oriented ISCC  in DSIN remains under-explored in several key areas, with much of the research in preliminary phases. As shown in Figure \ref{ISCC}, the goal-oriented ISCC mechanisms hold the potential to redefine how we utilize heterogeneous space-air-ground resources to provide on-demand services for diverse applications or tasks, bridging gaps between sensing, communication, and computation to create autonomous and resilient information network.
Nevertheless, integrating goal-oriented ISCC into DSIN faces several significant challenges, primarily stemming from the limitations of heterogeneous resources and the inherent difficulties of distributed collaborative scheduling within DSIN.
%Unlike traditional satellite networks, where sensing, communication, and computation function as discrete processes, ISCC integrates these elements into a cohesive and adaptive framework. This integration enables more efficient resource utilisation, real-time decision-making, and enhanced system intelligence, revolutionising how data is collected, transmitted, and processed within DSIN \cite{10480327}, as shown in Figure \ref{ISCC}.
%While recent studies have proposed several approaches to address these challenges of ISCC in satellite networks, i.e., ISCC in air-ground integrated networks \cite{fei2023air} and S-IoT \cite{10480327}, goal-oriented ISCC mechanisms in DSIN remain under-explored in several key areas, with much of the research in preliminary phases. The goal-oriented ISCC mechanisms hold the potential to redefine how we utilize heterogeneous space-air-ground resources to provide on-demand services for diverse tasks, bridging gaps between sensing, communication, and computation to create autonomous and resilient systems. Integrating goal-oriented ISCC into DSIN faces serveral significant challenges primarily from the limited hardware capacity, stringent power budgets, and high latency inherent to heterogeneous satellite networks:

%As shown in Figure \ref{ISCC}, the integration of ISCC in DSINs addresses several key roles, from high-precision data collection and effective resource utilization to adaptive task scheduling. In other words, the ISCC needs to optimize satellite capabilities across multiple applications, including remote sensing, disaster management, and environmental monitoring \cite{10480327}. For example, in disaster response scenarios, ISCC enables rapid data acquisition and immediate on-orbit data processing, which minimizes the delay in actionable information reaching ground control centers. This capability enhances satellite network responsiveness, making ISCC an essential component in DSINs.

From the computing perspective, the DSIN are restricted by the computational limitations of on-orbit hardware, which impacts the real-time processing capabilities.
Traditional satellites rely on GSs for data processing, causing large delays that undermine mission-critical applications.
Current solutions, such as deploying high-performance processors, are limited due to power constraints and radiating availability.
One promising direction is the use of edge computing \cite{xing2023task}, where data is processed closer to the sensing source (e.g., on the satellite itself or nearby satellites). This reduces latency and conserves bandwidth by minimizing the amount of data transmitted to GS.
Another promising approach involves hybrid edge-cloud solutions, where less time-sensitive tasks are processed in terrestrial cloud servers, while critical tasks are handled at the edge. However, this model necessitates robust coordination across network layers. Building on this, some research advocates for the use of distributed computing within the CCS system to manage resources more effectively. Despite its potential, this approach faces significant challenges, particularly due to the instability of inter-satellite communications \cite{fei2023air}.

From the communication perspective, the spectrum resources are actually limited in CCS system, leading to potential conflicts and interference. Besides, frequent handovers in CCS system further exacerbate the accuracy of information perception \cite{you2024ubiquitous}.
Thus, goal-oriented dynamic spectrum management, where satellites adaptively select frequency bands based on the availability of resources and the need by task, has been proposed as a solution.
However, goal-oriented dynamic spectrum management requires advanced low-complexity algorithms and policies to manage spectrum resources efficiently, particularly when coordinating with terrestrial networks.
To deal with this issue, some existing approaches utilize AI-driven or multi-agent cooperative scheduling approaches to maximize resource utilization \cite{sun2024integrated, sun2024integrated}.
The AI and ML techniques aim to allocate resources dynamically based on real-time satellite and mission status, allowing for enhancing the adaptability of ISCC in DSIN, but they are still in the early stages and remain susceptible to inaccuracies in rapidly changing environments, such as unexpected environmental conditions or equipment malfunctions.
In a multi-agent CCS system, each satellite is equipped with multiple onboard sensors, such as radars and cameras, enabling the capture of multi-modal, task-specific environmental data and its on-orbit processing. By integrating the dual functions of radio sensing and communication within a unified infrastructure, the goal-oriented ISCC in the CCS system can facilitate signal-level sensing information fusion, complementing information-level exchange among agents \cite{sun2024integrated}.

However, the mutual interactions among distributed satellites make the design of online ISAC strategies highly complex. While DRL has been explored for ISAC strategy development in multi-agent CCS systems, the dynamic nature of target determination and ISAC strategy design introduces substantial computational and operational complexity, limiting the applicability of traditional DRL algorithms.
In addition, the ISCC operations consume significant power, posing a challenge in maintaining energy efficiency. Energy-aware sensing, computation, and communication protocols have been developed to address this issue \cite{10437374,10159441}, yet they often involve trade-offs in performance, which can undermine the effectiveness of ISCC in real-time applications. Research into power-harvesting technologies in space, such as advancements in solar panel efficiency, shows promise but remains limited in its impact, particularly for low-power satellites.

\subsection{Direct Satellite-to-Device Communications}
With the breakthrough development in satellite communication techniques and the continuous reduction in manufacturing and launch costs of satellite platforms, the DSIN is expected to enable ubiquitous services and diverse applications to user terminals, i.e., direct satellite-to-device communications.
This application, as implied by its name, seeks to provide high-performance direct connectivity between satellites and common devices, and can further broaden its services to include mobile phones, and IoT terminals deployed in vehicles, aircraft, and maritime platforms \cite{5.5-1}.
Therefore, different from the conventional Mobile Satellite Communication Systems (MSCS) that focus solely on industrial users, direct satellite-to-device communications can offer diverse services for consumers in regions lacking coverage from terrestrial mobile networks with voice, messaging, and broadband Internet connectivity, which significantly expands the market size of SatCom \cite{5.5-1}.
Its commercial potential has also fostered deep collaboration across the entire industry chain, including satellite manufacturers and operators, cellular equipment manufacturers, mobile phone manufacturers, and mobile network operators, emerging as a focal point of interest among academia, industry, and research communities over the past three years \cite{5.5-3}.
Currently, various international organizations and telecommunication regulatory authorities from multiple countries have expressed their intention to promote the development of direct satellite-to-phone communication with substantial policy support.
Furthermore, industry leaders across related sectors have continued to increase their strategic investments in this area, making competition increasingly fierce.

\subsubsection{Technical Routes}
As of now, a unified route for direct satellite-to-phone communication has yet to be materialized, and it can be roughly categorized into three types based on its commercial progress.
\begin{itemize}
	\item \textit{Dual-Mode Mobile Device with Satellite-Terrestrial Independent Systems}: This route involves embedding dedicated SatCom chips or specific waveforms into a common mobile device and utilizing existing MSCS to provide services for these intelligent devices.
	This technical route is relatively mature, with products already available, notably Huawei's Mate60 Pro \cite{5.5-4} and Apple's iPhone 14 \cite{5.5-5}, which utilize the TianTong-1 and GlobalStar satellites to provide direct connectivity services, respectively.
	However, due to the non-standardized technologies of MSCS and their outdated air interface protocols, the communication capabilities of dual-mode devices are primarily restricted to voice and messaging, intended to provide emergency information transmission services for users in remote areas.
	Thus, it is mostly viewed as an interim solution for direct satellite-to-phone communication.
	
	\item \textit{Mobile Device with Terrestrial System Directly Connect to Satellites}: This route aims to provide direct connectivity services without altering the configurations of current ground-based mobile phones by manufacturing and launching satellite constellations specifically designed for this application.
	The primary challenge of this route lies in overcoming the significant attenuation associated with long-distance satellite-terrestrial transmission, particularly given the extremely limited performance of mobile device antenna.
	In response to this challenge, ASTS utilizes the BlueWalker-3  satellite equipped with a 64 m$^2$ phased array antenna to conduct 5G connectivity tests with existing mobile phones \cite{5.5-6}.
	Meanwhile, Lynk Global plans to deploy a complete LTE network in space and has currently completed the field experiments of the bidirectional voice communication between the test satellite and existing mobile phones \cite{5.5-7}.
	Furthermore, SpaceX introduces the Starlink V2 satellite, which features an additional 25 m$^2$ array antenna compared to V1, and achieves a downlink communication speed of 16.9 Mbit/s with unmodified mobile phones during tests \cite{5.5-8}.
	Nevertheless, while this route can provide broadband internet services, it places all modifications on the satellites, leading to higher engineering implementation costs.
	
	\item \textit{Integration of Satellite-Terrestrial Protocols based on 3GPP NTN}: This route requires the adoption of a unified protocol on both the satellite and terrestrial sides, enabling the next generation of mobile phones to have the capability to access both satellite and terrestrial networks simultaneously.
	To realize this vision, the NTN working group in 3GPP has completed the 5G New Radio (NR) standard for NTN in Release-17 \cite{5.5-9} and its enhancement technologies to support direct satellite-to-device service in Release-18 \cite{5.5-10}.
	Moreover, 3GPP plans to conduct research on supporting the standardization of spaceborne processing mode and spaceborne BS in Release-19 and Release-20 \cite{5.5-11}.
	In terms of standard implementation, satellite operators like Omnispace and EchoStar, equipment manufacturers such as ZTE, Unisoc and MediaTek, and mobile network operators like China Telecom, and China Mobile, have initiated validation work for spaceborne BSs based on 3GPP NTN standards and its performance for direct NTN satellite-to-device communications.
	However, the time required for standards to move from development to implementation is considerable, which hinders the rapid commercialization of this route.
\end{itemize}
\begin{figure}[t]
	\centering
	\includegraphics[width=0.9\textwidth]{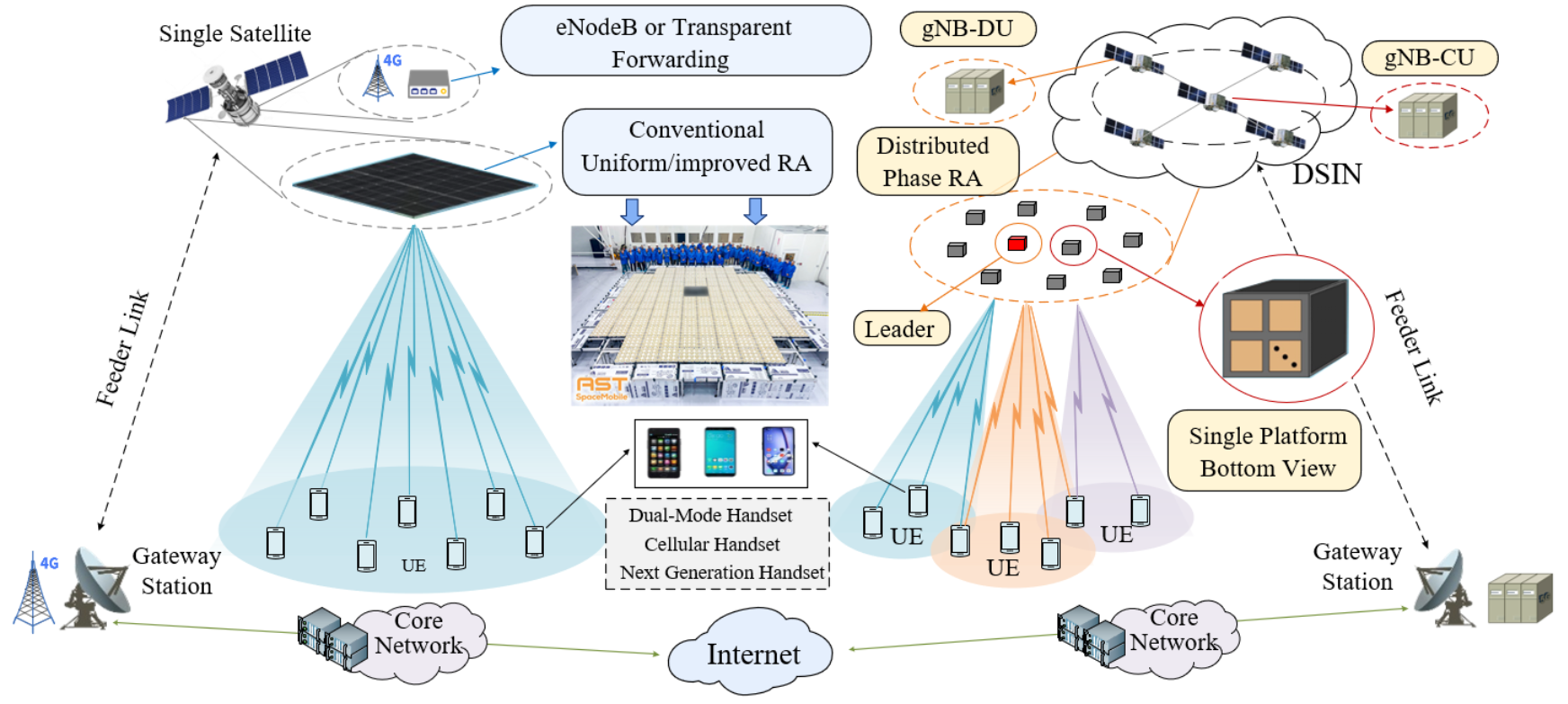}
	\caption{Comparison of direct satellite-to-device communication in DSIN and existing satellite communication system.}\label{phone}
\end{figure}

\subsubsection{Challenges and Solutions}
In practice, direct satellite-to-device communication encounters numerous technical challenges.
Firstly, due to the size and shape limitations, the typically low-gain omnidirectional antennas used in existing mobile phones struggle to establish communication with conventional satellites over long distances with significant signal loss.
As a result, improvements in aspects such as antenna design, power consumption, and weight constrain are needed for the next generation of satellites to enhance the transmission and reception capabilities of spaceborne antennas \cite{5.5-12}.

Another challenge lies in the spaceborne payload design.
Note that spaceborne BS deployment is a key trend in SatCom, as it significantly reduces network latency and leverages inter-satellite links to enable large-scale satellite networking, which can provide more efficient service for mass mobile users.
Following this, the weight and power limitations of satellite platforms impose stringent requirements on spaceborne BS payloads, necessitating advancements in lightweight design, high integration, thermal management and so on \cite{5.5-13}.
Although the partial BS deployment based on DU-CU separation is a potential scheme enabling effective tradeoffs between payload and cost, consensus on how to segment functional modules for optimal performance within acceptable cost remains elusive.

Additionally, to address the challenges posed by high Doppler shifts from satellite movement, time-varying network topologies, and uneven spatial and temporal distribution of traffic, etc, it is essential to further enhance the adaptability of existing SatCom protocols to the characteristics of satellite networks.
More precisely, the technology advancements in adaptive coding and modulation, time-frequency synchronization among DSIN, large-scale random access, intra-/inter-satellite beam switching, frequency-sharing between satellite and terrestrial network, and so forth \cite{5.5-14}, are necessary to meet the demand for high QoS to mobile users with direct connectivity.

Motivated by the above analysis, DSIN can be seen as an efficient transmission architecture to address the challenges as shown in Figure \ref{phone}.
\begin{itemize}
	\item \textit{Distributed Satellite-Based Array Antennas}: The satellites equipped with standard array antennas can tune with each other to create a larger equivalent aperture, providing substantial gain and narrow beam coverage \cite{5.5-15}.
	Alternatively, they can also operate as a virtual MIMO system through collaboration, allowing individual mobile users to receive signals from multiple satellites \cite{5.5-16}.
	The above configuration enables coherent signal transmission and reception, i.e.,  enhances the desired gain while suppressing interference, ultimately improving the SNR and spatial resolution.
	Moreover, the deployment of sub-arrays or units of a distributed array antenna across various satellite platforms cleverly circumvents the challenges posed by the conventional approach of loading an ultra-large multi-beam array antenna on a single satellite, which includes the technical demands of large deployable array antennas and the stringent design requirements for rocket fairings.
	As a result, the distributed satellite-based array antenna significantly reduces the costs of satellite production and launch while offering similar gains.
	
	\item \textit{Degree of Freedom and Fault Tolerance}: The distributed nature of DSIN introduces greater flexibility regarding inter-satellite spacing, number of satellites, payload configurations, etc \cite{5.5-17}.
	Considering the design of spaceborne payload, a BS can be categorized into different functional modules, and then be allocated to various satellite platforms within a CCS system formation to achieve the full functionality of a complete BS, while satisfying constraints such as weight and power consumption of a satellite through collaboration.
	On the other hand, leveraging the CU-DU separation architecture, we can position the CU on the leader satellite within a formation, which enables the CU satellite to manage and coordinate multiple follower DU satellites for access and forwarding, thus collectively functioning as a powerful spaceborne BS.
	Moreover, more refined network function slicing policies can be implemented, such as further dividing the CU into CU-CP and CU-UP modules \cite{3gpp.38.401}, with the CU-UP placed on select follower satellites to adapt to various communication goals or tasks.
	Furthermore, this distributed functional slicing or redundancy also enhances the overall fault tolerance of the CCS system, since the failure of one or more satellite platforms within the DSIN may result in performance degradation, but it does not lead to service interruption.
	
	\item \textit{Scalability with Adjustable Topology and Platform}: In DSIN, the number of satellites, the platform capabilities, and the topological configuration are all adjustable, which allows for low-cost maintenance and flexible application changes by simply replacing certain satellites within a CCS formation, showcasing its strong scalability \cite{5.5-19}.
	On this basis, by integrating the previously discussed air interface technologies and collaborative cross-layer optimization into individual satellite transmission mechanisms and multi-satellite coordinated management, respectively, the system can effectively tackle the challenges arising from the complex transmission environments, uneven network traffic, and diverse user demands in the satellite network.
	Consequently, the DSIN can provide continuous, reliable, and timely services to ground mobile users under limited payloads.
\end{itemize}

It is worth noting that the practical deployment of these technologies still faces substantial hurdles.
For example, the distributed satellite-based array antennas and spaceborne BS architectures remain largely at the theoretical and research stages, facing challenges like maintaining precise time-frequency synchronization, ensuring stable formation flying under limited power, and adapting beam management and resource optimization to complex scenarios.
Further development and rigorous validation are essential for advancing these technologies toward application readiness.

\section{Conclusion}
Benefiting from rapid progress in commercial space and satellite technologies, satellite networks are transitioning into a new era, evolving towards DSIN. 
This survey provides an in-depth exploration of the critical network architectures underpinning DSIN, focusing on distributed regenerative satellite network architecture, distributed satellite computing network architecture, and reconfigurable formation control.
A comprehensive review for enabling technologies of DSIN followed, covering essential air interface and collaborative transmission technologies (e.g., waveform design, GFRA, NOMA/RSMA multicast, cloud-native distributed MIMO cooperation) and cross-layer optimization techniques (e.g., mobility management, resource allocation, secure communication), alongside the testbeds of DSIN.
Finally, the survey identified several open research challenges and promising directions for future investigation, providing a valuable roadmap for advancing DSIN research.
By providing a thorough, multi-faceted overview, this work serves as a critical resource for researchers and developers, offering valuable insights to guide future innovations and propel the evolution of DSIN technologies at the forefront of satellite network research.

%%%%%%%%%%%%%%%%%%%%%%%%%%%%%%%%%%%%%%%%%%%%%%%%%%%%%%%
%%% Acknowledgements. ÖÂÐ»
%%%%%%%%%%%%%%%%%%%%%%%%%%%%%%%%%%%%%%%%%%%%%%%%%%%%%%%
%\Acknowledgements{This work was supported by the National Natural Science Foundation of China (Grant Nos. 00000000 and 11111111).}
\Acknowledgements{This work was supported in part by the Major Key Project of PCL under Grant PCL2024A01,
in part by the National Natural Sciences Foundation of China (NSFC) under Grant 62071141 and Grant 62027802, 
in part by the Shenzhen Science and Technology Program under Grant JCYJ20241202123904007, Grant GXWD20231127123203001 and Grant JSGG20220831110801003,
in part by the Fundamental Research Funds for the Central Universities under Grant HIT.OCEF.2024046,
in part by the National Key R\&D Program of China No. 2020YFB1806900, 
and in part by the Beijing Natural Science Foundation No. L242012.}

%%%%%%%%%%%%%%%%%%%%%%%%%%%%%%%%%%%%%%%%%%%%%%%%%%%%%%%
%%% Supplements. ²¹³ä²ÄÁÏ, ·Ç±ØÑ¡
%%%%%%%%%%%%%%%%%%%%%%%%%%%%%%%%%%%%%%%%%%%%%%%%%%%%%%%
\Supplements{Appendix A.}

%%%%%%%%%%%%%%%%%%%%%%%%%%%%%%%%%%%%%%%%%%%%%%%%%%%%%%%
%%% Reference section. ²Î¿¼ÎÄÏ×
%%% citation in the content using "some words~\cite{1,2}".
%%% ~ is needed to make the reference number is on the same line with the word before it.
%%%%%%%%%%%%%%%%%%%%%%%%%%%%%%%%%%%%%%%%%%%%%%%%%%%%%%%

%%%%%%%%%%%%%%%%%%%%%%%%%%%%%%%%%%%%%%%%%%%%%%%%%%%%%%%
%%% Appendix sections. ¸½Â¼ÕÂ½Ú, ·Ç±ØÑ¡
%%%%%%%%%%%%%%%%%%%%%%%%%%%%%%%%%%%%%%%%%%%%%%%%%%%%%%%
\begin{appendix}
\section{}
{
\scriptsize
\begin{longtable}{ll|ll}
	%\centering
	\caption{A list of abbreviations}
	\label{tab2} \\
	
	%\rowcolor{morelightGreen}
	\hline \textbf{Abbreviation} & \textbf{Definition} & \textbf{Abbreviation} & \textbf{Definition} \\
	\hline
	\endfirsthead
	\multicolumn{4}{l}{{(Continued)}} \\
	\hline \textbf{Abbreviation} & \textbf{Definition} & \textbf{Abbreviation} & \textbf{Definition} \\
	\hline
	\endhead
	\hline \multicolumn{4}{r}{{(Continued on next page)}} \\
	\endfoot
	\hline
	\endlastfoot
	
	3GPP & $3\text{rd}$ Generation Partnership Project & MEO & Medium Earth Orbit \\
	%\rowcolor{morelightGreen}
	%\hline
	5G & Fifth Generation & MIMO & Multiple-Input and Multiple-Output \\
	%\rowcolor{lightGreen}
	%\hline
	6G & Sixth Generation &  MIPv6 & Mobile IP version 6 \\
	%\rowcolor{morelightGreen}
	%\hline
	APD & Avalanche Photodiode & ML & Maximum Likelihood \\
	%\rowcolor{lightGreen}
	%\hline
	ACI & Adjacent Channel Interference & MMF & Mobility Management Function \\
	%\rowcolor{lightGreen}
	%\hline
	ACK & Acknowledgment & mMIMO & Massive MIMO \\
    AoA & Angles of Arrival & AoD & Angles of Departure \\
    AIMD & Additive Increase & MIMD &  Multiplicative Increase \\
     & Multiplicative Decrease & &  Multiplicative Decrease \\
	%\rowcolor{lightGreen}
	%\hline
	AMF & Access and Mobility Management Function & MMSE & Minimum Mean Square Error \\
	%\rowcolor{lightGreen}
	%\hline
	AoI & Age of Information & MPA & Message Passing Algorithm \\
	%\rowcolor{lightGreen}
	%\hline
	AoII & Age of Incorrect Information & MRB & Most Reliable Basis \\
	%\rowcolor{lightGreen}
	%\hline
	ARQ & Automatic Repeat Request & MRC & Maximum Ratio Combining \\
	%\rowcolor{lightGreen}
	%\hline
	AWGN & Additive White Gaussian Noise & MS & Min-Sum \\
	%\rowcolor{morelightGreen}
	%\hline
	BBSP & Best Beam Selection Policy & MSCS & Mobile Satellite Communication System \\
	%\rowcolor{lightGreen}
	%\hline
	BCH & Bose-Chaudhuri-Hocquenghem & MT & Mobile Terminal \\
	%\rowcolor{lightGreen}
	%\hline
	BER & Bit Error Rate &  NACK & Non-ACK \\
	%\rowcolor{morelightGreen}
	%\hline
	BLER & Block Error Rate &  NC & Network Coding  \\
	%\rowcolor{morelightGreen}
	%\hline
	BP & Belief Propagation &  NFV & Network Functions Virtualization \\
	%\rowcolor{morelightGreen}
	%\hline
	BPP & Binomial Point Process &  NOMA & Non-Orthogonal Multiple Access \\
	%\rowcolor{morelightGreen}
	%\hline
	CC & Chase Combining & NR & New Radio \\
    sCSI & Statistical CSI &  iCSI & Instantaneous CSI\\
    CCS & Cohesive Clustered Satellites & SatCom & Satellite Communication \\
	%\rowcolor{morelightGreen}
	%\hline
	CCI & Co-Channel Interference & NRV & Network Resource Virtualization \\
    6G  & Sixth Generation & DSIN & Distributed Satellite Information Networks \\
	ETSI & European Telecommunications & EO & Earth Observations \\
     & Standards Institute & MDI & Mirror Doppler Interference \\
    F6 & Fast, Flexible, Fractionated,  & DARPA & Defense Advanced Research \\
     &  Free-Flying & & Projects Agency \\
    BS & Base Station & OFDM & Orthogonal Frequency-Division Multiplexing\\
	CCSDS &  Consultative Committee for Space & NTN & Non-Terrestrial Network \\
	& Data Systems & NCR & Network Controllable Repeater \\
	%\rowcolor{lightGreen}
	%\hline
	CDMA & Code Division Multiple Access &  OSD & Ordered Statistics Decoding \\
	%\rowcolor{morelightGreen}
	%\hline
	CFO & Carrier Frequency Offset & OTFS & Orthogonal Time Frequency Space \\
	%\rowcolor{morelightGreen}
	%\hline
	CP &  Cyclic Prefix & PARP &  Peak-to-Average Power Ratio \\
	%\rowcolor{morelightGreen}
	%\hline
	CRC & Cyclic Redundancy Check & PGD & Pure Ground-based Deployment \\
	%\rowcolor{lightGreen}
	%\hline
	CS & Compressive Sensing & PHY & Physical Layer \\
    CU & Central Unit & DU & Distributed Unit \\
    RRC & Radio Resource Control & PDCP & Packet Data Convergence Protocol\\
    MAC & Medium Access Control & RLC & Radio Link Control\\
    RU & Radio Unit & NGSO & Non-Geostationary Satellite Orbits\\
	%\rowcolor{lightGreen}
	%\hline
	CSI & Channel State Information & PMIPv6 & Proxy MIPv6 \\
    ISL & Inter-Satellite Links & O-RAN & Open Radio Access Networks \\
    OTA & Over-the-Air & TDD & Time-Division Duplex \\
    OAI & OpenAirInterfac & UDP & User Datagram Protocol \\
	%\rowcolor{lightGreen}
	%\hline
	D3QN & Dueling Double DQN & POMDP & Partially Observable Markov \\
	DD & Delay-Doppler &  &  Decision Process \\
	%\rowcolor{lightGreen}
	%\hline
	Deep JSCC &  Deep Joint Source-Channel Coding & QC-LDPC & Quasi-Cyclic LDPC \\
    SAT-CU & Satellite Centralized Unit & SAT-DU & Satellite Distributed Units \\
    UDM & Unified Data Management & SMF & Session Management Function\\
    UPF & User Plane Function (UPF) &  PCF & Policy Control Function \\
    AUSF & Authentication Server Function & SAR & Synthetic Aperture Radar\\
    MPC & Model Predictive Control & RSFF & Reconfigurable Satellite Formation Flying \\
    ML & Machine Learning & LDS & Low-Density Signatures \\
    MTs &  Mobile Terminals & MMFs & Mobility Management Functions \\
    CR & Cognitive Radio & MADRL & Multi-Agent Deep Reinforcement Learning\\
	%\rowcolor{lightGreen}
	%\hline
	DQN & Deep Q-Network & RF &  Radio Frequency \\
	%\rowcolor{lightGreen}
	%\hline
	DRL & Deep Reinforcement Learning & RIS & Reconfigurable Intelligent Surface \\
	%\rowcolor{lightGreen}
	%\hline
	DTN &  Delay/Interrupt Tolerant Networking & RL & Reinforcement Learning \\
	%\rowcolor{lightGreen}
	%\hline
	DVB-S2 &  Digital Video Broadcasting-Satellite & RLNC & Random Linear NC \\
	& -Second Generation & RS & Reed-Solomon \\
	%\rowcolor{lightGreen}
	%\hline
	eMBB & Enhanced Mobile Broadband & RSMA &  Rate-Splitting Multiple Access \\
	%\rowcolor{lightGreen}
	%\hline
	FBMC & Filter Bank Multi-Carrier & SAGIN & Space-Air-Ground-Sea \\
	%\rowcolor{lightGreen}
	%\hline
	FDMMA &  Flexible and Distributed Mobility &  & Integrated Network  \\
        FSO & Free Space Optical & HAPs & High-Altitude Platforms\\ 
	& Management Architecture & SBL & Sparse Bayesian Learning\\
	%\rowcolor{lightGreen}
	%\hline
	FEC & Forward Error Correction & SC & Successive Cancellation \\
	%\rowcolor{lightGreen}
	%\hline
	FMIPv6 & Fast Handover for MIPv6 & SC-LDPC & Spatially-Coupled LDPC \\
	%\rowcolor{lightGreen}
	%\hline
	GE & Gaussian Elimination &  SCL & SC List  \\
    HTS & High-Throughput Satellite & IRS & Intelligent Reflecting Surfaces \\
    TCP/IP & Transmission Control Protocol/Internet & DBPR &  Distance-based Back-Pressure Routing \\
    &  Protocol & RTT & Round-Trip Time \\

	%\rowcolor{lightGreen}
	%\hline
	GFDM & Generalized Frequency Division Multiplexing & SCMA & Sparse Code Multiple Access \\
	%\rowcolor{lightGreen}
	%\hline
	GFRA & Grant-Free Random Access & SCPS & Space Communication Protocol Standard  \\
	%\rowcolor{lightGreen}
	%\hline
	GRAND & Guessing Random Additive  &  SDN &  Software-Defined Networking \\
     & Noise Decoding & LFNs & Long Fat Networks \\
	%\rowcolor{lightGreen}
	%\hline
	GS & Ground Station & SE &  Spectral Efficiency \\
    QKD & Quantum Key Distribution & QSDC &  Quantum Secure Direct Communication\\
	%\rowcolor{lightGreen}
	%\hline
	HARQ & Hybrid ARQ & SFF & Satellite Formation Flying \\
	%\rowcolor{lightGreen}
	%\hline
	HMIPv6 & Hierarchical MIPv6 & SFFT & Symplectic Finite Fourier Transform \\
	%\rowcolor{lightGreen}
	%\hline
	IAI & Inter-Antenna Interference & SGIMM & Space-Ground Integrated \\
	ICI & Inter-Carrier Interference &  & Mobility Management \\
	%\rowcolor{lightGreen}
	%\hline
	IDI & Inter-Doppler Interference & S-IoT & Satellite Internet of Things \\
	%\rowcolor{lightGreen}
	%\hline
	IETF & Internet Engineering Task Force & SIC & Successive Interference Cancellation \\
	%\rowcolor{lightGreen}
	%\hline
	IP & Internet Protocol &  SIGMA & Seamless IP diversity based  \\
	IQI & In-phase and Quadrature Imbalance &  & Generalized Mobility Architecture \\
	%\rowcolor{lightGreen}
	%\hline
	IR & Incremental Redundancy & SLA & Satellite LA \\
    ITU-R & Radio Communication Division of the & VITAL & Virtualized Hybrid Satellite-Terrestrial \\
     & International Telecommunication Union & & Systems \\
	%\rowcolor{lightGreen}
	%\hline
	IRS & Intelligent Reflecting Surfaces & SMFD & Space-based Management \\
	ISFFT &  Inverse SFFT &  & Function Deployment \\
	%\rowcolor{lightGreen}
	%\hline
	ISI & Inter-Symbol Interference & SNR &  Signal-to-Noise Ratio \\
	%\rowcolor{lightGreen}
	%\hline
	ISCC & Integrated Sensing, Communication, & SSIM &  Structural Similarity Index \\
	&  and Computation & TBCC & Tail-biting Convolutional Code \\
	%\rowcolor{lightGreen}
	%\hline
	JD & Joint Decoder & TCP & Transmission Control Protocol \\
	%\rowcolor{lightGreen}
	%\hline
	KB & Knowledge Base & TEP & Test Error Pattern \\
	%\rowcolor{lightGreen}
	%\hline
	LA & Location Area & TF & Time-Frequency \\
	%\rowcolor{lightGreen}
	%\hline
	LCMA & Lattice-Code Multiple Access & UFMC &  Universal Filtered Multi-Carrier \\
	%\rowcolor{lightGreen}
	%\hline
	LDPC & Low-Density Parity-Check & ULA & User LA \\
	%\rowcolor{lightGreen}
	%\hline
	LEC & Long Erasure Code & UoI & Utility of Information \\
	%\rowcolor{lightGreen}
	%\hline
	LEO & Low Earth Orbit & URLLC & Ultra-Reliable and Low Latency \\
	LM & Location Management & & Communications \\
	%\rowcolor{lightGreen}
	%\hline
	LMS & Land Mobile Satellite & VAC & Virtual Agent Cluster \\
	%\rowcolor{lightGreen}
	%\hline
	LNA & Low Noise Amplifier & VAD & Virtual Agent Domain \\
	%\rowcolor{lightGreen}
	%\hline
	LoS & Line of Sight & VAP & Virtual Attachment Point \\
	%\rowcolor{lightGreen}
	%\hline
	LSMR &  Least Squares Minimum Residual & VBI & Variational Bayesian Inference \\
	%\rowcolor{lightGreen}
	%\hline
    PAPR & Peak-to-Average Power Ratio & RAR & Random Access Response\\
    RAPID & Random Access Preamble Identifier & C-RNTI & Cell-Radio Network Temporary Identifier\\
    TA &  Timing Advance & TSC & Terrestrial-Satellite Communication\\
    ZC & Zadoff-Chu & DFT & Discrete Fourier Transform \\
	LSP &  Longer Side Priority & VMIPv6 & Virtual MIPv6 \\
    RMS & Root Mean Square & TDD &  Time Division Duplexing \\
    SNS3 & Satellite Network Simulator 3 & GA & Genetic Algorithms \\
    ACO & Ant Colony Optimization & SA & Simulated Annealing \\
	%\rowcolor{lightGreen}
	%\hline
	LTE &  Long-Term Evolution & VQA & Visual Question Answering\\
	%\rowcolor{lightGreen}
	%\hline
	LTP & Licklider Transmission Protocol & VoI &  Value of Information \\
	%\rowcolor{lightGreen}
	%\hline
	MBMS & Multimedia Broadcast/Multicast Service & WT & Whitening Transformation \\
    LOS & Line-of-Sight & IP & Internet Protocol \\
    UAVs & Unmanned Aerial Vehicles & LCRD & Laser Communications Relay Demonstration \\
    PAT & Pointing, Acquisition, and Tracking & FOV & Field-of-View, \\
    AoA & Angle-of-Arrival & OTA & Over-the-Air\\
    RRU &  Remote Radio Unit & FDD & Frequency-Division Duplex\\
    LLMs & Large Language Models & LPS & Low-Correlation-Zone Periodic Sequence\\
    RIS & Reconfigurable Intelligent Surfaces & & \\
	%\rowcolor{lightGreen}
	%\hline
	%\rowcolor{lightGreen}
	%\hline
\end{longtable}
}
\end{appendix}


\begin{thebibliography}{99}
%%%%%%%%%%%1.0 Introduction


%%%%%%%%%%%2.3 Reconfigurable Satellite Formation Flying
%%Àý×Ó£ºMao Y Y, You C S, Zhang J, et al. A survey on mobile edge computing: the communication perspective. IEEE Commun Surv Tut, 2017, 19: 2322¨C2358
%\bibitem{1} Author A, Author B, Author C. Reference title. Journal, 2024, 38: 13--28
%%%%%introduction
\bibitem{you6G} You X H, Wang C-X, Huang J, et al. Towards 6G wireless communication networks: Vision, enabling technologies, and new paradigm shifts. Sci China Inf Sci, 2021, 64: 110301

\bibitem{xu2024semantic} Xu L, Jiao J, Jiang S Y, et al. Semantic-aware coordinated transmission in cohesive clustered satellites: Utility of information perspective. Sci China Inf Sci, 2024, 67: 199301

\bibitem{9887918} Peng C, He Y Z, Zhao S H, et al. Integration of data center into the distributed satellite cluster networks: Challenges, techniques, and trends. IEEE Netw, 2023, 37: 52-58

\bibitem{9442378} Centenaro M, Costa C E, Granelli F, et al. A survey on technologies, standards and open challenges in satellite IoT. IEEE Commun Surv Tut, 2021, 23: 1693-1720

\bibitem{10400393} He Y Z, Wang C X, Qi C W, et al. Spatial ultra-sparse distributed antenna satellite-ground cooperative transmission architecture: Challenges, key technologies, and trends. IEEE Commun Mag, 2024, 62: 136-143

%%%%%%2.1
\bibitem{nan20225g} Zhang N, Li Z Y, Li R Y, et al. 5G-Advanced: network-controlled repeater (in Chinese). Telecommun Sci, 2022, 38: 169-176

\bibitem{5.5-19} Tuzi D, Delamotte T, Knopp A. Satellite swarm-based antenna arrays for 6G direct-to-cell connectivity. IEEE Access, 2023, 11: 36907--36928

\bibitem{10694119} Tuzi D, Delamotte T, Knopp A. Performance assessment of sparse satellite swarms for 6G direct-to-cell connectivity. In: Proceedings of IEEE International Workshop on Signal Processing Advances in Wireless Communications, Lucca, 2024. 616-620

\bibitem{3gpp.811} 3rd Generation Partnership Project (3GPP). Study on new radio (NR) to support nonterrestrial networks (NTN) (Release 16). TR 38.811. https://www.3gpp.org/ftp/Specs/archive/38\_series/38.811

\bibitem{10436159} Bhandari S, Vu T X , Chatzinotas S. User-centric flexible resource management framework for LEO satellites with fully regenerative payload. IEEE J Sel Areas Commun, 2024, 42: 1246-1261

\bibitem{10648786} O.B Y, Zineb G, Olivier B, et al. Evolution of high-throughput satellite systems: A vision of programmable regenerative payload. IEEE Commun Surv Tut, 2024. doi: 10.1109/COMST.2024.3450292.

\bibitem{10644023} Choi J, Li B, Al H, et al. Spectrum sharing through marketplaces for O-RAN based non-terrestrial and terrestrial networks. IEEE Internet Things Mag, 2024, 7: 128-134

\bibitem{c1} Yoshida Y, Mobile Xhaul Evolution: Enabling Tools for a Flexible 5G Xhaul Network. In: Proceedings of 2018 Optical Fiber Communications Conference and Exposition, San Diego, 2018. 1-85

\bibitem{3gpp.38.401} 3rd Generation Partnership Project (3GPP). NG-RAN; Architecture description (Release 17). TS 38.401. https://www.3gpp.org/ftp/Specs/archive/38\_series/38.401

\bibitem{3gpp.38.801} 3rd Generation Partnership Project (3GPP). Study on new radio access technology; Radio access architecture and interfaces (Release 14). TR 38.801. https://www.3gpp.org/ftp/Specs/archive/38\_series/38.801

\bibitem{9946423} Joda R, Pamuklu T, Iturria-Rivera P E, et al. Deep reinforcement learning-based joint user association and CU-DU placement in O-RAN. IEEE Trans Netw Serv Manag, 2022, 19: 4097-4110

%%%%%%%%2.2
\bibitem{Wang2023Satellite} Wang S G, Li Q. Satellite computing: Vision and challenges. IEEE Internet Things J, 2023, 10: 22514-22529

\bibitem{8051074} Li S Z, Maddah-Ali M A, Yu Q, et al. A fundamental tradeoff between computation and communication in distributed computing. IEEE Trans Inf Theory, 2018, 64(1): 109--128

\bibitem{9463425} Ng J S, Lim W Y B, Luong N C, et al. A comprehensive survey on coded distributed computing: Fundamentals, challenges, and networking applications. IEEE Commun Surv Tutorials, 2021, 23(3): 1800--1837

\bibitem{9933792} Duan S J, Wang D, Ren J, et al. Distributed artificial intelligence empowered by end-edge-cloud computing: A survey. IEEE Commun Surv Tutorials, 2023, 25(1): 591--624

\bibitem{10638529} Yang Y, He X Y, Lee J, et al. Collaborative deep reinforcement learning in 6G integrated satellite-terrestrial networks: Paradigm, solutions, and trends. IEEE Commun Mag, 2025, 63(1): 188--195
  
\bibitem{10433234} Deng S G, Zhao H L, Huang B B, et al. Cloud-native computing: A survey from the perspective of services. Proceedings of the IEEE, 2024, 112(1): 12-46

\bibitem{Jung2023Satellite} Jung D-H, Im G, Ryu J-G, et al. Satellite clustering for non-terrestrial networks: Concept, architectures, and applications. IEEE Veh Technol Mag, 2023, 18: 29-37

\bibitem{Pan2024Architecture} Pan X H, Lu L, Deng P K, et al. Architecture, technologies and experiment of satellite-terrestrial network based on space-based simplified core network (in Chinese). Telecommun Sci, 2024, 40: 49-59

\bibitem{Wang2024End} Wang Y C, Yang C, Lan S L, et al. End-edge-cloud collaborative computing for deep learning: A comprehensive survey. IEEE Commun Surv Tut, 2024, 26: 2647-2683.

\bibitem{Gao2023Onboard} Gao G, Yao L B, Li W F, et al. Onboard information fusion for multisatellite collaborative observation: Summary, challenges, and perspectives, IEEE Geosci Remote Sens Mag, 2023, 11: 40-59

\bibitem{Li2024FedFusion} Li D X, Xie W Y, Li Y S, et al. FedFusion: Manifold-driven federated learning for multi-satellite and multi-modality fusion. IEEE Trans Geosci Remote Sens, 2024, 62:1-13

\bibitem{Xi2023Multi} Xi S Y, Shang B D, Zhang H X, et al. Multi-satellite-enabled edge computing: An offloading and computation integration approach. In: Proceedings of International Conference on Intelligent Communications and Computing, Nanchang, 2023. 151-156

\bibitem{Cui2023Latency} Cui G F, Duan P F, Xu L X, et al. Latency optimization for hybrid GEO/CLEO satellite-assisted IoT networks. IEEE Internet Things J, 2023, 10: 6286-6297

\bibitem{Gong2024Intelligent} Gong Y K, Yao H P, Nallanathan A. Intelligent sensing, communication, computation, and caching for satellite-ground integrated networks. IEEE Netw, 2024, 38: 9-16

\bibitem{Zhang2023MDMA} Zhang P, Xu X D, Dong C, et al. Model division multiple access for semantic communications. Front Inf Technol Electron Eng, 2023, 24:801--812.


%%%%%%%%%%%%%%2.3
\bibitem{8293791} Liu G-P, Zhang S J. A survey on formation control of small satellites. Proc IEEE, 2018, 106: 440-457

\bibitem{yin2018review} Yin J F, Zhang Q J, Liu J, et al. A review on development of formation flying interferometric SAR satellite system. Spacecraft Eng, 2018, 27: 116-122

\bibitem{clohessy1960terminal} Clohessy W H, Wiltshire R S. Terminal guidance system for satellite rendezvous. J Aerospace Sci, 1960, 27: 653-658

\bibitem{tschauner1964optimale} Tschauner J, Hempel P. Optimale beschleunigungsprogramme fur das rendezvous-manover. Astron Acta, 1964, 10: 296

\bibitem{10679984} Lei C L, Feng W, Wei P, et al. Edge information hub: Orchestrating satellites, UAVs, MEC, sensing and communications for 6G closed-loop controls. IEEE J Sel Areas Commun, 2025, 43(1): 5--20

\bibitem{wang2016satellite} Wang D W, Wu B L, Poh E K. Satellite formation flying: relative dynamics, formation design, fuel optimal maneuvers and formation maintenance. Singapore: Springer Press, 2016

\bibitem{2.3.2} Cai H, Huang J. Leader-following adaptive consensus of multiple uncertain rigid spacecraft systems. Sci China Inf Sci, 2016, 59: 1-13

\bibitem{zou2012neural} Zou A-M, Kumar K D. Neural network-based distributed attitude coordination control for spacecraft formation flying with input saturation. IEEE Trans Neural Netw Learn Syst, 2012, 23: 1155--1162

\bibitem{balch1998behavior} Balch T, Arkin R C. Behavior-based formation control for multirobot teams. IEEE Trans Robot Autom, 1998, 14: 926--939

\bibitem{2.3.4} Schlanbusch R, Kristiansen R, Nicklasson P J. Spacecraft formation reconfiguration with collision avoidance. Automatica, 2011, 47: 1443-1449

\bibitem{2.3.5} Abbasi Y, Moosavian S A A, Novinzadeh A B. Formation control of aerial robots using virtual structure and new fuzzy-based self-tuning synchronization. IEEE Trans Inst Meas Control, 2017, 39: 1906-1919

\bibitem{2.3.6} Song Y Y, Zhou Q R, Chen Q W, et al. Optimal reconfiguration control of electromagnetic satellite formation. In: Proceedings of IEEE 41st Chinese Control Conference, Hefei, 2022. 1809-1814

\bibitem{2.3.7} Wang P K C, Hadaegh F Y. Minimum-fuel formation reconfiguration of multiple free-flying spacecraft. J Astronaut Sci, 1999, 47: 77-102

\bibitem{9761868} Marrero L M, Merlano-Duncan J C, Querol J, et al.  Architectures and synchronization techniques for distributed satellite systems: A survey. IEEE Access, 2022, 10: 45375-45409

\bibitem{graph} Blackmore L, Hadaegh F. Necessary and sufficient conditions for attitude estimation in fractionated spacecraft systems. In: Proceedings of AIAA Guidance, Navigation, and Control Conference, Chicago, 2009. 6095

\bibitem{li2014event} Li H Q, Liao X F, Huang T W, et al. Event-triggering sampling based leader-following consensus in second-order multi-agent systems. IEEE Trans Autom Control, 2014, 60: 1998-2003

\bibitem{lu2017distributed} L{\"u} Q G, Li H Q, Xia D W. Distributed optimization of first-order discrete-time multi-agent systems with event-triggered communication. Neurocomputing, 2017, 235: 255-263

\bibitem{garcia2014cooperative} Garcia E, Cao Y C, Casbeer D W. Cooperative control with general linear dynamics and limited communication: Centralized and decentralized event-triggered control strategies. In: Proceedings of American Control Conference, Portland, 2014. 159-164

\bibitem{zhu2014event} Zhu W, Jiang Z-P, Feng G. Event-based consensus of multi-agent systems with general linear models. Automatica, 2014, 50: 552-558

\bibitem{fan2014semi} Fan M-C, Chen Z Y, Zhang H-T. Semi-global consensus of nonlinear second-order multi-agent systems with measurement output feedback. IEEE Trans on Autom Control, 2014, 59: 2222-2227

\bibitem{seuret2016lq} Seuret A, Prieur C, Tarbouriech S, et al. LQ-based event-triggered controller co-design for saturated linear systems. Automatica, 2016, 74: 47-54

\bibitem{sconzo1962use} Sconzo P. The use of Lambert's theorem in orbit determination. Astron J, 1962, 67: 19-21

\bibitem{nelson1992alternative} Nelson S L, Zarchan P. Alternative approach to the solution of Lambert's problem. J Guid Control Dyn, 1992, 15: 1003-1009

\bibitem{avanzini2008simple} Avanzini G. A simple Lambert algorithm. J Guid Control Dyn, 2008, 31: 1587-1594

\bibitem{zhang2020terminal} Zhang G. A simple Lambert algorithm. J Guid Control Dyn, 2020, 43: 1529-1539

\bibitem{yang2015homotopic} Yang Z, Luo Y-Z, Zhang J, et al. Homotopic perturbed Lambert algorithm for long-duration rendezvous optimization. J Guid Control Dyn, 2015, 38: 2215-2223

\bibitem{zhang2020reachable} Zhang H Y, Zhang G. Reachable domain of ground track with a single impulse. IEEE Trans Aerosp Electron Syst, 2020, 57: 1105-1122

\bibitem{thorne2015convergence} Thorne J D. Convergence behavior of series solutions of the Lambert problem. J Guid Control Dyn, 2015, 38: 1821-1826

%%%%%%%%%%%%3.1.1 ´ýÐÞ¸Ä¸ñÊ½
\bibitem{6G} Wang C-X, You X H, Gao X Q, et al. On the road to 6G: Visions, requirements, key technologies, and testbeds. IEEE Commun Surv Tut, 2023, 25: 905-974

\bibitem{Sate} Xiang Z Y, Gao X Q, Li K-X, et al. Massive MIMO downlink transmission for multiple LEO satellite communication. IEEE Trans Commun 2024, 72: 3352-3364

\bibitem{test1} Milojevic M, Haardt M, Eberlein E et al. Channel modeling for multiple satellite broadcasting systems. IEEE Trans Broadcast, 2009, 55: 705-718

\bibitem{ISL} Tawfik M M, Sree M F A, Abaza M, et al. Performance analysis and evaluation of inter-satellite optical wireless communication system (IsOWC) from GEO to LEO at range 45000 km. IEEE Photon J, 2021, 13: 1-6

\bibitem{SG} Jung D-H, Ryu J-G, Byun W-J, et al. Performance analysis of satellite communication system under the shadowed-Rician fading: A stochastic geometry approach. IEEE Trans Commun, 2022, 70: 2707-2721

\bibitem{AICM} Homssi B A, Chan C C, Wang K, et al., Deep learning forecasting and statistical modeling for Q/V-Band LEO satellite channels. IEEE Trans Mach Learn Commun Netw, 2023, 1: 78-89

\bibitem{WCM} Fang X R, Feng W, Wei T, et al. 5G embraces satellites for 6G ubiquitous IoT: Basic models for integrated satellite terrestrial networks. IEEE Internet Things J, 2021, 8: 14399-14417

\bibitem{zCE} Zuo Y, Yue M Y, Zhang M C, et al. OFDM-based massive connectivity for LEO satellite Internet of Things. IEEE Trans Wirel Commun, 2023, 22: 8244-8258

\bibitem{Li} You L, Li K-X, Wang J H, et al. `Massive MIMO transmission for LEO satellite communications. IEEE J Sel Areas Commun, 2020, 38: 1851-1865

\bibitem{PCE} Li K-XX, Gao X Q, Xia X-G. Channel estimation for LEO satellite massive MIMO OFDM communications. IEEE Trans Wirel Commun, 2023, 22: 7537-7550

\bibitem{CSCE} Wang Y, Li Q W, Jiao J, et al. ARM: Adaptive random-selected multi-beamforming estimation scheme for satellite-based Internet of Things. IEEE Access, 2019, 7: 63264-63276

\bibitem{AICE} Zhang Y Y, Liu A J, Li P H, et al. Deep learning (DL)-based channel prediction and hybrid beamforming for LEO satellite massive MIMO system. IEEE Internet Things J, 2022, 9: 23705-23715

\bibitem{MCE} Lu S, Yu B G, Tang C K, et al. MIMO-OFDM channel detection algorithm in multi-station and multi-satellite uplink system based on deep learning. In: Proceedings of IEEE International Conference on Signal Processing, Communications and Computing, Xi'an, 2021. 1-4

\bibitem{8979256} Yuan J D, Ngo H Q, Matthaiou M. Machine learning-based channel prediction in massive MIMO with channel aging. IEEE Trans Wirel Commun, 2020, 19(5): 2960--2973

\bibitem{9439942} Zhang Y Y, Wu Y L, Liu A J, et al. Deep learning-based channel prediction for LEO satellite massive MIMO communication system. IEEE Wirel Commun Lett, 2021, 10(8): 1835--1839

\bibitem{9000850} Ma W Y, Qi C H, Zhang Z C, et al. Sparse channel estimation and hybrid precoding using deep learning for millimeter wave massive MIMO. IEEE Trans Commun, 2020, 68(5): 2838--2849

\bibitem{10550141} Ying M, Chen X M, Qi Q, et al. Deep learning-based joint channel prediction and multibeam precoding for LEO satellite Internet of Things. IEEE Trans Wirel Commun, 2024, 23(10): 13946--13960

\bibitem{10368353} Gizzini A K, Medjahdi Y, Ghandour A J, et al. Towards explainable AI for channel estimation in wireless communications. IEEE Trans Veh Technol, 2024, 73(5): 7389--7394
%%%%3.2新增分布式MIMO协同技术
\bibitem{XZY1} Xiang Z, Sun R, Gong X, et al. Massive MIMO uplink transmission for multiple LEO satellite communication. IEEE Trans Aerosp Electron Syst, 2024
\bibitem{multisatellitemimo} Bakhsh Z M, Omid Y, Chen G, et al. Multi-satellite MIMO systems for direct satellite-to-device communications: A Survey. IEEE Commun Surv Tut, 2024
\bibitem{XZY2} Xiang Z, Gao X, Li K X, et al. Massive MIMO downlink transmission for multiple LEO satellite communication. IEEE Trans Commun, 2024
\bibitem{r3x-1} Björnson E, Sanguinetti L. Making Cell-free massive MIMO competitive With MMSE processing and centralized implementation. IEEE Trans Wirel Commun, 2020, 19: 77-90
\bibitem{LEODMIMO} Abdelsadek M Y, Kurt G K, Yanikomeroglu H. Distributed massive MIMO for LEO satellite networks. IEEE Open J Comm Soc, 2022, 3: 2162-2177
\bibitem{r3x-3} Wang D M, You X H, Huang Y M, Xu W, et al. Full-spectrum cell-free RAN for 6G systems: system design and experimental results. Sci China Inf Sci, 2023, 66(3): 130305
\bibitem{cloud} Dalgitsis M, Cadenelli N, Serrano M A, et al. Cloud-native orchestration framework for network slice federation across administrative domains in 5G/6G mobile networks. IEEE Trans Veh Technol, 2024
\bibitem{synchronization} Cao Y, Wang P, Zheng K, et al. Experimental performance evaluation of cell-free massive MIMO systems using COTS RRU with OTA reciprocity calibration and phase synchronization. IEEE J Sel Areas Commun, 2023, 41(6): 1620-1634
\bibitem{synchronization1} Rogalin R, Bursalioglu O Y, Papadopoulos H, et al. Scalable synchronization and reciprocity calibration for distributed multiuser MIMO.  IEEE Trans Wirel Commun, 2014, 13(4): 1815-1831
\bibitem{synchronization2} Hamed E, Rahul H, Abdelghany M A, et al. Real-time distributed MIMO systems. In: Proceedings of the ACM SIGCOMM Conference. 2016. 412-425
\bibitem{synchronization4} Magounaki T, Kaltenberger F, Knopp R. Modeling the distributed MU-MIMO OAI 5G testbed and group-based OTA calibration performance evaluation. In: Proceedings of IEEE International Workshop on Signal Processing Advances in Wireless Communications. 2020. 1-5
\bibitem{r3x-5} Gharanjik A, Bhavani Shankar M.R., Arapoglou P. D., Bengtsson M, et al. Robust precoding design for multibeam downlink satellite channel with phase uncertainty. In: Proceedings of IEEE International Conference on Acoustics, Speech and Signal Processing. 2015. 3083-3087
\bibitem{DMIMO} Hong Z, Xu S, Li T, et al. Robust cascaded team MMSE precoding for cell-free distributed downlink under hierarchical fronthaul. IEEE Trans Wirel Commun, 2024

\bibitem{10234306} Wang C, Zhang Y R, Li Q, Zhou A, et al. Satellite computing: A case study of cloud-native satellites. In: Proceedings of IEEE International Conference on Edge Computing and Communications. 2023. 262-270
\bibitem{9183752} Wang D M, Zhang C, Du Y Q, et al. Implementation of a cloud-based cell-free distributed massive MIMO system. IEEE Commun Mag, 2020, 58(8): 61-67
\bibitem{9401305} Wang D M, Zhang C, Ji Z H, et al. Live demonstration: A cloud-based cell-free distributed massive MIMO system. In: Proceedings of IEEE International Symposium on Circuits and Systems, Daegu, Korea, 2021. 1--1

%%%%3.2
\bibitem{3.2-1} Hwang T, Yang C Y, Wu G, et al. OFDM and its wireless applications: A survey. IEEE Trans Veh Tech, 2009, 58: 1673--1694

\bibitem{3.2-2} Falconer D, Ariyavisitakul S, Benyamin-Seeyar B, et al. Frequency domain equalization for single-carrier broadband wireless systems. IEEE Commun Mag, 2002, 40: 58--66

\bibitem{3.2-3} Vakilian V, Wild T, Schaich F, et al. Universal-filtered multi-carrier technique for wireless systems beyond LTE. In: Proceedings of IEEE Global Communications Conference Workshops, Atlanta, 2013. 223--228

\bibitem{3.2-4} Zhang X, Jia M, Chen L, et al. Filtered-OFDM-enabler for flexible waveform in the 5th generation cellular networks. In: Proceedings of IEEE Global Communications Conference, San Diego, 2015. 1--6

\bibitem{3.2-5} Farhang-Boroujeny B. OFDM versus filter bank multicarrier. IEEE Signal Process Mag, 2011, 28: 92--112

\bibitem{3.2-6} Michailow N, Matth{\'{e}} M, Gaspar I S, et al. Generalized frequency division multiplexing for 5th generation cellular networks. IEEE Trans Commun, 2014, 62: 3045--3061

\bibitem{3.2-7} Sahin A, Arslan H. Edge windowing for OFDM-based systems. IEEE Commun Lett, 2011, 15: 1208--1211

\bibitem{3.2-8} Huang M N, Chen J, Feng S L. Synchronization for OFDM-based satellite communication system. IEEE Trans Veh Tech, 2021, 70: 5693--5702

\bibitem{3.2-9} 3rd Generation Partnership Project (3GPP). Study on narrow-band Internet of things (NB-IoT)/enhanced machine type communication (eMTC) support for non-terrestrial networks (NTN) (Release 17). TR 36.763. https://www.3gpp.org/ftp/Specs/archive/36\_series/36.763

\bibitem{3.2-10} Hadani R, Rakib S, Tsatsanis M, et al. Orthogonal time frequency space modulation. In: Proceedings of IEEE Wireless Communications and Networking Conference, San Francisco, 2017. 1--6

\bibitem{3.2-11} Chu T M C, Zepernick H -J, H{\"{u}}{\"{u}}k A, et al. OTFS modulation for non-terrestrial networks: Concepts, applications, benefits, and challenges. In: Proceedings of International Conference on Signal Processing and Communication System, Bydgoszcz, 2023. 1--10

\bibitem{3.2-12} Ma Y Y, Ma G Y, Ai B, et al. Characteristics of channel spreading function and performance of OTFS in high-speed railway. IEEE Trans Wirel Commun, 2023, 22: 7038--7054

\bibitem{3.2-13} 3rd Generation Partnership Project (3GPP). OTFS modulation waveform and reference signals for new RAT (Release 17). TSG RAN WG1, R1-162930. https://www.3gpp.org/ftp/tsg\_ ran/WG1\_RL1/TSGR1\_84/Docs

\bibitem{3.2-14} Farhang A, RezazadehReyhani A, Doyle L E, et al. Low complexity modem structure for OFDM-based orthogonal time frequency space modulation. IEEE Wirel Commun Lett, 2018, 7: 344--347

\bibitem{3.2-15} Shi J, Li Z, Hu J, et al. OTFS enabled LEO satellite communications: A promising solution to severe Doppler effects. IEEE Netw, 2024, 38: 203--209

\bibitem{3.2-16} Hedhly W, Musavian L, Thomos N. OTFS-NOMA system for MIMO communication networks with spatial diversity. In: Proceedings of IEEE International Conference on Communications, Denver, 2024. 569--574

\bibitem{3.2-17} Li S Y, Yuan W J, Wei Z Q, et al. Cross domain iterative detection for orthogonal time frequency space modulation. IEEE Trans Wirel Commun, 2022, 21: 2227--2242

\bibitem{3.2-18} Wei Z Q, Yuan W J, Li S Y, et al. Transmitter and receiver window designs for orthogonal time-frequency space modulation. IEEE Trans Commun, 2021, 69: 2207--2223

\bibitem{3.2-19} Qu H Y, Liu G H, Zhang L, et al. Low-complexity symbol detection and interference cancellation for OTFS system. IEEE Trans Commun, 2021, 69: 1524--1537

\bibitem{3.2-20} Huang K H, Qiu M, Tong J, et al. Performance of ODDM with imperfect channel estimation. In: Proceedings of IEEE International Workshop on Signal Processing Advances in Wireless Communications, Shanghai, 2023. 561--565

\bibitem{3.2-21} Li Q, Yuan J H, Qiu M, et al. Low complexity Turbo SIC-MMSE detection for orthogonal time frequency space modulation. IEEE Trans Commun, 2024, 72: 3169--3183

\bibitem{3.2-22} Tusha A, D{\u{o}}gan-Tusha S, Yilmaz F, et al. Performance analysis of OTFS under in-phase and quadrature imbalance at transmitter. IEEE Trans Veh Technol, 2021, 70: 11761--11771

\bibitem{3.2-23} Tusha A, D{\u{o}}gan-Tusha S, Yilmaz F, et al. Physical effect of in-phase and quadrature imbalance in delay-Doppler domain. In: Proceedings of IEEE Vehicular Technology Conference, Norman, 2021. 1--6

\bibitem{3.2-24} Neelam S G, Sahu P. Analysis, estimation and compensation of hardware impairments for CP-OTFS systems. IEEE Wirel Commun Lett, 2022, 11: 952--956

\bibitem{3.2-25} Murali K, Chockalingam A. On OTFS modulation for high-Doppler fading channels. In: Proceedings of Information Theory and Applications Workshop, 2018. 1--10

\bibitem{3.2-26} Raviteja P, Phan KT, Hong Y. Embedded pilot-aided channel estimation for OTFS in delay-Doppler channels. IEEE Trans Veh Tech, 2019, 68: 4906--4917

\bibitem{3.2-27} Guo Q, Jiang H Y, Xiang J H, et al. A CS-BEM OTFS channel estimation approach for Sparse continuous Doppler-spread channels. IEEE Wirel Commun Lett, 2024. doi: 10.1109/LWC.2024.3432090

\bibitem{3.2-28} Zhao L, Gao W-J , Guo W B. Sparse Bayesian learning of delay-Doppler channel for OTFS system. IEEE Commun Lett, 2020, 24: 2766--2769

\bibitem{3.2-29} Yuan W J, Li S Y, Wei Z Q, et al. Data-aided channel estimation for OTFS systems with a superimposed pilot and data transmission scheme. IEEE Wirel Commun Lett, 2021, 10: 1954--1958

\bibitem{3.2-30} Wang X Y, Shen W Q, Xing C W, et al. Joint Bayesian channel estimation and data detection for OTFS systems in LEO satellite communications. IEEE Trans Commun, 2022, 70: 4386--4399

\bibitem{3.2-31} Buzzi S, Caire G, Colavolpe G, et al. LEO satellite diversity in 6G non-terrestrial networks: OFDM vs. OTFS. IEEE Commun Lett, 2023, 27: 3013--3017

\bibitem{3.2-32} Caus C, Shaat M, P{\`{e}}rez-Neira A I, et al. Cooperative dual LEO satellite transmission in multi-user OTFS systems. In: Proceedings of IEEE International Conference on Acoustics, Speech, and Signal Processing Workshops, Rhodes Island, 2023. 1--5

\bibitem{3.2-33} Zhang Z Q, Wu Y C, Ma Z, et al. Coordinated multi-satellite transmission for OTFS-based 6G LEO satellite communication systems. IEEE J Sel Areas Commun, 2024. doi: 10.1109/JSAC.2024.3460108

\bibitem{3.2-34} Surabhi G D, Augustine R M, Chockalingam A. On the diversity of uncoded OTFS modulation in doubly-dispersive channels. IEEE Trans Wirel Commun, 2019, 18: 3049--3063

\bibitem{3.2-35} Ramachandran M K, Chockalingam A. MIMO-OTFS in high-Doppler fading channels: Signal detection and channel estimation. In Proceedings of IEEE Global Communications Conference, Abu Dhabi, 2018. 206--212

\bibitem{3.2-36} Loyka S L. Channel capacity of MIMO architecture using the exponential correlation matrix. IEEE Commun Lett, 2001, 5: 369--371

\bibitem{3.2-37} Thaj T, Viterbo E. Low-complexity linear diversity-combining detector for MIMO-OTFS. IEEE Wirel Commun Lett, 2022, 11: 288--292

\bibitem{3.2-38} Bora A S, Phan K T, Hong Y. Spatially correlated MIMO-OTFS for LEO satellite communication systems. In: Proceedings of IEEE International Conference on Communications Workshops, Seoul, 2022. 723--728

\bibitem{3.2-39} Qu H Y, Liu G H, Imran M A. Efficient channel equalization and symbol detection for MIMO OTFS systems. IEEE Trans Wirel Commun, 2022, 21: 6672--6686

\bibitem{3.2-40} Bora A S, Phan K T, Hong Y. Mitigating spatial correlation in MIMO-OTFS. IEEE Transactions on Vehicular Technology IEEE Trans Veh Technol, 2024, 73: 3608--3622

\bibitem{3.2-41} Shen B X, Wu Y P, An J P. Random access with massive MIMO-OTFS in LEO satellite communications. IEEE J Sel Areas Commun, 2022, 40: 2865--2881

\bibitem{3.2-42} Shen B X, Wu Y P, Gong S Q. Massive MIMO-OTFS-based random access for cooperative LEO satellite constellations. IEEE J Sel Areas Commun, 2024. doi: 10.1109/JSAC.2024.3460064

\bibitem{3.2-43} Gunjan G, Shrivastava S, Kashyap S. Modeling and analysis of physical layer security of OTFS systems under transmit antenna selection and passive eavesdropping. IEEE Commun Lett, 2024, 28: 483--487

\bibitem{3.2-44} Hu J F, Shi J, Ma S, et al. Secrecy analysis for orthogonal time frequency space scheme based uplink LEO satellite communication. IEEE Wirel Commun Lett, 2021, 10: 1623--1627

\bibitem{3.2-45} Niu H C, Hu J F, Shi J, et al. Secrecy performance analysis for OTFS modulation based downlink LEO satellite communication. In: Proceedings of IEEE Global Communications Conference Workshops, Kuala Lumpur, 2023. 2135--2139

%%%%%%%%%3.3
\bibitem{mudumbai2009distributed} Mudumbai R, Brown D R III, Madhow U, et al. Distributed transmit beamforming: Challenges and recent progress. IEEE Commun Mag, 2009, 47: 102-110

\bibitem{nanzer2017open} Nanzer J A, Schmid R L, Comberiate T M, et al. Open-loop coherent distributed arrays. IEEE Trans Microw Theory Tech, 2017, 65: 1662-1672

\bibitem{nanzer2021distributed} Nanzer J A, Mghabghab S R, Ellison S M, et al. Distributed phased arrays: Challenges and recent advances. IEEE Trans Microw Theory Tech, 2021, 69(11): 4893-4907

\bibitem{rashid2023high} Rashid M, Nanzer J A. High accuracy distributed Kalman filtering for synchronizing frequency and phase in distributed phased arrays. IEEE Signal Process Lett, 2023, 30: 688-692

\bibitem{marrero2022architectures} Marrero L M, Merlano-Duncan J C, Querol J, et al. Architectures and synchronization techniques for distributed satellite systems: A survey. IEEE Access, 2022, 10: 45375-45409

\bibitem{zhao2021millimeter} Zhao D X, Gu P, Zhong J C, et al. Millimeter-wave integrated phased arrays. IEEE Trans Circuits Syst I Reg Papers, 2021, 68: 3977-3990

\bibitem{zhao2024w} Zhao D X, Yu P G, Jiang S, et al. W-band CMOS beamforming ICs and integrated phased-array antennas with 20+ Gb/s data rates. Sci China Inf Sci, 2024, 67: 212301

\bibitem{liu202224} Liu H Q, Zhao D X, Yi Y R, et al. A 24.25-27.5 GHz 128-element dual-polarized 5G integrated phased array with 5.6\%-EVM 400-MHz 64-QAM and 50-dBm EIRP. Sci China Inf Sci, 2022, 65: 214301

\bibitem{zhao2021design} Zhao D X, Chen Z H, You X H. Design and implementation of CMOS millimeter-wave ICs and 4096 TX/4096 RX large-scale integrated phased-array antenna. Sci Sin Inf, 2021, 51: 505-519

\bibitem{luo2020scalable} Luo X, Ouyang J, Chen Z-H, et al. A scalable Ka-band 1024-element transmit dual-circularly-polarized planar phased array for SATCOM application. IEEE Access, 2020, 8: 156084-156095

\bibitem{fu2023low} Fu X, You D, Wang Y, et al. A low-power radiation-hardened Ka-band CMOS phased-array receiver for small satellite constellation. IEEE J Solid-State Circuits, 2024, 59: 349-363

\bibitem{zhao2022k} Zhao D X, Gu P, Yi Y R, et al. A K-band hybrid-packaged temperature-compensated phased-array receiver and integrated antenna array. IEEE Trans Microw Theory Tech, 2022, 71(1): 409-423

\bibitem{mudumbai2007feasibility} Mudumbai R, Barriac G, Madhow U. On the feasibility of distributed beamforming in wireless networks. IEEE Trans Wirel Commun, 2007, 6: 1754-1763

\bibitem{coleri2002channel} Coleri S, Ergen M, Puri A, et al. Channel estimation techniques based on pilot arrangement in OFDM systems. IEEE Trans Broadcast, 2002, 48: 223-229

\bibitem{yang2024k} Yang M R, Zhao D X, Xu C Y, et al. K/Ka-band hybrid-packaged four-element four-beam phased-array transmitter and receiver front-ends with optimized beamforming passive networks. IEEE J Solid-State Circuits, 2024, 59: 3142-3155

\bibitem{yeh2020multibeam} Yeh Y S, Floyd B A. Multibeam phased-arrays using dual-vector distributed beamforming: Architecture overview and 28 GHz transceiver prototypes. IEEE Trans Circuits Syst I Reg Papers, 2020, 67: 5496-5509

\bibitem{hu2020orthogonal} Hu Y, Zhan J, Jiang Z H, et al. An orthogonal hybrid analog-digital multibeam antenna array for millimeter-wave massive MIMO systems. IEEE Trans Antennas Propag, 2020, 69: 1393-1403

\bibitem{he2021review} He G L, Gao X, Sun L L, et al. A review of multibeam phased array antennas as LEO satellite constellation ground station. IEEE Access, 2021, 9: 147142-147154

\bibitem{hu2018digital} Hu Y, Hong W, Yu C, et al. A digital multibeam array with wide scanning angle and enhanced beam gain for millimeter-wave massive MIMO applications. IEEE Trans Antennas Propag, 2018, 66: 5827-5837

\bibitem{yu2019multibeam} Yu Y, Hong W, Jiang Z H, et al. Multibeam generation and measurement of a DDS-based digital beamforming array transmitter at Ka-band. IEEE Trans Antennas Propag, 2019, 67: 3030-3039

\bibitem{jang201816} Jang S, Jeong J, Lu R, et al. A 16-element 4-beam 1 GHz IF 100 MHz bandwidth interleaved bit stream digital beamformer in 40 nm CMOS. IEEE J Solid-State Circuits, 2018, 53: 1302-1312

\bibitem{peng2020ka} Peng N, Gu P, You X H, et al. A Ka-band CMOS 4-beam phased-array receiver with symmetrical beam-distribution network. IEEE Solid-State Circuits Lett, 2020, 3: 410-413

\bibitem{talisa2016benefits} Talisa S H, O'Haver K W, Comberiate T M, et al. Benefits of digital phased array radars. Proc IEEE, 2016, 104: 530-543

\bibitem{sohrabi2016hybrid} Sohrabi F, Yu W. Hybrid digital and analog beamforming design for large-scale antenna arrays. IEEE J Sel Top Signal Process, 2016, 10: 501-513

\bibitem{wan2021performance} Wan S J, Zhu H B, Kang K, et al. On the performance of fully-connected and sub-connected hybrid beamforming system. IEEE Trans Veh Technol, 2021, 70: 11078-11082

%%%%%%%%3.4
\bibitem{3.4-1} Shannon C E. A mathematical theory of communication. Bell Syst Tech J, 1948, 5: 379--423

\bibitem{3.4-2} Yang G H, He G N, Chen R R, et al. Progress and prospect of 6G wireless air-interface transmission technology research (in Chinese). Sci China Inf Sci, 2024, 54: 1078--1113

\bibitem{3.4-3} Blahut R E. Algebraic codes for data transmission. New York: Cambridge University Press, 2003

\bibitem{3.4-4} Viterbi A. Error bounds for convolutional codes and an asymptotically optimum decoding algorithm. IEEE Trans Inf Theory, 1967, 13: 260--269

\bibitem{3.4-5} Arikan E. Channel polarization: a method for constructing capacity-achieving codes for symmetric binary-input memoryless channels. IEEE Trans Inf Theory, 2009, 55: 3051--3073

\bibitem{3.4-6} Eslami A, Pishro-Nik H. A practical approach to polar codes. In: Proceedings of IEEE International Symposium on Information Theory Proceedings, St. Petersburg, 2011. 16--20

\bibitem{3.4-7} Wang R X, Liu R K. A novel puncturing scheme for polar codes. IEEE Commun Lett, 2014, 18: 2081--2084

\bibitem{3.4-8} 3rd Generation Partnership Project (3GPP). 5G; NR; Multiplexing and channel coding (Release 15). TS 38.212. https://www.3gpp.org/ftp/Specs/archive/38\_series/38.212

\bibitem{3.4-9} Gallager R. Low-density parity-check codes. IEEE Trans Inf Theory, 1962, 8: 21--28

\bibitem{3.4-10} Berrou C, Glavieux A, Thitimajshima P. Near shannon limit error-correcting coding and decoding: turbo-codes. In: Proceedings of IEEE International Conference on Communications, Geneva, 1993. 1064--1070

\bibitem{3.4-11} Xu H Z, Feng D, Luo R, et al. Construction of quasi-cyclic LDPC codes via masking with successive cycle elimination. IEEE Commun Lett, 2016, 20: 2370--2373

\bibitem{3.4-12} CCSDS. Synchronization and channel coding. 131.0-B-2-S, 2011. https://public.ccsds.org/Pubs/131x0b2ec1s.pdf

\bibitem{3.4-13} Costello D, Dolecek L, Fuja T, et al. Spatially coupled sparse codes on graphs: theory and practice. IEEE Commun Mag, 2014, 52: 168--176

\bibitem{3.4-14} Kudekar S, Richardson T J, Urbanke R L. Threshold saturation via spatial coupling: why convolutional LDPC ensembles perform so well over the BEC. IEEE Trans Inf Theory, 2011, 57: 803--834.

\bibitem{3.4-15} Wijekoon V B, Viterbo E, Hong Y. LDPC-staircase codes for soft decision decoding. In: Proceedings of IEEE Wireless Communications and Networking Conference, Seoul, 2020. 1--6

\bibitem{3.4-16} Felstrom J A, Zigangirov K S. Time-varing periodic convolutional codes with low-density parity-check matrix. IEEE Trans Inf Theory, 1999, 45: 2181--2191

\bibitem{3.4-17} Mithcell D G M, Lentmaier M, Costello D J. Spatially coupled LDPC codes constructed from protographs. IEEE Trans Inf Theory, 2015, 61: 4866--4889

\bibitem{3.4-18} Du J Y. A partially coupled LDPC coded scheme for the Gaussian wiretap channel. IEEE Commun Lett, 2020, 24: 7--10

\bibitem{3.4-19} Zhang Y S, Jiao J, Wang Y, et al. Soft OSD-sliding window decoding for staircase LDPC codes in deep space communications. In: Proceedings of IEEE/CIC International Conference on Communications in China, Dalian, 2023. 1-6

\bibitem{3.4-20} Ordentlich O, Polyanskiy Y. Low complexity schemes for the random access Gaussian channel. In: Proceedings of IEEE International Symposium on Information Theory, Aachen, 2017. 2528--2532

\bibitem{3.4-21} Yang T, Yang L, Guo Y J, et al. A non-orthogonal multiple-access scheme using reliable physical-layer network coding and cascade-computation decoding. Trans Wirel Commun, 2017, 16: 1633--1645

\bibitem{3.4-22} Kowshik S S, Andreev K, Frolov A, et al. Energy efficient coded random access for the wireless uplink. IEEE Trans Commun, 2020, 68: 4694--4708

\bibitem{3.4-23} Zamir R, Shamai S, Erez U. Nested linear/lattice codes for structured multiterminal binning. IEEE Trans Inf Theory, 2002, 48: 1250--1276

\bibitem{3.4-24} Yang T, Yu F T, Chen Q Z, et al. On the design of efficient lattice-code based multiple access. In: Proceedings of IEEE Global Communications Conference, Kuala Lumpur, 2023. 1--7

\bibitem{3.4-25} Tal I, Vardy A. List decoding of polar codes. IEEE Trans Inf Theory, 2015, 61: 2213--2226

\bibitem{3.4-26} Niu K, Chen K. CRC-aided decoding of polar codes. IEEE Commun Lett, 2012, 16: 1668--1671

\bibitem{3.4-27} Miloslavskaya V, Vucetic B. Design of short polar codes for SCL decoding. IEEE Trans Commun, 2020, 68: 6657--6668

\bibitem{3.4-28} Fossorier M P, Mihaljevic M, Imai H. Reduced complexity iterative decoding of low-density parity check codes based on belief propagation. IEEE Trans Commun, 1999, 47: 673--680

\bibitem{3.4-29} Cui H X, Ghaffari F, Le K. Design of high-performance and area-efficient decoder for 5G LDPC codes. IEEE Trans Circuits Syst I Reg Papers, 2021, 68: 879--891

\bibitem{3.4-30} Duffy K R, Li J, M{\'{e}}dard M. Guessing noise, not code-words. In: Proceedings of IEEE International Symposium on Information Theory, Vail, 2018. 671--675

\bibitem{3.4-31} Fossorier M P C, Lin S. Soft-decision decoding of linear block codes based on ordered statistics. IEEE Trans Inf Theory, 1995, 41: 1379--1396

\bibitem{3.4-32} Yue C T, Shirvanimoghaddam M, Vucetic B, et al. Ordered-statistics decoding with adaptive Gaussian elimination reduction for short codes. In: Proceedings of IEEE Global Communication Conference Workshops, Rio de Janeiro, 2022. 492--497

\bibitem{3.4-33} Liang J F, Wang Y W, Cai S H, et al. A low-complexity ordered statistic decoding of short block codes. IEEE Commun Lett, 2023, 2: 400-403

\bibitem{3.4-34} Yue C T, Shirvanimoghaddam M, Li Y, et al. Segmentation discarding ordered-statistic decoding for linear block codes. In: Proceedings of IEEE Global Communication Conference, Waikoloa, 2019. 1--6

\bibitem{3.4-35} Yue C T, Shirvanimoghaddam M, Park G, et al. Probability-based ordered-statistics decoding for short block codes. IEEE Commun Lett, 2021, 25: 1791--1795

\bibitem{3.4-36} Urman Y, Mogilevsky G, Burshtein D. Improving belief propagation list decoding of polar codes by post-processing. In: Proceedings of IEEE International Symposium on Information Theory, Espoo, 2022. 2571--2576

\bibitem{3.4-37} Nikopourand H, Baligh H. Sparse code multiple access. In: Proceedings of IIEEE Annual International Symposium on Personal, Indoor, and Mobile Radio Communications, London, 2013. 332--336

\bibitem{3.4-38} Yang T. Beyond integer-forcing receiver for lattice-code based multi-user MIMO system. IEEE Commun Lett, 2023, 27: 2553--2557

\bibitem{3.4-39} Rao Z G, Jiao J, Wang Y, et al. Code-domain collision resolution grant-free random access for massive access in IoT. IEEE Trans Wirel Commun, 2023, 22: 4611-4624

%%%%%%%3.5
\bibitem{KeIoTJ} Gao Z, Ke M L, Mei Y K, et al. Compressive-sensing-based grant-free massive access for 6G massive communication. IEEE Internet Things J, 2024, 1: 7411-7435

\bibitem{21Location} Wang W J, Chen T T, Ding R, et al. Location-based timing advance estimation for 5G integrated LEO satellite communications. IEEE Trans Veh Technol, 2021, 70: 6002-6017

\bibitem{21Golay} Yu N Y, Binary Golay spreading sequences and Reed-Muller codes for uplink grant-free NOMA. IEEE Trans Commun, 2021, 69: 276-290

\bibitem{24mask} Yu N Y, Yu W. Joint activity and data detection for massive grant-free access using deterministic non-orthogonal signatures. IEEE Trans Wirel Commun, 2024, 23: 9474-9487

\bibitem{23diff} Wang Y, Xu W J, Juntti M, et al.  Composite preambles based on differential phase rotations for grant-free random access systems. IEEE Internet Things J, 2023, 10: 17035-17046

\bibitem{10108023} Qi T, Lyu B, Hoang D T. Pilot sequences with low coherence and PAPR for grant-free massive access. IEEE Wirel Commun Lett, 2023, 12(7): 1254--1258

\bibitem{10246293} Xu L, Jiao J, Wang Y, et al. Low-correlation superimposed pilot grant-free massive access for satellite Internet of things. IEEE Trans Commun, 2023, 71(12): 7087--7101

\bibitem{9686735} Fengler A, Musa O, Jung P, et al. Pilot-based unsourced random access with a massive MIMO receiver, interference cancellation, and power control. IEEE J Sel Areas Commun, 2022, 40(5): 1522--1534

\bibitem{20IoTJ} Zhang Z J, Li Y, Huang C W, et al. User activity detection and channel estimation for grant-free random access in LEO satellite-enabled Internet of things. IEEE Internet Things J, 2020, 7: 8811-8825

%\bibitem{23OFDM_Rep}
%Y. Zuo, M. Yue, M. Zhang, S. Li, S. Ni, and X. Yuan, ``OFDM-based massive connectivity for LEO satellite Internet of things,'' \emph{IEEE Trans. Wireless Commun.}, vol.~22, no.~11, pp.~8244-8258, Nov. 2023.

\bibitem{23ZXY_TWC} Zhou X Y, Ying K K, Gao Z, et al. Active terminal identification, channel estimation, and signal detection for grant-free NOMA-OTFS in LEO satellite Internet-of-things. IEEE Trans Wirel Commun, 2023, 22: 2847-2866

\bibitem{24LDS} Zhang C, Liu Y S, Hu J, et al. Joint user identification, channel estimation and data Detection for grant-free NOMA in LEO satellite communications. IEEE J Sel Areas Commun, 2024. doi: 10.1109/JSAC.2024.3460059

\bibitem{24AFDM} Luo Q, Xiao P, Liu Z L, et al. AFDM-SCMA: A promising waveform for massive connectivity over high mobility channels. IEEE Trans Wirel Commun, 2024, 23: 14421-14436

\bibitem{10839280} Shen B X, Wu Y P, Zhang W J, et al. LEO satellite-enabled random access with large differential delay and Doppler shift. IEEE Trans Wirel Commun, 2025, pages: 1--1

\bibitem{10492466} Le T T T, Hassan N U, Chen X M, et al. A survey on random access protocols in direct-access LEO satellite-based IoT communication. IEEE Commun Surv Tutorials, 2025, 27(1): 426--462

\bibitem{3.7-27} Kaul S, Yates R, Gruteser M. Real-time status: How often should one update?. In: Proceedings of International Conference on Computer Communications, Orlando, 2012, 2731-2735

\bibitem{9766119} Yang T, Jiao J, Wu S H, Lu R X, et al. Grant free age-optimal random access protocol for satellite-based Internet of things. IEEE Trans Commun, 2022, 70(6): 3947-3961

\bibitem{10412105} Su S Y, Jiao J, Yang T, et al. Unequal timeliness protection massive access for mission critical communications in S-IoT. IEEE Trans Commun, 2024, 72(6): 3211--3226  

\bibitem{19MSC_RA} Zhao B, Ren G L, Zhang H N. Multisatellite cooperative random access scheme in low Earth orbit satellite networks. IEEE Syst J, 2019, 13: 2617--2628

\bibitem{KeJSAC} Ying K K, Gao Z, Chen S, et al. Quasi-synchronous random access for massive MIMO-based LEO satellite constellations. IEEE J Sel Areas Commun, 2023, 41: 1702-1722

%%%%%%%%%%3.6
\bibitem{7973146} Ding Z G, Lei X F, Karagiannidis G.K, et al. A survey on non-orthogonal multiple access for 5G networks: Research challenges and future trends. IEEE J Sel Areas Commun, 2017, 35: 2181-2195

\bibitem{8957684} Jiao J, Sun Y Y, Wu S H, et al. Network utility maximization resource allocation for NOMA in satellite-based Internet of Things. IEEE Internet Things J, 2020, 7: 3230-3242

\bibitem{8357810} Dai L L, Wang B C, Ding Z G, et al. A survey of Non-Orthogonal multiple access for 5G. IEEE Commun Surv Tut, 2018, 20: 2294-2323

\bibitem{7906532} Ding Z G, Zhao Z Y, Peng M G, et al. On the spectral efficiency and security enhancements of NOMA assisted multicast-unicast streaming. IEEE Trans Commun, 2017, 65: 3151-3163

\bibitem{9115278} Zhang H J, Zhang H S, Liu W, et al. Energy efficient user clustering, hybrid precoding and power optimization in Terahertz MIMO-NOMA systems. IEEE J Sel Areas Commun, 2020, 38: 2074-2085

\bibitem{7560605} Di B Y, Song L Y, Li Y H. Sub-channel assignment, power allocation, and user scheduling for non-orthogonal multiple access networks. IEEE Trans Wirel Commun, 2016, 15: 7686-7698

\bibitem{10336741} Hong H, Jiao J, Yang T, et al. Age of incorrect information minimization for semantic-empowered NOMA system in S-IoT. IEEE Trans Wirel Commun, 2024, 23: 6639-6652

\bibitem{kodheli2020satellite} Kodheli O, Lagunas E, Maturo N, et al. Satellite communications in the new space era: A survey and future challenges. IEEE Commun Surv Tut, 2020, 23: 70-109

\bibitem{perez2019non} Perez-Neira A I, Caus M, Vazquez M A. Non-orthogonal transmission techniques for multibeam satellite systems. IEEE Commun Mag, 2019, 57: 58-63

\bibitem{chu2021robust} Chu J H, Chen X M. Robust design for integrated satellite-terrestrial Internet of Things. IEEE Internet Things J, 2021, 57: 9072-9083

\bibitem{lin2019joint} Lin Z, Lin M, Wang J-B, et al. Joint beamforming and power allocation for satellite-terrestrial integrated networks with non-orthogonal multiple access. IEEE J Sel Top Signal Process, 2019, 13: 657-670

\bibitem{Liu2021Spectrum} Liu X, Lam K Y, Li F, et al. Spectrum sharing for 6G integrated satellite-terrestrial communication networks based on NOMA and CR. IEEE Netw, 2021, 35: 28-34

\bibitem{jiao2023age} Jiao J, Hong H, Wang Y, et al. Age-optimal downlink NOMA resource allocation for satellite-based IoT network. IEEE Trans Veh Technol, 2023, 72: 211575-11589

\bibitem{Zhao2024Dynamic} Zhao M H, Yu H X, Pan J X, et al. Dynamic resource allocation for multi-satellite cooperation networks: A decentralized scheme under statistical CSI. IEEE Access, 2024, 12: 15419-15437

\bibitem{Zhao2023Multi} Zhao M H, Ye N, Ouyang Q L, et al. Multi-satellite cooperative communication: Exploiting time asynchrony in non-orthogonal transmissions. IEEE Trans Veh Technol, 2023, 72: 6868-6873.

\bibitem{9093213} Huang H J, Yang Y C, Ding Z G, et al. Deep learning-based sum data rate and energy efficiency optimization for MIMO-NOMA systems. IEEE Trans Wirel Commun, 2020, 19(8): 5373--5388

\bibitem{10824682} Ding Z G. A study on the optimality of downlink hybrid NOMA. IEEE Signal Process Lett, 2025, 32: 511--515

\bibitem{10283937} Katwe M, Singh K, Li C P, et al. Spectral-efficient downlink systems under imperfect SIC and CSI: MC-NOMA or partial NOMA? IEEE Wirel Commun Lett, 2024, 13(1): 133--137

\bibitem{10185599} Shi Z Y, Lu H B, Xie X Z, et al. Active RIS-aided EH-NOMA networks: A deep reinforcement learning approach. IEEE Trans Commun, 2023, 71(10): 5846--5861

\bibitem{10103152} Deshpande R, Katwe M V, Singh K, et al. Resource allocation design for spectral-efficient URLLC using RIS-aided FD-NOMA system. IEEE Wirel Commun Lett, 2023, 12(7): 1209--1213

\bibitem{mao2022rate} Mao Y J, Dizdar O, Clerckx B, et al. Rate-splitting multiple access: Fundamentals, survey, and future research trends. IEEE Commun Surv Tut, 2022, 24: 2073-2126

\bibitem{mishra2021rate} Mishra A, Mao Y J, Dizdar O, Clerckx, B. Rate-splitting multiple access for downlink multiuser MIMO: Precoder optimization and PHY-layer design. IEEE Trans Commun, 2021, 70: 874-890

\bibitem{9217326} Dizdar O, Mao Y J, Han W, et al. Rate-splitting multiple access for downlink multi-antenna communications: Physical layer design and link-level simulations. In: Proceedings of IEEE Annual International Symposium on Personal, Indoor and Mobile Radio Communications, London, 2020. 1-6

\bibitem{de2022rate} de Sena A S, Nardelli P H J, da Costa D B, et al. Rate-splitting multiple access and its interplay with intelligent reflecting surfaces. IEEE Commun Mag, 2022, 60: 52-57

\bibitem{9420034} Yin L F, Clerckx B, Mao Y J. Rate-splitting multiple access for multi-antenna broadcast channels with statistical CSIT. In: Proceedings of IEEE Wireless Communications and Networking Conference Workshops, Nanjing, 2021. 1-6

\bibitem{zhuo2023ris} Zhuo B T, Gu J P, Duan W, et al. RIS-IoE for data-driven networks: New mentalities, trends and preliminary solutions. IEEE Internet Things Mag, 2023, 6: 102-107

\bibitem{Huang2024} Huang J F, Yang Y, Lee J, et al. Deep reinforcement learning based resource allocation for RSMA in LEO satellite-terrestrial networks. IEEE Trans Commun, 2024, 72: 1341-1354

\bibitem{xu2024distributed} Xu Y N, Yin L F, Mao Y J, et al. Distributed rate-splitting multiple access for multilayer satellite communications. IEEE Trans Commun, 2024, 72: 6131-6144

\bibitem{8385504} Vazquez M, Caus M, Perez-Neira A. rate splitting for MIMO multibeam satellite systems. In: Proceedings of International ITG Workshop on Smart Antennas, Bochum, 2018. 1-6

\bibitem{9684855} Si Z W, Yin L F, Clerckx B. Rate-splitting multiple access for multigateway multibeam satellite systems with feeder link interference. IEEE Trans Commun, 2022, 70: 2147-2162


%%%%3.7
\bibitem{3.7-1} Liu J J, Shi Y P, Fadlullah Z M. Space-air-ground integrated network: A survey. IEEE Commun Surv Tut, 2018, 20: 27-14

\bibitem{3.7-2} Leung K-C, Li V O K. Transmission control protocol (TCP) in wireless networks: Issues, approaches, and challenges. IEEE Commun
Surv Tut, 2006, 8: 64--79

\bibitem{3.7-3} CCSDS. Overview of space communications protocols, report concerning space data system standards. 130.0-G-2, 2007. https://public.ccsds.org/Pubs/130x0g2s.pdf

\bibitem{3.7-4} Fall K. A delay-tolerant network architecture for challenged Internets. In: Proceedings of ACM Conference on Applications, Technologies, Architectures, and Protocols for Computer Communications, Karlsruhe, 2003. 27-34

\bibitem{3.7-5} Burleigh S, Ramadas M, Farrell S. Licklider Transmission Protocol-Specification. IETF RFC 5325, 2008. https://datatracker.ietf.org/doc/html/rfc5325

\bibitem{3.7-6} Shi L L, Jiao J, Sabbagh A, et al. Integration of Reed-Solomon codes to licklider transmission protocol (LTP) for space DTN. IEEE Aerosp Electron Syst Mag, 2017, 32: 48--55

\bibitem{3.7-7} Hasegawa Y, Ito T, Ono Y, et al. A throughput model of TCP-FSO/ADFR for free-space optical satellite communications. In: Proceedings of IEEE Global Communications Conference, Waikoloa, 2019. 1--6

\bibitem{3.7-8} Le H D, Mai V V, Nguyen C T, et al. Design and analysis of sliding window ARQ protocols with rate adaptation for burst transmission over FSO turbulence channels. J Opt Commun Netw, 2019, 11: 151--163

%\bibitem{3.7-9} Chen H, Maunder G R, Hanzo L. A survey and tutorial on low
%complexity turbo coding techniques and a holistic hybrid ARQ design example. IEEE Commun Surv Tutor, 2013, 15: 1546--1566

\bibitem{3.7-9} Rosas F, Souza R D, Pellenz M E, et al. Optimizing the code rate of energy-constrained wireless communications with HARQ. IEEE Trans Wirel Commun, 2016, 15: 191--205

\bibitem{3.7-10} Lee W, Simeone O, Kang J, et al. HARQ buffer management: An information-theoretic view. IEEE Trans Commun, 2015, 63: 4539--4550

\bibitem{3.7-11} Ahmed A, Al-Dweik A, Iraqi Y, et al. Hybrid automatic repeat request (HARQ) in wireless communications systems and standards: A contemporary survey. IEEE Commun Surv Tut, 2021, 23: 2711--2752

\bibitem{3.7-12} Makki B, Svensson T, Caire G, et al. Fast HARQ over finite block length codes: A technique for low-latency reliable communication. IEEE Trans Wirel Commun, 2019, 18: 194--209

\bibitem{3.7-13} Shirvanimoghaddam M, Khayami H, Li Y. Dynamic HARQ with guaranteed delay. In: Proceedings of IEEE Wireless Communications and Networking Conference, Seoul, 2020. 1--6

\bibitem{3.7-14} Berardinelli G, Khosravirad S R, Pedersen K I, et al. Enabling early HARQ feedback in 5G networks. In: Proceedings of IEEE Vehicular Technology Conference, Nanjing, 2016. 1--5

\bibitem{3.7-15} Strodthoff N, G{\"{o}}ktepe B, Schierl T, et al. Enhanced machine learning techniques for early HARQ feedback prediction in 5G. IEEE J Sel Areas Commun, 2109, 37: 2573--2587

\bibitem{3.7-16} CCSDS. Erasure correcting codes for use in near-earth and deep-space communications. 131.5-O-1, 2014. https://public.ccsds.org/Pubs/1 31x 5o1.pdf

\bibitem{3.7-17} Shirvanimoghaddam M, et al., Short block-length codes for ultra-reliable low latency communications. IEEE Commun Mag, 2019, 57: 130--137

\bibitem{3.7-18} Luby M. LT codes. In: Proceedings of IEEE Symposium on Foundations of Computer Science, Vancouver, 2002. 271-280

\bibitem{3.7-19} Shokrollahi A. Raptor codes. IEEE Trans Inf Theory, 2006, 21: 2551--2567

\bibitem{3.7-20} 3rd Generation Partnership Project (3GPP). Technical specification group services and system aspects; Multimedia broadcast/multicast service (MBMS); Protocols and codecs (Release 6). TS 26.346. https://www.3gpp.org/ftp/Specs/archive/26\_series/26.346

\bibitem{3.7-21} S{\o}rensen J H, Koike-Akino T, Orlik P. Rateless feedback codes. In: Proceedings of IEEE International Symposium on Information Theory, Cambridge, 2012. 1767--1771

\bibitem{3.7-22} Jiao J, Nie S X, Yang Y, et al. Distributed systematic Raptor coding scheme in deep space communications (in Chinese). J Astronaut, 2016, 37: 1232-1238

\bibitem{3.7-23} Ho T, Medard M, Koetter R, et al. A random linear network coding approach to multicast. IEEE Trans Inf Theory, 2006, 52: 4413-4430

\bibitem{3.7-24} Tsimbalo E, Tassi A, Piechocki R J. Reliability of multicast under random linear network coding. IEEE Trans Commun, 2018, 66: 2547-2559

\bibitem{3.7-25} Yang S H, Yeung R W. Batched Sparse Codes. IEEE Trans Inf Theory, 2014, 60: 5322-5346
%Yang S, Yeung R W. Coding for a network coded fountain. In: Proceedings of IEEE International Symposium on Information Theory, St. Petersburg, 2011. 2647--2651

\bibitem{3.7-26} Jiao J, Ni Z N, Wu S H, et al. Energy efficient network coded HARQ transmission scheme for S-IoT. IEEE Trans Green Commun Netw, 2021, 5: 308-321


\bibitem{3.7-28} Jiao J, Liu S Q, Ding J, et al. Age-optimal network coding HARQ transmission scheme for dual-hop satellite-integrated Internet. IEEE Trans Veh Technol, 2022, 71: 10666-10682

\bibitem{3.7-29} Ding J, Jiao J, Huang J H, et al. Age-optimal network coding HARQ scheme for satellite-based Internet of things. IEEE Internet Things J, 2022, 9: 21984-21998

\bibitem{3.7-30} Huang J H, Jiao J, Wang Y, et al. Age-critical long erasure coding-CCSDS file delivery protocol for dual-hop S-IoT. IEEE Internet Things J, 2023, 10: 17070-17084

\bibitem{3.7-31} Li D Q, Wu S H, Jiao J, et al. Towards age-optimal transmission in satellite-integrated IoT: A two-layer coding approach. IEEE Trans Veh Technol, 2023, 72: 1137-1148

\bibitem{3.7-32} Le H D, Trinh P V, Pham T V, et al. Throughput analysis for TCP over the FSO-based satellite-assisted Internet of vehicles. IEEE Trans Veh Technol, 2022, 71: 1875-1890

\bibitem{3.7-33} Hu C B, Yang H J, Li B, et al. A high-throughput cooperative network coding HARQ transmission scheme for integrated satellite-terrestrial networks. In: Proceedings of IEEE Vehicular Technology Conference, Hong Kong, 2023. 1-5

\bibitem{3.7-34} Cai D H, Ding Z G, Fan P Z, et al. On the performance of NOMA with hybrid ARQ. IEEE Trans Veh Technol, 2018, 67: 10033-10038

\bibitem{3.7-35} Shi Z, Zhang C M, Fu Y R, et al. Achievable diversity order of HARQ-aided downlink NOMA systems. IEEE Trans Veh Technol, 2020, 69: 471-487

\bibitem{3.7-36} Ghanami F, Hodtani G A, Vucetic B, et al. Performance analysis and optimization of NOMA with HARQ for short packet communications in Massive IoT. IEEE Internet Things J, 2021, 8: 4736-4748

\bibitem{3.7-37} Marasinghe D, Rajatheva N, Latva-Aho M. Block error performance of NOMA with HARQ-CC in finite blocklength. In: Proceedings of IEEE International Conference Communications Workshops, Dublin, 2020, 1-6

\bibitem{3.7-38} Wu S H, Deng Z H, Li A M, et al. Minimizing age-of-information in HARQ-CC aided NOMA systems. IEEE Trans Wirel Commun, 2023, 22: 1072-1086

\bibitem{3.7-39} Liu K P, Li A M, Wu S H. Deep reinforcement learning-assisted age-optimal transmission policy for HARQ-aided NOMA networks. In: Proceedings of IEEE Conference on Computer Communications Workshops, Hoboken, 2023. 1-6

%\bibitem{3.7-33} Ceran E T, G{\"{u}}nd{\"{u}}z D, Gy{\"{o}}rgy A. A reinforcement learning approach to age of information in multi-user networks with HARQ. IEEE J Sel Areas Commun, 2021, 39: 1412--1426
%\bibitem{3.7-35} Vilni S S, Moltafet M, Leinonen M. Multi-source AoI-constrained resource minimization under HARQ: Heterogeneous sampling processes. IEEE Trans Veh Technol, 2024, 73: 1084--1099

%%%%%%%%%3.8
\bibitem{samy2022space} Samy R, Yang H-C, Rakia T, Alouini M-S. Space-air-ground FSO networks for high-throughput satellite communications. IEEE Commun Mag, 2022, 60: 82-87
\bibitem{5.5-2} Al-Hraishawi H, et al. A survey on nongeostationary satellite systems: The communication perspective. IEEE Commun Surv Tut, 2023, 25: 101--132
\bibitem{le2022link} Le H D, Pham A T. Link-layer retransmission-based error-control protocols in FSO communications: A survey. IEEE Commun Surv Tut, 2022, 24(3): 1602-1633
\bibitem{10477428} Feng W, Wang Y M, Chen Y F, et al. Structured satellite-UAV-terrestrial networks for 6G Internet of Things. IEEE Netw, 2024, 38(4): 48--54
\bibitem{arum2020review} Arum S C, Grace D, Mitchell P D. A review of wireless communication using high-altitude platforms for extended coverage and capacity. Computer Communications, 2020, 157: 232-256
\bibitem{kaymak2018survey} Kaymak Y, Rojas-Cessa R, Feng J H, Ansari N, et al. A survey on acquisition, tracking, and pointing mechanisms for mobile free-space optical communications. IEEE Commun Surv Tut, 2018, 20(2): 1104-1123
\bibitem{moon2024pointing} Moon H-J, Chae C-B, Wong K-K, Alouini M-S. Pointing-and-acquisition for optical wireless in 6G: From algorithms to performance evaluation. IEEE Commun Mag, 2024, 62(3): 32-38
\bibitem{2019USAConference} Velazco J E, Wernicke D, Griffin J, et al. Inter-spacecraft omnidirectional optical communicator for swarms. In: Proceedings of Annual AIAA/USU Conference, 2019


\bibitem{sundararaman2005clock} Sundararaman B, Buy U, Kshemkalyani A D. Clock synchronization for wireless sensor networks: A survey. Ad hoc networks, 2005, 3(3): 281-323

\bibitem{9348867} Jeong S, Farhang A, Flanagan M. Collaborative Vs. non-collaborative CFO estimation for distributed large-scale MIMO systems. In: Proceedings of IEEE Vehicular Technology Conference, Victoria, BC, 2020. 1-6
\bibitem{wu2019deep} Wu H M, Sun Z, Zhou X. Deep learning-based frame and timing synchronization for end-to-end communications. Journal of Physics: Conference Series, 2019, 1169(1): 012060
\bibitem{wang2013learning} Wang Y Y, Zhang C, Peng Q, Wang Z H. Learning to detect frame synchronization. In: Proceedings of ICONIP, Daegu, 2013. 570-578

\bibitem{qi2021using} Qi X X, Zhang B, Qiu Z L, et al. Using inter-mesh links to reduce end-to-end delay in walker delta constellations. IEEE Communications Letters, 2021, 25: 3070-3074

\bibitem{dong2023novel} Dong Y Y, Xu X F, Zhang Y Y, et al. A novel virtual node-based multi-controller management architecture for LEO mega-constellation satellite networks. In: Proceedings of IEEE International Conference on Wireless Communications and Signal Processing, Hangzhou, 2023. 701-706

\bibitem{hu2022software} Hu M L, Li J, Cai C, Deng T P, et al. Software defined multicast for large-scale multi-layer leo satellite networks. IEEE Trans Netw Serv Manag, 2022, 19: 2119-2130

\bibitem{jiang2023software} Jiang W W. Software defined satellite networks: A survey. Digit Commun Netw, 2023, 9: 1243-1264

\bibitem{chen2019topology} Chen Q, Guo J M, Yang L, et al. Topology virtualization and dynamics shielding method for LEO satellite networks. IEEE Commun Lett, 2019, 24: 433-437

\bibitem{wang2023reliability} Wang R B, Kishk M A, Alouini M-S. Reliability analysis of multi-hop routing in multi-tier leo satellite networks. IEEE Trans Wirel Commun, 2024, 23: 1959-1973

\bibitem{hu2022qos} Hu M L, Yang R H, Hu Y, et al. QoS-aware software-defined multicast in LEO satellite networks. IEEE Trans Aerosp Electron Syst, 2022, 58: 5307-5317

\bibitem{roth2022distributed} Roth M M, Brandt H, Bischl H. Distributed SDN-based load-balanced routing for low earth orbit satellite constellation networks. In: Proceedings of IEEE Advanced Satellite Multimedia Systems Conference and Signal Processing for Space Communications Workshop, Graz, 2022. 1-8

\bibitem{deng2022distance} Deng X, Chang L, Zeng S Y, et al. Distance-based back-pressure routing for load-balancing leo satellite networks. IEEE Trans Veh Technol, 2022, 72: 1240-1253

\bibitem{kumar2021fybrrlink} Kumar P, Bhushan S, Halder D, et al. fybrrlink: efficient QoS-aware routing in SDN enabled future satellite networks. IEEE Trans Netw Serv Manag, 2021, 19: 2107-2118

\bibitem{soret2024q} Soret B, Leyva-Mayorga I, Lozano-Cuadra F, et al. Q-learning for distributed routing in LEO satellite constellations. In: Proceedings of IEEE International Conference on Machine Learning for Communication and Networking, Stockholm, 2024. 208-213

\bibitem{10623789} Pachler N, Crawley E F, Cameron B G. Robust beam-to-satellite routing strategies for megaconstellations. IEEE Wirel Commun Let, 2024, 13: 3040-3043

\bibitem{ji2021mega} Ji S J, Zhou D, Sheng M, et al. Mega satellite constellation system optimization: From a network control structure perspective. IEEE Trans Wirel Commun, 2021, 21: 913-927

\bibitem{wei2006fast} Wei D X, Jin C, Low S H, et al. FAST TCP: Motivation, architecture, algorithms, performance. IEEE/ACM Trans Netw, 2006, 14: 1246-1259

\bibitem{ahmad2019enhancing} Ahmad S, Arshad M J. Enhancing fast TCP's performance using single TCP connection for parallel traffic flows to prevent head-of-line blocking. IEEE Access, 2019, 7: 148152-148162

\bibitem{deutschmann2023cubic} Deutschmann J, Hielscher K-S, German R. CUBIC local loss recovery vs. BBR on (satellite) Internet paths. In: Proceedings of IEEE International Symposium on Local and Metropolitan Area Networks, London, 2023. 1-3

\bibitem{claypool2021comparison} Claypool S, Chung J, Claypool M. Comparison of TCP congestion control performance over a satellite network. In: Proceedings of International Conference on Passive and Active Network Measurement, 2021. 499-512

\bibitem{liu2022effects} Liu M X, Liu Y C, Ma Z F, et al. The effects of a performance enhancing proxy on TCP congestion control over a satellite network. In: Proceedings of IEEE International Performance, Computing, and Communications Conference, Austin, 2022. 325-331

\bibitem{9484567} Shreedhar T, Kaul S K, Yates R D. An empirical study of ageing in the Cloud. In: Proceedings of IEEE Conference on Computer Communications Workshops, Vancouver, 2021. 1-6

\bibitem{9562140} Guloglu U, Baghaee S, Uysal E. Evaluation of age control protocol (ACP) and ACP+ on ESP32. In: Proceedings of IEEE International Symposium on Wireless Communication Systems, Berlin, 2021. 1-6

\bibitem{chiu1989analysis} Chiu D-M, Jain R. Analysis of the increase and decrease algorithms for congestion avoidance in computer networks.  Comput Netw ISDN Syst, 1989, 17: 1-14

\bibitem{10228914} Cao X Y, Zhang X Y. SaTCP: Link-layer informed TCP adaptation for highly dynamic LEO satellite networks. In: Proceedings of IEEE Conference on Computer Communications, New York, 2023. 1-10


%%%%%%%%%4.1
\bibitem{4.1-1} Ma T, Qian B, Qin X H, et al. Satellite-terrestrial integrated 6G: An ultra-dense LEO networking management architecture. IEEE Wirel Commun, 2024, 31: 62--69

\bibitem{4.1-2} Sun Y H, Peng M G, Zhang S J, et al. Integrated satellite-terrestrial networks: Architectures, key techniques, and experimental progress. IEEE Netw, 2022, 36: 191--198

\bibitem{4.1-3} 3rd Generation Partnership Project (3GPP). Solutions for NR to support non-terrestrial networks (NTN) (Release 16). TR 38.821. https://www.3gpp.org/ftp/Specs/archive/38\_series/39.821

\bibitem{4.1-4} Portillo I, Cameron B, Crawley C. Ground segment architectures for large LEO constellations with feeder links in EHF-bands. In: Proceedings of IEEE Aerospace Conference, Big Sky, 2018. 1-14

\bibitem{4.1-5} Papa A, Cola T, Vizarreta P, et al. Design and evaluation of reconfigurable SDN LEO constellations. IEEE Trans Netw Serv Manage, 2020, 17: 1432--1445

\bibitem{4.1-6} Ji S J, Zhou D, Sheng M, et al. Mega satellite constellation system optimization: From a network control structure perspective. IEEE Trans Wirel Commun, 2022, 21: 913--927
%Kim J, Lee J, Ko H, et al. Space mobile networks: Satellite as core and access networks for B5G. IEEE Commun Mag, 2022, 60: 58--64

\bibitem{4.1-7} Du P P, Bai W G, Li J D, et al. Dynamic hierarchical VAP-based location management for mega satellite networks. IEEE Internet Things J, 2024, 11: 19749--19761

\bibitem{4.1-8} 3rd Generation Partnership Project (3GPP). 5G; System architecture for the 5G system (5GS) (Release 16). TS 23.501. https://www.3gpp.org/ftp/Specs/archive/23\_series/23.501

\bibitem{4.1-9} Ji S J, Zhou D, Sheng M, et al. Dynamic space-ground integrated mobility management strategy for mega LEO satellite constellations. IEEE Trans Wirel Commun, 2024, 23: 11043--11060

\bibitem{4.1-10} Ji S, Sheng M, Zhou D, et al. Flexible and distributed mobility management for integrated terrestrial-satellite networks: Challenges, architectures, and approaches. IEEE Netw, 2021, 35:  73--81

\bibitem{4.1-11} Chowdhury P K, Atiquzzaman M, Ivancic W. Handover schemes in satellite networks: state-of-the-art and future research directions. IEEE Commun Surv Tut, 2006, 8: 2--14.

\bibitem{4.1-12} Re E D, Fantacci R, Giambene G. Efficient dynamic channel allocation techniques with handover queuing for mobile satellite networks. IEEE J Sel Areas Commun, 1995, 13: 397--405

\bibitem{4.1-13} Maral G, Restrepo J, Re E, et al. Performance analysis for a guaranteed handover service in an LEO constellation with a ``satellite-fixed cell" system. IEEE Trans Veh Technol, 1998, 47: 1200--1214

\bibitem{4.1-14} Wu Z F, Jin F L, Luo J X, et al. A graph-based satellite handover framework for LEO satellite communication networks. IEEE Commun Lett, 2016, 20: 1547--1550

\bibitem{4.1-15} Feng L, Liu Y F, Wu L, et al. A satellite handover
strategy based on MIMO technology in LEO satellite networks. IEEE Commun Lett, 2020, 24: 1505--1509

\bibitem{4.1-16} Zhang S B, Liu A J, Han C, et al. A network-flows-based satellite handover strategy for LEO satellite networks. IEEE Wirel Commun Lett, 2121, 10: 2669--2673.

\bibitem{4.1-17} Xu H H, Li D S, Liu M L, et al. QoE-driven intelligent handover for user-centric mobile satellite networks. IEEE Trans Veh Technol, 2020, 69: 10127--10139

\bibitem{4.1-18} He S, Wang T, Wang S. Load-aware satellite handover strategy based on multi-agent reinforcement learning. In: Proceedings of IEEE Global Communications Conference, Taipei, 2020. 1--6

\bibitem{4.1-19} Cao Y, Lien S-Y, Liang Y-C. Deep reinforcement learning For multi-user access control in non-terrestrial networks. IEEE Trans Commun, 2021, 69: 1605--1619

\bibitem{4.1-20} Yang F, Wu W J, Gao Y, et al. Multi-agent fingerprints-enhanced distributed intelligent handover algorithm in LEO satellite networks. IEEE Trans Veh Technol, 2024, 73: 15255--15269

\bibitem{4.1-21} Liu H T, Wang Y C, Li P X, et al. A multi-agent deep reinforcement learning-based handover scheme for mega-constellation under dynamic propagation conditions. IEEE Trans WireL Commun, 2024, 23: 13579-13596

\bibitem{4.1-22} Pacheco-Paramo D, Akyildiz I F, Casares-Giner V. Local anchor based location management schemes for small cells in HetNets. IEEE Trans Mobile Comput, 2016, 15: 883--894
%Chaurasia S N, Singh A. A hybrid evolutionary approach to the registration area planning problem. Appl Intell, 2014, 41: 1127--1149

\bibitem{4.1-23} Deng T, Wang X, Fan P Z, et al. Modeling and performance analysis of a tracking-area-list-based location management scheme in LTE
networks. IEEE Trans Veh Technol, 2016, 65: 6417--6431

\bibitem{4.1-24} Johnson D, Perkins C, Arkko J. Mobility support in IPv6. IETF, RFC 3775, 2004. https://www.rfc-editor.org/pdfrfc/rfc6275.txt.pdf

\bibitem{4.1-25} Darwish T, Kurt G K, Yanikomeroglu H, et al. Location management in Internet protocol-based future LEO satellite networks: A review. IEEE Open J Commun Soc, 2022, 3: 1035--1062

\bibitem{4.1-26} Shahriar A Z M, Atiquzzaman M, and Rahman S. Mobility management protocols for next-generation all-IP satellite networks. IEEE Wirel Commun, 2008, 15: 46--54

\bibitem{4.1-27} Du P P, Li J D, Bai W G, et al. Dual location area based distributed location management for hybrid LEO/MEO mega satellite networks. IEEE Trans Veh Technol, 2023, 72: 2307--2321

\bibitem{4.1-28} Li D G, Li H Y, Zhang S, et al. Virtual agent clustering based mobility management over the satellite networks. In: Proceedings of IEEE International Conference on Wireless Communications and Signal Processing, Hangzhou, 2018. 1--5

\bibitem{4.1-29} Jeon S, Figueiredo S, Aguiar R L, et al. Distributed mobility management for the future mobile networks: A comprehensive analysis of key design options. IEEE Access, 2017, 5: 11423--11436

\bibitem{4.1-30} Zhang X S, Shi K Y, Zhang S, et al. Virtual agent clustering based mobility management over the satellite networks. IEEE Access, 2019, 7: 89544--89555

%%%%%%4.2
\bibitem{9500539} Xu X B, Zhao H, Liu C, et al. On the Aggregated Resource Management for Satellite Edge Computing. In: Proceedings of IEEE International Conference on Communications, Montreal, 2021. 1-6

\bibitem{jia2020intelligent} Jia M, Zhang X M, Sun J T, et al. Intelligent resource management for satellite and terrestrial spectrum shared networking toward B5G. IEEE Wirel Commun, 2020, 27: 54-61

\bibitem{deng2022dynamic} Deng D, Wang C W, Pang M L, et al. Dynamic resource management in integrated NOMA terrestrial-satellite networks using multi-agent reinforcement learning. IEEE Wirel Commun Lett, 2022, 12: 75-79

\bibitem{cao2022edge} Cao X L, Yang B, Shen Y L, et al. Edge-assisted multi-layer offloading optimization of LEO satellite-terrestrial integrated networks. IEEE J Sel Areas Commun, 2022, 41: 381-398

\bibitem{he2023hierarchical} He H M, Zhou D, Sheng M, et al. Hierarchical cross-domain satellite resource management: An intelligent collaboration perspective. IEEE Trans Commun, 2023, 71: 2201-2215

\bibitem{jia2022joint} Jia M, Zhang L, Wu J, et al. Joint computing and communication resource allocation for edge computing towards Huge LEO networks. China Commun, 2022, 71: 73-84

\bibitem{guo2024network} Guo B Q, Chang Z, Han Z, Yang W T, et al. Network slicing strategy for real-time applications in large-scale satellite networks with heterogeneous transceivers. IEEE Wirel Commun Lett, 2024, 13: 2195-2199

\bibitem{he2024digital} He M C, Wu H Q, Zhou C H, et al. Digital twin-assisted robust and adaptive resource slicing in LEO satellite networks. https://arxiv.org/pdf/2411.03635v1

\bibitem{10371362} Feng W, Lin Y S, Wang Y M, et al. Radio map-based cognitive satellite-UAV networks towards 6G on-demand coverage. IEEE Trans Cognit Commun Netw, 2024, 10: 1075-1089
  
%%%% 4.3
\bibitem{diffie1976new} Diffie W, Hellman M. New directions in cryptography. IEEE Trans Inf Theory, 1976, 22: 644-654

\bibitem{ingemarsson1981encryption} Ingemarsson I, Wong C. Encryption and authentication in on-board processing satellite communication systems. IEEE Trans Commun, 1981, 29: 1684-1687

\bibitem{liu2017physical} Liu Y L, Chen H-H, Wang L M. Physical layer security for next generation wireless networks: Theories, technologies, and challenges. IEEE Commun Surv Tut, 2017, 19: 347-376

\bibitem{han2022challenges} Han S, Li J X, Meng W X, et al. Challenges of physical layer security in a satellite-terrestrial network. IEEE Netw, 2022, 36: 98-104

\bibitem{zeng2015physical} Zeng K. Physical layer key generation in wireless networks: Challenges and opportunities. IEEE Commun Mag, 2015, 53: 33-39

\bibitem{chen2023covert} Chen X Y, An J P, Xiong Z H, et al. Covert communications: A comprehensive survey. IEEE Commun Surv Tut, 2023, 25: 1173-1198

\bibitem{Yu2024covert} Yu H P, Yu J H, Liu J H, et al. Covert satellite communication over overt channel: A randomized Gaussian signalling approach. IEEE Trans Aerosp Electron Syst, 2024. doi: 10.1109/TAES.2024.3475994

\bibitem{feng2024covert} Feng S H, Lu X, Sun S M, et al. Covert communication in large-scale multi-tier LEO satellite networks. IEEE Trans Mob Comput, 2024, 23: 11576-11587

\bibitem{pan2024evolution} Pan D, Long G-L, Yin L G, et al. The evolution of quantum secure direct communication: On the road to the Qinternet. IEEE Commun Surv Tut, 2024, 26: 1898-1949

\bibitem{yin2016measurement} Yin H L, Chen T Y, Yu Z W, et al. Measurement-device-independent quantum key distribution over a 404 km optical fiber. Phys Rev Lett, 2016, 117: 190501.

\bibitem{liao2017satellite} Liao S K, Cai W Q, Liu W Y, et al. Satellite-to-ground quantum key distribution. Nature, 2017, 549: 43-47.

%%%%%%4.4
\bibitem{quan2017enhancing} Quan W, Liu Y N, Zhang H K, et al. Enhancing crowd collaborations for software defined vehicular networks. IEEE Commun Mag, 2017, 55: 80-86

\bibitem{SAT-5G} European Commission. SAT-5G Project. CORDIS, 2020. https:// www.sat5g-project.eu/

\bibitem{cheng2020comprehensive} Cheng N, Quan W, Shi W S, et al. A comprehensive simulation platform for space-air-ground integrated network. IEEE Wirel Commun, 2020, 27: 178-185

\bibitem{6934571} Erl S, De Cola T. Deep learning for joint source-channel coding of text. In: Proceedings of Advanced Satellite Multimedia Systems Conference, Livorno, 2014. 382-389

\bibitem{9814560} Afhamisis M, Barillaro S, Palattella M R. A Testbed for LoRaWAN satellite backhaul: Design principles and validation. In: Proceedings of IEEE International Conference on Communications Workshops, Seoul, 2022. 1171-1176

\bibitem{spiridonov2022small} Spiridonov A A, Saetchnikov V A, Ushakov D V, et al. Small satellite orbit determination using single pass doppler measurements. IEEE J Miniaturization Air Space Syst, 2022, 3: 162-170

\bibitem{10560107} Tiwari A K, Mehto R, Tiwari S, et al. 3D satellite visualization using SGP4. In: Proceedings of IEEE International Conference on Advances in Electronics, Computers and Communications, Bengaluru, 2023. 1-5

\bibitem{9148757} Storek K-U, Schwarz R T, Knopp A. Multi-satellite multi-user MIMO precoding: Testbed and field trial. In: Proceedings of IEEE International Conference on Communications, Dublin, 2020. 1-7

\bibitem{10154319} Minardi M, Drif Y, Vu T X, et al. SDN-based Testbed for emerging use cases in beyond 5G NTN-terrestrial networks. In: Proceedings of IEEE Network Operations and Management Symposium, Miami, 2023. 1-6

\bibitem{qiao2024orbit} Qiao Y, Teng S Y, Luo J, et al. On-orbit DNN distributed inference for remote sensing images in satellite Internet of Things. IEEE Internet of Things Journal, 2024. doi: 10.1109/JIOT.2024.3488076

\bibitem{minardi2022virtual} Minardi M, Vu T X, Lei L, et al. Virtual network embedding for NGSO systems: Algorithmic solution and SDN-testbed validation. IEEE Trans Netw Serv Manag, 2022, 20: 3523-3535


%%%%%%%5.1
\bibitem{10236381} Wang C, An W Y, Li X. On-board Hybrid Heterogeneous Distributed Computing Resource Virtualization. In: Proceedings of International Conference on Optical Communications and Networks, Qufu, 2023. 1-3

\bibitem{10275860} Syamala M, Kaliappan S, M P H, et al. Performance Analysis of Lightweight Virtualization for Environments with Edge Computing Based on NFV. In: Proceedings of International Conference on Smart Electronics and Communication, Trichy, 2023. 687-691

\bibitem{xuRP} Xu X D, Tao X F, Wu C L, et al. Interference analysis of OFDMA based distributed network architecture and resource pooling (in Chinese). J Beijing Univ Post Telecommun, 2007, 30: 19--22

\bibitem{xu2013resource} Xu X D, Wang D, Tao X F, et al. Resource pooling for frameless network architecture with adaptive resource allocation. Sci China Inf Sci, 2013, 56: 1-12

\bibitem{zhu2020two} Zhu X M, Jiang C X, Kuang L L, et al. Two-layer game based resource allocation in cloud based integrated terrestrial-satellite networks. IEEE Trans Cognit Commun Netw, 2020, 6: 509-522

\bibitem{qin2022application} Qin J X, Guo X Y, Ma X T, et al. Application and performance evaluation of resource pool architecture in satellite edge computing. Aerospace, 2022, 9: 451

\bibitem{shenxuemin} Shen X M, Cheng N, Zhou H B, et al. Space-air-ground integrated networks: review and prospect (in Chinese). Chinese J Int Things, 2020, 4: 3-19

\bibitem{8647843} Papa A, De Cola T, Vizarreta P, et al. Dynamic SDN controller placement in a LEO constellation satellite network. In: Proceedings of IEEE Global Communications Conference, Montreal, 2018. 206-212

\bibitem{8460139} Ahmed T, Alleg A, Ferrus R, er al. On-demand network slicing using SDN/NFV-enabled satellite ground segment systems. In: Proceedings of IEEE Conference on Network Softwarization and Workshops, Abu Dhabi, 2018. 242-246

\bibitem{ku2017study} Ku H-J, Jung J-H, Kwon G-I. A study on reinforcement learning based SFC path selection in SDN/NFV. Int J Appl Eng Res, 2017, 12: 3439-3443

%%%%%%%%5.2
\bibitem{pachler2021updated} Pachler N, del Portillo I, Crawley E F, et al. An updated comparison of four low earth orbit satellite constellation systems to provide global broadband. In: Proceedings of IEEE International Conference on Communications Workshops, Montreal, 2021. 1-7

\bibitem{chen2021analysis} Chen Q, Giambene G, Yang L, et al. Analysis of inter-satellite link paths for LEO mega-constellation networks. IEEE Trans Veh Technol, 2021, 70: 2743-2755

\bibitem{mcmahan2017communication} McMahan B, Moore E, Ramage D,  et al. Communication-efficient learning of deep networks from decentralized data. In: Proceedings of Artificial Intelligence and Statistics, Ft. Lauderdale, 2017. 1273-1282

\bibitem{wu2023split} Wu W, Li M S, Qu K G, et al. Split learning over wireless networks: Parallel design and resource management. IEEE J Sel Areas Commun, 2023, 41: 1051-1066

\bibitem{li2020federated} Li T, Sahu A K, Zaheer M, et al. Federated optimization in heterogeneous networks. In: Proceedings of Machine Learning and Systems, Austin, 2020. 429-450

\bibitem{wang2020tackling} Wang J Y, Liu Q H, Liang H, et al. Tackling the objective inconsistency problem in heterogeneous federated optimization. In: Proceedings of Advances in Neural Information Processing Systems, Vancouver, 2020. 7611-7623

\bibitem{10092560} Wu C R, Zhu Y F, Wang F X. DSFL: Decentralized satellite federated learning for energy-aware LEO constellation computing. In: Proceedings of IEEE International Conference on Satellite Computing, Shenzhen, 2022. 25-30

\bibitem{razmi2022ground} Razmi N, Matthiesen B, Dekorsy A, et al. Ground-assisted federated learning in LEO satellite constellations. IEEE Wirel Commun Lett, 2022, 11: 717-721

\bibitem{razmi2022boardhuiyi} Razmi N, Matthiesen B, Dekorsy A, et al. On-board federated learning for dense LEO constellations. In: Proceedings of IEEE International Conference on Communications, Seoul, 2022. 4715-4720

\bibitem{so2022fedspace} So J H, Hsieh K, Arzani B,et al. Fedspace: An efficient federated learning framework at satellites and ground stations. https://arxiv.org/pdf/2202.01267

\bibitem{10021101} Elmahallawy M, Luo T. AsyncFLEO: Asynchronous federated learning for LEO satellite constellations with high-altitude platforms. In Proceedings of IEEE International Conference on Big Data, Osaka, 2022. 5478-5487

\bibitem{10304545} Wu L L, Zhang J J. FedGSM: Efficient federated learning for LEO constellations with gradient staleness mitigation. In Proceedings of IEEE International Workshop on Signal Processing Advances in Wireless Communications, Shanghai, 2023. 356-360

\bibitem{Nguyen2021FederatedLW} Nguyen J, Malik K, Zhan H Y, et al. Federated learning with buffered asynchronous aggregation. https://arxiv.org/pdf/2106.06639v4

\bibitem{10042025} Rodrigues T K, Kato N. Hybrid centralized and distributed learning for MEC-equipped satellite 6G networks. IEEE J Sel Areas Commun, 2023, 41: 1201-1211

%%%%%%%5.3
\bibitem{5.3-1} Uysal E, Kaya O, Ephremides A, et al. Semantic communications in networked systems: A data significance perspective. IEEE Net, 2022, 36: 233--240

\bibitem{5.3-2} G{\"{u}}nd{\"{u}}z D, et al. Beyond transmitting bits: Context, semantics, and task-oriented communications. IEEE J Sel Areas Commun, 2023, 41: 5--41

\bibitem{5.3-3} Shi G Y, Xiao Y, Li Y Y, et al. From semantic communication to semantic-aware networking: Model, architecture, and open problems.
IEEE Commun Mag, 2021, 59: 44--50

\bibitem{10485510} Peng H X, Zhang Z H, Liu Y L, et al. Semantic communication in non-terrestrial networks: A future-ready paradigm. IEEE Netw, 2024, 38(4): 119--127

\bibitem{10486973} Deng D H, Wang C W, Xu L X, et al. Semantic communication empowered NTN for IoT: Benefits and challenges. IEEE Netw, 2024, 38(4): 32--39
 
\bibitem{5.3-4} Weaver W. Recent contributions to the mathematical theory of communication. ETC, 1953, 10: 261-281

\bibitem{5.3-5} Yang W T, Du H Y, Liew Z Q, et al. Semantic communications for future Internet: Fundamentals, applications, and challenges. IEEE Commun Surv Tut, 2023, 2: 215-250

\bibitem{5.3-6} Strinati E C, Barbarossa S. 6G networks: Beyond Shannon towards semantic and goal-oriented communications. Comput Netw, 2021, 190: 107930

\bibitem{5.3-7} Qin Z J, Ye H, Li G Y, et al. Deep learning in physical layer communications. IEEE Wirel Commun, 2019, 26: 93--99

\bibitem{5.3-8} Farsad N, Rao M, Goldsmith A. Deep learning for joint source-channel coding of text. In: Proceedings of IEEE International Conference on Acoustics, Speech and Signal Processing, Calgary, 2018. 2326--2330

\bibitem{5.3-9} Bourtsoulatze E, Kurka D B, G{\"{u}}nd{\"{u}}z D. Deep joint source-channel coding for wireless image transmission. IEEE Trans Cognit Commun Netw, 2019, 5: 567--579

\bibitem{5.3-10} Weng Z Z, Qin Z J. Semantic communication systems for speech transmission. IEEE J Sel Areas Commun, 2021, 39: 2434--2444

\bibitem{5.3-11} Jankowski J, G{\"{u}}nd{\"{u}}z D, Mikolajczyk K. Wireless image retrieval at the edge. IEEE J Sel Areas Commun, 2021, 39: 89--100

\bibitem{5.3-12} Weng Z Z, Qin Z J, Tao X M, et al. Deep learning enabled semantic communications with speech recognition and Synthesis. IEEE Trans Wirel Commun, 2023, 22: 6227--6240

\bibitem{5.3-13} Xie H Q, Qin Z J, Tao X M, et al. Task-oriented multi-user semantic communications. IEEE J Sel Areas Commun, 2022, 40: 2584--2597

\bibitem{5.3-14} Xie H Q, Qin Z J, Li G Y. Semantic communication with memory. IEEE J Sel Areas Commun, 2023, 41: 2658--2669

\bibitem{5.3-15} Yang Y, Guo C L, Liu F F, et al. Semantic communications with AI tasks. https://arxiv.org/pdf/2109.14170

\bibitem{5.3-16} Wang M, Wang H F, Qi G L, et al. Richpedia: A large-scale, comprehensive multi-modal knowledge graph. Big Data Res, 2020, 22: 100159

\bibitem{5.3-17} Kountouris M, Pappas N. Semantics-empowered communication for networked intelligent systems. IEEE Commun Mag, 2021, 59: 96--102

\bibitem{5.3-18} Luo X W, Chen H-H, Guo Q. Semantic communications: Overview, open Issues, and future research directions. IEEE Wirel Commun, 2022, 29: 210--219

\bibitem{5.3-19} Yates R, Sun Y, Brown D, et al. Age of Information: An introduction and survey. IEEE J Sel Areas Commun, 2021, 39: 1183--1210

\bibitem{5.3-20} Kosta A, Pappas N, Ephremides A, et al. The cost of delay in status updates and their value: Non-linear ageing. IEEE Trans Commun, 2020, 68: 4905--4918

%Soleymani T. Value of information analysis in feedback control. Ph.D. dissertation, Technical University of Munich, 2019.

\bibitem{5.3-21} Wang Z, Badiu M-A, Coon J P. A framework for characterizing the value of information in hidden Markov models. IEEE Trans Inform Theory, 2022, 68: 5203--5216

\bibitem{5.3-22} Holm J, Chiariotti F, Kal\o r A E, et al. Goal-oriented scheduling in sensor networks with application timing awareness. IEEE Trans Commun, 2023, 71: 4513--4527

%Sun Y, Polyanskiy Y, Uysal E. Sampling of the Wiener process for remote estimation over a channel with random delay. IEEE Trans Inf Theory, 2020, 66: 1118--1135

\bibitem{5.3-23} Maatouk A, Assaad M, Ephremides A. The age of incorrect information: An enabler of semantics-empowered communication. IEEE Trans Wirel Commun, 2023, 22: 2621-2635

\bibitem{5.3-24} Lu M Y, Huang J H, Jiao, J, et al. Utility loss of information-optimal for semantic empowered RSMA in satellite-integrated Internet. IEEE Internet Things J, 2024. doi: 10.1109/JIOT.2024.3452317

\bibitem{5.3-25} Xu L, Jiao J, Yang T, et al. Semantic utility loss of information for energy efficient semantic status update communications. IEEE Trans Cognit Commun Netw, 2024. doi: 10.1109/TCCN.2024.3417627

%%%%%5.4
\bibitem{10480327} Zuo Y, Yue M Y, Yang H Y, et al. Integrating communication, sensing and computing in satellite Internet of Things: Challenges and opportunities. IEEE Wirel Commun, 2024, 31: 332-338

\bibitem{you2024ubiquitous} You L, Zhu Y X, Qiang X Y, et al. Ubiquitous integrated sensing and communications for massive MIMO LEO satellite systems. IEEE Internet Things Mag, 2024, 7: 30-35

\bibitem{zhang2024joint} Zhang Y B, Wang J J, Li Q, et al. Joint communication, sensing, and computing in space-air-ground integrated networks: System architecture and handover procedure. IEEE Veh Technol Mag, 2024, 19: 70-78

\bibitem{kaushik2024toward} Kaushik A, Singh R, Dayarathna S, et al. Toward integrated sensing and communications for 6G: Key enabling technologies, standardization, and challenges. IEEE Commun Stand Mag, 2024, 8: 52-59

\bibitem{fei2023air} Fei Z S, Wang X Y, Wu N, et al. Air-ground integrated sensing and communications: Opportunities and challenges. IEEE Commun Mag, 2023, 61: 55-61

\bibitem{xing2023task} Xing H, Zhu G X, Liu D Z, et al. Task-oriented integrated sensing, computation and communication for wireless edge AI. IEEE Netw, 2023, 37: 135-144

\bibitem{sun2024integrated} Sun Z, Yu Z W, Guo B, et al. Integrated sensing and communication for effective multi-agent cooperation systems. IEEE Commun Mag, 2024, 62: 68-73

\bibitem{10437374} Khalili A, Rezaei A, Xu D F, et al. Energy-aware resource allocation and trajectory design for UAV-enabled ISAC. In: Proceedings of IEEE Global Communications Conference, Kuala Lumpur, 2023. 4193-4198

\bibitem{10159441} Liu Y M, Liu S, Liu X, et al. Sensing fairness-based energy efficiency optimization for UAV enabled integrated sensing and communication. IEEE Wirel Commun Lett, 2023, 12: 1702-1706

\bibitem{he2024integrating} He D X, Hou H Z, Jiang R K, et al. Integrating sensing and communication for IoT systems: Task-oriented control perspective. IEEE Internet Things Mag, 2024, 7: 76-83

\bibitem{qi2024architecture} Qi Y L, Zhou Y Q, Cai Q, et al. Architecture, characteristics, and resource management of integration of sensing, communications, and computing in 6G. IEEE Netw, 2024, 38: 54-61

\bibitem{wen2024integrated} Wen D Z, Li X Y, Zhou Y, et al. Integrated sensing-communication-computation for edge artificial intelligence. IEEE Internet Things Mag, 2024, 7: 14-20

%%%%%%%5.5
\bibitem{5.5-1} Brandon W, Mahle C. Key technology trends-ground terminals. Space Commun, 2000, 16: 125--137


\bibitem{5.5-3} Zhang S J, Zhao X T, Zhao Y F, et al. Integrated satellite-terrestrial networks: Integrated mode, frequency usage and application prospects (in Chinese). Radio Commun Technol, 2023, 49: 775--787

\bibitem{5.5-4} Johnson A. The first phone maker to add satellite texting to its devices is. . . Huawei. The Verge, 2022. https://www.theverge.com/2022/9/6/23339717/huawei-mate-50-pro-satellite-text-china-beidou

\bibitem{5.5-5} Jewett R. Apple to debut iPhone with emergency messaging enabled by Globalstar satellites. Via Satellite, 2022.
https://www.satellitetoday.com/mobile-connectivity/2022/09/07/apple-to-debut-iphone-with-emergency-messaging-enabled-by-globalstar-satellites/

\bibitem{5.5-6} Bissinger B. AST SpaceMobile announces summer launch date of BlueWalker-3 for direct-to-cell phone connectivity testing AST science. AST SpaceMobile, 2022. https://ast-science.com/2022/06/13/bluewalker-3-launch-date

\bibitem{5.5-7}Robert S T, Miller C E. Cellular core network and radio access network infrastructure and management in space. Lynk Global, EP3830980A1.

\bibitem{5.5-8} Wall M. SpaceX Starlink satellites to beam service straight to smartphones. Future PLC, 2022. https://www.space.com/spacex-starlink-direct-service-smartphones-t-mobile

%\bibitem{5.5-4} ITU. Direct satellite connectivity to mobile. 2023. https://www.itu.int/hub/2023/10/direct-satellite-connectivity-to-mobile

\bibitem{5.5-9} 3rd Generation Partnership Project (3GPP). Study on architecture aspects for using satellite access in 5G (Release 17). TR 23.737. https://www.3gpp.org/ftp/Specs/archive/23\_series/23.737

\bibitem{5.5-10} 3rd Generation Partnership Project (3GPP). NR NTN (non-terrestrial networks) enhancements (Release 18). TSG RAN, RP-223534. https://www.3gpp.org/ftp/tsg\_ran/TSG\_RAN/TSGR\_98e/Docs

\bibitem{5.5-11} 3rd Generation Partnership Project (3GPP). New WID: non-terrestrial networks (NTN) for NR phase 3 (Release 19). TSG RAN, RP-234078. https://www.3gpp.org/ftp/tsg\_ran/TSG\_RAN/TSGR\_102/Docs

\bibitem{5.5-12} He Y Z, Xiao Y W, Zhang S J. et al. Direct-to-smartphone for 6G NTN: Technical routes, challenges, and key technologies. IEEE Netw, 2024, 38: 128--135

\bibitem{5.5-13} Sun Y H, Xu H T, Peng M G. Direct-to-mobile low earth orbit satellite communication: Architecture, key technologies, and future perspective (in Chinese). Mobile Commun, 2024, 48: 103--110

\bibitem{5.5-14} Bakhsh Z M, Omid Y, Chen G, et al. Multi-satellite MIMO systems for direct satellite-to- device communications: A survey. IEEE Commun Surv Tut, 2024. doi: 10.1109/COMST.2024.3449430

\bibitem{5.5-15} Abdelsadek M Y, et al. Broadband connectivity for handheld devices via LEO satellites: Is distributed massive MIMO the answer?. IEEE Open J Commun Soc, 2023, 4: 713--726

\bibitem{5.5-16} Heo J, Sung S, Lee H, et al. MIMO satellite communication systems: A survey from the PHY layer perspective. IEEE Commun Surv Tut, 2023, 25: 1543--1570

\bibitem{5.5-17} Tuzi D, Aguilar E F, Delamotte T, et al. Distributed approach to satellite direct-to-cell connectivity in 6G non-terrestrial networks. IEEE Wirel Commun, 2023, 30: 28--34
%\bibitem{5.5-19} Tuzi D, Delamotte T, Knopp A. Satellite swarm-based antenna arrays for 6G direct-to-cell connectivity. IEEE Access, 2023, 11: 36907--36928

%\bibitem{2} Author A, Author B, Author C, et al. Reference title. In: Proceedings of Conference, Place, 2024. 6--12

\end{thebibliography}
\end{document}